\documentclass[12pt, a4paper]{article}
\usepackage{hyperref}
\usepackage{geometry}
\usepackage{graphicx}
\usepackage{subcaption}
\usepackage[numbers]{natbib}
\usepackage{amssymb}                                                                                                                                                                                                                                                                                                                                                                                                                                                                                                                                                                                                                                                                                                                                                                                                                                                                                                                                                                                                                                                                                                                                                                                                                                                                                                                                                                                                                                                                                                                                                                                                                                                                                                                                                                                                                                                                                                                                                                                                                                                                                                                                                                                                                                                                                                                                                                                                                                                                                                                                                                                                                                                                                                                                                                                                                                                                                                                                                                                                                                                                                                                                                                                                                                                                                                                                                                                                                                                                                                                                                                                                                                                                                                                                                                                                                                                                                                                                                                                                                                                                                                                                                                                                                                          
\usepackage{amsmath}
\usepackage{slashed}
\usepackage{anyfontsize}
\usepackage{bm}
\usepackage{slashed}
\usepackage{multirow}
\usepackage{float}
\usepackage[font=small]{caption}
\usepackage{listings}
\usepackage{color}
\definecolor{codegreen}{rgb}{0,0.5,0}
\definecolor{codeblue}{rgb}{0,0,0.5}
\definecolor{maroon}{rgb}{0.675,0.1,0.1}
\usepackage{upgreek}
\usepackage{tensor}

\usepackage[utf8]{inputenc}
\usepackage[english]{babel}
\usepackage{amsthm}
\usepackage{slashed}
\usepackage{xcolor}
\definecolor{ForestGreen}{RGB}{0, 176, 80}
\usepackage{hyperref}
\newgeometry{lmargin=1.1in, rmargin=1.1in, tmargin=1in, bmargin=1in}
\title{The Theory of Fundamental Duality, Quantum Dualiton and Topological Dual Invariance}
\author{B.T.T.Wong\footnote{CERN, The University of Hong Kong,  u3500478@connect.hku.hk}}
\date{}
\newtheorem{definition}{Definition}[section]
\theoremstyle{definition}

\begin{document}

\maketitle
\begin{abstract}
Fundamental duality is a concept which refers to two irreducible, heterogeneous principles which are in opposite and complementary of each other. The complementary principle in quantum mechanics is also praised by Bohr. This important concept is known to appear in a lot of places in our physical universe, however a rigorous mathematical definition and physics theory has not yet ever developed in a formal way. In this paper, we establish a formalism for fundamental duality and study its various properties and theorems. One of the most profound results is that we establish  a relation between dual invariance and topological invariance, and we find that the topological Chern-Simons form is a dual invariant action. Finally we apply the concept of duality to study dual state oscillation, and predict a theoretical new matter of state of dualiton, which is the particle excitation of the dual field by second quantization. This new exotic quasi-particle state is expected to have an impact in particle physics and condensed matter physics.
\end{abstract}


\section{Introduction}
Fundamental duality \footnote{The duality we refer here is different from the duality in common physics terms, such as the duality in particle-wave duality and T-duality, S-duality in string theory. To avoid confusion, we use the word ``fundamental''.}, conceptually is simply the study of elements with opposite nature. For example, $+$ and $-$ is a dual pair in which $+$ and $-$ are opposite; the two spin states of an electron, spin-up $\uparrow$ and spin-down $\downarrow$ is also a dual pair. Macroscopic and microscopic world are dual to each other that the former follows the law of general relativity and it is absolutely deterministic while the latter follows the law of quantum mechanics which is probabilistic. There are numerous examples that duality shows its appearance.

A fundamental dual system contains two opposite elements, and the two opposite elements form a fundamental dual pair. One element is dual to the other, and vice versa. We can see that such duality is a fundamental property in nature and it appears everywhere in our universe. Although the concept of fundamental duality is simple, it can be extremely profound and far more difficult than what it seems. It is a no easy task to define rigorously by means of mathematical definition for each duality system. 

It is well-agreed that two things which are dual to each other have opposite properties, and they are never identical. We will never say an element is same as the dual element. However, we will like to introduce a very important idea for observable frame or observation perspective, that allows us to make equivalent statement of dual elements in a system.

The next question for fundamental duality is very deep, it addresses the problem of whether fundamental duality is always conserved. In most of the cases, we may agree that duality is generally conserved where we have two sides equally. However, in reality, especially from the perspective of physics, this is not true. Here we will give profound examples from particle physics. We know that a fermion (spin-$\frac{1}{2}$) must have its own antiparticle. The electron particle $e^-$ has its own antiparticle positron $e^+$ \citep{DiracAntiMatter, Dirac, Dirac2}. This particle-antiparticle pair share the same mass but opposite charge. However, in our universe, matter dominates over antimatter naturally by an order of parts per $\sim 10^9$, and this is the long-lasting unsolved problem of matter antimatter asymmetry problem in physics \cite{MAMasymmetry}. Next, we know that parity is conserved in electromagnetic force and strong force, but is violated in weak force in a maximal manner due to the vertex minus axial vector $V-A$ structure \cite{parityviolation, Wu}. In the Higgs mechanism, the process of spontaneous symmetry breaking is to pick a positive vacuum  $+v$ over a $-v$ one in which both happens to be equally probable, such that this allows particles to gain mass, i.e. `to bring them into existence' \citep{Higgs1, Higgs2, Englert, Weinberg, Salam, Kibble}. If all particles have no mass, basically our universe literally cannot exist realistically. Hence, in nature, fundamental duality sometimes conserves, but sometimes not. And it is very difficult to answer when it is conserved and when it is not. So often if it is not conserved it is maximally violated.

The context of this subject becomes even more difficult when there are different layers of fundamental duality superimposing all at once in one collective framework, known as multi-duality. Then the extraction of information in the framework is highly non-trivial.

In the first part of this paper, it is written in the aim of developing formal mathematical formalism that can define fundamental duality properly, in conjunction with observation perspective. We would like to transform the abstract concept of fundamental duality into the language of rigorous mathematics. The construction is based on the parity group $\mathbb{Z }_2$ , which is a dual symmetry. Much work is also established for $\mathbb{Z}_2 \times \mathbb{Z}_2 $, which is the double duality symmetry group, mathematically the Klein-4 group. It is then followed to study the general multi-duality symmetry group $\mathbb{Z}_2 \times \mathbb{Z}_2 \times \cdots \times \mathbb{Z}_2$. Therefore, groups and representation theory are used throughout the text. Secondly, the idea of dual symmetry is integrated with quantum mechanics. In particular, in simple words we can formulate the 2 dual elements in a fundamental duality system as two states $|0\rangle$ and $|1 \rangle$. We will give a thorough establishment  for a novel theoretical development, and construct a number of theorems on this subject. To study how we can interpret information of dual systems, ideas from entropy in information theory is  used, and this give a useful way to study fundamental dual systems.

In the second part of the paper, we will apply the concept of duality and multi-duality symmetry quantum field theories. First, for scalar field theories, there are remarkable consequences of duality symmetries in high-order interaction terms. This enriches the study of quantum field theory on behalf to the traditional ones. Next, we discover a very important equivalent relation between topological invariance and dual invariance in the Chern-Simons field theory. Finally, we study the subject of duality wave, which is a wave oscillation of dual states. Upon quantization, it promotes to the duality field and gives a prediction of a new state of matter-the quanta of duality wave known as \emph{dualiton}. This quasi particle is expected to give rise a new exotic matter state in particle physics and condensed matter physics.

\section{The Theory of Duality and Fundamental Formalism}
\subsection{Single Duality Structure}

We will begin by introducing a full set of definitions for duality. 
\begin{definition}\end{definition}
(I) Let $U$ be an element set and its dual $U^*$, where there exists a one-to-one bijective map on $U$ and $U^*$. Define the duality map $*$, as a function $* : U \rightarrow U^*$ which is a representation of the parity group $\mathbb{Z}_2$. The inverse is just the map itself $* : U^* \rightarrow U$. The double duality map $** $ is the identity map $I_d$, that  $** : U \rightarrow U$ and $** : U^* \rightarrow U^*$. $U$ and $U^*$ are said to be dual if $U \cap U^* = \emptyset$ under the $*$ map. Define the zero set as $\{0 \}$. The complete duality set $W$ is defined by $W = U \cup U^* \cup \{0 \}$. The concept of zero is introduced such that $U=U^*$ is dual invariant if and only if $U=U^* = \{0\}$. For  $u_i\in U$ and $u_i^*\in U^*$, there exist a map $\cdot$ such that $u_i\cdot u_i^* = 0$, where $0$ here is the zero element in the zero set.

(II) The duality set embedded in an extrinsic observer frame in $k$-dimension is said to be a complete single  duality structure. The observer's frame of in $k$ dimension forms a duality $S_k$ and $S_k^{\star}$. Let the duality operator for observer's frame be a map $\star : S_k \rightarrow S_k^{*}$, which is a is a representation of the parity group $\mathbb{Z}_2$. The inverse is the operator itself, $\star : S_k^{\star} \rightarrow S_k$, and $\star\star$ is the identity map $I_d$. We define the zero set for observer as $\{0 \}_{S_k}$. For observer at $\{0 \}_{S_k}$, it is defined as the intrinsic observer of the duality system. We concern the extrinsic frame, and define the complete observer's as $B = S_k \cup S_k^{\star}$.  There exists a set in the duality set which is independent of the observer's frame, which is the zero set $\{ 0 \} \in W$. The complete duality structure is defined as $\{W, B\}$. 

(III) Each set or dual set have to be observer's frame specific. In the $S_k$ observer's frame, we specify the set and dual set under the observer frame is denoted by $(U|S_k )$ and $(U^{*} | S_k )$ respectively. In the $S_k^*$ observer's frame, we have $(U| S_k^\star)$ and $(U^{*}|S_k^\star )$ respectively.

(IV) The dual equivalence of two elements $a,b$, denoted by $a\equiv b$ (or $a:=: b , a \,*=* \, b$ ) is defined by
\begin{equation}
a \equiv b \,\text{if}  \,
\begin{cases}
a = b \\
\text{or} \\
a \neq b \,\, \text{but}\, a,b \,\text{are equivalent by some relation establishment}
\end{cases}
\end{equation} 

(V) In complete duality, we have the following identity for dual equivalence,
\begin{equation}
( U | S_k) \equiv ( U^{*} | S_k^\star)\,\,\,\,{\rm and}\,\,\,\, ( U | S_k^\star) \equiv ( U^{*}| S_k )\,.
\end{equation}
Element-wise, let $u \in U$ and $u^* \in  U^*$, we have
\begin{equation} \label{eq:equivalentt}
( u | S_k ) \equiv (u^{*} | S_k^{\star} )\,\,\,\,{\rm and}\,\,\,\, (u| S_k^{\star}) \equiv ( u^{*} | S_k )\,.
\end{equation}

(VI) The duality operator of the element set $*$ acts as the following:
\begin{equation}
*( u | S_k ) = ( u^{*} | S_k )\,\,\,\, , \,\,\,\, ** ( u | S_k )  = * (u^{*} | S_k ) = (u| S_k )\,.
\end{equation}

(VII) The duality operator of observer set $*$ acts as the following:
\begin{equation}
\star ( u| S_k ) = ( u | S_k^{\star} )\,\,\,\, , \,\,\,\,\star \star ( u | S_k ) = \star ( u |S_k^{\star}) =  ( u | S_k )\,.
\end{equation}

(VIII) The dual map of the set and the dual operator can act together, which is an identity map.
The two different duality map commutes. In other words, $\star \circ * = * \circ \star = I_d$. From example, we have,
\begin{equation}
\star *( u | S_k ) \equiv \star ( u^{*} | S_k ) \equiv ( u^{*} | S_k^{\star} )\,\,\,\,{\rm and }\,\,\,\, *\star (u | S_k ) = * (u | S_k^{\star}) = ( u^{*} | S_k^{\star})\,.
\end{equation}

(IX) The $\{I_d , * \}$ and $\{I_d^\prime , \star \}$ are elements of two parity groups $\mathbb{Z}_2$ under multiplication.  The Klein-4 group, which is the called the 4-duality group, is $\mathbb{Z}_2 \times \mathbb{Z}_2 = \{   I , * , \star, *\circ \star\}$. The $( u | S_k ), ( u^{*} | S_k ), ( u | S_k^\star ) $ and $( u^* | S_k^\star ) $ form a 4-representation of $\mathbb{Z}_2 \times \mathbb{Z}_2$, and can be represented by a 4-tableau diagram,

\begin{figure}[H]
\centering
\includegraphics[trim=0cm 0cm 0cm 0cm, clip, scale=1]{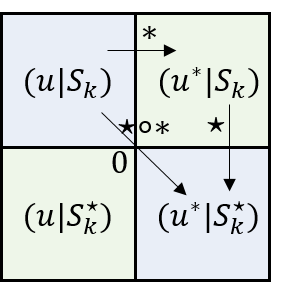}
\caption[]
{ The 4-tableau representation of a complete duality structure. Boxes with the same colour denote dual equivalence relation of each other. \label{fig:Duality}}
\end{figure}
The last definition (IX) is achieved by constructing an isomorphism from the four cases to the basis of $\mathbb{Z}_2 \times \mathbb{Z}_2$, such we have a one-to-one map identification as
\begin{equation} \label{eq:basiss}
(u|S_k ) \rightarrow |00\rangle \,,\quad \, (u|S_k^\star ) \rightarrow |01\rangle \,,\quad \, (u^*|S_k )\rightarrow |10\rangle\,,\quad \, (u^*|S_k^\star ) \rightarrow |11\rangle ,.
\end{equation}
The $(u|S_k ) $ is called an identity element, in which no dual operation is acted upon on it. Without loss of generality, we can also pick $(u^* | S_k^{\star} ) $ as the identity. 
In terms of the number of dual operation that act on the element and observer, we can write 
\begin{equation}
0 + 0 \equiv 1 +1 \quad\text{and}\quad 0+1 \equiv 1+0 \,.
\end{equation}

The element-observer composite $( \, | \,)$ can also be viewed as a tensor product, i.e. $(a|b)\equiv (a|\otimes| b) \equiv a\otimes b $. Consider the basis of representation of $\mathbb{Z}_2$ be $u \oplus u^*$, and the basis of representation of another $\mathbb{Z}_2$ as $S_{k}\oplus S_{k}^\star$. Then we have
\begin{equation}
\begin{aligned}
(u\oplus u^*) \otimes (S_k\oplus S_k^\star) &= (u\otimes S_k)\oplus  (u\otimes S_k^\star) \oplus(u^*\otimes S_k) \oplus (u^* \otimes S_k^\star) \\
&= (u|S_k ) \oplus (u|S_k^\star ) \oplus (u^*|S_k ) \oplus (u^*|S_k^\star ) \,,
\end{aligned}
\end{equation}
which is the basis of representation of $\mathbb{Z}_2 \otimes \mathbb{Z}_2$ direct product group. But since $\mathbb{Z}_2 \otimes \mathbb{Z}_2 \cong \mathbb{Z}_2 \times \mathbb{Z}_2$, therefore the above serves as the basis of the 4-duality group.
For simplicity, using \ref{eq:basiss} we can write it as
\begin{equation}
\psi = |00\rangle \oplus |01\rangle \oplus |10\rangle \oplus |11\rangle
\end{equation}
Since by recalling that \ref{eq:equivalentt}, $( u | S_k ) \equiv (u^{*} | S_k^{\star} )\,\,\,\,{\rm and}\,\,\,\, (u| S_k^{\star}) \equiv ( u^{*} | S_k )\,$, we have $|00\rangle \equiv |11\rangle \,\,\,\,{\rm and}\,\,\,\,|01\rangle \equiv |10\rangle$. This can be viewed as
\begin{figure}[H]
\centering
\includegraphics[trim=0cm 0cm 0cm 0cm, clip, scale=0.4]{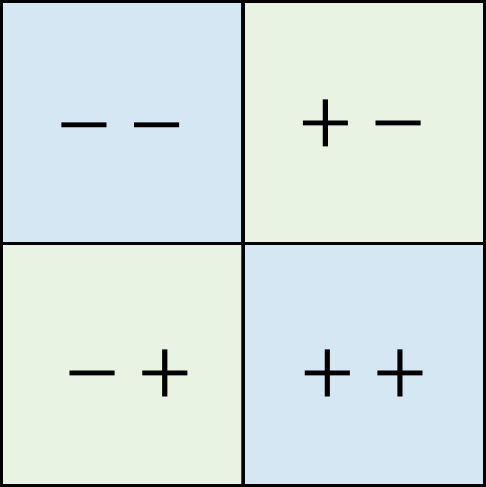}
\caption[]
{ The two equivalence relationship can be viewed as $++ , --\rightarrow +$ and $+-,-+ \rightarrow -$ In other words, considering the backward argument, the conventional relationship of arithmetic can be viewed as the equivalent relationships of tensor product of dual sets as demonstrated above. . \label{fig:plusminus}}
\end{figure}
Explicitly in the tensor product representation, we have the operators as
\begin{equation}
\begin{aligned}
&(1\otimes 1)(u\otimes S_k) = ( u\otimes S_k ) \\
&(*\otimes 1)(u\otimes S_k) = (u^*\otimes S_k ) \\
&(1\otimes \star)(u\otimes S_k) = (u\otimes S_k^\star ) \\
& (*\otimes 1)(1\otimes \star)(u\otimes S_k)= (*\otimes \star)( u\otimes S_k )=(u^* \otimes S_k^\star )
\end{aligned}
\end{equation}
We have the Klein-4 group as $\mathbb{Z}_2 \otimes \mathbb{Z}_2 =\{ 1\otimes 1 , *\otimes 1 , 1\otimes \star , *\otimes\star \} $.

We group the terms as
\begin{equation}
\psi = \big[ ( u | S_k ) \oplus (u^{*} | S_k^{\star} )  \big]_D \oplus \big[ (u| S_k^{\star}) \oplus ( u^{*} | S_k ) \big]_{D^*} \,,
\end{equation}
or
\begin{equation}
\psi = \big[ |00\rangle \oplus |11\rangle\big]_D \oplus \big[ |01\rangle \oplus |10\rangle\big]_{D^*} = | \psi_D \rangle \oplus |\psi_{D^*} \rangle \,, 
\end{equation}
where the subscript $D$ and $D^*$ indicate the two dual partitions, and the two states are orthogonal to each other $\langle \psi_D | \psi_{D^*} \rangle =0 .$ Now we will show that in fact the basis of $\mathbb{Z}_2 \otimes \mathbb{Z}_2 $ can be separated into two dual partitions, such that the blue boxes are dual to the green boxes in figure 2.1. Define the parity operator of the partition $D$ as follow,
\begin{equation}
\hat{P}_D = ( 1 \otimes \star ) \oplus ( 1 \otimes \star ) \,.
\end{equation}
Then we have
\begin{equation}
\begin{aligned}
\hat{P}_D \big[ |00\rangle \oplus |11\rangle\big]_D &= \big[( 1 \otimes \star ) \oplus ( 1 \otimes \star )\big]\big[ |00\rangle \oplus |11\rangle\big]_D \\
&=( 1 \otimes \star )(|0\rangle  \otimes |0\rangle   ) \oplus ( 1 \otimes \star )(|1\rangle  \otimes |1 \rangle   ) \\
&= (|0\rangle  \otimes |1\rangle ) \oplus (|1 \rangle \otimes |0\rangle ) \\
&= \big[ |01\rangle \oplus |10\rangle\big]_{D^*} \,.
\end{aligned}
\end{equation}
It follows that 
\begin{equation}
\hat{P}_D \big[ |01\rangle \oplus |10\rangle\big]_{D^*} = \hat{P}_D^2  |00\rangle \oplus |11\rangle\big]_D = [|00\rangle \oplus [|11\rangle\big]_D  \,.
\end{equation}
It can be also easily checked that
\begin{equation}
\begin{aligned}
\hat{P}_D^2 &=[( 1 \otimes \star ) \oplus ( 1 \otimes \star )][( 1 \otimes \star ) \oplus ( 1 \otimes \star )] \\
&= ( 1 \otimes \star )( 1 \otimes \star ) \oplus ( 1 \otimes \star )( 1 \otimes \star ) \\
&= (1\otimes 1)\oplus (1 \otimes 1) \\
&= 1\oplus 1 \\
&= I
\end{aligned}
\end{equation}
which is the identity matrix. 
Therefore, the two bases $ \big[ |00\rangle \oplus |11\rangle\big]_D \,\,\,\,\text{and}\,\,\,\, \big[ |01\rangle \oplus |10\rangle\big]_{D^*}$ are the basis of of the duality group $\mathbb{Z}_2$. Therefore, $\psi$ can be decomposed into two EPR basis pair. Symbolically we can write
\begin{equation}
\psi = 2 \oplus 2 \,.
\end{equation}
Note that the choice of $\hat{P}_D$ is not unique, we can also define,
\begin{equation}
\hat{Q}_D = (*\otimes 1) \oplus (* \otimes 1) \,.
\end{equation}
Then we have
\begin{equation}
\begin{aligned}
\hat{Q}_D \big[ |00\rangle \oplus |11\rangle\big]_D &= \big[( * \otimes 1 ) \oplus ( * \otimes 1 )\big]\big[ |00\rangle \oplus |11\rangle\big]_D \\
&=( * \otimes 1)(|0\rangle  \otimes |0\rangle   ) \oplus ( * \otimes 1 )(|1\rangle  \otimes |1 \rangle   ) \\
&= (|1\rangle \rangle \otimes |0\rangle \rangle) \oplus (|0 \rangle \otimes |1\rangle ) \\
&= \big[ |01\rangle \oplus |10\rangle\big]_{D^*} \,.
\end{aligned}
\end{equation}
And similarly we have $\hat{Q}_D^2 = I$ which is the identity map. 

Furthermore, we can have rectangular duality. We consider $\psi$ with the following partitions,
\begin{equation}
\psi = \big[ |00\rangle \oplus |01\rangle\big]_P \oplus \big[ |11\rangle \oplus |10\rangle\big]_{P^*} = |\psi_P \rangle \oplus |\psi_{P^*} \rangle  \,.
\end{equation}
Clearly partitions $P$ and $P^*$ are dual to each other, and the two dual basis are orthogonal to each other $\langle \psi_P | \psi_{p^*} \rangle =0$. This is referred as the the vertical rectangular duality. Similarly, we can have
\begin{equation}
\psi = \big[ |00\rangle \oplus |10\rangle\big]_Q \oplus \big[ |11\rangle \oplus |01\rangle\big]_{Q^*} = |\psi_Q \rangle \oplus |\psi_{Q^*} \rangle\,, 
\end{equation}
where $Q$ and $Q^*$ are dual to each other, and the two dual basis are orthogonal to each other $\langle \psi_Q | \psi_{Q^*} \rangle =0$. This is referred as the horizontal rectangular duality. The idea is illustrated as follow.
\begin{figure}[H]
\centering
\includegraphics[trim=0cm 0cm 0cm 0cm, clip, scale=0.5]{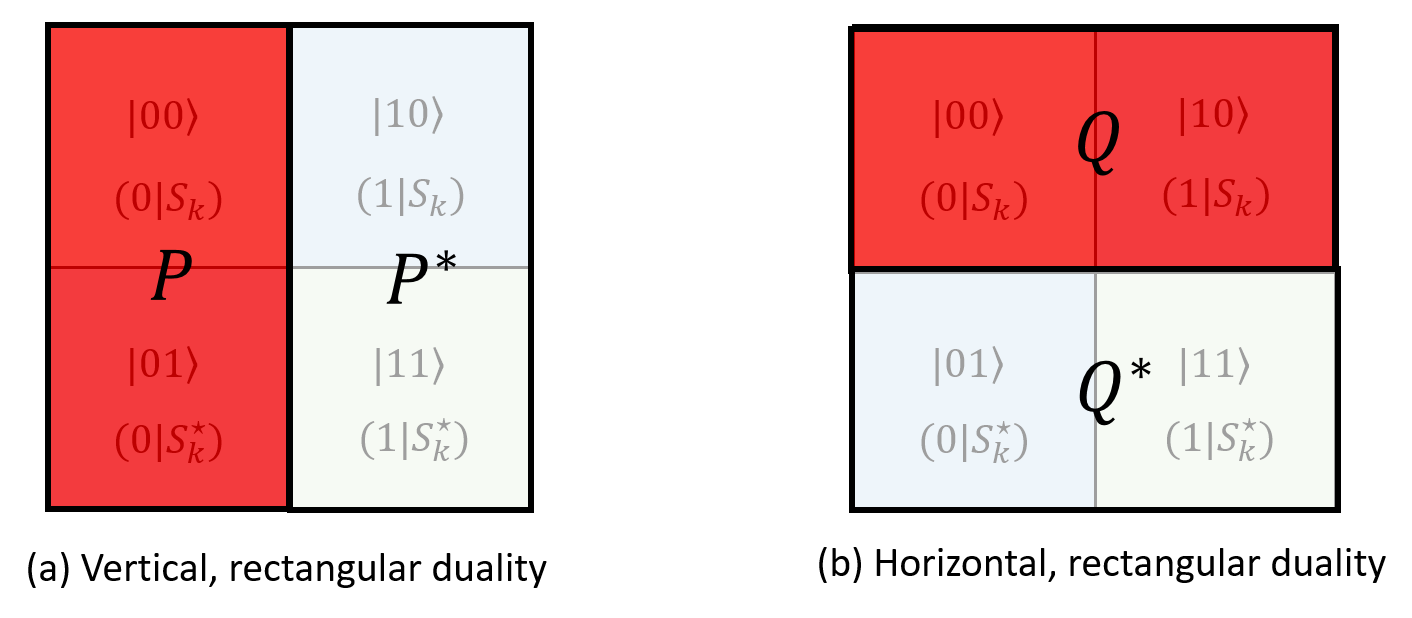}
\caption[]
{ \label{fig:vdduality}}
\end{figure}

Explicitly, we can draw out the whole idea of duality for illustration. The dual elements and dual observers form a generic 4-dual diagram as follow.
\begin{figure}[H]
\centering
\includegraphics[trim=0cm 0cm 0cm 0cm, clip, scale=1]{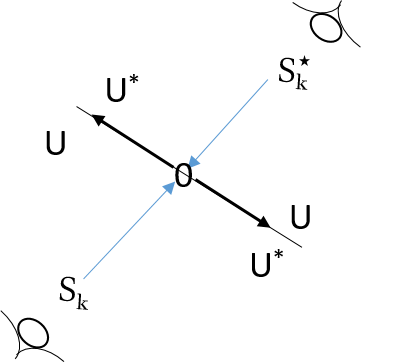}
\caption[]
{In the diagram, we can see that $U$ observed by $S_k$ is equivalent to $U^*$ observed by $S_k^\star$ such that $(U|S_k ) \equiv (U^*|S_k^\star )$; and $U^*$ observed by $S_k$ is equivalent to $U$ observed by $S_k^\star$ such that $(U^* |S_k ) \equiv (U|S_k^\star )$. \label{fig:genericdiagram4dual}}
\end{figure}

\begin{figure}[H]
\centering
\includegraphics[trim=0cm 0cm 0cm 0cm, clip, scale=1]{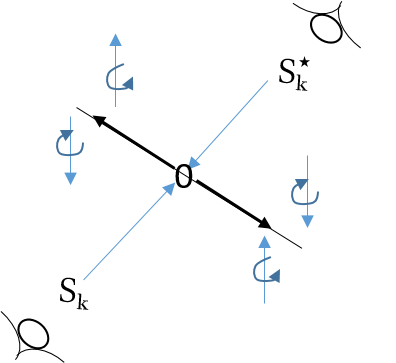}
\caption[]
{ \label{fig:spindual}}
\end{figure}

The above abstract definition can be easily understood by some examples. The simplest case for a dual system would be positive and negative numbers. Let start from the most fundamental case. Let $U=\{-1\}$ and $U^* = \{+1\}$, and the zero set $\{0\}$. Consider a pair of dual observers living on a 2D manifold, the one in front of the two numbers is $S_2$, and the one behind the two numbers is $S_2^{*}$. Let's use the normal convention of a number line, the left is -1 and the right is +1, then we have,
\begin{equation}
(-1 |S_2) \equiv (+1 | S_2^{*} )\,\,\,\,{\rm and}\,\,\,\,(-1 | S_2^* ) \equiv (+1 | S_2 )\,.
\end{equation}
And we have the map $\cdot = +$ such that $(-1) + (+1) = 0 $. If we let $U= \mathbb{R}^-$ and $U^*= \mathbb{R}^+$ and the zero set, the you have the duality for the real number system. Another example would be spin. Let $U=\{ \downarrow \} = \begin{pmatrix} 
0 \\  
1 
\end{pmatrix}
$ 
and $U^* = \{\uparrow \}=
\begin{pmatrix} 
1 \\  
0 
\end{pmatrix}
$, and $*=
\begin{pmatrix} 
0 & 1 \\  
1 & 0 
\end{pmatrix}
$ with $**=\mathbb{I}$, where we observe in a 3 dimensional space. Then we have
\begin{equation}
(\downarrow |S_3) \equiv ( \uparrow | S_3^{*} )\,\,\,\,{\rm and}\,\,\,\,( \downarrow | S_3^* ) \equiv (\uparrow | S_3 )\,.
\end{equation}
This is demonstrated in figure \ref{fig:spindual}. And we have the $\cdot$ map as the inner product $\langle \uparrow | \downarrow \rangle =0$.

Although not obviously noticed, spontaneously duality symmetry breaking of choice is always implicitly 
inferred. For example we define left-hand side as negative in our observation perspective, but this is equivalently to a positive right-hand side in the dual perspective. However we often make a particular choice of representation so that at the end only one representation out of the two equivalence is used. Without the loss of generality we can pick the dual one, but for realistic observable we must pick a particular one. In quantum mechanics terms, this is a state collapse of a dual state. Explicitly,
\begin{equation}
|\Psi \rangle = \frac{1}{\sqrt{2}} \Big( |(U,U^* |S_k )\rangle  + | ( U^*,U |S_k^\star ) \rangle \Big) \,,
\end{equation}
where we can map $|(U,U^* |S_k )\rangle \rightarrow |01\rangle$ and $| ( U^*,U |S_k^\star \rangle \rightarrow |10\rangle )$ respectively, with equal probability of $1/2$. This is an EPR pair and an entangled state. In general we can write
\begin{equation}
|\Psi \rangle = \cos\theta |(U,U^* |S_k )\rangle  + \sin\theta | ( U^*,U |S_k^\star \rangle \,,
\end{equation}
When the phase is at $\pi/4$, we have both the probabilities as $1/2$. At $\theta = 0 $ or $2\pi$, we have a deterministic state for $|(U,U^* |S_k )\rangle$ and at $\theta = \pi$, we have a deterministic state for $| ( U^*,U |S_k^\star) \rangle$.

Next we would like to promote the idea into a more abstract way. We can call $U$ as a left dual $^{*}U$. In figure \ref{fig:genericdiagram4dual}, if we slice along the $S_k , S_k^\star$ frame, we can see the pair of element $U^{*}{}^{*}U$ and its dual $^{*}U U^* $. We identify as follow:
\begin{equation}
U^{*}{}^{*}U \text{as}\,\, RL \quad \text{and} \quad ^{*}U U^* \text{as}\,\, LR\,.
\end{equation}
In the case of of RL, the two $*$s are in the inner side and we term this as ``bonding" denoted as $\rightarrow \leftarrow$, while in the case of LR, the two $*$s are at the outer side and we term this as ``anti-bonding" denoted as $\leftarrow\rightarrow$. Thus the "bonding" and "anti-bonding" representation is a dual representation. And the two objects $U^{*}{} ^{*}U $ and $^{*}U U^*$ form a basis of irreducible representation of $\mathbb{Z}_2 $. We can go in the other way that if there exist such a dual pair, then the notion of observation frame is implied.

The role of element and observer is interchangeable in a 4-duality system. Now we can treat the element as observer and observer as element, this is known as element-observer duality.

There are several more important examples for duality. We would like to show that the set of odd numbers and even numbers are duality. Let the odd number set be $O = \{ -(2k-1) , \cdots ,-5,-3,-1,1,3,5,\cdots ,(2k-1)   \}$ and the even number set be $\{ -2k, -4,-2,0,2,4,\cdots ,2k \,   \}$. There is a one-to-one bijective map from the even set to the odd set. We also have $O\cap E =\emptyset$ Consider the $*$ function as adding $+1$ to each of the number in the set. We have
\begin{equation}
*O = O+1 = E\,.
\end{equation}
It follows that
\begin{equation}
**O = (O+1)+1 = E+1 = \{  -(2k+1) , \cdots ,-3,-1,1,3,5,7,\cdots ,(2k+1)  \} = O\,.
\end{equation}
Therefore $**=1$ is the identity map. Hence the odd number set and the even number is a duality.

Next we would like to show that momentum and position are duality. Let $U = \{x \}$ and $U^* = \{ 1/x \}$ excluding $x=1$. We see $U\cap U^* = \emptyset$. There is a one-to-one correspondence between the two sets. Now consider
\begin{equation}
*(x) = x^{-1} \,,
\end{equation}
then
\begin{equation}
**(x) = *(x^{-1}) = (x^{-1})^{-1} =x \,.
\end{equation}
Therefore $**=1$ is the identity map. Hence $x$ and $1/x$ are dual to each other. Also we have the $\cdot$ map as multiplication $\times$ such that $\frac{1}{x}\times x =1$. One important consequence is that for $x=0$, we have $1/0 = \infty$ thus $0$ and $\infty$ are dual to each other. Mathematically we write
\begin{equation}
*0=\infty \quad \text{and} \quad *\infty = 0 \,.
\end{equation} 
Meaning-wise, we say nothing $(0)$ is dual to everything $(\infty)$, or extremely small is dual to extremely large. One important property is that we see when $x=1$, we get the same values that $*1=1$. This is the dual invariant number. This serves as the zero number $\{0 \}$. Therefore we have $W=U\cup U^* \cup \{1\}$ as the complete set of duality. Now returning to physics, consider $x$ as the wavelength $\lambda$, the momentum is $p=h/ \lambda$. where $h$ is the planck's constant and can be regarded as 1 in the natural unit. Hence it follows  that $x$ and $p$ are dual to each other, therefore position and momentum are dual to each other. 

We can construct a general dual invariant function. The following function
\begin{equation}
f(x) = \bigg(x + \frac{1}{x} \bigg)^n
\end{equation}
for $n$ is any positive integer greater than 1 is dual invariant that $f(x) = f(\frac{1}{x})$, so this function remains the same for the exchange of $x \leftrightarrow \frac{1}{x}$.  A special attention goes to the case for $n=2$, for which
\begin{equation}
f(x) = x^2 + \frac{1}{x^2} + 2 = x^2 + \frac{1}{x^2} + \text{constant} \,. 
\end{equation}
In string theory, the mass spectrum for a closed bosonic string with $26$ dimensions has a mass spectrum as \citep{string1, string2}
\begin{equation} \label{eq:stringmass}
M^2 = \frac{n^2}{R^2} + \frac{m^2 R^2}{\alpha^{\prime 2}} + \frac{2}{\alpha^\prime}(N_L + N_R -2) \,,
\end{equation}
where $N_L$ is the number of left-moving modes, $N_R$ is the number of right moving modes, $n$ is the quantized number of Kaluza-Kelin momentum mode and $m$ is the winding number. The $\alpha^\prime$ is the string's length scale and is related to the tension of the string. Also, $N_R - N_L = nm.$ The mass spectrum in \ref{eq:stringmass} is invariant under the interchange $n\leftrightarrow m$ and $R \leftrightarrow \tilde{R}=\frac{\alpha^\prime}{R}$. This is known as the T-duality \citep{string1, string2}. In particular, when $n=m$ and in generic natural length unit $\alpha^\prime =1$, we have 
\begin{equation}
M^2 = \frac{n^2}{R^2} + n^2 R^2 + \text{constant} \,,
\end{equation}
which is a $\mathbb{Z}_2 $ invariant.

\subsubsection*{The dual equation} 
Finally we would like to write out the duality theory in a compact form. Let $a= ( u | S_k )$, and the full negation operator be $! = \star *$, then we nicely obtain the follow equation,
\begin{equation} \label{eq:inv1}
a\equiv\, !a \,.
\end{equation}
The negation of this equation is 
\begin{equation} \label{eq:inv2}
!a\equiv \,a \,.
\end{equation}
This is because 
\begin{equation}
!! =\star* \star* = (\star\otimes*)(\star\otimes*)= (\star\star\otimes **) = (I\otimes I) = I. 
\end{equation}
Equation \ref{eq:inv1} is the same as \ref{eq:inv2}. Thus both equations \ref{eq:inv1} and \ref{eq:inv2} are dual invariant. 
On the other hand, let $\bar{a}= ( u^* | S_k )$ then we have
\begin{equation} \label{eq:inv3}
\bar{a} \equiv \,! \bar{a} \,.
\end{equation}
The negation of this equation is
\begin{equation} \label{eq:inv4}
!\bar{a}\equiv \,a
\end{equation}
which is same as \ref{eq:inv3}. Thus both equations \ref{eq:inv3} and \ref{eq:inv4} are dual invariant. 
Therefore, the dual equation consists of 4 equations \ref{eq:inv1}, \ref{eq:inv2}, \ref{eq:inv3} , \ref{eq:inv4},
\begin{equation}
\begin{cases}
& a\equiv\, !a  \,\, \text{and} \,\, !a\equiv \,a \,\,\text{for partition} \,\, D \\
& \bar{a}\equiv\, !\bar{a}  \,\, \text{and} \,\, !\bar{a}\equiv \,\bar{a} \,\,\text{for partition} \,\,D^*
\end{cases} \,.
\end{equation}
Diagramatically,
\begin{figure}[H]
\centering
\includegraphics[trim=0cm 0cm 0cm 0cm, clip, scale=0.6]{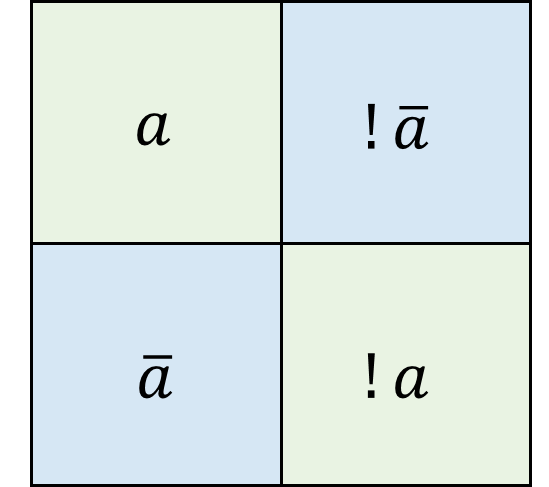}
\caption[]
{ \label{fig:table7}}
\end{figure}

\subsection{Representation of dual operators}
In the section we will construct matrix representations of the dual operators. Let $|0 \rangle$ and $|1 \rangle$ be a pair of dual states, satisfying $*|0\rangle = |1\rangle$ and $*|1\rangle = |0\rangle$ where $*=\hat{P}$ is the dual operator, satisfying the orthogonal relation $\langle 0 |1 \rangle = \langle 1 |0 \rangle =0$. 

The identity element can be constructed by
\begin{equation}
\pmb{1} = |0\rangle \langle 0 | + |1\rangle \langle 1 | \,,
\end{equation}
and the parity element can be constructed by\begin{equation}
\hat{P} = |0\rangle \langle1 | + |1\rangle \langle 0 | \,. 
\end{equation}
This can be proven easily by
\begin{equation}
\begin{aligned}
\hat{P}^2 &= ( |0\rangle \langle1 | +|1\rangle \langle 0 | )( |0\rangle \langle1 | +|1\rangle \langle 0 | ) \\
&= |0\rangle \langle  1 | 0\rangle \langle 1 | + |0\rangle \langle 1 | 1 \rangle \langle 0| + |1\rangle \langle 0 | 0\rangle \langle 1 | + |1\rangle \langle 0 | 1 \rangle \langle 0  | \\
&= |0\rangle \langle 0 | + |1\rangle \langle 1 | \\
&= \pmb{1} \,.
\end{aligned}
\end{equation}
It can also be checked that
\begin{equation}
\begin{aligned}
\pmb{1}^2 &= (|0\rangle \langle 0 | + |1\rangle \langle 1 |)(|0\rangle \langle 0 | + |1\rangle \langle 1 |) \\
&=|0\rangle \langle 0 | 0\rangle \langle 0 | + |0\rangle \langle 0 | 1\rangle \langle 1 | + |1\rangle \langle 1 | 0\rangle \langle 0 | + |1\rangle \langle 1 | 1\rangle \langle 1 | \\
&= |0\rangle \langle 0 | +|1\rangle \langle 1 | \\
&= \pmb{1}  \,.
\end{aligned}
\end{equation}
Next, to construct matrix representation for $\pmb{1}$ and $\hat{P}$,
\begin{equation}
\pmb{1}_{ab} = \langle a | \pmb{1} | b \rangle \,.
\end{equation}
Explicitly,
\begin{equation}
\pmb{1} = \begin{pmatrix}
 \langle 0| \pmb{1} | 0 \rangle &  \langle 0| \pmb{1} | 1 \rangle \\
  \langle 1| \pmb{1} | 0 \rangle & \langle 1| \pmb{1} | 1 \rangle 
\end{pmatrix} \,.
\end{equation}
To evaluate the matrix element, we compute
\begin{equation}
\begin{aligned}
\pmb{1}_{ab} &=  \langle a |0\rangle \langle 0 | b\rangle + \langle a |1\rangle \langle 1 | b \rangle \\
&= \delta_{a0}\delta_{0b} = \delta_{a1}\delta_{1b}\,.
\end{aligned}
\end{equation}
Therefore as expected,
\begin{equation}
\pmb{1} = \begin{pmatrix}
 1 &  0 \\
  0 & 1 
\end{pmatrix} \,.
\end{equation}
For $\hat{P}$, 
\begin{equation}
\hat{P}_{ab} = \langle a | \hat{P} | b\rangle \,.
\end{equation}
Explicitly,
\begin{equation}
\hat{P} = \begin{pmatrix}
 \langle 0| \hat{P} | 0 \rangle &  \langle 0| \hat{P} | 1 \rangle \\
  \langle 1| \hat{P} | 0 \rangle & \langle 1| \hat{P} | 1 \rangle 
\end{pmatrix} \,.
\end{equation}
To evaluate the matrix element,
\begin{equation}
\begin{aligned}
\hat{P}_{ab} &= \langle a |0\rangle \langle 1 | b\rangle + \langle a |1\rangle \langle 0 | b \rangle \\
&=\delta_{a0}\delta_{1b} + \delta_{a1}\delta_{0b} \,.
\end{aligned}
\end{equation}
Therefore we have
\begin{equation} \label{eq:dualmatrix}
\hat{P} =   \begin{pmatrix}
 0 & 1 \\
  1 & 0 
\end{pmatrix} \,.
\end{equation}
which is called the $\pmb{M}$ matrix. The matrix representation in \ref{eq:dualmatrix} can be diagonalized, since the eigenvalue is $1$ or $-1$, we have
\begin{equation}
\hat{P}^\prime =   \begin{pmatrix}
 1 & 0 \\
  0 & -1 
\end{pmatrix} \,.
\end{equation}
which is another representation of the parity operator. And there exists a similarity transformation such that $U\hat{P}U^{-1} =\hat{P}^\prime$.

One can observe that because
\begin{equation}
\pmb{1} \cdot \pmb{1} =  \pmb{1} \quad , \quad \hat{P}\cdot \hat{P} = \pmb{1} \quad , \quad  \pmb{1} \cdot \hat{P} = \hat{P} \quad , \quad  \hat{P} \cdot \pmb{1} = \hat{P} \,,
\end{equation}
this is isomorphic to the law of arithmetic multiplication that $++ =+ , --=+ , +-=-, -+=-$. Thus we can identify $+ \rightarrow \pmb{1},\,\,-\rightarrow \hat{P}$. Also, this structure is isomorphic to $\mathrm{even} +\mathrm{even} = \mathrm{even}$, $\mathrm{odd} +\mathrm{odd} = \mathrm{even}$, $\mathrm{even} +\mathrm{odd} = \mathrm{odd}$ , $\mathrm{odd} +\mathrm{even} = \mathrm{odd}$.   In addition, notice that $|00\rangle$ and $|11\rangle $ states are dual invariant (D.I) states, as no matter you look from the left or right they remains the same,
\begin{equation}
(|00\rangle | L) \equiv (|00\rangle | R) \quad \text{and} \quad (|11\rangle | L) \equiv (|11\rangle | R)  \,,
\end{equation} 
we call them full states. While $|01\rangle$ and $|10\rangle$ states are non-dual invariant (non-D.I) states, we have
\begin{equation}
(|01\rangle |L) \equiv |10\rangle |R ) \quad \text{and} \quad (|10\rangle |L) \equiv (|01\rangle |R ) \,.
\end{equation}
We regard these two states as the half states. Therefore the full states are dual even invariant, while the half states are non-dual invariant. Now we see that
\begin{equation}
\begin{aligned}
|00\rangle \,(\mathrm{D.I}) + |11\rangle \,(\mathrm{D.I}) &= \,(\mathrm{D.I})  \\
|01\rangle \,(\mathrm{non-D.I}) + |10\rangle \,(\mathrm{non-D.I}) &= \,(\mathrm{D.I})  \\
|00\rangle \,(\mathrm{D.I}) + |10\rangle \,(\mathrm{non-D.I}) &= \,(\mathrm{non-D.I})  \\
|01\rangle \,(\mathrm{non-D.I}) + |11\rangle \,(\mathrm{D.I}) &= \,(\mathrm{non-D.I})  \\
\end{aligned}
\end{equation} 
In words, we can write,
\begin{equation}
\begin{aligned}
\mathrm{full} + \mathrm{full} &= \mathrm{full} \\
\mathrm{half} + \mathrm{half} &= \mathrm{full} \\
\mathrm{half} + \mathrm{full} &= \mathrm{half} \\
\mathrm{full} + \mathrm{half} &= \mathrm{half} \\
\end{aligned}
\end{equation}
With such isomorphism, we can refer $\mathrm{full} \cong \mathrm{even} $ and $\mathrm{half} \cong \mathrm{odd}$.

\subsection{Mirror duality and identities}
It is interesting to ask a question: does there exist a structure which is dual to the above, such  that $++ =-, +- = +, -+=+, -- = -$? The answer is yes when we consider matrix operation towards the left. Normally we multiply matrix towards the right, here we introduce a new idea-left matrix multiplication. Consider
\begin{equation}
\begin{pmatrix}
 0 & 1 \\
  1 & 0 
\end{pmatrix} \overset{\mathrm{left\,\,multiplication}}{=} 
 \begin{pmatrix}
 0 & 1 \\
  1 & 0 
\end{pmatrix}
 \begin{pmatrix}
 0 & 1 \\
  1 & 0 
\end{pmatrix}
\overset{\mathrm{right\,\, multiplication}}{=} 
\begin{pmatrix}
 1 & 0 \\
 0 & 1 
\end{pmatrix} \,,
\end{equation}
similarly, we have
\begin{equation}
\begin{pmatrix}
 1 & 0 \\
  0 & 1 
\end{pmatrix} \overset{\mathrm{left\,\,multiplication}}{=} 
 \begin{pmatrix}
 0 & 1 \\
  1 & 0 
\end{pmatrix}
 \begin{pmatrix}
 1 & 0 \\
  0 & 1 
\end{pmatrix}
\overset{\mathrm{right\,\, multiplication}}{=} 
\begin{pmatrix}
 0 & 1 \\
 1 & 0 
\end{pmatrix}\,
\end{equation}
and similar for the others. We have the following,
\begin{equation}
(\pmb{1} \cdot \pmb{1})_L =  \hat{P} \quad , \quad (\hat{P}\cdot \hat{P})_L = \hat{P} \quad , \quad  (\pmb{1} \cdot \hat{P})_L = \pmb{1} \quad , \quad  (\hat{P} \cdot \pmb{1})_L = \pmb{1} \,,
\end{equation}
where the subscript label refers that we are multiplying matrix to the left. The dualities involving left matrix multiplication is called the mirror duality.
 
\subsubsection{Association dualities}

The general dualities here are set on the following grounds on dual studies:
\begin{equation}\hat{P}_+ =
\begin{pmatrix}
1 & 0 \\
0 & 0 
\end{pmatrix} \,,\,
\hat{P}_- =
\begin{pmatrix}  
 0 & 0 \\
 0  & 1  
\end{pmatrix} \,,\,
\hat{P}_+^* =
\begin{pmatrix}
0 & 1 \\
0 & 0
\end{pmatrix}  \,,\,
\hat{P}_-^* =
\begin {pmatrix}
0 & 0 \\
1 & 0
\end{pmatrix}  \,,\,
\end{equation}
They can be constructed by the following,
\begin{equation} \label{eq:construction1}
\hat{P}_+ = |0\rangle \langle 0 | \,,\quad \hat{P}_- = |1\rangle \langle 1 | \,,\quad
\hat{P}^*_+ = |0\rangle \langle 1 | \,,\quad
\hat{P}^*_- = |1\rangle \langle 0 | \,. 
\end{equation}
The generic form is
\begin{equation}
\hat{P}_{ab} = |a\rangle \langle b | \quad \text{where} \,\, a, b = 0 ,1 \,,
\end{equation}
the matrix element of the association dualities are 
\begin{equation}
\langle c | \hat{P}_{ab}| d\rangle = \langle c |a\rangle \langle b| d \rangle = \delta_{ca}\delta_{bd} \quad \text{where}\,\, a, b ,c ,d = 0 ,1 \,.
\end{equation}

We have the dual operations of
\begin{equation}
\hat{P}_+^* =\hat{P}_+ * = \hat{P}_+ \pmb{M}\,, \quad , \quad \hat{P}_-=\hat{P}_{+}^{(R)} \,\,,\,\, \hat{P}_-^* = *\hat{P}_+ = \pmb{M}\hat{P}_+ \,.
\end{equation}
Now we compute the product of these operators. $L$ means we multiply the matrix towards the left direction, while $R$ means we multiply the matrix towards the normal right direction. 
\begin{equation}
(\hat{P}_+\hat{P}_- | L ) =P_-^* \quad \text{and} \quad (\hat{P}_-\hat{P}_+ |L  )=\hat{P}^*_+ \,,
\end{equation}
\begin{equation}\label{eq:333}
(\hat{P}_+ \hat{P}_-|R)= ( \hat{P}_- \hat{P}_+|R)=  \pmb{0}   \,,
\end{equation}
where for the last equation we have dual invariance for the operator element over the right action perspective. Since we have $(\hat{P}_+\hat{P}_-)^\star=(\hat{P}_-\hat{P}_+),$ which is left-land action invariant. Next we have
\begin{equation}
(\hat{P}^*_+ \hat{P}_-^*|R)= \hat{P}_+  \quad \text{and} \quad  (\hat{P}^*_- \hat{P}_+^*|R)= \hat{P}_-  \,,
\end{equation}
\begin{equation}
(\hat{P}_+^* P_-^* | L) = (\hat{P}_-^* P_+^* | L) = \pmb{0} \,.
\end{equation}
To simply the expression for clear demonstration process, let $A=\hat{P}_+\hat{P}_-$, and its dual $A^\star =\hat{P}_-\hat{P}_+$, we have
\begin{equation}
(A|R) \equiv (A^\star |R) \,. 
\end{equation}
Thus $A=A^\star = \pmb{0}$\,, thus the zero matrix is a dual invariant under left or right matrix operation.
We further have the following properties, consider the anticommutators of the operators:
\begin{equation}
(\{ \hat{P_+} , \hat{P}_- \} | L ) =\pmb{M} \quad , \quad (\{ \hat{P_+} , \hat{P}_- \} | R ) =\pmb{0} \quad ; \quad
\end{equation}
while
\begin{equation}
(\{ \hat{P}_+^* , \hat{P}^*_- \}\ | L ) = \pmb{0} \quad,\quad (\{ \hat{P}_+^* , \hat{P}^*_- \}| R ) = \pmb{1}\,.
\end{equation}
Now consider the following comparison,
\begin{equation} \label{eq:set1}
\begin{aligned}
&(\hat{P}_+ \hat{P}_+ + \hat{P}_- \hat{P}_- |R) = \pmb{1} \quad \text{and} \quad  (\hat{P}_+ \hat{P}_- + \hat{P}_- \hat{P}_+ |R) = \pmb{0} \,.\\
&(\hat{P}_+ \hat{P}_+ + \hat{P}_- \hat{P}_- |L) = \pmb{0} \quad \text{and} \quad  (\hat{P}_+ \hat{P}_- + \hat{P}_- \hat{P}_+ |L) = \pmb{M} \,.
\end{aligned}
\end{equation}
And their dual, 
\begin{equation} \label{eq:set2}
\begin{aligned}
&(\hat{P}_+^* \hat{P}_+^* + \hat{P}_-^* \hat{P}_-^* |R) = \pmb{0} \quad \text{and} \quad  (\hat{P}_+^* \hat{P}_-^* + \hat{P}_-^* \hat{P}_+^* |R) = \pmb{1} \,.\\
&(\hat{P}_+^* \hat{P}_+^* + \hat{P}_-^* \hat{P}_-^* |L) = \pmb{M} \quad \text{and} \quad  (\hat{P}_+^* \hat{P}_-^* + \hat{P}_-^* \hat{P}_+^* |L) = \pmb{0} \,.
\end{aligned}
\end{equation}
Here we define the sum dualities, as we can see
\begin{equation}
\begin{aligned}
&*(\hat{P}_+ \hat{P}_- + \hat{P}_- \hat{P}_+ |R) = (\hat{P}_+^* \hat{P}_-^* + \hat{P}_-^* \hat{P}_+^* |R) \,. \\
&*(\hat{P}_+ \hat{P}_+ + \hat{P}_- \hat{P}_- |R)= (\hat{P}_+^* \hat{P}_+^* + \hat{P}_-^* \hat{P}_-^* |R) \,.
\end{aligned} 
\end{equation}
Therefore we have for $R$,
\begin{equation} \label{eq:dual10}
(*\pmb{0} =\pmb{1}|R) \quad \text{and} \quad (*\pmb{1} =\pmb{0} |R)\,.
\end{equation}
Hence in fact the zero matrix and the identity matrix is dual to each other under the right matrix operation. 
Then we also have the perspective duality
\begin{equation}
\begin{aligned}
&\star(\hat{P}_+ \hat{P}_+ + \hat{P}_- \hat{P}_- |R) = (\hat{P}_+ \hat{P}_+ + \hat{P}_- \hat{P}_- |L) \,. \\
& \star(\hat{P}_+^* \hat{P}_-^* + \hat{P}_-^* \hat{P}_+^* |R) = (\hat{P}_+^* \hat{P}_-^* + \hat{P}_-^* \hat{P}_+^* |L) \,.
\end{aligned}
\end{equation}
Therefore we have
\begin{equation}
\star \pmb{1} = \pmb{0} \quad \text{and} \quad \star \pmb{0} = \pmb{1} \,.
\end{equation}
Now for $L$,
\begin{equation}
\begin{aligned}
& *(\hat{P}_+ \hat{P}_- + \hat{P}_- \hat{P}_+ |L) = (\hat{P}_+^* \hat{P}_-^* + \hat{P}_-^* \hat{P}_+^* |L) \,.\\
& *(\hat{P}_+ \hat{P}_+ + \hat{P}_- \hat{P}_- |L)= (\hat{P}_+^* \hat{P}_+^* + \hat{P}_-^* \hat{P}_-^* |L) \,.
\end{aligned}
\end{equation}
Therefore we have for $L$,
\begin{equation}
(*\pmb{M} =\pmb{0}|L) \quad \text{and} \quad (*\pmb{0} =\pmb{M} |L)\,.
\end{equation}
Hence in fact the zero matrix and the identity matrix is dual to each other under the left matrix operation. Then we also have the perspective duality
\begin{equation}
\begin{aligned}
& \star(\hat{P}_+ \hat{P}_- + \hat{P}_- \hat{P}_+ |R) = (\hat{P}_+ \hat{P}_- + \hat{P}_- \hat{P}_+ |L) \,. \\
&\star(\hat{P}_+^* \hat{P}_+^* + \hat{P}_-^* \hat{P}_-^* |R) = (\hat{P}_+^* \hat{P}_+^* + \hat{P}_-^* \hat{P}_-^* |L) \,,
\end{aligned}
\end{equation}
Therefore for the $*$ case we have
\begin{equation}
\star\pmb{M} = \pmb{0} \quad \text{and} \quad \star\pmb{0} = \pmb{M}\,.
\end{equation}
If we consider the collection representation matrices of \ref{eq:set1} and   \ref{eq:set2}
\begin{equation}
\begin{pmatrix}
\pmb{1} & \pmb{0} \\
\pmb{0} & \pmb{M} 
\end{pmatrix} \quad \text{and} \quad 
\begin{pmatrix}
\pmb{0} & \pmb{1} \\
\pmb{M} & \pmb{0} 
\end{pmatrix}\,,
\end{equation}
which is left-right dual to each other.

Next we define the following 4 association matrices
\begin{equation}
\hat{-}=\begin{pmatrix}
 0 & 0 \\
1 &1 
\end{pmatrix} \quad , \quad
\hat{+}=\begin{pmatrix}
1 & 1 \\
0 &0 
\end{pmatrix} \quad , \quad
\hat{\#}=\begin{pmatrix}
1 & 0 \\
1 &0
\end{pmatrix} \quad , \quad
\hat{\flat}=\begin{pmatrix}
 0 & 1 \\
0 &1 
\end{pmatrix} \quad , \quad
\end{equation}

Next we can check that the following identities hold:
\begin{equation}
\begin{aligned}
&\hat{-} = (\hat{P}_+ \hat{P}_+^\star | R ) + (\hat{P}_+ \hat{P}_+^\star | L ) \\
&\hat{+} = (\hat{P}_- \hat{P}_-^\star | R ) + (\hat{P}_- \hat{P}_-^\star | L )\\ 
&\hat{\#} = (\hat{P}_-^* \hat{P}_+ | R ) + (\hat{P}^*_- \hat{P}_+ | L ) \\
&\hat{\flat} = (\hat{P}_+^* \hat{P}_- | R ) + (\hat{P}^*_+ \hat{P}_- | L )
\end{aligned} \,.
\end{equation} 

Note that $\{\hat{+},\hat{-}\}$ and $\{\hat{\#},\hat{\flat}\}$ are dual pair such that 
\begin{equation}
*\hat{-} =\hat{+} \quad ,\quad \hat{*}+ =\hat{-} \quad ; \quad \hat{\#}=\flat* \quad , \quad \flat=\hat{\#}*\,. 
\end{equation}

Now we will investigate representations of $\mathbb{Z}_2 \times \mathbb{Z}_2$ based on these operators under left or right matrix operations. The matrix operators are regarded as the perspectives. For $\{\hat{+},\hat{-}\}$ It is easy to prove the following,
\begin{equation} \label{eq:sys1}
\begin{aligned}
&(\hat{-}\hat{+}|R) \equiv (\hat{+}\hat{-}|L) = \hat{+} \\
&(\hat{-}\hat{+}|L) \equiv  (\hat{+}\hat{-}|R) =\hat{-}
\end{aligned} \,\,\,\,.
\end{equation}
And for $\{\hat{\#},\hat{\flat}\}$,
\begin{equation} \label{eq:sys2}
\begin{aligned}
&( \hat{\#} \hat{\flat} | R ) \equiv (\hat{\flat}\hat{\#}  |L ) = \hat{\flat} \\
&( \hat{\#} \hat{\flat} |L ) \equiv (\hat{\flat}\hat{\#}  |R) = \hat{\flat}
\end{aligned} \,\,\,\,.
\end{equation}
Therefore for $\{\hat{+} ,\hat{-}\}$, and $\{\hat{\#} ,\hat{\flat}\}$ two dual sets respectively, we have \ref{eq:sys1} and \ref{eq:sys2}

It takes the generic form of 
\begin{equation}
\begin{aligned}
(AB|R) \equiv (BA|L)
\end{aligned}
\end{equation}
If $A$ and $B$ are dual to each other, i.e. $B=*A$ and $A=*B$. then the Klein-4 group $\mathbb{Z}_2 \times \mathbb{Z}_2$ is preserved. However if they are not, we will see that this leads to breaking down of the Klein-4 group to its subgroup $\mathbb{Z}_2$. Let's us check,
\begin{equation}
(\hat{-}\hat{\#} |R ) \equiv (\hat{\#}\hat{-} |L ) = 2\hat{P}_+ \,. 
\end{equation}
However,
\begin{equation} \label{eq:dual1}
(\hat{-}\hat{\#}|L) = \begin{pmatrix}
0 & 0\\
0 & 0
\end{pmatrix} \quad \text{and} \quad (\hat{\#}\hat{-}|R) = \begin{pmatrix}
1 & 1\\
1 & 1
\end{pmatrix} 
\end{equation}
in which they are not equivalent, 
\begin{equation}
(\hat{-}\hat{\#}|L) \neq(\hat{\#}\hat{-}|R) \,.
\end{equation}
Similarly we have
\begin{equation}
(\hat{-}\hat{\flat}|R) \equiv (\hat{\flat}\hat{-}|L) = 2\hat{P}_+^* \,,
\end{equation}
but
\begin{equation} \label{eq:dual2}
(\hat{-}\hat{\flat}|L) = \begin{pmatrix}
1 & 1\\
1&1 \end{pmatrix} \quad \text{and} \quad (\hat{\flat}\hat{-}|R) = \begin{pmatrix}
0 & 0\\
0 & 0
\end{pmatrix} \,.
\end{equation}
Therefore
\begin{equation}
(\hat{-}\hat{\flat}|R)  \neq (\hat{\flat}\hat{-}|L).
\end{equation}

We also have
\begin{equation}
(\hat{+}\hat{\#} |R ) \equiv (\hat{\#}\hat{+} |L ) = 2\hat{P}_-^* \,. 
\end{equation}
but
\begin{equation} \label{eq:dual3}
(\hat{+}\hat{\#} |L) = \begin{pmatrix}
1 & 1\\
1&1 \end{pmatrix} \quad \text{and} \quad  (\hat{\#}\hat{+} |R)=\begin{pmatrix}
0 & 0\\
0 & 0
\end{pmatrix} \,.
\end{equation}
Therefore,
\begin{equation}
(\hat{+}\hat{\#} |L) \neq (\hat{\#}\hat{+} |R)\,.
\end{equation}
Finally we also have 
\begin{equation}
(\hat{+}\hat{\flat} |R ) \equiv (\hat{\flat}\hat{+} |L ) = 2\hat{P}_-
\end{equation}
but
\begin{equation} \label{eq:dual4}
(\hat{+}\hat{\flat} |L ) =\begin{pmatrix}
0 & 0\\
0 & 0
\end{pmatrix} \quad \text{and} \quad (\hat{\flat}\hat{+} |R ) =  \begin{pmatrix}
1 & 1\\
1&1  \end{pmatrix} \,.
\end{equation}
Therefore we have,
\begin{equation}
(\hat{+}\hat{\flat} |L ) \neq (\hat{\flat}\hat{+} |R )\,.
\end{equation}

\begin{figure}[H]
\centering
\includegraphics[trim=0cm 0cm 0cm 0cm, clip, scale=0.45]{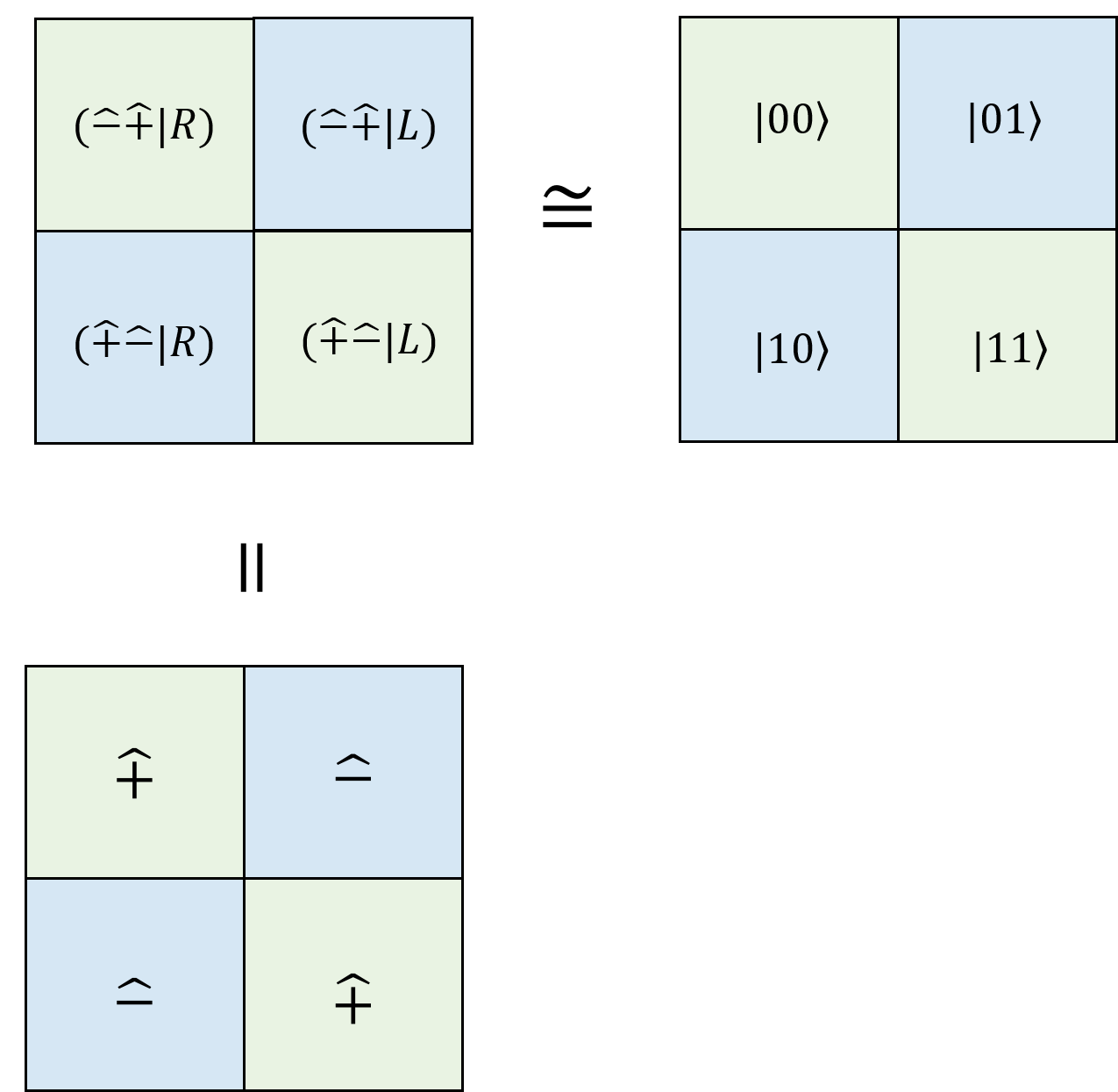}
\caption[]
{} \label{fig:left1}
\end{figure}

\begin{figure}[H]
\centering
\includegraphics[trim=0cm 0cm 0cm 0cm, clip, scale=0.45]{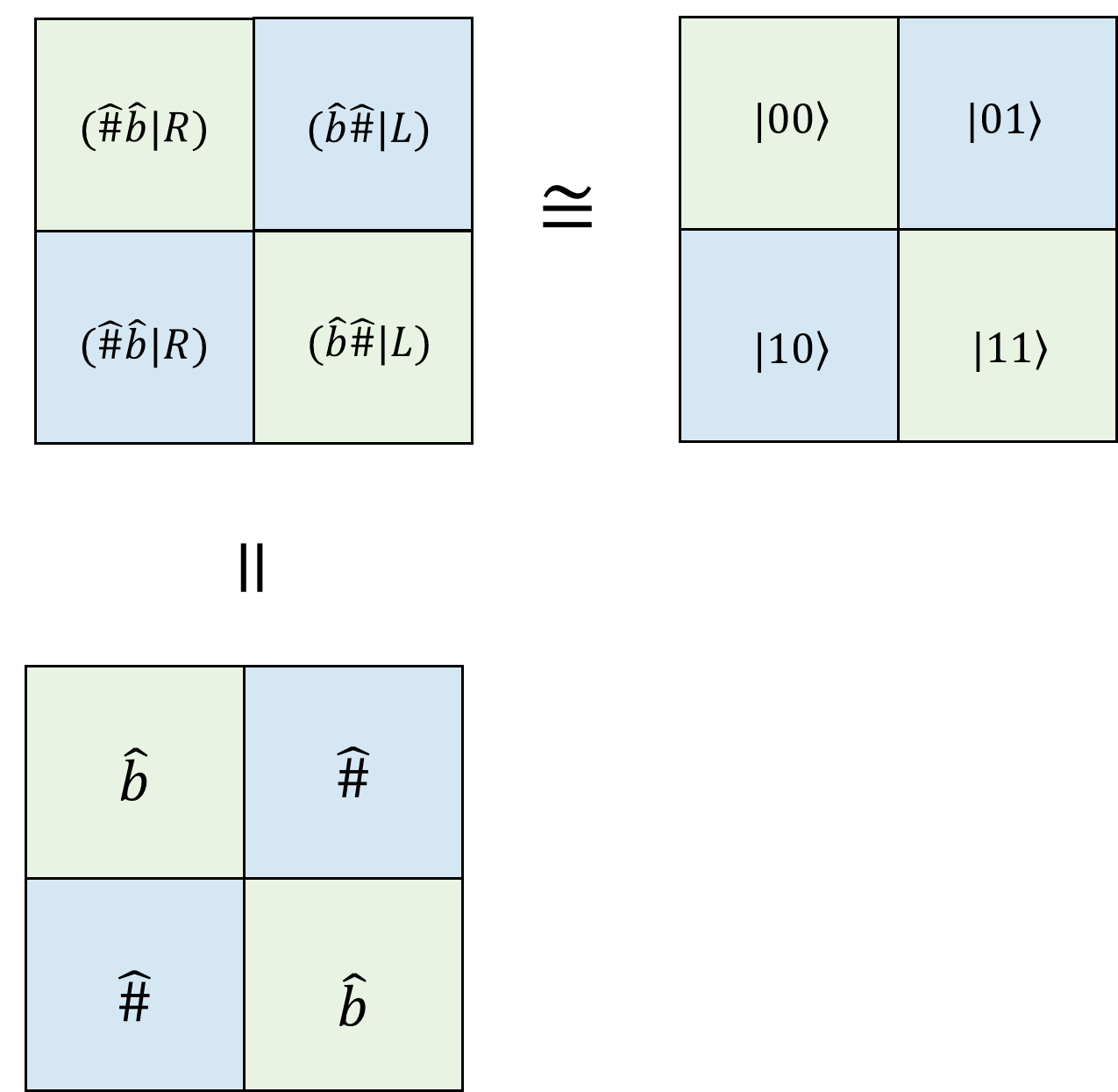}
\caption[]
{} \label{fig:left2}
\end{figure}

The whole idea is fairly abstract, we can construct diagramatic approach to illustrate the full idea.  Let's divide them into two big cases, for $A,B$ that are dual to each other, and $A,B$ which are non-dual to each other. There are two sub-cases for the case which $A,B$ are dual to each other, while there are 4 sub-cases for the case which $A,B$ are not dual to each other. In other words, there are six sub-cases. We first consider the case with $\{\hat{+}, \hat{-} \}$, we map $R\rightarrow 0$ and $L\rightarrow 0$, ; $\hat{-}\hat{+}\rightarrow 0 $ and $\hat{+}\hat{-}\rightarrow 1$ (see figure \ref{fig:left1}). Similarly for the $\{\hat{\#}, \hat{\flat} \}$, we have in figure \ref{fig:left2}.


However, for the remaining case for which $A$ and $B$ are not dual to each other, things are much more complicated, the structural of $\mathbb{Z}_2 \times \mathbb{Z}_2 $ duality breaks down, we only have one statement for duality equivalence out of two. 

The full idea can be represented by a diagram as follow:
\begin{figure}[H]
\centering
\includegraphics[trim=0cm 0cm 0cm 0cm, clip, scale=0.8]{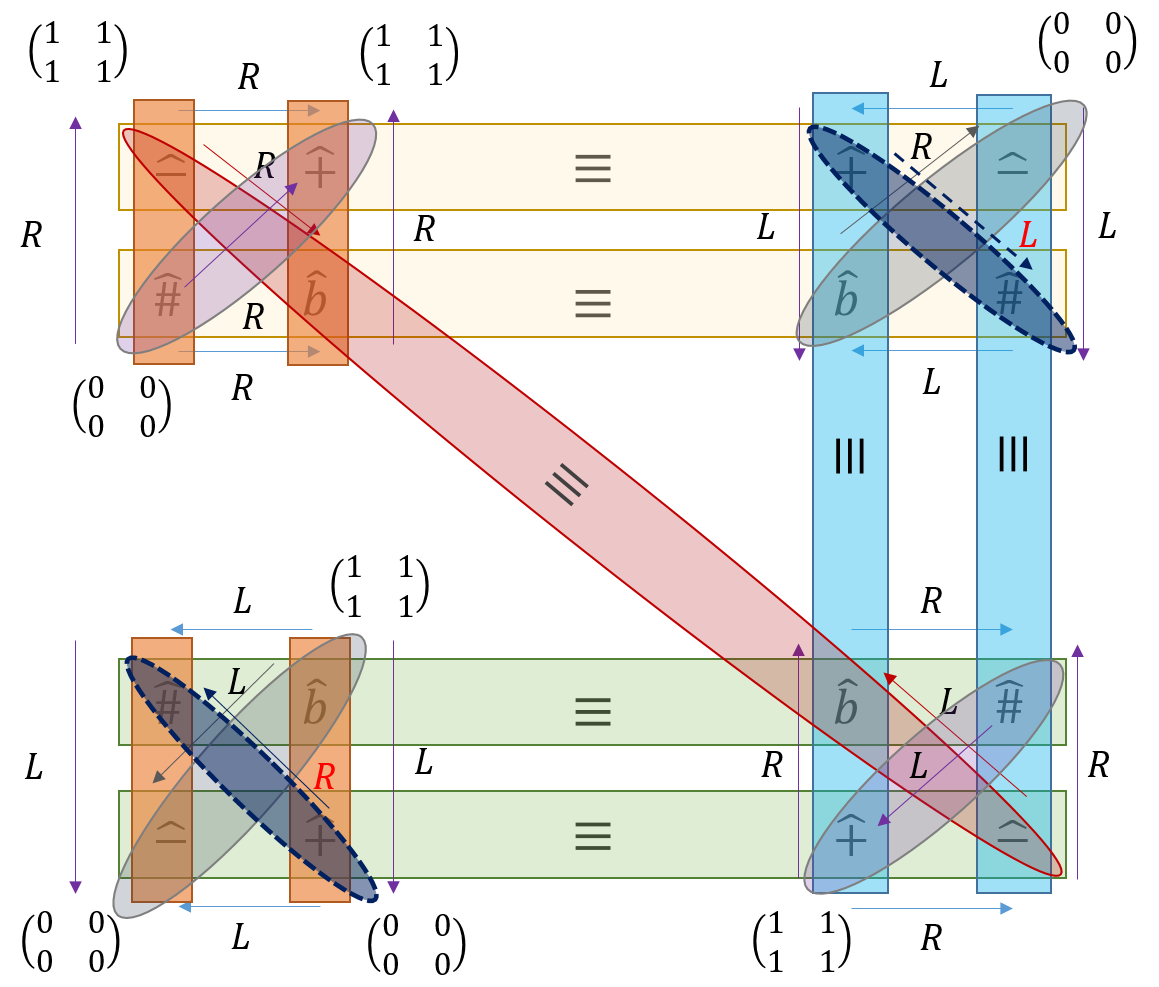}
\caption[]
{In the diagram, the arrow denotes the direction of the matrix multiplication, $R$ denotes right and $L$ denotes left. A special note is remarked for the navy-blue oval with dotted boundary, this corresponds to the dual equivalence of $(\hat{+}\hat{\#} |R ) \equiv (\hat{\#}\hat{+} |L )$. Here for this case we read from the back of the page, so we have the red $R$ and red $L$. For the others we read the direction from above the page. } \label{fig:fullillustration}
\end{figure}

From equation \ref{eq:dual1}, \ref{eq:dual2}, \ref{eq:dual3} and \ref{eq:dual4}, we see that,
\begin{equation}
\begin{aligned}
& (\hat{\#}\hat{-}|R) = *\star (\hat{-}\hat{\#}|L) = !(\hat{-}\hat{\#}|L) \,,\\
& (\hat{\flat}\hat{-}|R) = *\star (\hat{-}\hat{\flat}|L) = !(\hat{-}\hat{\flat}|L) \,,\\
& (\hat{\#}\hat{+}|R) = *\star (\hat{+}\hat{\#}|L) = !(\hat{+}\hat{\#}|L) \,, \\
& (\hat{\flat}\hat{+}|R) = *\star (\hat{+}\hat{\flat}|L) = !(\hat{+}\hat{\flat}|L) \,,
\end{aligned}
\end{equation}
where $!=*\star$. Therefore from these results we have
\begin{equation} \label{eq:dual!}
\begin{pmatrix}
1  & 1 \\
1 &  1
\end{pmatrix} = !\begin{pmatrix}
0 & 0 \\
0 &  0
\end{pmatrix} \quad \text{and} \quad 
\begin{pmatrix}
0  & 0 \\
0 &  0
\end{pmatrix} = !\begin{pmatrix}
1 & 1 \\
1 &  1
\end{pmatrix}
\end{equation}
with $!!= \pmb{1}$. Therefore, $\begin{pmatrix}
0  & 0 \\
0 &  0
\end{pmatrix}$ and $\begin{pmatrix}
1 & 1 \\
1 &  1
\end{pmatrix}$
is dual to one another. Define 
$
\mathbb{I} = \begin{pmatrix}
1 & 1 \\
1 &  1
\end{pmatrix}
$
as the dual invariant identity matrix, we write \ref{eq:dual!} as
\begin{equation}
\mathbb{I} = !\,\pmb{0} \quad \text{and} \quad \pmb{0} = !\,\mathbb{I} \,.
\end{equation}
In addition, we also have
\begin{equation}
\begin{aligned}
& (\hat{-}\hat{\#}|L) = (*_1\hat{+}*_2\hat{\flat}|L) =(\hat{+}\hat{\flat} |L) = \pmb{0} \\
& (\hat{\flat}\hat{-}|R) = (*_1\hat{\flat}*_2\hat{-}|R) =(\hat{\#}\hat{+} |R) = \pmb{0} \,, \\
& (\hat{-}\hat{\flat}|L) = (*_1\hat{-}*_2\hat{\flat}|L) =(\hat{+}\hat{\#} |L) = \mathbb{I} \,, \\
& (\hat{\#}\hat{-}|R) = (*_1\hat{\#}*_2\hat{-}|R) =(\hat{\flat}\hat{+}|R) = \mathbb{I} \,.
\end{aligned} 
\end{equation}
This can be further grouped into $L$ and $R$ category as follows :
\begin{figure}[H]
\centering
\includegraphics[trim=0cm 0cm 0cm 0cm, clip, scale=0.45]{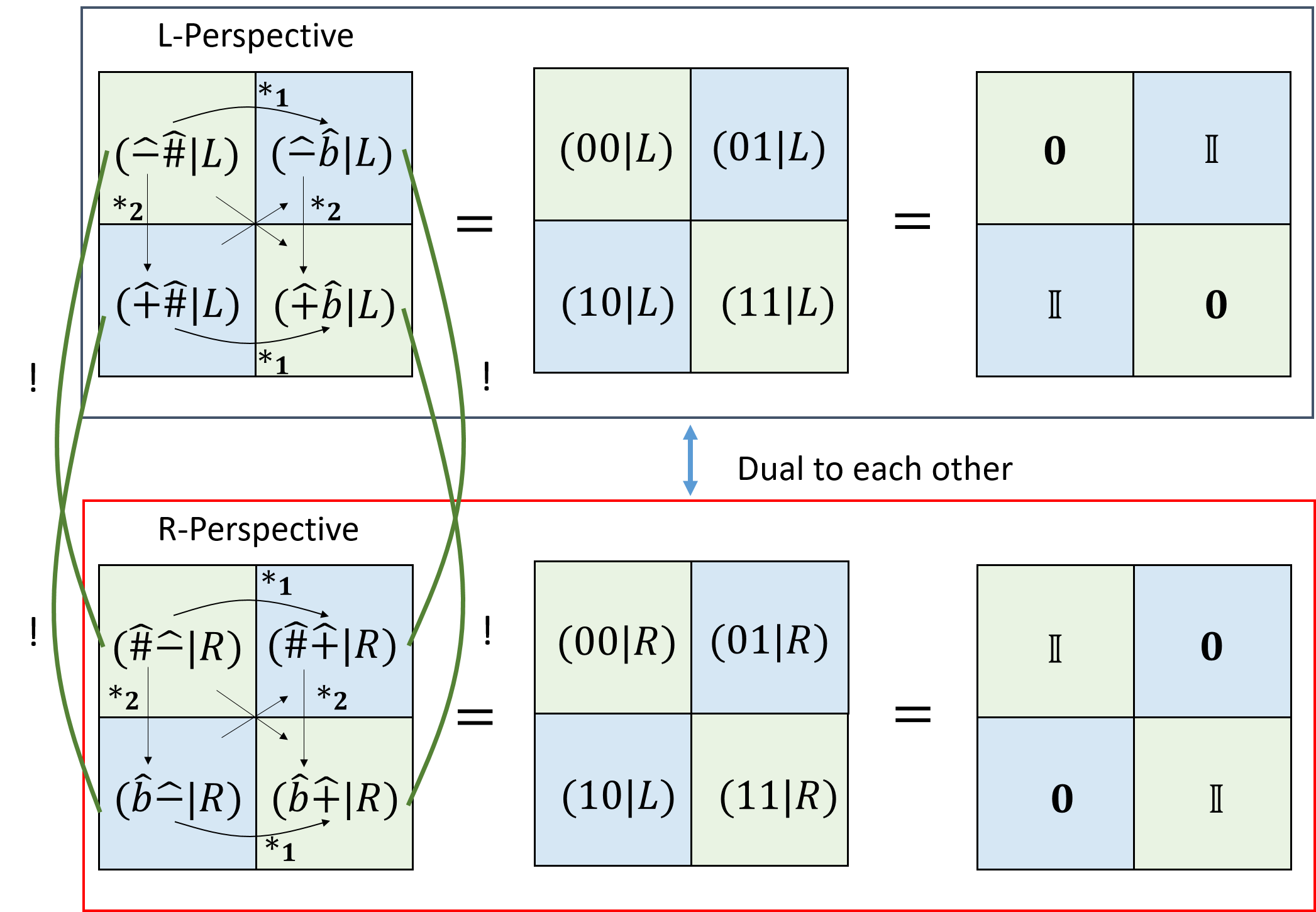}
\caption[]
{} \label{fig:bigpicture}
\end{figure}

\subsection{Addition and pointwise multiplication duality}
Since we know that $\pmb{0}$ and $\mathbb{I}$ is dual to each other, now we would like to see how to construct these dual matrices from their progenitor dual matrices. First consider
\begin{equation}
\pmb{0} + \mathbb{I} = \mathbb{I} + \pmb{0} = \mathbb{I} \quad \text{and} \quad \pmb{0} \bullet\mathbb{I} = \mathbb{I} \bullet \pmb{0} = \pmb{0} \,,
\end{equation}
where $\bullet$ is the pointwise matrix product (Hadamard product). Therefore in terms of operator, we say the matrix addition and the matrix pointwise operator is dual to each other. Let $\hat{O}$ be $+$ and $\hat{O}^*$ be $\bullet$, we have the dual operator set as $\{+ , \bullet\}$. Element-wise, we define
\begin{equation}
\begin{pmatrix}
1  & 1 \\
1 &  1
\end{pmatrix} = !\begin{pmatrix}
0 & 0 \\
0 &  0
\end{pmatrix}
=
\begin{pmatrix}
!0  & !0 \\
!0 &  !0
\end{pmatrix} \quad\text{and}\quad
\end{equation}
\begin{equation}
\begin{pmatrix}
0  & 0 \\
0 &  0
\end{pmatrix} = !\begin{pmatrix}
1 & 1 \\
1 &  1
\end{pmatrix}
=
\begin{pmatrix}
!1  & !1 \\
!1 &  !1
\end{pmatrix} \quad\,,
\end{equation}
therefore 
\begin{equation} \label{eq:01dual}
1= !\,0  \quad\text{and}\quad  0=!\,1 \,.
\end{equation}
Hence 0 and 1 are dual to each other, we can form the dual set $\{0,1 \}$. This 0,1 duality has many applications, for example,
\begin{equation}
\begin{pmatrix}
1   \\
0 
\end{pmatrix} =
!\begin{pmatrix}
0\\ 
1   
\end{pmatrix}
=\begin{pmatrix}
!0\\ 
!1  
\end{pmatrix} \,,
\end{equation}
\begin{equation}
\begin{pmatrix}
1  & 0 \\
0 &  1
\end{pmatrix} =
!\begin{pmatrix}
0  & 1 \\
1&   0
\end{pmatrix}
=\begin{pmatrix}
!0  & !1 \\
!1&   !0
\end{pmatrix} \,,
\end{equation}
\begin{equation}
\begin{pmatrix}
1  & 1 \\
0 &  0
\end{pmatrix} =
!\begin{pmatrix}
0  & 0 \\
1&   1
\end{pmatrix}
=\begin{pmatrix}
!0  & !0 \\
!1&   !1
\end{pmatrix} \,.
\end{equation} 
So we have duality sets, for examples$
\{
\begin{pmatrix}
1 & 0 \\ 
0 & 1
\end{pmatrix} ,
\begin{pmatrix}
0 & 1 \\ 
1 & 0
\end{pmatrix}
\}
\quad , \quad
\{\begin{pmatrix}
1 &1 \\ 
0 & 0
\end{pmatrix} ,
\begin{pmatrix}
0 & 0 \\ 
1 & 1
\end{pmatrix}
\}
\,
$ , etc.

\subsection{Column and row duality for $\hat{+}$ and $\hat{-}$ operators}
Finally, we would like to investigate the column and row duality for the $\hat{+}$ and $\hat{-}$ operators. Consider their products
\begin{equation}
\begin{aligned}
& (\hat{+}\hat{+}|R) = \hat{+} \quad \text{and} \quad (\hat{+}\hat{+}|L) = \hat{+} \\
& (\hat{-}\hat{-}|R) = \hat{-} \quad \text{and} \quad (\hat{-}\hat{-}|L) = \hat{-} \\
&  (\hat{+}\hat{-}|R) = \hat{+} \quad \text{and} \quad (\hat{+}\hat{-}|L) = \hat{-} \\
& (\hat{-}\hat{+}|R) = \hat{-} \quad \text{and} \quad (\hat{-}\hat{+}|L) = \hat{+} 
\end{aligned} \,\,\,\,.
\end{equation}
Diagramatically,
\begin{figure}[H]
\centering
\includegraphics[trim=0cm 0cm 0cm 0cm, clip, scale=0.6]{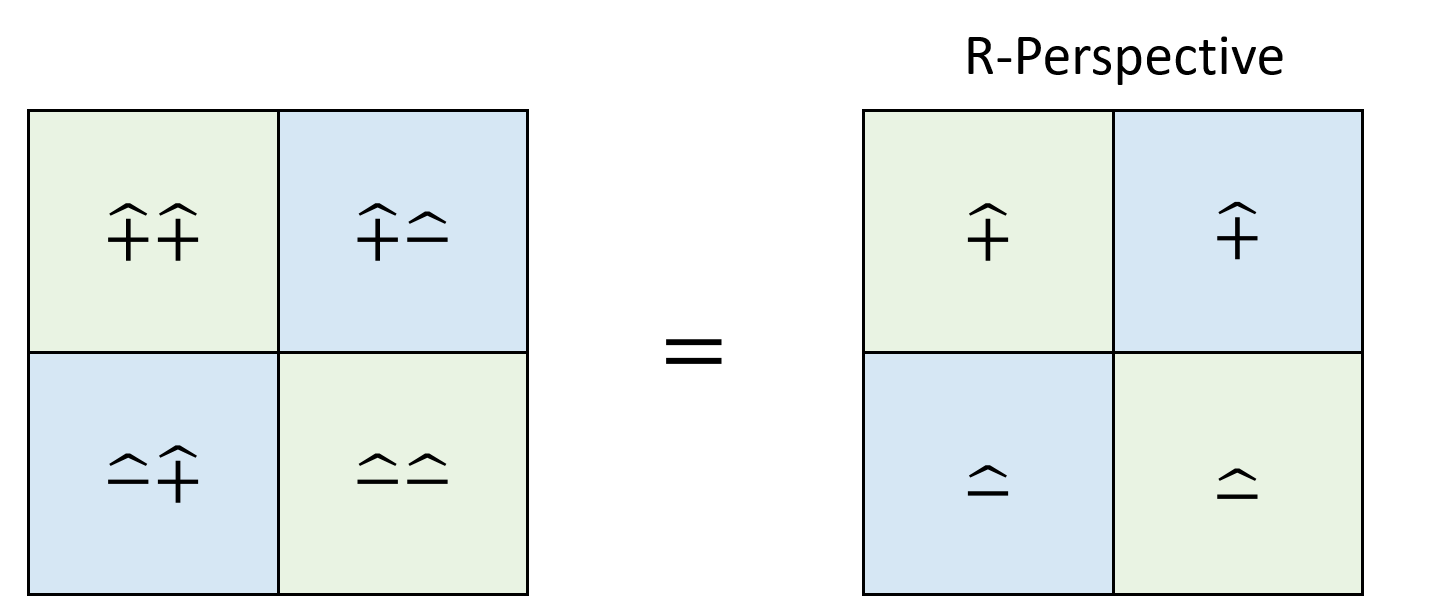}
\caption[]
{This is from the R-Perspective, which is a row duality.} \label{fig:RP}
\end{figure}
\begin{figure}[H]
\centering
\includegraphics[trim=0cm 0cm 0cm 0cm, clip, scale=0.6]{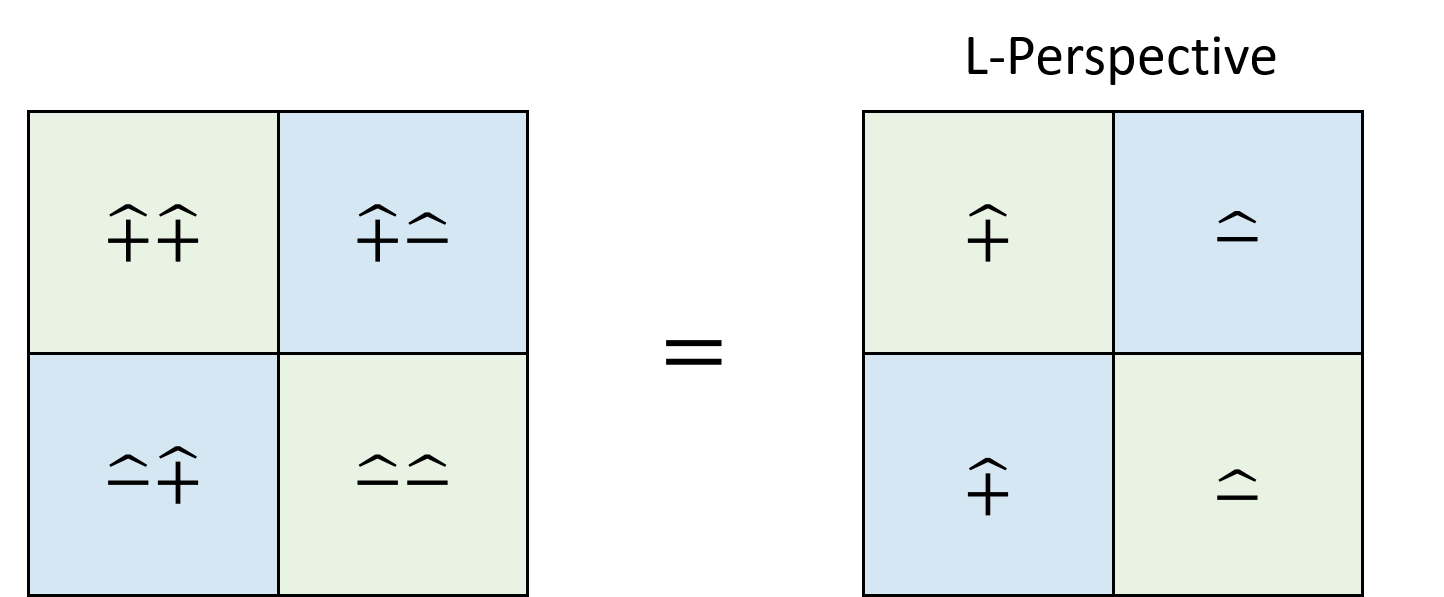}
\caption[]
{This is from the L-Perspective, which is a column duality.} \label{fig:LP}
\end{figure}

\subsubsection{Duality equivalence in left and right transpose}
Next we will introduce the concept of transpose and dual transpose. Transpose of a matrix is given by the usual definition, which swap elements by the diagonal. Given a matrix $A$, $(A^T)_{ij} = A_{ji}$. We call this left transpose. Now we also define its dual operation of right transpose, which swap elements along the off-diagonal, for example,
\begin{equation}
\begin{pmatrix}
a & b \\
c & d
\end{pmatrix}^R =
\begin{pmatrix}
d & b \\
c & a
\end{pmatrix} \,.
\end{equation}
We have
\begin{equation}
\hat{+}^T = \hat{\flat} \quad , \quad  \hat{-}^T = \hat{\sharp} \quad , \quad \hat{+}^R = \hat{\sharp} \quad , \quad  \hat{-}^R = \hat{\flat}\,. 
\end{equation}
Therefore we have
\begin{equation}
\hat{+}^T = \hat{-}^R \quad \text{and} \quad \hat{+}^R = \hat{-}^T \,.
\end{equation}
In perspective representation, we can write it as
\begin{equation}
(\hat{+}|T) \equiv (\hat{-} | R) \quad \text{and} \quad (\hat{+}|R) \equiv (\hat{-} | T)
\end{equation}
For simplicity, if we write $\hat{+}\rightarrow 0, R\rightarrow 0$ and $\hat{-}\rightarrow 1 , L\rightarrow 1$ , we have
\begin{equation}
(0|0) \equiv (1|1) \quad \text{and} \quad (0|1) \equiv (1|0) ,
\end{equation}
which are representations of $\mathbb{Z}_2 \times \mathbb{Z}_2$ as before. 
We also have the following
\begin{equation}
\hat{+}^{(TR)} = \hat{+}^{(RT)} = \hat{-} \quad \text{and} \quad \hat{-}^{(TR)} = \hat{-}^{(RT)} = \hat{+} \,.
\end{equation}
Therefore the $RT$ (or $TR$) operations act as the dual matrix $\pmb{M}$.

\subsubsection{Complementary matrices}
Next we define complementary matrices, these matrices have three 1 entries and one 0 entry, denoted by $A_{ij}$, where $ij$ is the position of the 0,
\begin{equation}
A_{11} =\begin{pmatrix}
0 & 1\\
1 & 1
\end{pmatrix} \quad,\quad
A_{22} =\begin{pmatrix}
1 & 1\\
1 & 0 
\end{pmatrix}\quad,\quad
A_{12} =\begin{pmatrix}
1 & 0\\
1 & 1 
\end{pmatrix} \quad ,\quad
A_{21} =\begin{pmatrix}
1 & 1\\
0 & 1 
\end{pmatrix}\quad\,.
\end{equation}
and we have their pointwise-element dual,
\begin{equation}
!A_{11} = \hat{P}_+ = \begin{pmatrix}
1 & 0 \\
0 & 0
\end{pmatrix}
\quad, \quad
!A_{22} = \hat{P}_- = \begin{pmatrix}
0 & 0 \\
0 & 1
\end{pmatrix}
\end{equation}
\begin{equation}
!A_{12} = \hat{P}_+^* = \begin{pmatrix}
0 & 1 \\
0 & 0
\end{pmatrix}
\quad, \quad
!A_{21} = \hat{P}_-^* = \begin{pmatrix}
0 & 0 \\
1 & 0
\end{pmatrix} \,.
\end{equation}
Since 
\begin{equation}
!A_{ij} + A_{ij} = A_{ij}+!A_{ij}= \mathbb{I} \quad\text{and}\quad !A_{ij} \bullet A_{ij} = A_{ij}\bullet !A_{ij}= \pmb{0} \,,
\end{equation}
therefore it follows that $!A_{ij} , A_{ij}$ are dual to each other. Now we have the following identities,
\begin{equation}
\begin{aligned}
& (A_{11} \hat{P}_+ |R) \equiv (\hat{P}_- A_{22} |R ) = \hat{P}^*_- \,, \\
& (\hat{P}_+ A_{11} |R) \equiv (A_{22} \hat{P}_-|R ) = \hat{P}^*_+   \,,
\end{aligned}
\end{equation}
or
\begin{equation}
\begin{aligned}
& (A_{11} !A_{11} |R) \equiv (!A_{22} A_{22} |R ) = !A_{21} \,, \\
& (!A_{11} A_{11} |R) \equiv (A_{22} ! A_{22} |R) = !A_{12} \,. 
\end{aligned}
\end{equation}
Since $\hat{P}_-^*$ and $\hat{P}_+^*$ are dual to each other (by right transpose),
\begin{equation}
*^\prime (\hat{P}_-^*) = (\hat{P}_-^*)^R = \hat{P}_+^* \,. 
\end{equation}
We then have
\begin{equation}
\begin{aligned}
& *^\prime (A_{11} \hat{P}_+|R) = ( \hat{P}_+ A_{11}|R) \,,\\
& *^\prime (\hat{P}_- A_{22} |R ) = (A_{22} \hat{P}_-|R ) \,. 
\end{aligned}
\end{equation}
We can also see that 
\begin{equation} \label{eq:eq11}
A_{11} = \hat{*}_R A_{22} = A^R_{22} \quad \text{and} \quad \hat{*}_R A_{11}= A^R_{11} = A_{22} \,. 
\end{equation}
Next we have the following identities
\begin{equation}
\begin{aligned}
&(A_{21} \hat{P}_-^*|L) \equiv (P_+^* A_{12}|L) = \hat{P}_-  \,,\\
& (\hat{P}_-^* A_{21} |L) \equiv (A_{12} P_+^* |L) = \hat{P}_+ \,,
\end{aligned}
\end{equation}
or
\begin{equation}
\begin{aligned}
&(A_{21}  !A_{21} |L) \equiv (!A_{12} A_{12}|L) = !A_{22}  \,,\\
& (!A_{21} A_{21} |L) \equiv (A_{12} !A_{12} |L) =!A_{11} \,,
\end{aligned}
\end{equation}
Since $\hat{P}_-$ and $\hat{P}_+$ are dual to each other (by left transpose),
\begin{equation}
*^{\prime\prime}\hat{P}_- = \hat{P}_-^L = \hat{P}_+ \,.
\end{equation}
We then have
\begin{equation}
\begin{aligned}
*^{\prime\prime}(A_{21} \hat{P}_-^*|L) = (\hat{P}_-^* A_{12} |L) \,,  \\
*^{\prime\prime}(\hat{P}_+^* A_{21} |L) = (A_{12} P_+^* |L) \,.
\end{aligned}
\end{equation}
We can also see that
\begin{equation} \label{eq:eq22}
A_{21} = \hat{*}_L A_{12} = A^T_{12} \quad \text{and} \quad \hat{*}_L A_{21}= A^T_{21} = A_{12} \,. 
\end{equation}
Therefore from \ref{eq:eq11} and \ref{eq:eq22} we obtain the following generalization. For $i,j = 1,2$, $*i=j$ and $*j=i$ with $**=\pmb{1}$. We have
\begin{equation}
*A_{ij} = A_{*i*j} = A_{ji}\quad \text{and} \quad *A_{ii} = A_{*i*i} = A_{jj}\,.
\end{equation}
Next, we continue to have the following identities,
\begin{equation}
\begin{aligned}
&(A_{22} \hat{P}_- |L ) \equiv ( \hat{P}_-^* A_{21} | R ) = \hat{+} \,,\\
&(A_{11}\hat{P}_+ |L  )  \equiv (\hat{P}^*_+ A_{12} | R ) =\hat{-} \,,
\end{aligned} 
\end{equation}
or
\begin{equation}
\begin{aligned}
&(A_{22} !A_{22} |L ) \equiv ( !A_{21} A_{21} | R ) = \hat{+} \,,\\
&(A_{11}!A_{11} |L  )  \equiv (!A_{12} A_{12} | R ) =\hat{-} \,,
\end{aligned}
\end{equation}
Therefore we have
\begin{equation}
\begin{aligned}
&\hat{*}_R *^{\prime\prime}(A_{22} \hat{P}_- |L )=(A_{11}\hat{P}_+ |L  ) \,,\\
& *^\prime \hat{*}_L ( \hat{P}_-^* A_{21} | R )=(\hat{P}^*_+ A_{12} | R )\,.
\end{aligned} 
\end{equation}
Hence, $\hat{*}_R *^{\prime\prime}$ and $ *^\prime \hat{*}_L$ are dual to each other. Finally we have the identities of,
\begin{equation}
\begin{aligned}
&(\hat{P}_- A_{22} | L ) \equiv ( A_{12} \hat{P}_+^* | R) =\hat{\flat} \,, \\
&(\hat{P}_+ A_{11} | L ) \equiv (A_{21} \hat{P}_-^* | R) =\hat{\#}  \,,
\end{aligned}
\end{equation}
or
\begin{equation}
\begin{aligned}
&(!A_{22} A_{22} | L ) \equiv ( A_{12} !A_{12} | R) =\hat{\flat} \,,\\
&(!A_{11} A_{11} | L ) \equiv (A_{21} !A_{21} | R) =\hat{\#}  \,.
\end{aligned}
\end{equation}
Therefore we have
\begin{equation}
\begin{aligned}
&*^{\prime\prime}\hat{*}_R (\hat{P}_- A_{22} | L )=(\hat{P}_+ A_{11} | L ) \,,\\
&\hat{*}_L *^\prime ( A_{12} \hat{P}_+^* | R) = (A_{21} \hat{P}_-^* | R) \,.
\end{aligned}
\end{equation}
Hence, $*^{\prime\prime}\hat{*}_R$ and $\hat{*}_L *^\prime $ are dual to each other.

\subsubsection{Non-dual invariant identity and dual invariant identity}
In matrix operation, we are already familiar with the usual identity matrix $\pmb{1}$. However as we know, it is only the identity under right matrix operation, we say it is non-dual invariant identity. Let's see for the following illustration
\begin{figure}[H]
\centering
\includegraphics[trim=0cm 0cm 0cm 0cm, clip, scale=1]{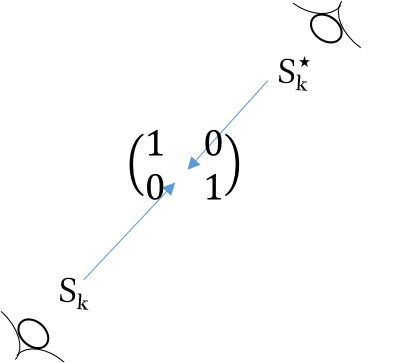}
\caption[]
{} \label{fig:nondualidentity}
\end{figure}
It is only the identity matrix in the $S_k$ perspective, but if we view from the other side, i.e. $S_k^\star$ perspective, it is the dual matrix $\pmb{M}$. Mathematically we have
\begin{equation}
(\pmb{1}|S_k ) \equiv (\pmb{M}|S_k^\star ) \,.
\end{equation}
And similarly we also have
\begin{equation}
(\pmb{M}|S_k ) \equiv (\pmb{1}|S_k^\star ) \,.
\end{equation}
To construct a dual invariant identity matrix $\mathbb{I}$, we can simply add up the identity matrix and its dual, so that
\begin{equation}
\pmb{1} + \pmb{M} = *(\pmb{1} + \pmb{M}) = \begin{pmatrix}
1 & 1 \\
1 & 1
\end{pmatrix} = \mathbb{I}\,,
\end{equation}
in which this matrix remains the same no matter which perspective do we look at. Also we have
\begin{equation}
\hat{+} + \hat{-} = * (\hat{+} + \hat{-}) = \mathbb{I} \,,
\end{equation}
where $\hat{-} = *\hat{+}=!\hat{+}$. Also,
\begin{equation}
\hat{\#} + \hat{\flat} = !(\hat{\#} + \hat{\flat})  = \mathbb{I} \,,
\end{equation}
where $\hat{\#} =!\hat{\flat}$. Also
\begin{equation}
A_{ij} + !A_{ij} = !(A_{ij} + !A_{ij} )=\mathbb{I} \,.
\end{equation}
And while $\pmb{1}^n = \pmb{1}$, but for $\mathbb{I}$ we have
\begin{equation}
\mathbb{I}^n = 2^{n-1} \mathbb{I}\,.
\end{equation}
In addition, since
\begin{equation}
\hat{+} \bullet \hat{-} = ! (\hat{+} \bullet \hat{-}) =\hat{-} \bullet\hat{+=}\pmb{0} \,, 
\end{equation}
\begin{equation}
\hat{\#} \bullet \hat{\flat} = !(\hat{\#} \bullet \hat{\flat})  = \hat{\flat}\bullet\hat{\#}=\pmb{0} \,,
\end{equation}
\begin{equation}
A_{ij} \bullet !A_{ij} =!(A_{ij} \bullet !A_{ij} ) = \pmb{0} \,.
\end{equation}
and $\mathbb{I}$, $\pmb{0}$ are dual to each other, thus this further generalize $+,\bullet$ are dual to each other for all cases.

\subsection{General Analysis and Topological duality}
In this section, we wil carry out a combined analysis of duality of the previous sections discussed above. Notice that from \ref{eq:construction1} we assume working on right operation. The result is different if we work on the left operation. The following diagram illustrates the different outcomes:
\begin{figure}[H]
\centering
\includegraphics[trim=0cm 0cm 0cm 0cm, clip, scale=0.45]{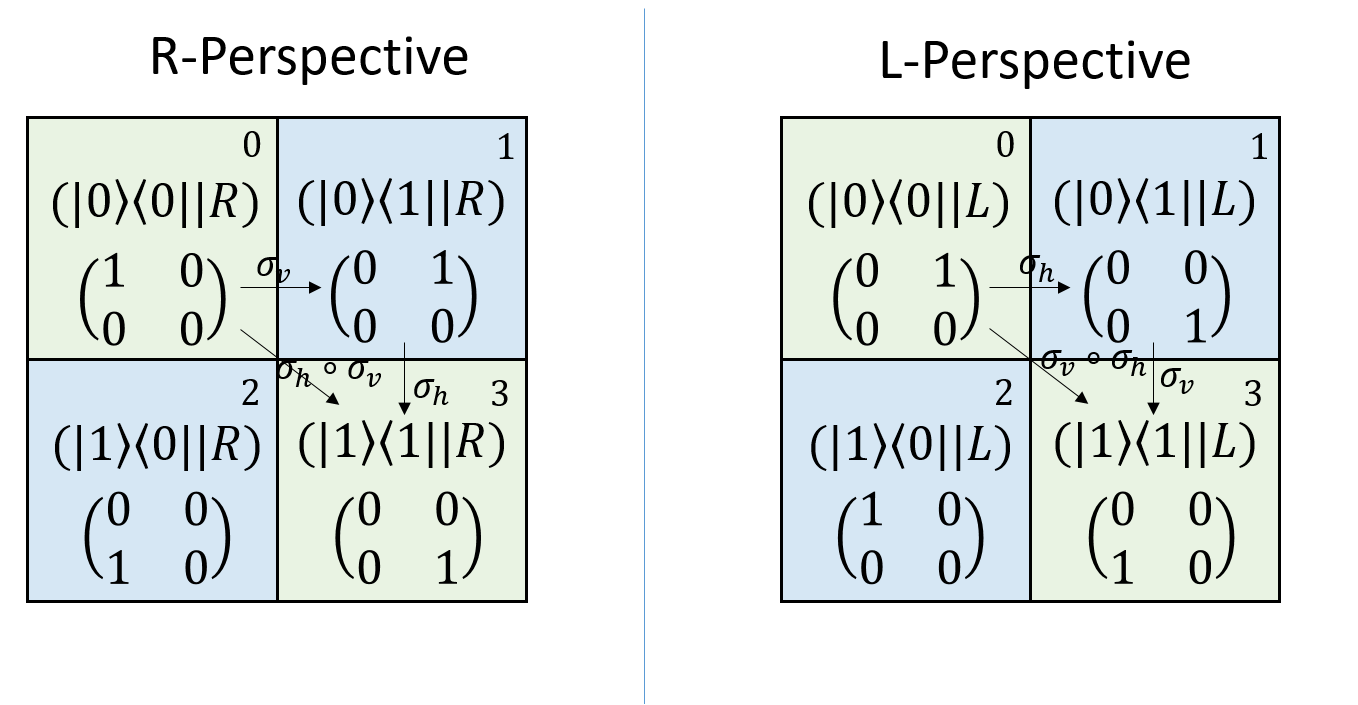}
\caption[]
{The association dualities in $R$ and $L$ matrix operation perspective. The $\sigma_v$ is the reflection operator along the vertical plane and the $\sigma_h$ is the reflection operator along the horizontal plane. The $\hat{P}_+ , \hat{P}_- , \hat{P}_+^* , \hat{P}_-^* $ form the matrix basis of the $\mathbb{Z}_2 \times \mathbb{Z}_2$ group in the matrix space. } \label{fig:dualx}
\end{figure}
We see that 
\begin{equation}
(|0\rangle\langle 0| +|1\rangle\langle 1||R) \equiv (|1\rangle\langle 0| +|0\rangle\langle 1||L) = \pmb{1}
\end{equation}
and
\begin{equation}
(|0\rangle\langle 1| +|1\rangle\langle 0||R) \equiv (|0\rangle\langle 0| +|1\rangle\langle 1||L) = \pmb{M}
\end{equation}
Therefore we see that in fact $u=|0\rangle\langle 0| +|1\rangle\langle 1|$ and $u^* = |0\rangle\langle 1| +|1\rangle\langle 0|$ are dual to each other. Also as $L$ and $R$ is dual to each other, this shows the follow structure of
\begin{equation}
(u|S_k) =(u^* | S_k^\star) = \pmb{1} \quad \text{and} (u^*|S_k) =(u | S_k^*) = \pmb{M}
\end{equation}
for $S_k  =R$ and $S_k^\star = L$. Notes that $u$ is a dual invariant, while $u^*$ is another dual invariant, this is simply both $u$ and $u^*$ are complete,
\begin{equation}
u = \hat{P_+} + \hat{P_-}\,,\quad u^R = \hat{P_-} + \hat{P_+} = u \,,
\end{equation}
and
\begin{equation}
u^* = \hat{P^*_+} + \hat{P^*_-}\,,\quad u^T = \hat{P^*_-} + \hat{P^*_+} = u^* \,.
\end{equation} 
So we can see that two different dual invariants can be dual to each other. 
 
Next, we consider the row and column dualities in both $R$ and $L$ perspectives. First for convenience, we address the decimal number representation on the upper-right corner in each box of \ref{fig:dualx} for both $R$ and $L$ perspective. For example we denote $(0|R) = (|0\rangle \langle 0||R),\,\,(1|R) = (|0\rangle \langle 1||R) $, etc. Then we have the following equivalence,
\begin{equation} \label{eq:eqa}
\begin{aligned}
& (0|R) = (2|L) = \hat{P}_+ \\
& (1|R) = (0|L) = \hat{P}_+^* \\
& (2|R) = (3|L) = \hat{P}_-^* \\
& (3|R) = (1|L) = \hat{P}_- \,.
\end{aligned}
\end{equation}
Note that $\{\hat{P}_+ , \hat{P}_- \}$ is a dual pair and $\{\hat{P}_+^* , \hat{P}_-^* \}$, this is arise from the fact that $00 ,11$ are dual pair and $01, 10$ is another dual pair. It follows that 0 is dual to 3 and 1 is dual to 2. Now we add up up two association dualities to form $\hat{-},\hat{+},\hat{\sharp},\hat{\flat}$. We have the following, for the right perspective,
\begin{equation}
\begin{aligned}
\hat{+} &=(|0\rangle\langle 0||R) + (|0\rangle\langle 1||R) =(0|R)+(1|R) = (1|R) \\
\hat{-} &=(|1\rangle\langle 1||R) + (|1\rangle\langle 0||R) =(3|R)+(2|R) = (5|R)
\end{aligned}
\end{equation}
We know that $\{\hat{+},\hat{-} \}$ is a dual pair under the $\sigma_h$ reflection and this can be confirmed by seeing $0,1$ duality, i.e. \begin{equation}
*[(|0\rangle\langle 0||R) + (|0\rangle\langle 1||R)] = (|1\rangle\langle 1||R) + (|1\rangle\langle 0||R)\,.
\end{equation}
Therefore, 1 and 5 are dual to each other. Since 1 and 5 are both odd, we call $\{\hat{+},\hat{-} \}$ the odd association dual pair under the right perspective. Next we also have
\begin{equation}
\begin{aligned}
\hat{\sharp} &= (|0\rangle\langle 0||R) + (|1\rangle\langle 0||R) =(0|R)+(2|R) =  (2|R) \\
\hat{\flat} &= (|1\rangle\langle 1||R) + (|1\rangle\langle 0||R) =(3|R)+(1|R)= (4|R)
\end{aligned} \,.
\end{equation}
We know that $\{\hat{\sharp},\hat{\flat} \}$ is a dual pair under the $\sigma_v$ reflection and this can be confirmed by seeing $0,1$ duality, i.e. \begin{equation}
*[(|0\rangle\langle 0||R) + (|1\rangle\langle 0||R)] = (|1\rangle\langle 1||R) + (|0\rangle\langle 1||R)\,.
\end{equation}
Therefore, 2 and 4 are dual to each other. Since 2 and 4 are both even, we call $\{\hat{\sharp},\hat{\flat} \}$ the even association dual pair.

For the left perspective,
\begin{equation}
\begin{aligned}
\hat{+} &=(|0\rangle\langle 0||L) + (|1\rangle\langle 0||L) =(0|L) + (2|L) = (2|L) \\
\hat{-} &=(|1\rangle\langle 1||L) + (|0\rangle\langle 1||L) =(3|L) + (1|L) = (4|L) \,.
\end{aligned}
\end{equation}
Again, we see that these two matrices are dual to each other by $0 \leftrightarrow 1$ duality. As $2$ and $4$ are even, we say the dual pair $\{\hat{+}, \hat{-}\}$ is even association dual pair under the left perspective. Next we also have
\begin{equation}
\begin{aligned}
\hat{\sharp} &= (|1\rangle\langle 1||L) + (|1\rangle\langle 0||L) = (3|L) + (2|L)=(5|L) \\
\hat{\flat} &= (|0\rangle\langle 0||L) + (|0\rangle\langle 1||L) = (0|L) + (1|L)=(1|L) \,.
\end{aligned}
\end{equation}
Again, we see that these two matrices are dual to each other by $0 \leftrightarrow 1$ duality. As $1$ and $5$ are odd, we say the dual pair $\{\hat{\sharp}, \hat{\flat}\}$ is odd association dual pair under the left perspective. 
Therefore, for a dual pair, whether it is odd or even parity subjects to $L/R$ operation perspective. The above demonstrates nicely the following fact that
\begin{equation}
\begin{aligned}
( \mathrm{even}| R  ) &\equiv ( \mathrm{odd}| R  ) \\
( \mathrm{odd}| R  ) &\equiv ( \mathrm{even}| R  ) \,.
\end{aligned}
\end{equation}

Diagramatically, we can express $\{\hat{+} , \hat{-}\}$ in rows and $\{\hat{\sharp} , \hat{\flat}\}$ in columns in the right perspective, vice versa. This will give the concept of row and column dualities, 
\begin{figure}[H]
\centering
\includegraphics[trim=0cm 0cm 0cm 0cm, clip, scale=0.45]{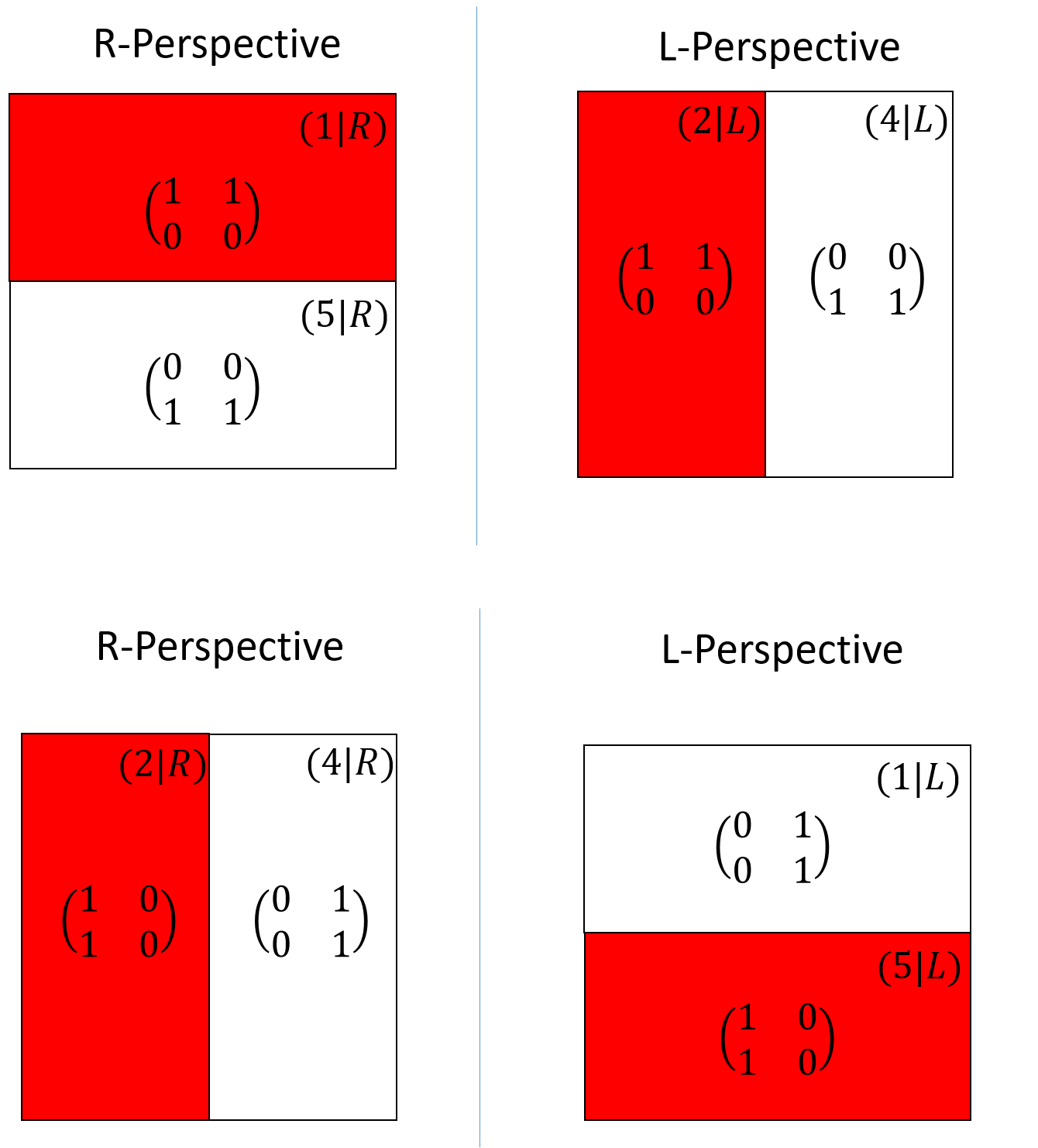}
\caption[]
{ } \label{fig:dualx2}
\end{figure}
In general, the following rule is satisfied,
\begin{equation}
\begin{aligned}
(\mathrm{row}|R) &\equiv (\mathrm{column}|L) \\
(\mathrm{column}|R) &\equiv (\mathrm{row} |L)
\end{aligned}
\end{equation}
Therefore, in general a row is dual to a column and a column is dual to a row. However, there is still a degree of freedom when we carry out transformation between row and columns. A row (column) can be converted to a a column (row) by either left or right transpose. In here, for the $\{\hat{+} , \hat{-} \}$ dual pair in right respective, it is brought to the left perspective by left transpose operation $T$ (anti-clockwise rotation). Also, from the fact that $(1|R) \equiv (2|L)$, the change of perspective involves an increment of decimal representation by 1; while $(5|R) \equiv (4|L)$, the change of perspective involves an decrement of decimal representation by 1. For a general form of $(a_R | R) \equiv (a_L | L)$, in the $\hat{+}$ case, we have $\Delta_+ = a_L -a_R = 2-1 = + 1$. For the $\hat{-}$ case,we have $\Delta_- = a_L -a_R = 4-5 = - 1$. Therefore we have $\Delta_- = -\Delta_+$. So in fact $\Delta_-$ is dual to $\Delta_+$. It follows that
\begin{equation}
(\Delta_+ |\hat{+} ) \equiv (\Delta_- |\hat{-}) 
\end{equation}
which is just
\begin{equation}
(+1 |\hat{+} ) \equiv (-1 |\hat{-}) \,. 
\end{equation}
The same idea goes for the $\{\hat{\sharp}, \hat{\flat}\}$ dual pair, where we have
\begin{equation}
(\Delta_\sharp |\hat{\sharp} ) \equiv (\Delta_\flat |\hat{\flat}) 
\end{equation}
which is just
\begin{equation}
(+3 |\hat{\sharp} ) \equiv (-3 |\hat{\flat}) \,. 
\end{equation}
In here, for the $\{\hat{\sharp} , \hat{\flat} \}$ dual pair in right respective, it is brought to the left perspective by right transpose operation $R$ (clockwise rotation), which is dual to $T$ (anti-clockwise direction) in $\{\hat{+} , \hat{-} \}$. This affirms $\{\hat{+} , \hat{-} \}$ and $\{\hat{\sharp} , \hat{\flat} \}$ are indeed dual to each other.

Generally speaking, the difference between $\Delta$ values of the dual representations characterize the specific dual representation. For the above case, $\{\hat{+}, \hat{-} \} $ has $\{\Delta_+ , \Delta_- \} = \{ +1 , -1 \}$, and representation $R$ is brought to representation $L$ by left transpose $T$. $\{\hat{\sharp}, \hat{\flat} \} $ has $\{\Delta_\sharp , \Delta_\flat \} = \{ +3 , -3 \}$, and representation $R$ is brought to representation $L$ by right transpose $R$. And  we must have $|\Delta| = |\Delta^*|$ so the $\Delta$ values of the two dual representation must be conserved.

Now we consider a special case,
\begin{equation}
\pmb{1} = (|0\rangle \langle 0| + |1\rangle \langle 1||R) = (0|R) + (3|R) = (3|R) \,,
\end{equation}
the same goes for
\begin{equation}
\pmb{1} = (|1\rangle \langle 0| + |0\rangle \langle 1||L) = (2|L) + (1|L) = (3|L) \,.
\end{equation}
Therefore we have
\begin{equation}
(3|R) = (3|L) \,.
\end{equation}
Therefore $3=3^*$, 3 is a dual invariant number. Therefore we have $\Delta_{\pmb{1}} = 0$. Similarly we also have the case of its dual $\mathbb{M}$, 
\begin{equation}
\pmb{M} = (|0\rangle \langle 1| + |1\rangle \langle 0||R) = (1|R) + (2|R) = (3|R) \,,
\end{equation}
the same goes for
\begin{equation}
\pmb{M} = (|0\rangle \langle 0| + |1\rangle \langle 1||L) = (0|L) + (3|L) = (3|L) \,.
\end{equation}
Again $\Delta_{\pmb{M}} =0$. This implies that both $\pmb{1}$ and $\pmb{M}$ are dual invariant. However, what does it mean? We know that $\{ \pmb{1}, \pmb{M} \}$ forms a dual pair, how can they be dual invariant? It seems that we arrive at some contradiction. Yet if we consider the diagonal perspective, dual invariance is conserved.  
\begin{figure}[H]
\centering
\includegraphics[trim=0cm 0cm 0cm 0cm, clip, scale=0.45]{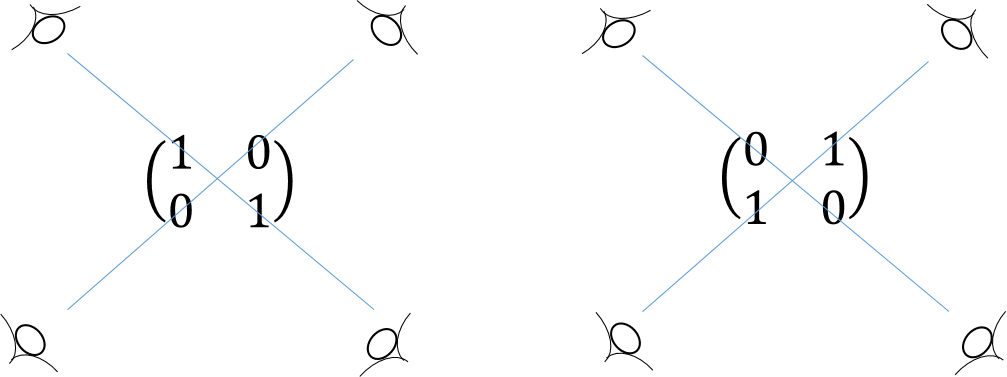}
\caption[]
{ } \label{fig:dualx2}
\end{figure}
In general, we expect that for some representation $X$ and its dual $X^*$ if $\Delta_X = \Delta_{X^*} =0$, then they are dual invariant under the diagonal perspective. 

Finally, we can construct the full dual invariant $\mathbb{I}$. Notice that for the $R$-perspective
\begin{equation}
\mathbb{I} = (1|R) + (5|R) = (2|R) + (4|R) = (3|R) + (3|R) = (6|R)\,.
\end{equation}
And for the $L$ perspective,
\begin{equation}
\mathbb{I} = (2|L) + (4|L) = (1|L) + (5|L) = (3|L) + (3|L) = (6|L)\,.
\end{equation}
Therefore we have
\begin{equation}
(6|R) = (6|L) \,.
\end{equation}
Thus $6$ is also a dual invariant number, and it represents the $\mathbb{I}$ matrix. Hence the above expression $6= 1 + 5 = 2+4 = 3+3$ can be expressed in matrix form. 

From the above, we have shown that 1 is dual to 5, 2 is dual to 4, and 3 is dual invariants. This means in the number line 1,2,3,4,5 we have $3$ that partitions this set into two halves. The number 3 acts like a mirror such that the number in two opposite sides are dual to each other. The following illustrates this important result
\begin{figure}[H]
\centering
\includegraphics[trim=0cm 0cm 0cm 0cm, clip, scale=0.45]{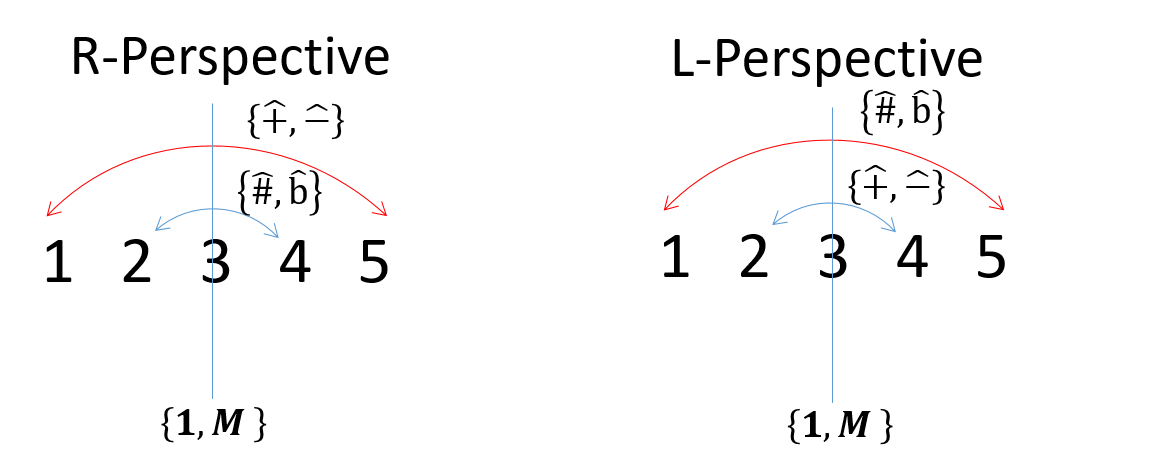}
\caption[]
{ } \label{fig:dualx4}
\end{figure}
In addition, we know that 6 is a dual invariant, and of course 0 is a dual invariant number. To complete the whole construction, we assign $0$ as the zero matrix $\pmb{0}$. This makes sense because as in section 2.1.4, we have $!\pmb{0}=\mathbb{I}$, so since $3$ is the duality mirror, and each opposite side of $3$ have matrices exchanging $0 \leftrightarrow 1$. The following illustration  shows the whole idea:
\begin{figure}[H]
\centering
\includegraphics[trim=0cm 0cm 0cm 0cm, clip, scale=0.45]{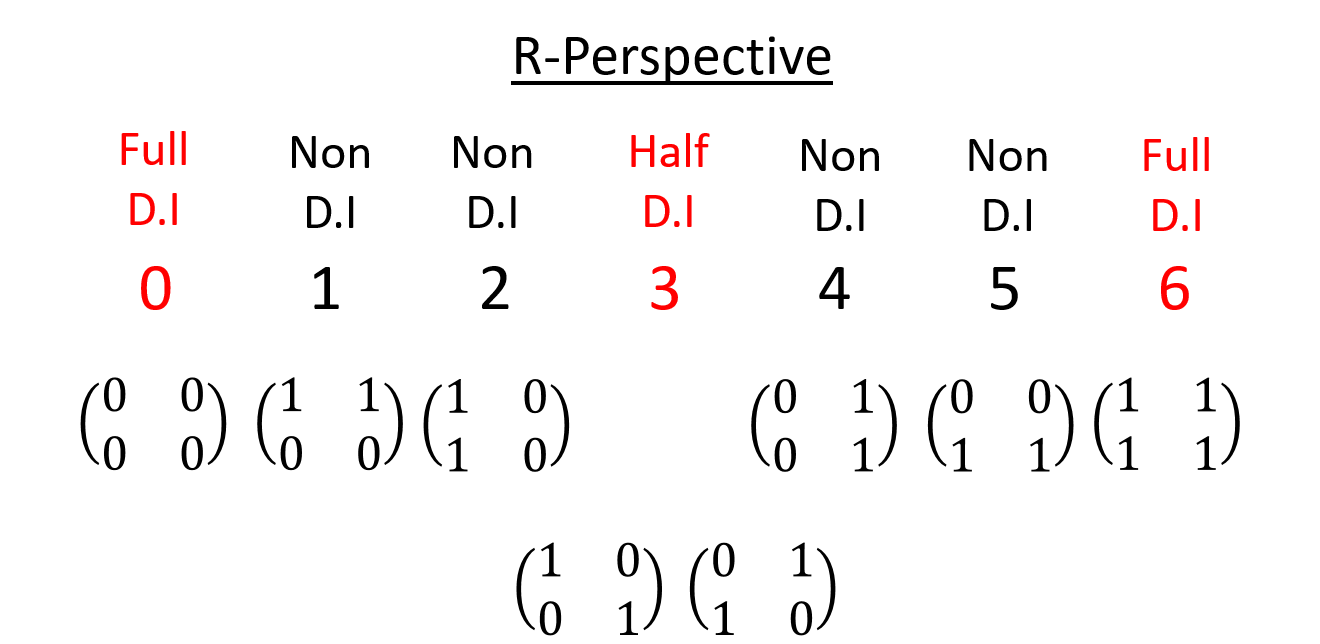}
\caption[]
{Representation of numbers 0 to 6 by duality matrices. 0, 3,6 are dual invariant matrices in $R$ and $L$ operation perspective. Others are non-dual invariant under $R$ and $L$ operation perspective.} \label{fig:dualx5}
\end{figure}

Finally, we would like to express the $R$ and $L$ perspective graphically. According to \ref{eq:eqa}, if we arrange $0-4$ to be the corners of a square, we would like to work out the path order of $L$-perspective and $R$- perspective. For the $R$-perspective, let's start from the order of 0123, then it is a square with one hole with zero node. For $L$-perspective, it is 2031, which is a twisted square with two holes and one node. Hence, a square representation in $R$-perspective is a twisted square representation in $L$-perspective. On the other way, starting from the $L$-perspective, if we choose the order of 0123, then it is a square; while it would be 1302 for the $R$-perspective, which is a twisted square. Thus we have the following duality equations,
\begin{equation}
\begin{aligned}
\bigg( 
\begin{gathered}
 \includegraphics[trim=0cm 0cm 0cm 0cm, clip, scale=0.5]{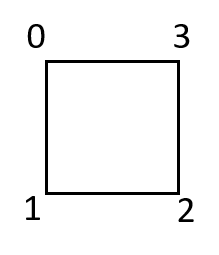} 
\end{gathered}
\bigg| R \bigg)  &\equiv \bigg( \begin{gathered}
 \includegraphics[trim=0cm 0cm 0cm 0cm, clip, scale=0.5]{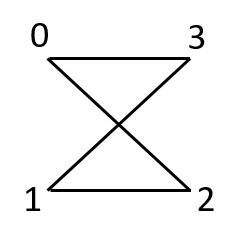} 
\end{gathered} \bigg| L \bigg) \\
\bigg( 
\begin{gathered}
 \includegraphics[trim=0cm 0cm 0cm 0cm, clip, scale=0.5]{topo2.png} 
\end{gathered}
\bigg| R \bigg)  &\equiv \bigg( \begin{gathered}
 \includegraphics[trim=0cm 0cm 0cm 0cm, clip, scale=0.5]{topo1.png} 
\end{gathered} \bigg| L \bigg) \,.
\end{aligned}
\end{equation}
It is easier to visualize in 2D. If we consider 2D, the square corresponds a topological torus, and the twisted square corresponds to a topological surface of genius equal to 2. 
\begin{figure}[H]
\centering
\includegraphics[trim=0cm 3cm 0cm 0cm, clip, scale=0.9]{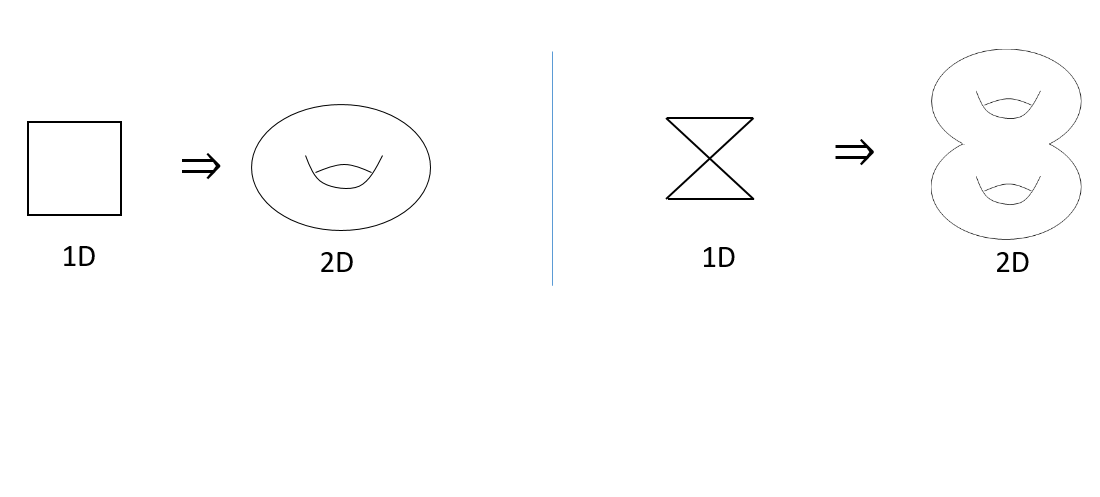}
\caption[]
{} \label{fig:topo5}
\end{figure}
Therefore, a surface of genus 1 is dual to a surface of genus 2. A surface of one hole is dual to the surface to two holes. This can be further entails that, if $u=u^*$, it should be a topological invariant. This has important implication in the application of topological quantum field theory in later sections.

\subsection{Arithmetic Duality}
In this section, we will study dualities in arithmetic operators. First let $x$ and $x^\prime$ be elements, then we define a function $f(x, x^\prime)$. If $f(x , x^\prime) = f(x^\prime , x  )$ then it is symmetric and it is dual invariant. In particle physics term, it is bosonic. If  $f(x , x^\prime) = *f( x , x^\prime   )$ where $*$ is some dual operator with $** = 1$, then it is anti-symmetric, or non-dual invariant. In particle physics term, it is fermonic.

First, we study the addition operator $+$. We have the following,
\begin{equation}
f(x , x^\prime) = x + x^\prime =  x^\prime + x = f( x^\prime ,x ) \,.
\end{equation} 
Thus the addition function is dual invariant (which is same when looking from the left or looking from the right ) , it is symmetric and hence addition corresponds to bosonic operator. Explicitly,
\begin{equation}
 (x + x^\prime |R) \equiv  (x^\prime +x  |L) = (x + x^\prime |L) \equiv (x^\prime +x  |R) \,.
\end{equation}
An implication for this result is that the cosine function is symmetric, and hence bosonic,
\begin{equation}
\cos x = \frac{e^{ix} + e^{-ix}}{2} = \frac{e^{-ix} + e^{ix}}{2}  \,. 
\end{equation}

Next, we study the multiplication operator $\times$, We have the following,
\begin{equation}
f(x , x^\prime) = x \times x^\prime =  x^\prime \times x = f( x^\prime ,x ) \,.
\end{equation} 
Thus the multiplication function is dual invariant,
\begin{equation}
 (x \times x^\prime |R) \equiv  (x^\prime \times x  |L) = (x \times x^\prime |L) \equiv (x^\prime \times x  |R) \,.
\end{equation}
It is also symmetric and bosonic.

Now, consider the subtraction operator $-$. We have the following, 
\begin{equation}
f(x , x^\prime) = x - x^\prime =  -(x^\prime - x) = -f( x^\prime ,x ) \,.
\end{equation} 
Thus the subtraction function is anti-symmetric , it is not dual invariant in the left/right observer perspective, thus subtraction is fermonic. We have $*=-$ and $--=+1$. Notice that 
\begin{equation}
(x-x^\prime |R) = x-x^\prime \quad,\quad (x-x^\prime |L) = x^\prime -x \quad ,\quad (x^\prime - x |R) = x^\prime -x \quad ,\quad (x^\prime - x |L) = x-x^\prime \,.
\end{equation}
Therefore,
\begin{equation}
(x-x^\prime |R) \equiv (x^\prime - x |L) \quad \text{and} \quad (x-x^\prime |L) \equiv (x^\prime - x |R) \,.
\end{equation}
We can express this in terms of the standard $\mathbb{Z}_2 \times \mathbb{Z}_2$ 4-tableau, 
\begin{figure}[H]
\includegraphics[trim=0cm 0cm 0cm 0cm, clip, scale=0.6]{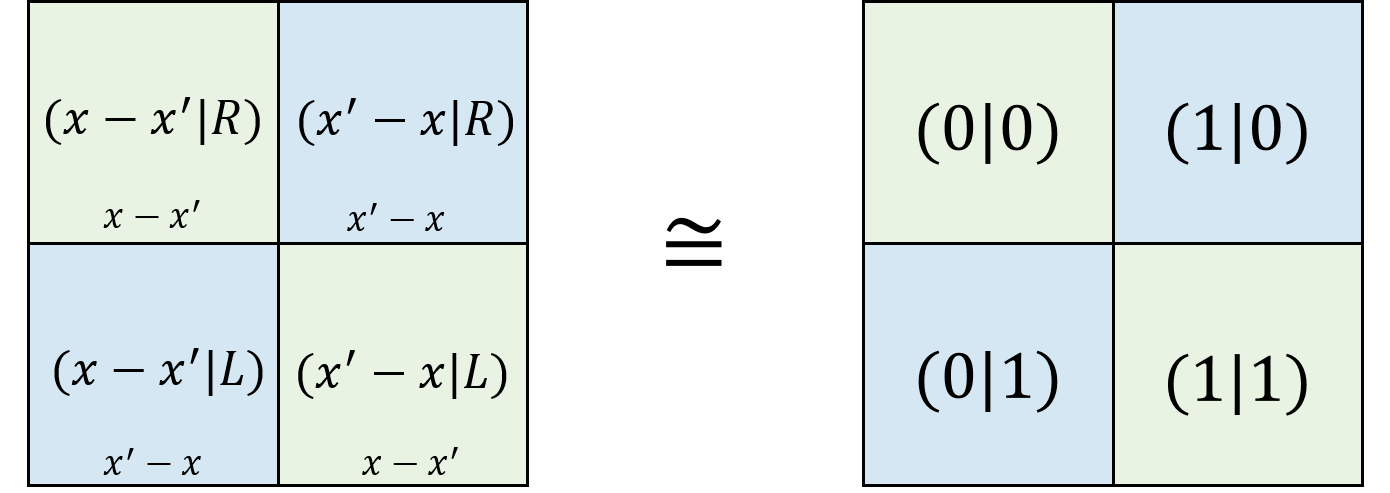}
\caption[]
{} \label{fig:substractdual}
\end{figure}
in which we map $x-x^\prime \rightarrow 0,\,\,x^\prime -x \rightarrow 1$ and $R\rightarrow 0,\,\, L \rightarrow 1$.
An implication for this result is that the sine function is symmetric, and hence fermonic,
\begin{equation}
\sin x = \frac{e^{ix} - e^{-ix}}{2i} = - \frac{e^{-ix} - e^{ix}}{2i}  \,. 
\end{equation}

Finally we consider the division operator $\div$. We have the following,
\begin{equation}
f(x, x^\prime) = x \div x^\prime = (x^\prime \div x)^{-1} =f^{-1} (x^\prime , x)
\end{equation}
Thus the subtraction function is anti-symmetric, it is not dual invariant in the left/right observer perspective, We have $*f = f^{-1}$ and $**f= (f^{-1})^{-1} =f$ thus $**=1$. Notice that 
\begin{equation}
(x\div x^\prime |R) = x\div x^\prime \quad,\quad (x\div x^\prime |L) = x^\prime \div x \quad ,\quad (x^\prime \div x |R) = x^\prime \div x \quad ,\quad (x^\prime \div x |L) = x\div x^\prime \,.
\end{equation}
Therefore,
\begin{equation}
(x\div x^\prime |R) \equiv (x^\prime \div x |L) \quad \text{and} \quad (x\div x^\prime |L) \equiv (x^\prime \div x |R) \,.
\end{equation}

We can express this in terms of the standard $\mathbb{Z}_2 \times \mathbb{Z}_2$ 4-tableau, 
\begin{figure}[H]
\includegraphics[trim=0cm 0cm 0cm 0cm, clip, scale=0.6]{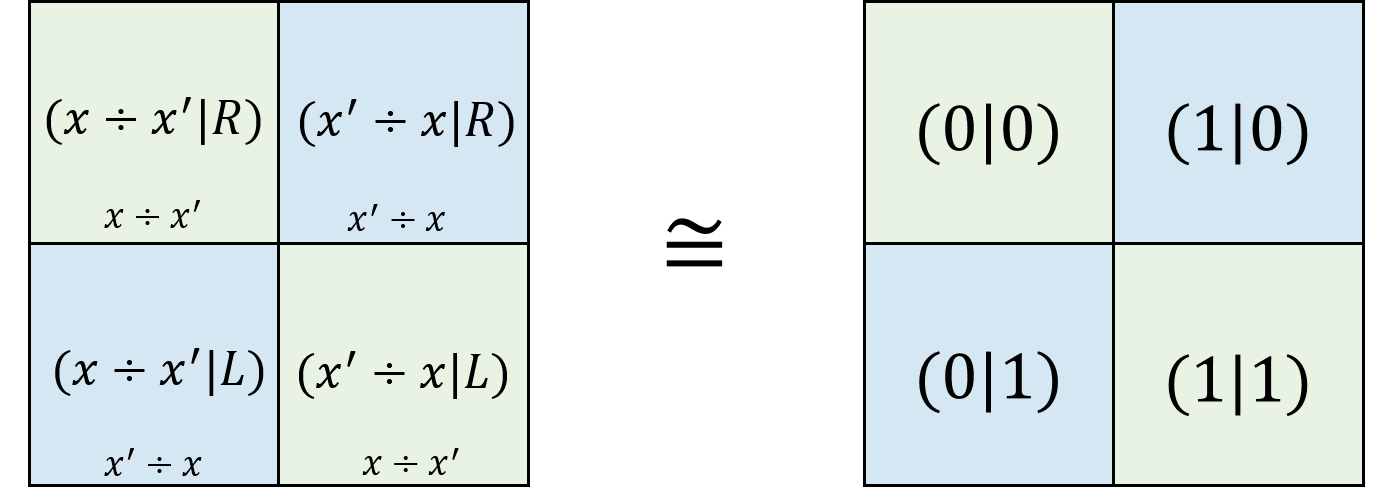}
\caption[]
{} \label{fig:divdual}
\end{figure}
in which we map $x\div x^\prime \rightarrow 0,\,\,x^\prime \div x \rightarrow 1$ and $R\rightarrow 0,\,\, L \rightarrow 1$.

Therefore, all in all, we have both addition and multiplication $+,\times$ as one category-the dual category; and subtraction and division as the other category, the non-dual category. We can define $W = \{ +,\times\}$ and $W^*=\{ - , \div \}$.

Next, we will show that $+$ and $\times$ operators are dual to each other, despite that they are both symmetric and bosonic. This can be shown by the $0,1$ duality we showed in previous section. We consider two function $F(0,1)$ and $F(1,0)$. We then define the perspective as a map such that for some $a,b$,
\begin{equation}
L:F(a,b)\rightarrow a \quad \text{and} \quad R:F(a,b)\rightarrow b \,, 
\end{equation}
where $L:F(a,b) =: (F(a,b)|L)$ and $R:F(a,b) =: (F(a,b)|R)$. We then have the following,
\begin{equation}
(F(0,1)|L) = (F(1,0)|R) = 0 \quad \text{and} \quad (F(0,1)|R) = (F(1,0)|L) = 1\,.
\end{equation}
First we have
\begin{equation}
(F(0,1)|L) = 0 \times 1 =0= 1 \times 0 = (F(1,0)|R) \,,
\end{equation}
then we have
\begin{equation}
(F(0,1)|R) = 0 + 1 =1= 1 + 0 = (F(1,0)|L) \,.
\end{equation}
But by the result in \ref{eq:01dual}, $!0 = 1$ and $!1 =0$ where 0 and 1 are dual to each other, so $(F(0,1)|L)$ and $ (F(1,0)|R)$ corresponds to the multiplication operation $\times$; while  $(F(0,1)|R)$ and $(F(1,0)|L)$ corresponds to the addition operation $+$. In terms of the $\mathbb{Z}\times \mathbb{Z}$ tableau, 
\begin{figure}[H]
\includegraphics[trim=0cm 0cm 0cm 0cm, clip, scale=0.6]{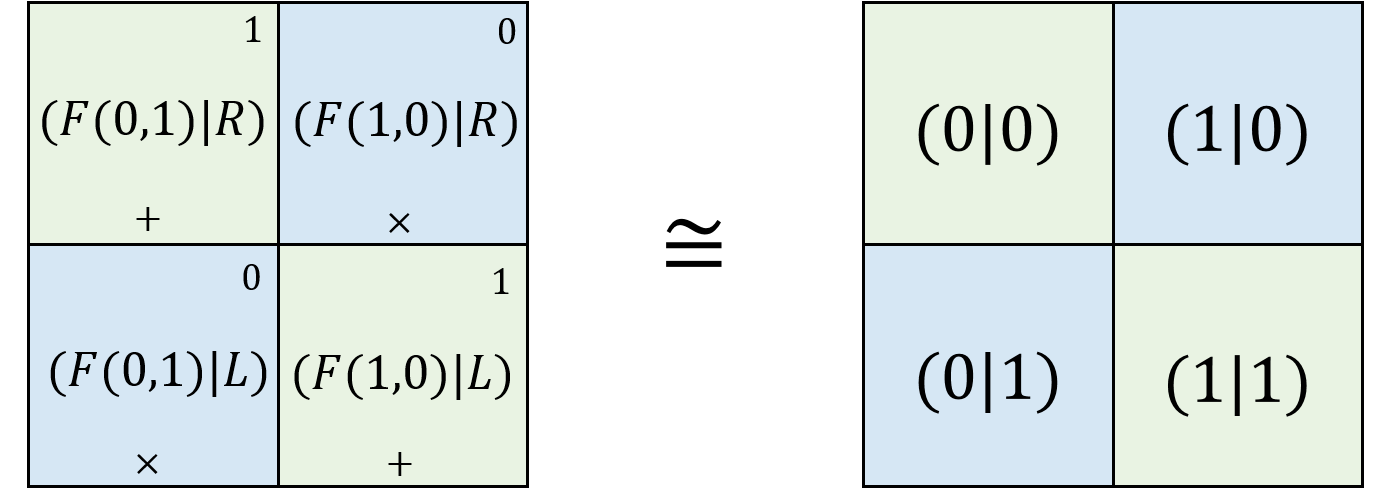}
\caption[]
{} \label{fig:+xdual}
\end{figure}
Therefore $+$ and $\times$ are dual to each other as $0$ and $1$ are dual to each other.

\subsection{Arithmetic Expression Duality and the Cyclic Rule}
Now, we establish the duality on arithmetic expression under left, right observation perspective. We will show that positive numbers are left-right dual invariant, while negative numbers are not. Recalling that expression of $f(x, x^\prime) = x+ x^\prime $, which is bosonic, which implies $f(x, x^\prime) = f( x^\prime ,x)$ . This also means that when we look from the $R$ perspective, this is $ x+ x^\prime$. If we look from the $L$ prespective, this is $ x^\prime + x$. Thus bosonic also means that the expression is right-left perspective dual invariant,
\begin{equation}
(x + x^\prime | R ) \equiv (x + x^\prime | L ) \,.
\end{equation}
Now consider another bosonic expression $f(x, x^\prime) = -x-x^\prime $. This expression is simply $-x-x^\prime$ when looking from the right perspective, but how about looking from the left perspective? We are running into a problem as we have $x^\prime - x -$ which is an operator but not an element. However, as we can obviously see that $f(x, x^\prime) = f(x^\prime ,x) $ for this case, so we expect it is left right dual invariant.

Let's see for the fermionic case. For $f(x , x^\prime) = x-x^\prime$, when we look from the right perspective, it is $x-x^\prime$. If we look from the left perspective, it becomes $x^\prime -x$. This is totally fine. However, if we have $f(x , x^\prime) = -x+x^\prime$, we know that again it satisfies $f(x , x^\prime) = -f(x^\prime ,x) $, which is not dual invariant. But when we look from the right perspective, we get $x^\prime + x -$ which is problematic. To counter this problem we have to rearrange the terms as $-x + x^\prime = x^\prime -x$, such that when we look from the left perspective, it becomes $x-x^\prime$ which is $-(-x + x^\prime)=-f(x,x^\prime)$.
Although the $-+$ case can be solved by rearranging the terms, we see that the $--$ case, $-x-x^\prime$ cannot be solved by rearranging the terms.  This urges us that we demand a more generic rule to define arithmetic expression duality. 

The generic rule is as follow, which we call the cyclic rule. We wind up any arithmetic expressions in a close ring. 
\begin{center}
\begin{figure}[H]
\includegraphics[trim=0cm 0cm 0cm 0cm, clip, scale=0.6]{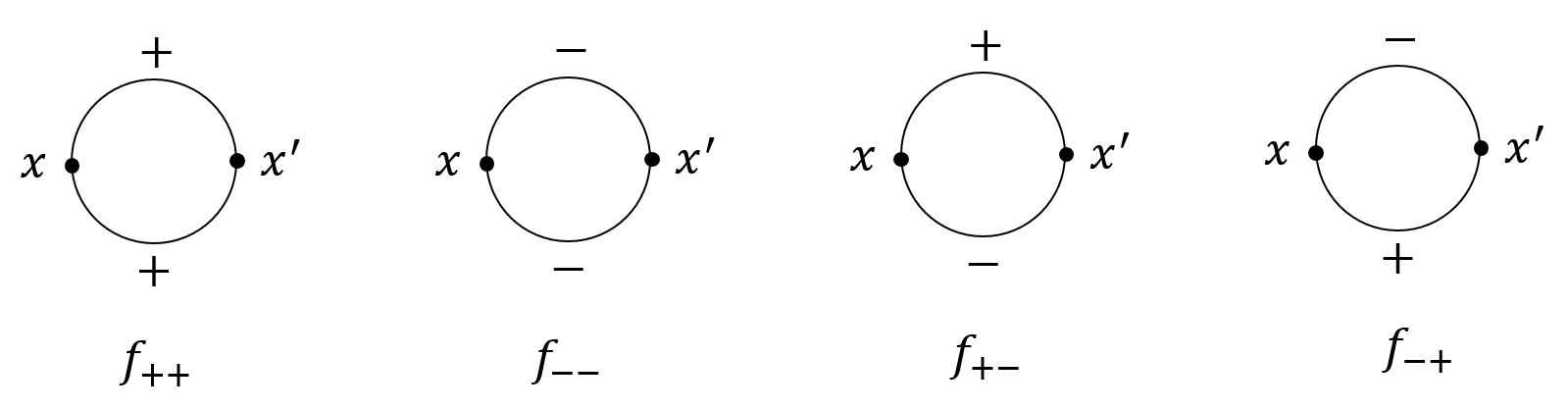}
\caption[]
{We define $f_{++}(x, x^\prime) = + x + x^\prime$, $f_{--}(x, x^\prime) = - x - x^\prime$, $f_{+-}(x, x^\prime) = + x - x^\prime$ and $f_{-+}(x, x^\prime) = - x + x^\prime$.} \label{fig:cyclic}
\end{figure}
\end{center}
To read these ring diagrams we have the following rules:
\begin{enumerate}
\item The order goes with operator(sign)-element-operator(sign)-element.
\item The value obtained along a particular cyclic direction must be the same, no matter  the starting point.
\item The anti-clockwise direction denotes the one of the perspective, while the clockwise direction denotes the dual perspective. 
\end{enumerate}
Using the $f_{--}$ as the example, using the first point , we begin with the $-$ sign on the top, going towards the anti-clockwise direction we have $-x-x^\prime$ for one$(R)$ perspective. For the dual$(L)$ perspective, start from the $-$ sign on the top, going with the clockwise direction, we have $-x^\prime -x$ for one$(L)$ perspective. The similar idea holds all for $f_{++}, f_{+-}$ and $f_{-+}$ cases. In terms of the 4-tableau, we have
\begin{figure}[H]
\begin{center}
\includegraphics[trim=0cm 0cm 0cm 0cm, clip, scale=0.6]{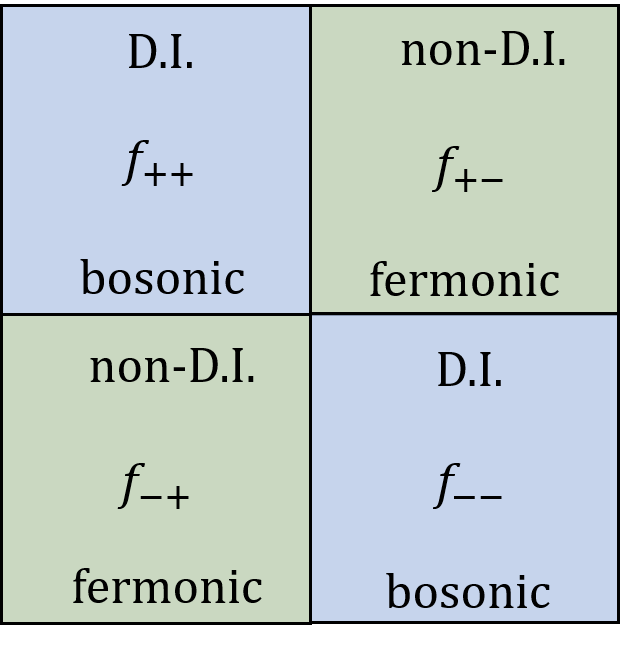}
\caption[]
{In this diagram, D.I. stands for left-right perspective dual invariant.} \label{fig:cyclic2}
\end{center}
\end{figure}
From the $f_{++}$ case, if we pick $x=0$, then we have $(0 + x^\prime |R) \equiv ( + x^\prime + 0 |L) \implies (+x^\prime |R) \equiv (+x^\prime |L) $. Therefore positive numbers are dual invariant. However, for the $f_{+-}$ case and the $f_{-+}$ case, as $(f_{+-}|R)\equiv (f_{-+}|L)$, i.e. $(x-x^\prime | R) \equiv (x^\prime -x |L)$, if we pick $x =0$, this gives $(-x^\prime |R) \equiv (x^\prime |L)$, so negative numbers are not dual invariant.

\subsection{Cyclic rule on tensor product states}
In this section, we will investigate the power of cyclic rule on tensor product state. First notice the following identity:
\begin{equation}
\begin{aligned}
|0\rangle \otimes \langle 0 | &= \langle 0 | \otimes |0\rangle = \hat{P}_+ \\
|1\rangle \otimes \langle 1 | &= \langle 1 | \otimes |1\rangle = \hat{P}_- \,. 
\end{aligned}
\end{equation}
On the other hand,
\begin{equation}
\begin{aligned}
|0\rangle \otimes \langle 1 | &=  \langle 1 | \otimes |0\rangle = \hat{P}^*_+ \\
|1\rangle \otimes \langle 0 | &=  \langle 0 | \otimes |1\rangle = \hat{P}^*_- \,.
\end{aligned}
\end{equation}
We can illustrate such results by the cyclic rule.
\begin{figure}[H]
\begin{center}
\includegraphics[trim=0cm 2cm 0cm 0cm, clip, scale=0.55]{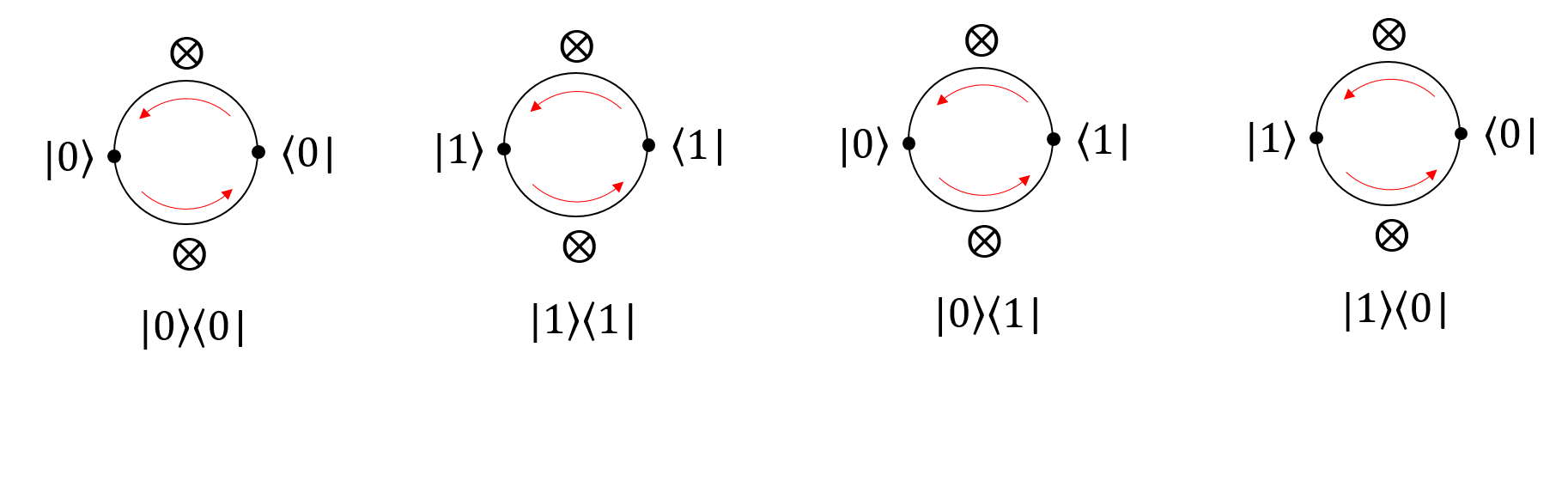}
\caption[]
{Tensor product states with the anticlockwise perspective} \label{fig:cyclic3}
\end{center}
\end{figure}
For the first diagram $|0\rangle \langle 0 |$, following the lower half anti-clockwise direction, we have $|0\rangle \otimes \langle 0|$. Following the upper anti-clockwise direction, we have $\langle 0| \otimes |0\rangle$. Since along the same direction, the expression must be the same, thus by the cyclic rule we must have $ |0\rangle \otimes \langle 0| = \langle 0| \otimes |0\rangle $. The same idea goes with the second digram for the $|1\rangle \langle 1|$ case. For the third diagram, following the lower half anti-clockwise direction, we have $|0\rangle \otimes \langle 1|$. Following the upper anti-clockwise direction, we have $\langle 1| \otimes |0\rangle$. Since along the same direction, the expression must be the same, thus by the cyclic rule we must have $ |0\rangle \otimes \langle 1| = \langle 1| \otimes |0\rangle $. The same idea goes with the fourth digram for the $|1\rangle \langle 0|$ case.

Now consider the dual perspective, which is the clockwise operation.
\begin{figure}[H]
\begin{center}
\includegraphics[trim=0cm 2cm 0cm 0cm, clip, scale=0.55]{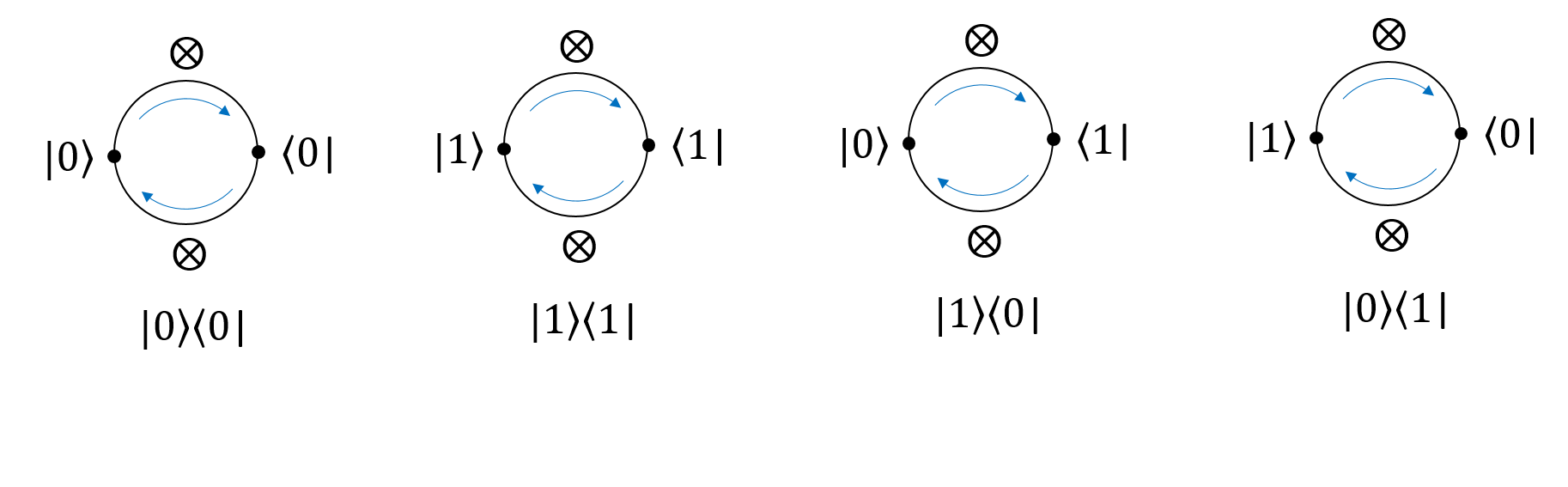}
\caption[]
{Tensor product states with the dual clockwise perspective} \label{fig:cyclic4}
\end{center}
\end{figure}
For the first two diagrams, they remain the same. Therefore, as expected, $|0\rangle \langle 0 |$ and $|1\rangle \langle 1 |$ states are dual invariant. We have 
\begin{equation}
\begin{aligned}
(|0\rangle \otimes \langle 0 | | \mathrm{anticlockwise}) &\equiv (|0\rangle \otimes \langle 0 | | \mathrm{clockwise}) \\
(|1\rangle \otimes \langle 1 ||  \mathrm{anticlockwise}) &\equiv (|1\rangle \otimes \langle 1 | | \mathrm{clockwise} ) \,,
\end{aligned}
\end{equation}
while the $|0\rangle \langle 1 | $ case and $|1\rangle \langle 0 | $ case are non-dual invariant.

\subsection{Element-Operator duality}
In the final part, we study element and operator duality. Let $a$ be an element (a number), $+$ be an operator. We have $+a =a$ is an element, but $a+$ is an operator. Define also the $L$ and $R$ perspective, then we have
\begin{equation}
(+a|R) =\text{element}\,\,,\,\,(+a|L) =\text{operator} \,,\,\,,(a+|R) = \text{operator}\,\,\,,\,\,(a+|L) = \text{element}\,.
\end{equation}
Therefore we have
\begin{equation}
(+a|R) \equiv (a+|L) \quad \text{and} \quad (a+|R) \equiv (+a|L) \,. 
\end{equation}
In terms of $\mathbb{Z}_2 \times \mathbb{Z}_2$ 4-tableau, we have
\begin{figure}[H]
\includegraphics[trim=0cm 0cm 0cm 0cm, clip, scale=0.6]{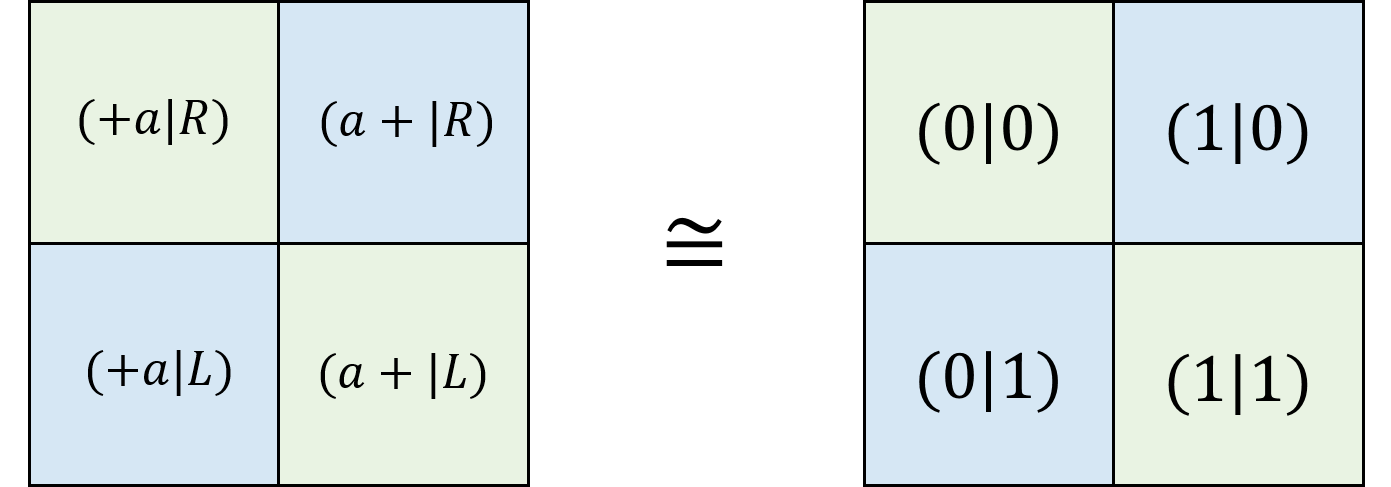}
\caption[]
{} \label{fig:eodual}
\end{figure}

Next, we would study the fundamental element-operator duality in multiplication. Consider the  multiplication of two elements, $A_1 \cdot A_2 $. Now consider inserting an identity element in the middle, $A_1 (1) A_2 = A_1 A_2$. Explicitly, there are two ways to for insertion and two forms to insert. Let $1$ be the element and the multiplication $\cdot$ be the operator. Define the element operator (EO) form  as $1\cdot$, and the operator element form (OE) form as $\cdot 1$. We immediately have
\begin{equation}
(1\cdot |R) \equiv (\cdot 1 | L) \quad\text{and}\quad (1\cdot |L) \equiv (\cdot 1 | R) \,.
\end{equation}
Therefore $1\cdot$ and $\cdot 1$ are dual to each other,
\begin{equation}
1\cdot = *(\cdot 1) \quad\text{and} \quad \cdot 1 = * (1 \cdot) \,.
\end{equation}
Or equivalently EO form and OE form are dual to each other, $\mathrm{EO} = *\mathrm{OE}$ and $\mathrm{OE} = *\mathrm{EO}$. 
The two ways for insertion are
\begin{equation}
A_1 \,_\wedge \,\,\,\,_\wedge \cdot A_2 \quad\text{and}\quad A_1  \cdot \,_\wedge \,\,\,\,_\wedge A_2 \,\,\,,
\end{equation}
where the first one is the left insertion denoted by $l$ perspective and the second one is the right insertion by $r$ perspective. To restore the product of $A_1 A_2$ , we need to have
\begin{equation}
A_1 \,_\wedge (\cdot 1)\,_\wedge \cdot A_2 \quad\text{and}\quad A_1  \cdot \,_\wedge ( 1\cdot)\,_\wedge A_2 \,\,\,,
\end{equation}
so both of them looks like $A_1 \cdot 1 \cdot A_2$, we notice that the OEO form $\cdot 1 \cdot$ is dual invariant with respect to the right observer and left observer, i.e. $(\cdot 1 \cdot | L)\equiv (\cdot 1 \cdot | R) $. The OEO form here is referred as the connector.  Therefore we have
\begin{equation}
(\cdot 1 | l) \equiv (1\cdot | r ) \,,
\end{equation}
or
\begin{equation}
(\mathrm{OE} | l) \equiv (\mathrm{EO} | r )\,.
\end{equation}
Next, we have another form of insertion, 
\begin{equation}
A_1 \,_\wedge ( 1\cdot)\,_\wedge \cdot A_2 \quad\text{and}\quad A_1  \cdot \,_\wedge ( \cdot1)\,_\wedge A_2 \,\,\,,
\end{equation}
This gives,
\begin{equation}
A_1 \,\,\, 1\cdot\cdot A_2 \quad\text{and}\quad A_1 \cdot\cdot 1\,\,\, A_2 \,.
\end{equation}
The EOO form $1\cdot\cdot$ and OOE form $\cdot\cdot 1$ is not dual invariant, in particular we see that
\begin{equation}
(1\cdot\cdot |R) \equiv (\cdot\cdot 1 |L) \quad \text{and} \quad (1\cdot\cdot |L) \equiv (\cdot\cdot 1 |R) \,.
\end{equation}
Such EOO form and OOE form are referred as the disconnector, they will break the product into two separate elements $A_1 , A_2$ and does not return to a value. In terms of $\mathbb{Z}_2 \times \mathbb{Z}_2$ tableau, we have the following diagram representation,
\begin{figure}[H]
\includegraphics[trim=0cm 0cm 0cm 0cm, clip, scale=0.6]{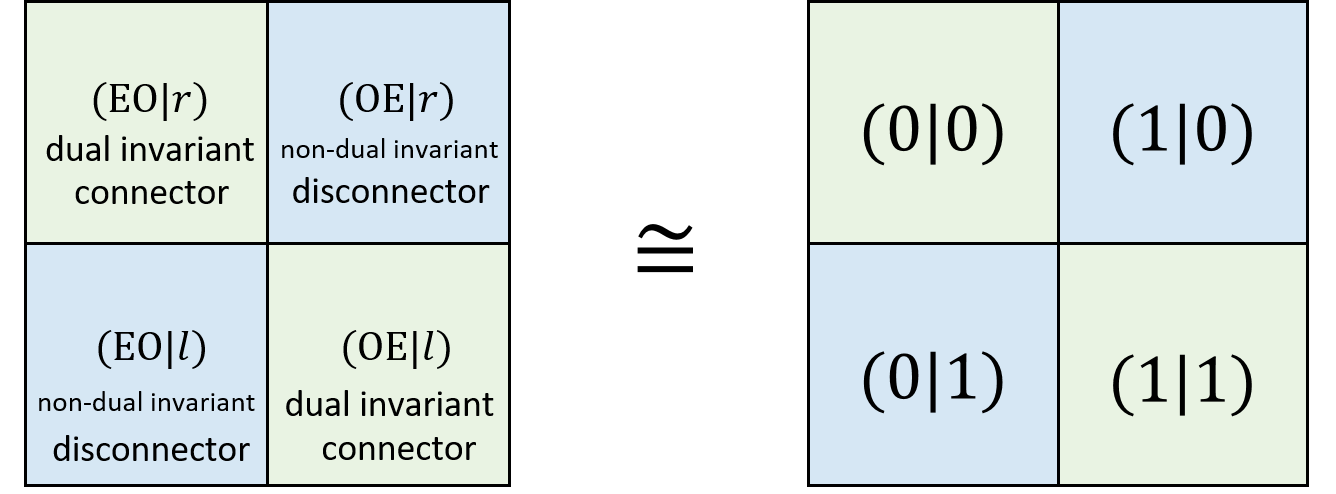}
\caption[]
{} \label{fig:eooe}
\end{figure}

\subsection{Multi-duality Structure}
The theory of a single duality system can be generalized into multi-duality system, in which more than one complete duality structure is concerned. 
\begin{definition}
Let the $i^{\rm{th}}$ unit of complete duality be $D_i =\{W_i , B_i\}$. The full set of total duality is defined as $D_{tot}$ with $|D_{tot}|=n$, which is the union of all $U_i$. Each $D_i$ is a partition set of $D_{tot}$. Mathematically,
\begin{equation}
D_{tot} = \bigcup_{i=1}^n D_i \,\,\,\, {\mathrm{and}} \,\,\,\,W_{tot} = \bigcup_{i=1}^n W_i = \bigcup_{i=1}^n U_i \cup U_i^{*} \cup \{0\}_i \,\, , \,\, B_{tot} = \bigcup_{i=1}^n B_i = \bigcup_{i=1}^n S_{k,i} \cup S_{k,i}^{\star}\,. 
\end{equation}

\begin{equation}
\bigcap_{i=1}^n D_i = \emptyset\,\,\,\,{\mathrm{and}}\,\,\,\,  \bigcap_{i=1}^n U_i \cup U_i^*= \emptyset\,\, , \,\, \bigcap_{i=1}^n S_{k,i} \cup S_{k,i}^{\star} = \emptyset\,. 
\end{equation}
\end{definition}

Next we would like to study the algebra of the multi-duality. We promote the definition of dual set to dual space. (Remark the dual space here is not referring to that in differential Geometry). The dual spaces are related by tensor product.

\begin{definition}\end{definition}
(I) Let $\mathcal{V}_i$ be a $i^{\mathrm{th}}$ of two dimensional space which can be partitioned into two one dimensional sub-space $V_i$ and its dual $V^*_i$ such that $\mathcal{V}_i= V_i \oplus V_i^*$. The multi-dual spaces satisfy the following algebra. Define each partition as
\begin{equation}
\bigotimes^{p,q}_{i,j} v_{i,j} := \bigotimes_{i}^p V_i \otimes \bigotimes_{j}^q V_j^{*}\,.
\end{equation}

(II) Let $\mathcal{W}$ be the full multi-duality space. The multi-duality system if multiplicity $n$ possess the natural decomposition algebra,
\begin{equation} \label{eq:DualityAlgebraFirstClassElement}
\mathcal{W}=\bigotimes_{i=1}^n \mathcal{V}_i = \bigoplus_{p+q = n} \bigotimes^{p,q}_{i,j} v_{i,j}\,.
\end{equation}

(III) Let $\mathcal{S}_{k,i}$ be the $i^{\rm th}$ observer's space corresponding to $\mathcal{V}_i$ which can be partitioned into $\mathcal{S}_{k,\,\,i}$ and its dual $\mathcal{S}_{k,\,\,i}^{\star}$. The multi-dual observer's space satisfies
\begin{equation}
\bigotimes^{p,q}_{i,j} s_{k,\,\,i,j} := \bigotimes_{i}^p S_{k\,\,i} \otimes \bigotimes_{j}^q S_{k\,\,j}^{\star}\,.
\end{equation}
and
\begin{equation} \label{eq:DualityAlgebraFirstClassObserver}
\mathcal{B}=\bigotimes_{i=1}^n \mathcal{S}_{k,\,\,i} = \bigoplus_{p+q = n} \bigotimes^{p,q}_{i,j} s_{k,\,\,i,j}\,.
\end{equation}

(IV) The algebra of the element space and the algebra of the observer space is an isomorphism $\bigotimes_{i=1}^n \mathcal{V}_i  \cong \bigotimes_{i=1}^n \mathcal{S}_{k,\,\,i}$ for each $i$. There exists a bijective map for the two algebras.  

(V) The complete dual space of multi-duality $\mathcal{C}$ is a natural duality if $\mathcal{D}=\mathcal{W}\oplus \mathcal{B}$, where $\mathcal{D}$ induces a new observer's space $\mathcal{O}_l$ and its dual $\mathcal{O}_l^{\star}$ in dimension $l$ such that 
\begin{equation} \label{eq:InducedDualityFirstClass}
( \mathcal{W} |\mathcal{O}_l ) \equiv (\mathcal{B} |\mathcal{O}_l^{\star}) \,\,\,\,{\mathrm{and}}\,\,\,\, (\mathcal{W} | \mathcal{O}_l^{\star}) \equiv (\mathcal{B}| \mathcal{O}_l )
\end{equation} 
where
\begin{equation}
\mathcal{B}=\mathcal{W}^{*}\,\,\,\,{\mathrm{and}}\,\,\,\,\mathcal{W}=\mathcal{B}^{*}\,.
\end{equation}
The role of element space and observer's space is interchangeable under such duality system.

(VI) Let the dual map for element space $*$ such that $*_{V_i} : V_i \rightarrow V_i^*$ and $*_{V_i^*} : V_i^* \rightarrow V_i$ where $**$ is the identity map $I_d (V_i)$. Similarly define dual map for observer space $\star$ such that ${\star}_{S_{k,\,\,i}} : S_{k\,\,i}  \rightarrow S_{k,\,\,i}^{\star}$ and ${\star}_{S_{k,\,\,i}^{\star}} : S_{k\,\,i}^{\star}  \rightarrow S_{k,\,\,i}$ where $\star \star$ is the identity map $I_d (S_{k,i})$. The maps $*_{V_i} \circ \star_{S_{k,\,\,i}} = I_d (V_i, S_{k,\,\,i})$ and $\star_{S_{k,\,\,i}} \circ *_{V_i}  = I_d (S_{k,\,\,i}, V_i)$ are identity maps. For each partition, we define the collective duality map as the following:
\begin{equation}
*_{p,q} = \prod_i^p *_{V_i} \prod_j^q *_{V_j^*}\,\,\,\,{\mathrm{and}}\,\,\,\,
\star_{p,q} = \prod_i^p \star_{S_{k,\,\,i}} \prod_{j}^q \star_{S_{k,\,\,j}^{\star}}\,,  
\end{equation}
where the product sign here for notation simplicity denotes the operation of composite maps. 

(VII) With all these maps defined we have the following theorems.
 \label{DualityMapsFirstClass}
The arbitrary number of $r \leq p$ $*_{V_i}$ maps and the arbitrary number of $s \leq q$ $*_{V_{j}^*}$ maps acting on any partition must return to any other partition.
\begin{equation} \label{DualtiyOperationsParition}
*_{r,s}\bigotimes_{i,j}^{p,q} v_{i,j} = \bigotimes_{i,j}^{p-r+s, q+r-s} v_{i,j}
\end{equation} 
for $p-r+s = p^{\prime}$ and $q+r-s = q^{\prime}$. The same theorem holds for observer spaces and its dual operators.

(VIII) \label{IdentityMapsFirstClass}
The identity map $*_{r,s} \circ \star_{r,s} = \star_{r,s} \circ *_{r,s} = I_d$ acts on the partition as
\begin{equation}
*_{r,s} \star_{r,s} \left( \bigotimes_{i,j}^{p,q} v_{i,j} \bigg\vert \bigotimes_{i,j}^{p,q} s_{k,\,\,i,j} \right) = \left(\bigotimes_{i,j}^{p-r+s, q+r-s} v_{i,j} \bigg\vert \bigotimes_{i,j}^{p-r+s, q+r-s} s_{k,\,\,i,j} \right)= \left( \bigotimes_{i,j}^{p,q} v_{i,j} \bigg\vert \bigotimes_{i,j}^{p,q} s_{k,\,\,i,j} \right)\,.
\end{equation}

The duality operators can be viewed as discrete parity symmetry and they form a parity group. Define the parity group for elements as $\rho_i(V_i) = \{I_d(V_i),\,*_{V_i}\}$ and for observer as $\rho_i(S_{k,\,\,i}) = \{I_d(S_{k,\,\,i}), \star_{S_{k,\,\,i}} \}$. These parity groups are isomorphic to the group $\mathbb{Z}_2$. The multi-duality of the first class is the study of tensor product representations of the parity groups of elements and observers. The $\mathcal{V}_i$ and $\mathcal{S}_{k,\,\,i}$ spaces are representation vector spaces. Equations \ref{eq:DualityAlgebraFirstClassElement} and \ref{eq:DualityAlgebraFirstClassObserver} show the reducible representation tensor product vector spaces as the direct sum of each irreducible representations. 

The irreducible tensor product vector spaces can recombine to form duality systems. One grouping criteria is according to binomial coefficients. It can be seen that the number of ways of the $p$ and $q$ indices combine follow the binomial distribution. The total number of irreducible representation vector spaces is,
\begin{equation}
\sum_{i=0}^{n} {\mathrm{C}}_{i}^n =\sum_{i=0}^{n}\frac{n!}{i!(n-i)!} = 2^n\,.
\end{equation}
If viewing the binomial coefficients as the pascal triangle, one sees that ${\mathrm{C}}_{i}^{n}$ and ${\mathrm{C}}_{n-i}^n$ is symmetric. Thus the irreducible representation spaces can be naturally group as $2^{n-1}$ sub-duality systems, and they can be further re-group into one large  duality system. The one large duality system is as follow,
\begin{equation}
\mathcal{W}=\mathcal{V}_1 \otimes \cdots \otimes \mathcal{V}_n = \mathcal{A} \oplus \mathcal{A^*} = \bigoplus_{l=1}^{2^{n-1}} (\mathcal{K}_l \oplus \mathcal{K}_l^{*})
\end{equation}
where
\begin{equation}
\mathcal{A}= \bigoplus_{p+q=n, \,\,p \geq q} \bigotimes_{i,j}^{p,q} v_{i,j}\,\,\,\,{\mathrm{and}}\,\,\,\, \mathcal{A}^* = * \mathcal{A} = \bigoplus_{p+q=n, \,\, p \leq q} \bigotimes_{i,j}^{p,q} v_{i,j}\,,
\end{equation}
and this is the one large duality system. For the sub-$2^{n-1}$ duality systems,
\begin{equation}
\mathcal{K}_l =\bigoplus_{i,j}^{p,q} (v_{i,j})_l\,\,\,\,{\mathrm{and}}\,\,\,\,\mathcal{K}_l^* = *_{p,q} \mathcal{K}_l =\bigoplus_{i,j}^{q,p} (v_{i,j})_l\,.
\end{equation}
Any partition with its full dual forms the $K$ space. Alternatively, the $K_l$ dual spaces can be explicitly defined through the pair-wise dual operations $*_{V_j} *_{V_j^*}$. The $K_{l^{\prime}} \oplus K_{l^{\prime}}^*$ dual space serves as the role of generating other $K$-dual spaces under the pair-wise dual operators.

(IX) Each $K_l \oplus K_l^*$ dual space is a generator space, which can generate other $K_{l^{\prime}} \oplus K_{l^{\prime}}^*$ dual spaces under the pair-wise dual operators  $*_{V_j} *_{V_j^*}$. Mathematically,

\begin{equation} \label{eq:DualityKspacePairwiseOperator}
*_{V_j} *_{V_j^*} (K_{l} \oplus K_{l}^*) = (K_{l^{\prime}} \oplus K_{l^{\prime}}^*)\,. 
\end{equation}
If we apply fully the $*_{V_j} *_{V_j^*}$ operators for all $j$ on a $(K_{l} \oplus K_{l}^*)$, then we obtain all other $(K_{l^{\prime}} \oplus K_{l^{\prime}}^*)$ except  itself.
\begin{equation} \label{eq:DualitySumKspacePairwiseOperator}
\sum_{j=1}^n *_{V_j} *_{V_j^*} (K_{l} \oplus K_{l}^*) = \bigoplus_{l^{\prime} \neq l}(K_{l^{\prime}} \oplus K_{l^{\prime}}^*)\,.
\end{equation}

(X) The $\mathcal{W}$ space is the representation space of the multi-duality group $\mathbb{Z}_2 \otimes \cdots \otimes \mathbb{Z}_2$. The map $\rho$  
\begin{equation}
\rho : \mathbb{Z}_2 \otimes \mathbb{Z}_2 \otimes \cdots \otimes \mathbb{Z}_2 \rightarrow \mathcal{V} \otimes \mathcal{V} \otimes \cdots \otimes \mathcal{V} \,.
\end{equation}

Next we would like to study some conserved dual operations under some circumstances of invariance in duality system. 
\subsection{Duality Transformation and Duality Symmetry}
Since a complete duality system bases on both the element space of observer's space, when we consider dual action we have the following circumstances. (1) The element space is transformed while keeping the observer space constant. (2) The observer space is transformed while keeping the element space constant. (3) Both the element space and observer space transform. First we consider the (1) case.
\subsubsection{Local duality transformation}
\begin{definition}
If $\mathcal{W}=\mathcal{V}_1 \otimes \cdots \otimes \mathcal{V}_n$ is an invariant, then any dual operations on a particular partition or re-partitions in $\mathcal{W}$  must induce a simultaneous same dual operation on that original partition, such that $\mathcal{W}$ remains unchanged.  
\end{definition}
The $\mathcal{W}$ has $2^n$ distinct partitions. Suppose $P_1 = \bigotimes_{i,j}^{p_1,q_1} v_{i,j}$ where $p_1+q_1=n$ is one of the partitions. Now we act dual operators $*_{r_1 , s_1}$ where $r_1 \leq p_1$ and $s_1 \leq q_1$ on $P_1$. Then the original $P_1$ will become another partition $P_2^{\prime}$ in $\mathcal{W}$,
\begin{equation}
*_{r_1 , s_1} P_1 = P_2^{\prime}\,.
\end{equation}  
The prime on $P_2$ denotes that $P_2$ is transformed from $P_1$. However, if $\mathcal{W}$ is an invariant under any dual operations, then $P_1 \rightarrow P_2^{\prime} = *_{r_1 , s_1} P_1$ breaks the invariant as $P_1 \notin \mathcal{W}$, and now we have two $P_2$s, one from the original one in $\mathcal{W}$ and the new one $P_2^{\prime}$ from $P_1$. To preserve $\mathcal{W}$ we have to transform the original $P_2$ to $P_1^{\prime}$, $*_{r_2, s_2}^{-1}P_2 = P_1^{\prime}$. But since the inverse is just the same as itself in the parity group, thus in fact $*_{r_2, s_2}^{-1} = *_{r_1, s_1}$. Therefore we just demand $*_{r_1, s_1}P_2 = P_1^{\prime}$ using the same dual operator. Hence, before dual transformation, we have $P_1$ and $P_2$ $\in \mathcal{W}$, after local dual transformation, we have $P_2^\prime$ and $P_1^\prime$ $\in \mathcal{W}$ such that $\mathcal{W}$ remains unchanged. We call such dual transformation local as it just operates on a particular partition. 
One important note is the instantaneous induction on $P_2$ transforming back to $P_1^{\prime}$. The two operations must have to be synchronized, as the invariance of $\mathcal{W}$ must be conserved at any time. This will have essential physics interpretation later.

The definition for the $K_l$ space in \ref{eq:DualityKspacePairwiseOperator} is also naturally a local duality transformation. The above concept applies similarly.

\subsubsection{Global duality transformation}
\begin{definition}
Global duality transformation is a dual transformation of all partitions  and all elements in the partition in $\mathcal{W}$. The full transformation is simply denoted as $*$. 
\end{definition}

\subsubsection{The Dual Symmetry}
We define a system to have dual symmetry, or called $D$-symmetry if the system is invariant under the transformation of the dual operator.
\begin{definition}
Let $U$ be some element space which $U \subseteq W$. If $U$ can be partitioned into one space and is dual space, $U=X \oplus X^{*}$, then $U$ possesses dual symmetry such that $*U = U$. 
\end{definition}
Since $\mathcal{W} = \mathcal{A} \oplus \mathcal{A}^*$, it is trivial to see that $* \mathcal{W} = \mathcal{W}$, the full element space is global dual symmetry invariant. The partition spaces $\mathcal{K}_l \oplus \mathcal{K}_l^{*}$ for each $l$ is also a dual symmetry invariant.

Next, we would like to show that the full $\mathcal{W}$ space is invariant under the sub-dual symmetry of full dual operations $*_{V_j} \oplus *_{V_j^*}$, i.e.,
\begin{equation}
(*_{V_j} \oplus *_{V_j^*}) W = W\,.
\end{equation}
This is equivalent to say, the $*_{V_i} \oplus *_{V_i^*}$ acting on all partitions remain the same, which is an identity map $I_d$. We also need
\begin{equation}
*_{V_i} V_i^* = 0\,\,\,\,{\text{and}}\,\,\,\, *_{V_i^*} V_i = 0\,.
\end{equation}

The proof is straight forward by going the opposite way,
\begin{equation}
\begin{aligned}
(*_{V_j} \oplus *_{V_j^*} )\mathcal{W} &= (*_{V_j} \oplus *_{V_j^*} )\bigoplus_{p+q=n}\bigotimes_{s,t}^{p,q} v_{s,t}\\
 & = (*_{V_j} \oplus *_{V_j^*}) \bigotimes_{i=1}^n (V_i \oplus V_i^* ) \\
 & = (*_{V_j} \oplus *_{V_j^*}) \big( (V_1 \oplus V_1^*) \otimes \cdots \otimes (V_j \oplus V_j^*) \otimes \cdots \otimes (V_n \oplus V_n^*) \big)\\
 & =  (*_{V_j} \oplus *_{V_j^*})  \big( (V_1 \oplus V_1^*) \otimes \cdots \otimes V_j  \otimes \cdots \otimes (V_n \oplus V_n^*) \big)\\
 & \,\,\,\,\,\, \oplus  (*_{V_j} \oplus *_{V_j^*})  \big( (V_1 \oplus V_1^*) \otimes \cdots \otimes V_j^*  \otimes \cdots \otimes (V_n \oplus V_n^*) \big)\\
 & =  \big( (V_1 \oplus V_1^*) \otimes \cdots \otimes V_j^*  \otimes \cdots \otimes (V_n \oplus V_n^*) \big)\\
 & \,\,\,\,\,\, \oplus  \big( (V_1 \oplus V_1^*) \otimes \cdots \otimes V_j  \otimes \cdots \otimes (V_n \oplus V_n^*) \big)\\
 & = (V_1 \oplus V_1^*) \otimes \cdots \otimes (V_j^* \oplus V_j) \otimes \cdots \otimes (V_n \oplus V_n^*) \\
&= \mathcal{W} \,. \\
\end{aligned}
\end{equation}

All of the above theorems apply to the observer space, since the element space and the observer space themselves are a duality system. By definition, the two spaces are isomorphic. Thus all theorems for one space apply to the other. Therefore the (2) case would be the same as the (1) case, but just a change of notations.

\subsubsection{Local duality transformation}
\begin{definition}
If $(\mathcal{W}|\mathcal{B})$ is an invariant, then any dual operations on a particular partition or re-partitions in $(\mathcal{W}|\mathcal{B})$ must be invariant. If a partition in the element space is transformed by $*_{r,s}$, the corresponding observer space is transformed by $\star_{r,s}$, such that the overall change $*_{r,s}\circ \star_{r,s} = I_d$ is an identity, vice versa.
\end{definition}
This is just the consequence of \ref{IdentityMapsFirstClass}. One may think of whether we need to do the same transformation for the original partition back to a new one just like the case (1). The answer is no, because the change of observer's space at the same has compensated the issue. Let $\mathcal{P_1}$ be the corresponding partition for the element space $P_1$, together as $(P_1|\mathcal{P}_1)$. For case (1) we are doing $*_{r_1, s_1}(P_1|\mathcal{P}_1) = (P_2^{\prime}|\mathcal{P}_1)$, holding $\mathcal{P}_1$ constant. But in this case $*_{r_1, s_1}\star_{r_1, s_1}(P_1|\mathcal{P}_1) = (P_2^{\prime}|\mathcal{P}_2^{\prime})$ but this new $(P_2^{\prime}|\mathcal{P}_2^{\prime})$ is the same as the original $(P_1|\mathcal{P}_1)$. The $*_{r_1, s_1}\star_{r_1, s_1}$ is just the identity map.

\subsubsection{Global duality transformation}
\begin{definition}
Define the complete global duality transformation as dual transformation of all partitions in $\mathcal{W}$, denoted as $*$, and dual transformation of all partitions in $\mathcal{B}$, denoted as $\star$. The composite map is an identity map.
\begin{equation}
* \star (\mathcal{W}|\mathcal{B}) = (\mathcal{W}^*|\mathcal{B}^{\star}) \equiv (\mathcal{W}|\mathcal{B})\,.
\end{equation}
\end{definition}

\subsubsection{The Duality Symmetry}
The idea of duality symmetry for case (3) would be similar to case (1). The $(\mathcal{W}| \mathcal{B})$ is a full duality invariant under the action of pair-wise operators for both element space,
\begin{equation}
(*_{V_j^*} \oplus *_{V_{j}^*})( \star_{S_{k,\,\,j}} \oplus \star_{S_{k,\,\,j}^{\star}}) (\mathcal{W}|\mathcal{B}) = (\mathcal{W}|\mathcal{B})\,.
\end{equation} 
The proof will be the same as case(1) but just include the observer's space. Note that since the two different maps commute, we can write,
\begin{equation}
(*_{V_j^*} \oplus *_{V_{j}^*})(\star_{S_{k,\,\,j}} \oplus\star_{S_{k,\,\,j}^{\star}}) (\mathcal{W}|\mathcal{B}) =  (\star_{S_{k,\,\,j}} \oplus\star_{S_{k,\,\,j}^{\star}})(*_{V_j^*} \oplus *_{V_{j}^*}) (\mathcal{W}|\mathcal{B})\,.
\end{equation}

We will demonstrate all the above abstract definitions of the duality space above using a duality system with multiplicity $n=3$ as an example. Let there be three duality units, so we have three parity groups and three representation vector spaces together with three corresponding observer spaces. Then the tensor product space is,
\begin{equation} \label{eq:FirstClassDualityExampleN3}
\begin{aligned}
\,\,\,\,\,\, & \big(\mathcal{V}_1 \otimes \mathcal{V}_2 \otimes \mathcal{V}_3 | \mathcal{S}_{k,\,\,1} \otimes \mathcal{S}_{k,\,\,2} \otimes \mathcal{S}_{k,\,\,3}\big)\\
= & \big( (V_1 \oplus V_1^*) \otimes (V_2 \oplus V_2^*) \otimes (V_3 \oplus V_3^*) | (S_{k,\,\,1} \oplus S_{k,\,\,1}^{\star}) \otimes (S_{k,\,\,2} \oplus S_{k,\,\,2}^{\star}) \otimes (S_{k,\,\,3} \oplus S_{k,\,\,3}^{\star})  \big) \\
= & \big( (V_1 \otimes V_2 \otimes V_3) \oplus (V_1^* \otimes V_2 \otimes V_3) \oplus (V_1 \otimes V_2^* \otimes V_3) \oplus (V_1 \otimes V_2 \otimes V_3^*)\,\,|\,\, S_k \cdots \big) \oplus\\
\,\,\,\,& \big( (V_1^* \otimes V_2^* \otimes V_3^*) \oplus (V_1 \otimes V_2^* \otimes V_3^*) \oplus (V_1^* \otimes V_2 \otimes V_3^*) \oplus (V_1^* \otimes V_2^* \otimes V_3) |S_k \cdots)\,,
\end{aligned}
\end{equation}
where the $S_k \cdots$ is the expansion counterparts for observer spaces (as we are running out of space). We can see by definition \ref{DualityMapsFirstClass} any dual maps on a particular partition will give you another partition. For example,
\begin{equation}
*_{V_1} *_{V_2^*} (V_1 \otimes V_2^* \otimes V_3) = V_1^* \otimes V_2 \otimes V_3
\end{equation}
is another partition. By definition \ref{IdentityMapsFirstClass} we see that for example,
\begin{equation}
\begin{aligned}
\,\,\,\,\,\, & *_{V_1} \star_{S_{k,\,\,1}} *_{V_2^*} \star_{S_{k,\,\,1}^{\star}} \big( V_1 \otimes V_2^* \otimes V_3 | S_{k,\,\,1} \otimes S_{k,\,\,2}^{\star} \otimes S_{k,\,\,3}\big) \\
=& \big( V_1^*\otimes V_2 \otimes V_3 | S_{k,\,\,1}^{\star} \otimes S_{k,\,\,2} \otimes S_{k,\,\,3}\big)\\
\equiv & \big( V_1 \otimes V_2^* \otimes V_3 | S_{k,\,\,1} \otimes S_{k,\,\,2}^{\star} \otimes S_{k,\,\,3}\big)\,. 
\end{aligned}
\end{equation}
We identify the the second last line of \ref{eq:FirstClassDualityExampleN3} as $(A|A_{S_k})$ and the last line as $(A^{\star}|A^{\star}_{S_k})$. Finally as $n=3$ then we have $2^{3-1} = 4$ sub-duality system, which is identified as follow:
\begin{equation}
\begin{aligned}
& (\mathcal{K}_1 | \mathcal{K}_{S_k}) = \big( V_1 \otimes V_2 \otimes V_3 |S_{k,\,\,1} \otimes S_{k,\,\,2} \otimes S_{k,\,\,3}\big)\\
& (\mathcal{K}_1^{*} | \mathcal{K}^{\star}_{S_k^{\star}} \big) = \big( V_1^{*} \otimes V_2^* \otimes V_3^* |S_{k,\,\,1}^{\star} \otimes S_{k,\,\,2}^{\star} \otimes S_{k,\,\,3}^{\star}\big) \\
& \vdots
\end{aligned}
\end{equation}
and similarly for the 2,3 and 4 cases. We can see that,
\begin{equation}
\begin{aligned}
(*_{V_1} \oplus *_{V_1^*}) (K_1 \oplus K_1^*) & = (K_2 \oplus K_2^*) \\
(*_{V_2} \oplus *_{V_2^*}) (K_1 \oplus K_1^*) & = (K_3 \oplus K_3^*) \\
(*_{V_3} \oplus *_{V_3^*}) (K_1 \oplus K_1^*) & = (K_4 \oplus K_4^*) \\
\end{aligned}
\end{equation}
thus this is an example demonstration of \ref{eq:DualityKspacePairwiseOperator} and \ref{eq:DualitySumKspacePairwiseOperator}.

\section{Construction of the diagramatic basis representation of 4-duality group}
In this chapter we study the construction of basis of irreducible representation of the 4-duality group $\mathbb{Z}_2 \times \mathbb{Z}_2$. We would extensively use the diagramatic representation of the 4-box tableaux, which is called the 4-fundamental tableaux representation,
\begin{figure}[H]
\centering
\includegraphics[trim=0cm 0cm 0cm 0cm, clip, scale=1]{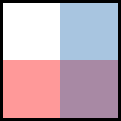}
\caption[4-box tableaux representation of the 4-duality group]
{4-box tableaux representation of the 4-duality group} \label{fig:4fundamentalBox}
\end{figure}

Recalling the definition of the dual space, which consists of two element vector space of $V$ and $V^*$ which are isomorphic to each other \footnote{In the most general general definition of dual space the isomorphism is not necessarily imposed.}. Here each coloured box, (red (R), blue (B), magenta (M) and white (W) ) represent the following,
\begin{equation} \label{eq:Dspace}
{\rm R}:= V , \quad {\rm{ B}} := V^* , \quad {\rm{ M}}:= V \oplus V^*  , \quad W = 0 .
\end{equation}
If one consider dual set then
\begin{equation} \label{eq:DSet}
{\rm R}:=U , \quad {\rm B}:= U^*, \quad \rm{M}:= U \cup U^* , \quad \rm{W} = U \cap U^* = \emptyset \,.
\end{equation}
And in particular, we have $W= M^* = * (U \cup U^* ) = U^* \cap U = \emptyset$ and $*\emptyset = *(U^* \cap U) = U\cup U^* = W$, where we have used de-Morgan's theorem. Thus $W$ and $M$ is dual to each other.  The full union is sometimes written as `All' , while the null intersection is sometimes written as `Null' or `None'.
This can be understood diagramatically by the four colour in the 4-fundamental tableaux representation. The magenta is the mixing of red and blue, while the white has no overlap between them. The origin, which is defined as the central zero, can be omitted at the moment. We also define each coloured box to have unity unit of area, thus the 4-fundamental tableau is a 4-unit object.  Next we define the 4 quadrants for the  4-fundamental tableaux representation. The four quadrants correspond to the 4 boxes, for which each quadrant is a vector space $Q$. The quadrant number is defined by indexing the coloured boxes by the following binary number $q_Q$:
\begin{equation} \label{eq:QuandrantIndex}
\text{W}: (00) , \,0  ; \quad \text{R}: (01),\,1  ; \quad \text{B}: (10),\,2 ; \quad \text{M}: (11) , 3 \,.
\end{equation}

In terms of vectors,
\begin{equation} \label{eq:Dspace}
{\rm R}:= v , \quad {\rm{ B}} :=  v^* , \quad {\rm{ M}}:= v + v^*  , \quad W =  \pmb{0} \,,
\end{equation}

Suppose $|0\rangle \in V$ and $|1\rangle \in V^*$ where $V$ and $V^*$ are one dimensional vector spaces. Now consider the qubit $|\psi(\theta)\rangle = \cos\theta |0\rangle + \sin\theta |1\rangle$. For convenience, we consider the labelling of the zero vector,
\begin{equation}
0|u\rangle = \pmb{0}_u =  \pmb{0} \,,\quad \,0|u^*\rangle = \pmb{0}_{u^*}  =  \pmb{0} \,.
\end{equation}
We also take
\begin{equation}
*\pmb{0}_u = \pmb{0}_{u^*} \,,\quad \, *\pmb{0}_{u^*} = \pmb{0}_{u}
\end{equation}
But note that the zero vector itself has no parity, we just introduce it for convenience, so notice that
\begin{equation}
\pmb{0}_u = \pmb{0}_{u^*} = \pmb{0} \,.
\end{equation}
Then we have
\begin{equation}
|\psi(0)\rangle = |0\rangle + 0|1\rangle = |0\rangle + \pmb{0}_1 \,,
\end{equation}
\begin{equation}
|\psi (\pi/2)\rangle = 0|0\rangle + |1\rangle = \pmb{0}_0 + |1\rangle \,,
\end{equation}
The fundamental four-box tableaux now reads
\begin{figure}[H]
\centering
\includegraphics[trim=0cm 0cm 0cm 0cm, clip, scale=0.6]{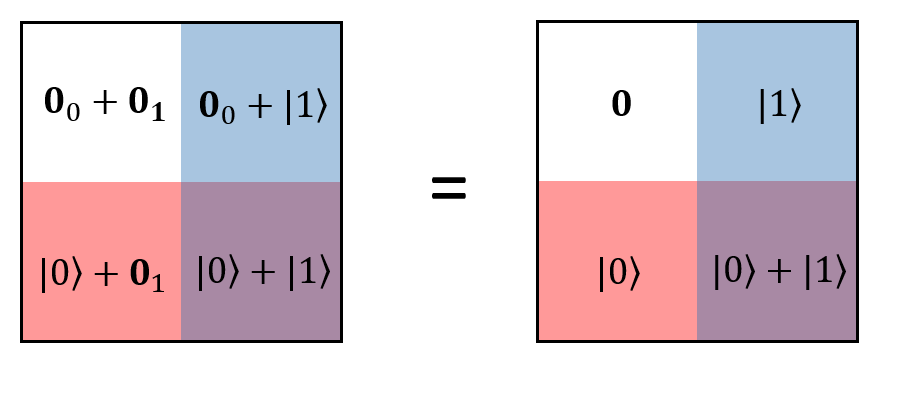}
\caption[]
{} \label{fig:table3}
\end{figure}
We can see that
\begin{equation}
|\psi (\pi/2)\rangle = *|\psi(0)\rangle
\end{equation}
The Null state and the All state are dual invariant,
\begin{equation}
*\pmb{0} = \pmb{0}_1 + \pmb{0}_0 = \pmb{0} \quad \text{and} \quad *( |0\rangle + |1\rangle)=  |1\rangle + |0\rangle = |0\rangle + |1\rangle \,,
\end{equation}
while the half states are not. We can further check that states in the dual partition are orthogonal. For R and B, we have $\langle v |v^*\rangle = \langle 0 |1\rangle =0$ which satisfies. So clearly they are orthogonal. For M and W, it is clearly that $\langle \pmb{0}|(|0\rangle + |1 \rangle   ) = 0$\, so the full vector and the null vector are orthogonal and dual to each other.
Notice that the role of zero vector is to ensure $|v\rangle +\pmb{0} =\pmb{0}+|v\rangle =|v\rangle  $, this is in analogy to the role of empty set that $U\cup \emptyset = \emptyset \cup U = U$.

Then in dimensional representation we have 
\begin{figure}[H]
\centering
\includegraphics[trim=0cm 0cm 0cm 0cm, clip, scale=0.6]{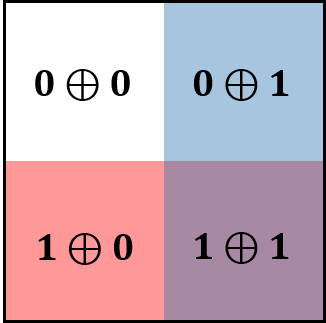}
\caption[]
{} \label{fig:table4}
\end{figure}
The dimensional representation is isomorphic to the basis of $\mathbb{Z}_2 \times \mathbb{Z}_2$. Thus we have constructed the basis representation of $\mathbb{Z}_2 \times \mathbb{Z}_2$ by the dimensional representation of the 4-box tableaux.

Now we would like to construct another new representation in terms of direct-sum vector space.
\begin{equation} \label{eq:Dspace}
{\rm R}:= V \oplus \pmb{0}_{V^*}, \quad {\rm{ B}} := \pmb{0}_V \oplus V^* , \quad {\rm{ M}}:= V \oplus V^*  , \quad W = \pmb{0}_{V} \oplus \pmb{0}_{V^*} \,,
\end{equation}
where
\begin{equation}
\pmb{0}_{V} = \pmb{0}_{V^*}= \pmb{0} = |\pmb{0}\rangle\,.
\end{equation}
In terms of element,
\begin{figure}[H]
\centering
\includegraphics[trim=0cm 0cm 0cm 0cm, clip, scale=0.6]{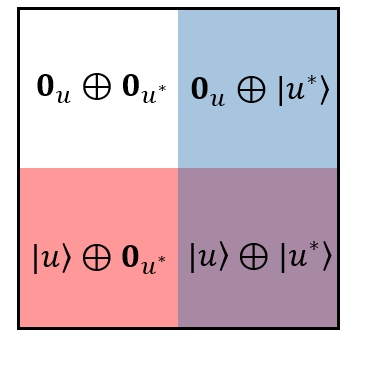}
\caption[]
{} \label{fig:tablele}
\end{figure}
The zero vector is the only dual invariant element in the representation, such that
\begin{equation}
*\pmb{0} = \pmb{0} \,.
\end{equation}

Next we can check that the vectors in the same dual partition are orthogonal. We need to use the following identity,
\begin{equation}
\bigg\langle \bigoplus_{i=1}^n v_i \bigg\vert \bigoplus_{i=1}^n u_i \bigg\rangle = \sum_{i=1}^n \langle v_i | u_i \rangle \,.
\end{equation}
Let $|w_{\mathrm{full}}\rangle= |u\rangle \oplus |u^*\rangle$, $|w_{\mathrm{+}}\rangle= |u\rangle \oplus |\pmb{0}_{u^*}\rangle$, $|w_{\mathrm{-}}\rangle= | \pmb{0}_u \rangle \oplus |u^*\rangle$ and $|w_{\mathrm{null}}\rangle= \pmb{0}$, we have
\begin{equation}
\begin{aligned}
&\langle w_{\mathrm{null}}|w_{\mathrm{full}} \rangle = \langle \pmb{0} | u\oplus u^* \rangle = \langle \pmb{0}_u |  u \rangle +  \langle \pmb{0}_{u^*} |  u^* \rangle = 0 \\
&\langle w_{\mathrm{+}} |  w_{\mathrm{-}}\rangle  = \langle u \oplus  \pmb{0}_{u^*} |\pmb{0}_{u} \oplus u^* \rangle = \langle u |\pmb{0}_{u} \rangle + \langle \pmb{0}_{u^*} | u^*\rangle = 0 \,.
\end{aligned}
\end{equation}

Furthermore, we can check that we can generate dual vector simply by acting the $*$ on the vector that satisfies orthogonality. The dual construction of \ref{fig:tablele} is illustrated as follows in \ref{fig:dualtab}: 
\begin{figure}[H]
\centering
\includegraphics[trim=0cm 0cm 0cm 0cm, clip, scale=0.6]{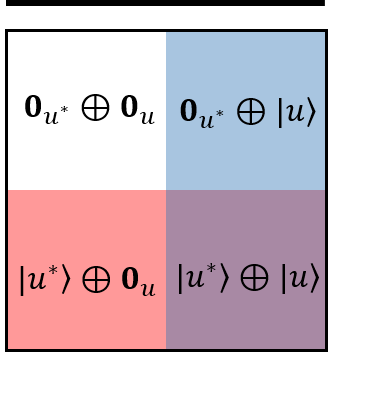}
\caption[]
{} \label{fig:dualtab}
\end{figure}

We can check that, 
\begin{equation}
\begin{aligned}
&\langle w^*_{\mathrm{full}} | w_{\mathrm{full}}\rangle = \langle u^* \oplus u| u\oplus u^*\rangle = \langle u^* | u\rangle + \langle u|u^*\rangle = 0 \,.\\
&\langle w^*_{+} | w_{+} \rangle = \langle u^*  \oplus  \pmb{0}_u |u   \oplus\pmb{0}_{u^*} \rangle= \langle u^* | u\rangle + \langle \pmb{0}_u | \pmb{0}_{u^*} \rangle = 0 \,. \\
&\langle w^*_{-} | w_{-} \rangle = \langle \pmb{0}_{u^*} \oplus u |  \pmb{0}_{u} \oplus u^* \rangle =  \langle \pmb{0}_{u^*} | \pmb{0}_{u} \rangle +\langle u | u^*\rangle = 0 \,.\\
& \langle w^*_{\mathrm{null}} | w_{\mathrm{null}} \rangle = \langle \pmb{0} | \pmb{0} \rangle = 0\,.
\end{aligned}
\end{equation}

Next, we would like to construct a larger basis for the $\mathbb{Z}_2 \times \mathbb{Z}_2$ group from the 4-fundamental tableaux representation. We construct another 4 representations from the 4-fundamental tableaux representation by reflections along the horizontal and vertical axes, and define the abelian  $\mathbb{Z}_4 $ group (which is also the cyclic group $C_4$) with elements $\{ I, \sigma_L ,\sigma_D , \sigma_d \}$(where $\sigma_d = \sigma_L \sigma_D$) over the 4 representations as follow. (Here $L$ means reflect left/right-wise and $U$ means reflect up/down-wise, $d$ means reflect diagonal-wise. It is illustrated as follow.
\begin{figure}[H]
\centering
\includegraphics[trim=0cm 0cm 0cm 0cm, clip, scale=0.6]{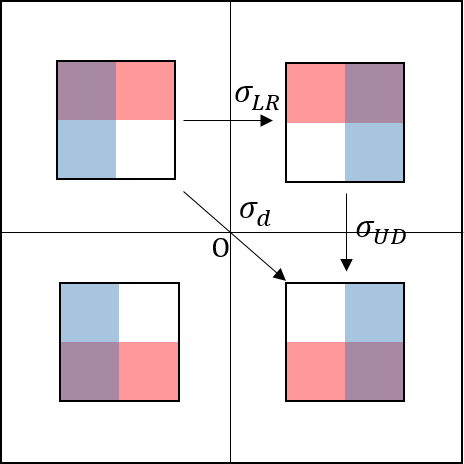}
\caption[Extensive 4-network. formed by]
{Extensive 4-network} \label{fig:4diagram}
\end{figure}

Such construction allows us to define all other larger interesting objects. The joining of the four individual tableau glued by the original $O$ defines the repeating unity of the 4-duality network, and we suppose the network is defined infinitely. 
\begin{figure}[H]
\centering
\includegraphics[trim=0cm 0cm 0cm 0cm, clip, scale=0.5]{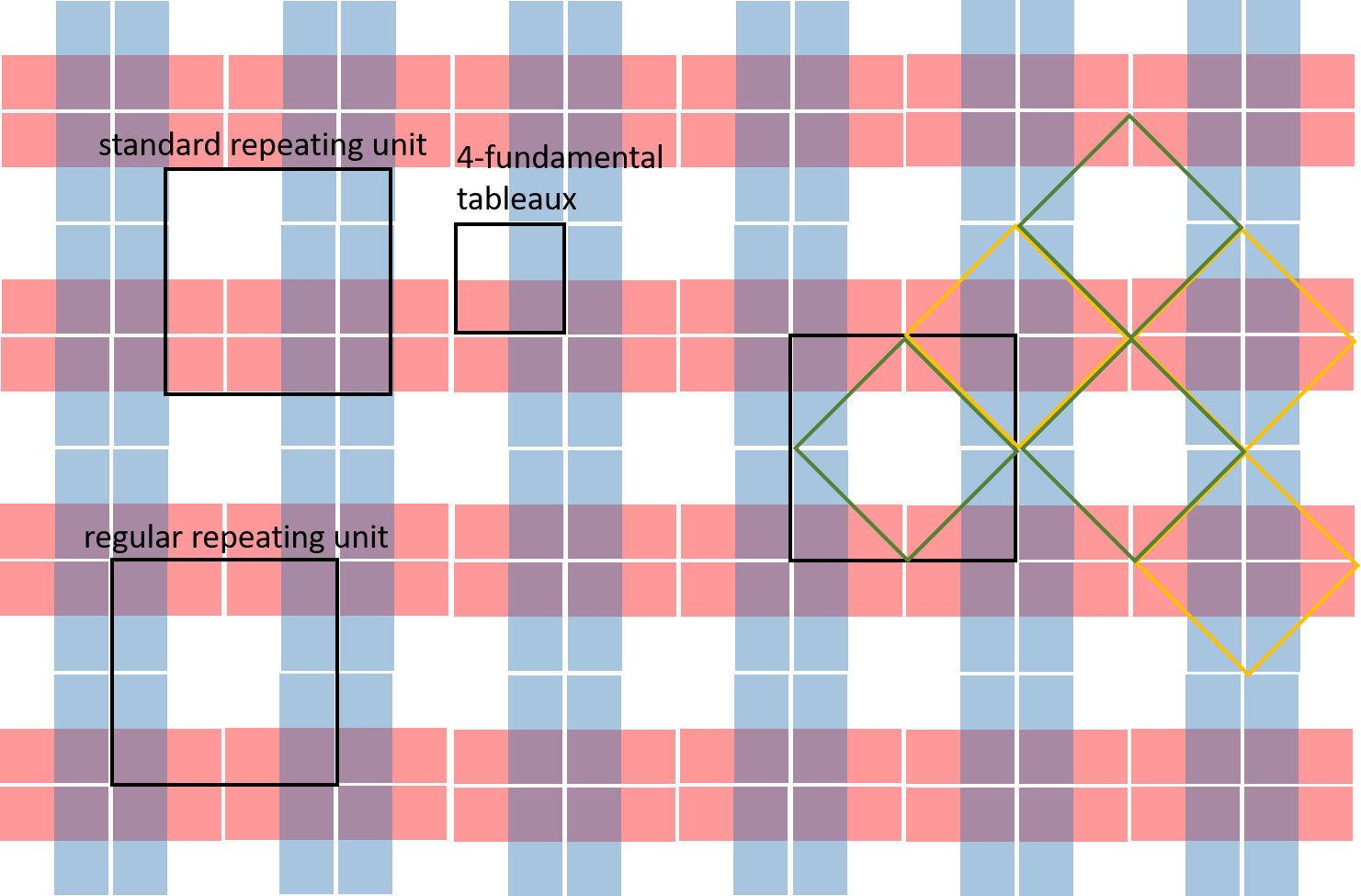}
\caption[Extensive 4-network. ]
{Extensive 4-network formed by the \ref{fig:4diagram}. Note that the white gaps in between do not exist but is left for clear demonstration. } \label{fig:4network}
\end{figure}

Note that the choice of repeating unit is not unique at least locally, but it has to be a 16-unit square. We can see that the extended version of the 4-fundamental tableau reappears as a 16-unit repeating unit of the network, shown in the upper left corner of \ref{fig:4network}.

We can have different sub-diagrams in the network, and the sub-diagram can have different units of red, blue, magenta and white boxes. We defined the standard notation as $(\mathrm{C_1 m_1 :C_2 m_2 : C_3 m_3 : C_4 m_4})$, where $\mathrm{C_i}$ refers to the colour and $\mathrm{m_i}$ refers to the number of units possessed by that colour. If $m_i =0$ we can choose to omit the $\mathrm{C_i m_i}$ term. 

Now we need to introduce some formal definitions in a rigorous manner for the repeating units. 
\begin{definition}
In a 4-network, the standard repeating unit is a repeating unit of 16 area-units which is the extension of the 4-fundamental tableau with no reflectional symmetry.
\end{definition}
\begin{definition}
The regular repeating unit is a 16 area-unit that is formed by the horizontal, vertical and diagonal reflections of the 4-fundamental tableau by \ref{fig:4network} with 4 reflectional symmetries (1 horizontal, 1 vertical and 2 diagonals). 
\end{definition}
There are 4 possible regular repeating units in total. Starting from the lower-left one in \ref{fig:4network}, translation in the horizontal direction by 2 units, translation in the vertical direction by 2 units, and their composition would give the remaining 3 repeating units. The total 4 repeating units form the basis representation of $\mathbb{Z}_2 \times \mathbb{Z}_2$ 4-dual group. 
\begin{definition}
The diamond representation a shrinked or extended representation with a rotation by $\pi/4$, which is defined as the dual representation of the square representation, which defines the deficit colour representation.
\end{definition}
The reason for why the diamonds are dual to the square would be apparent in the moment. It is noted that the diamond representation is not a repeating unit of the network, ad we require two set of different diamond representations in order to cover the whole network.  

Next we would introduce the concept of level $n$. 
\begin{definition}
The level $n \in \mathbb{Z}$ of the box or diamond representation defines the index of the layers. The $n=0$ is the lowest level and defined as the ground level in which the box cannot be further split into other colour except for itself. 
\end{definition}
It is easy to see that for $n \in  \mathbb{Z}^{-}$ being negative, the results would be the same $n=0$ case, as we just continue to split within a same coloured box. Therefore we would only have non-trivial results when $n > 0$.

The following shows the illustration,
\begin{figure}[H]
\centering
\includegraphics[trim=0cm 0cm 0cm 0cm, clip, scale=0.6]{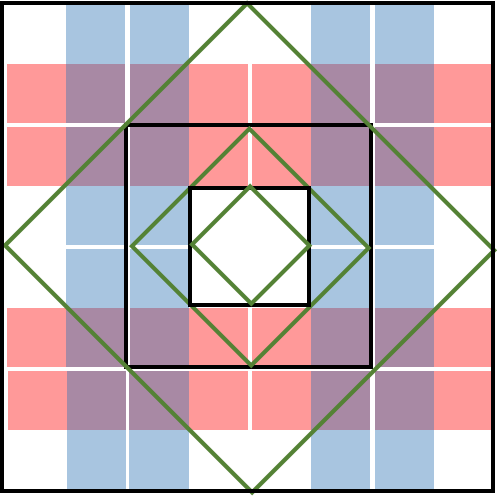}
\caption[Levels of square or diamond representation ]
{Levels of square or diamond representation } \label{fig:levels}
\end{figure}

Next we will introduce the concept of observation frame. We will define two frames, one the interior frame and its dual, exterior frame.
\begin{definition}
Let $R$ be the vector space that the box is situated in, while $S$ be the vector space that the diamond is situated in, where $R, S \subset \mathbb{R}^2$. The interior frame is defined by the space enclosed by $R$ or $S$. Denote $A(R)$ and $A(S)$ the area of the box and diamond respectively defined by $A : R, S \rightarrow \mathbb{R}$. The interior space $I$ is defined by
\begin{equation}
\mathrm{I}=
\begin{cases}
R & \text{if} \quad A(R) < A(S) \\
S & \text{if}\quad A(S) < A(R)
\end{cases} \,.
\end{equation}
\end{definition}
The exterior space $E$ is defined by
\begin{equation}
\mathrm{E}=
\begin{cases}
R\setminus S & \text{if} \quad A(R) > A(S) \\
S\setminus R & \text{if}\quad A(S) > A(R)
\end{cases}
\,.
\end{equation}
Define the interior frame ${\mathrm{I}}=F$ and the exterior frame as its dual ${\mathrm{E}}=F^*$. If we have an isomorphism between $R$ and $S$, $R \cong S$, then $A(R) = A(S)$. if follows that $A(\mathrm{E}) =A(\mathrm{I}) $. Then we have $\mathrm{E} \cong \mathrm{I}$. We say $\mathrm{E}$ and $\mathrm{I}$ is dual invariant $F= F^*$, writing $F\equiv F^*$, in which the condition is the same area of both representations.

We would also call the interior frame normal (existing) frame and the exterior frame the null frame. The reason for this terminology is because, the interior is considered as an ownership while the exterior is considered as things belonging to "outside". 

Finally we have to define the translational operation on the network. 
\begin{definition}
Let $k \in \mathbb{Z}$ be the units to be translated across the 4-network. Define the map $\rho^{(k)}$ as the translation of $k$ units along the horizontal direction $LR$ or vertical direction $UD$, with $\rho^{(k)}_{LR}$ and $\rho^{(k)}_{UD}$.  The $\rho^{(k)}$ is independent of layer $n$, in which $\rho^{(k)} = \rho^{(k)}_n$ for all $n \in \mathbb{Z}$.
\end{definition}

It is easy to see that the map has a periodicity of $4$,
\begin{equation}
\rho^{(k)} = \rho^{(k+4)} \,.
\end{equation}

\begin{definition}
The map of translation $\rho^{k}$ for $k=0,1,2,3$ forms an abelian group which is isomorphic to the cyclic group $C_4$ (or $\mathbb{Z}_4 $), explicitly $C_4 = \{ I, \rho^{1} , \rho^{2} , \rho^{3}   \}$ with $I =\rho^{0} $ the identity element. 
\end{definition}

We have naturally three of these groups $C_{4\,LR}$ , $C_{4\,UD}$ and $C_{4\,d}$  for the horizontal, vertical and diagonal translations respectively.

Now let's first study the ground level $n=0$. It can be illustrated by the following diagram.
\begin{figure}[H]
\centering
\includegraphics[trim=0cm 0cm 0cm 0cm, clip, scale=0.65]{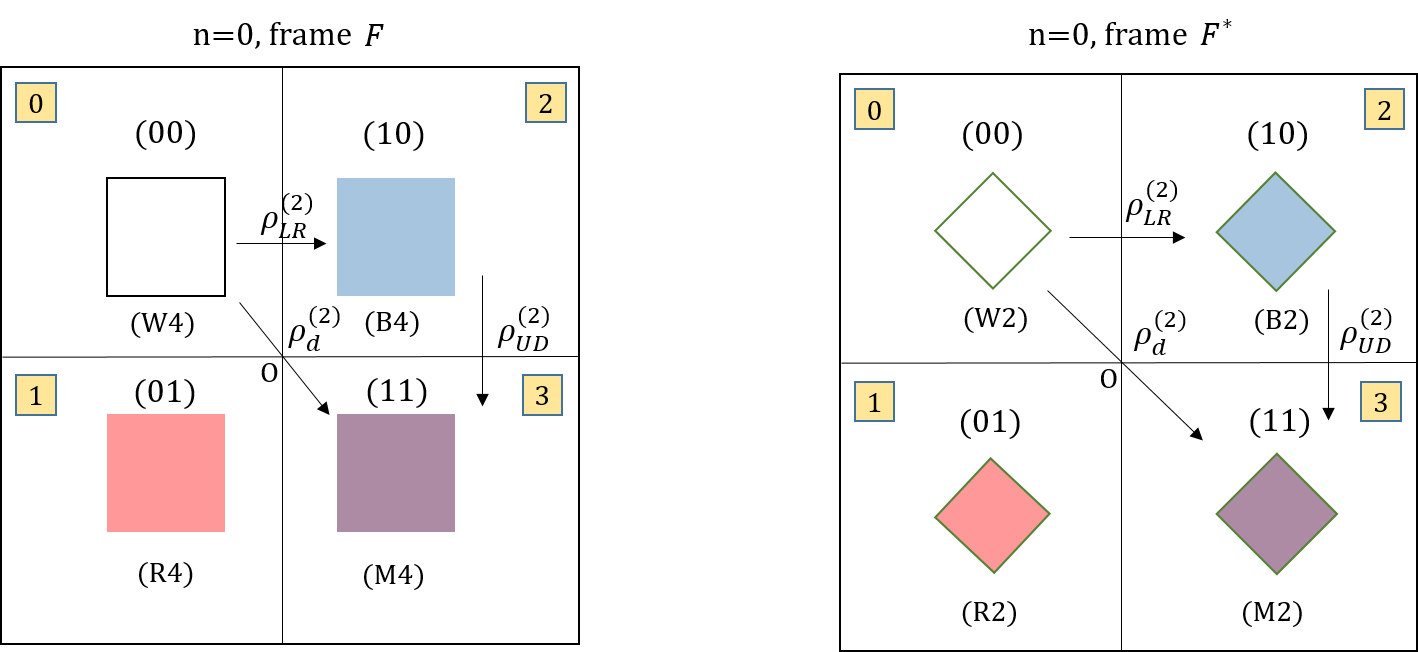}
\caption[The study of ground level]
{The study of ground level. The quandrant number is labelled by the binary number.} \label{fig:groundlevel}
\end{figure}
For the $n=0$ case, both the box and diamond representations share the same quadrant number. It is noted that in general for $n >0$, the quadrant number will not be invariant for both representations as we will see. We can see that the only difference between the box the diamond representation is the number of coloured area units it enclose, for which is halved for the diamond case.

The most important feature for the $n=0$ ground level is that it is interior and exterior dual invariant by definition 2.0.5 . It is a very special property for the ground level. It is easy to see that cases for $n <0$ is also dual invariant. Thus for all $n \leq 0$, the box representation and the diamond representation is dual invariant under the condition of same area.

Thus in summary, the $\mathrm{E}-\mathrm{I}$ dual invariant for the $n=0$ ground level case, which means $F^* \equiv F^{*\star}$ implies
\begin{enumerate}
\item No further colour splitting possible
\item Unchanging quadrant number
\item Same area  $A(\mathrm{E)} = A{\mathrm(I)}$
\end{enumerate} 
Also, we can define weak dual invariant if only some items of the above is satisfied. For our $n=0$ case, $F\equiv F^{*}$ is a weak dual invariant as it just satisfies two items, as we have a change from $\mathrm{C_i}4$ to $\mathrm{C_i}2$ when we go from box representation to diamond representation.

\begin{figure}
\centering
\includegraphics[trim=0cm 0cm 0cm 0cm, clip, scale=0.42]{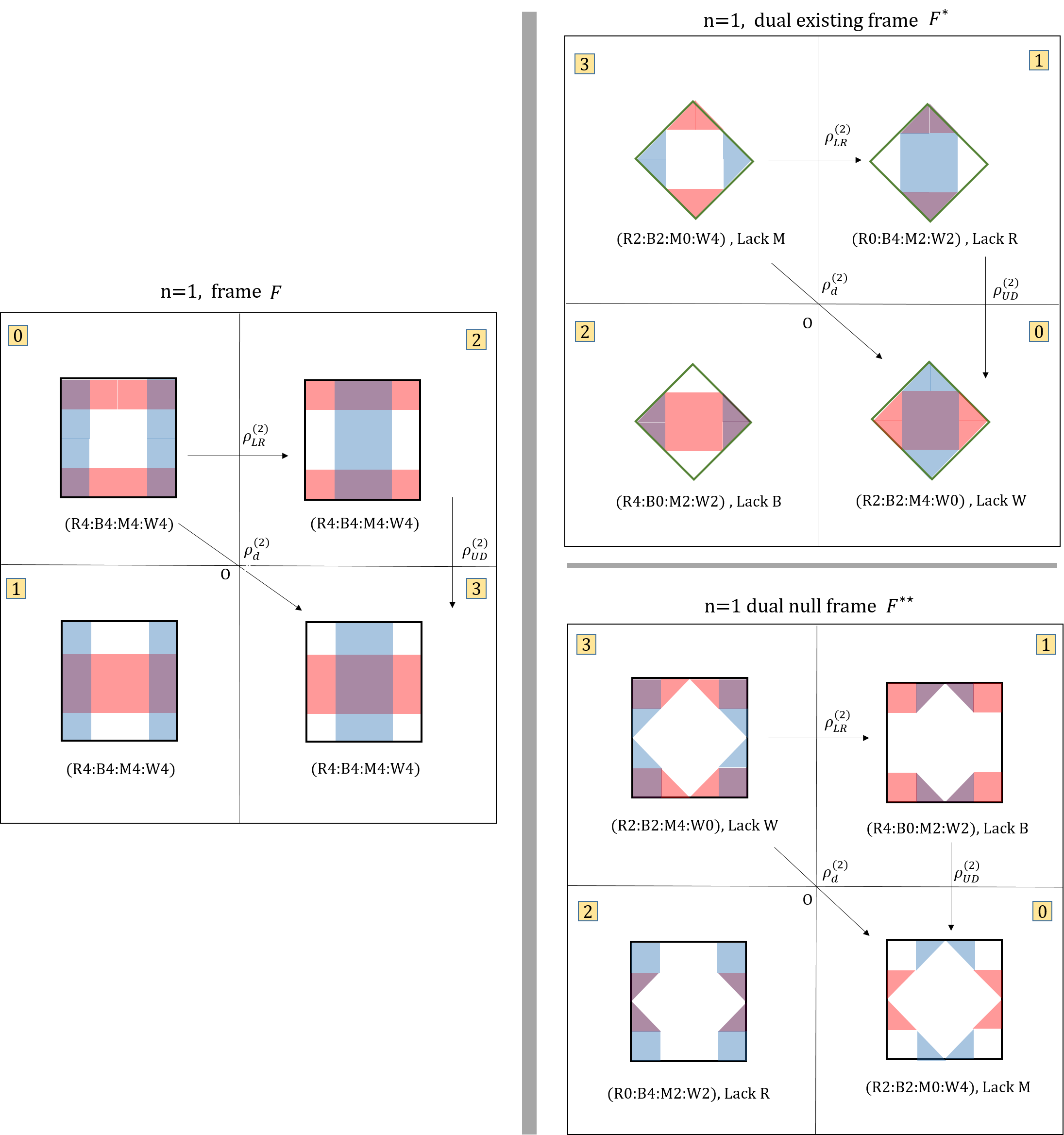}
\caption[The study of first level]
{The study of first level \label{fig:n1level}}
\end{figure}

Now for $n=1$ case, things become much more complicated. First  remember there are two levels of duality. First, it is the duality between the box representation and diamond representation. Next it the duality for the interior and exterior frame. There are sub-duality inside a duality structure. In both cases they are not dual invariant. For $n=1$ case, the concept of lacking colour units for the dual diamond representation is clearly demonstrated in figure \ref{fig:n1level}.

\begin{figure}
\centering
\includegraphics[trim=0cm 0cm 0cm 0cm, clip, scale=0.42]{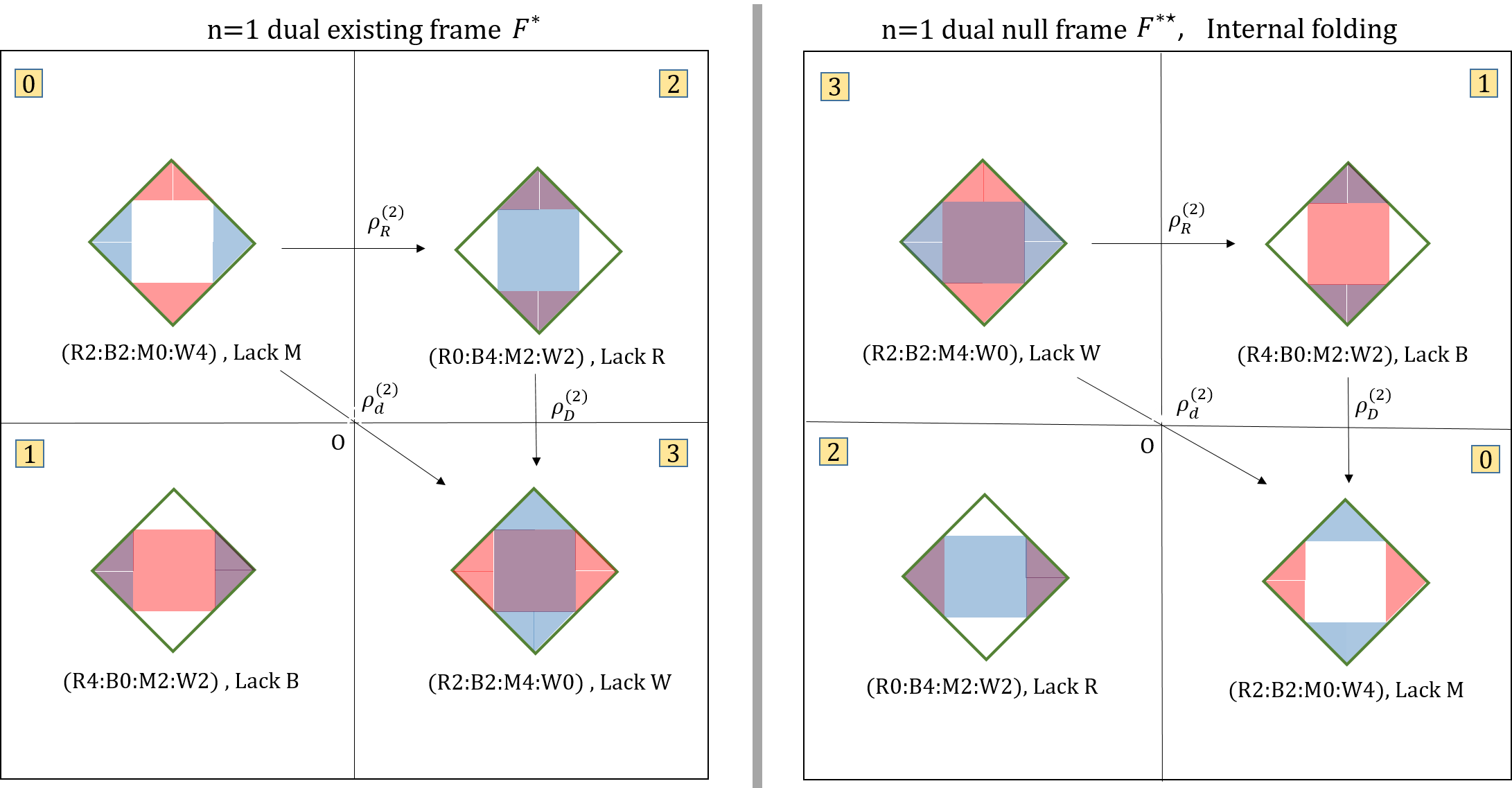}
\caption[The study of first level with internal folding for $F^{*\star} $ frame and the comparison to $F^{*}$.]{
The study of first level with internal folding for $F^{*\star} $ frame and the comparison to $F^{*}$.} \label{fig:n1InternalFolding}
\end{figure}

Note that all of them are basis of irreducible representation of $\mathbb{Z}_2 \times \mathbb{Z}_2$. 
Let's first consider the general differences between the box representation and diamond  representation. In the box case, all bases have the same number or red, blue, magenta and white boxes, which are 4; while in the diamond case, all bases have different coloured units. However for each of them we can figure out one particular colour is missing. Thus this is the reason we call it the dual representation. (Note that in the $F^{*\star}$ the white diamond square do not count as we only contribute the exterior part by definition. ) Due to the obvious difference from the box representation, the quadrant indexes have to be relabelled. Using \ref{eq:QuandrantIndex}, the new quadrant index in $F^*$ frame is given by
\begin{equation}
q_Q^{* } = ({\mathrm{max}}\,q_Q ) - Q = 3-q_Q \,.
\end{equation}
For the ease of comparison, we define internal folding for the diagramatic basis for $F^{*\star}$ basis. The internal folding is defined by joining the four corners to the center. The result is shown in \ref{fig:n1InternalFolding}.
Thus we can see that they are apparently different, not just in position for the colour of the square but the colour in the triangles have swapped.

It is remarked that these colour representations are just denoting the dual set or dual space, in the end the would translate back the language of colour back to the elements of dual set or dual space. Let's define the lacking of an element in dual space by the action `!' \footnote{Not to confuse with the use of negation $NOT$ in programming languages}. 

Next we introduce the concept of perspective. We can have two perspectives, the perspective of presence and the perspective of absence. For example !$W$ is in the absence perspective, while R,B,M is in the presence perspective. The full analysis is shown below.

\begin{figure}[H]
\centering
\includegraphics[trim=0cm 0cm 0cm 0cm, clip, scale=0.5]{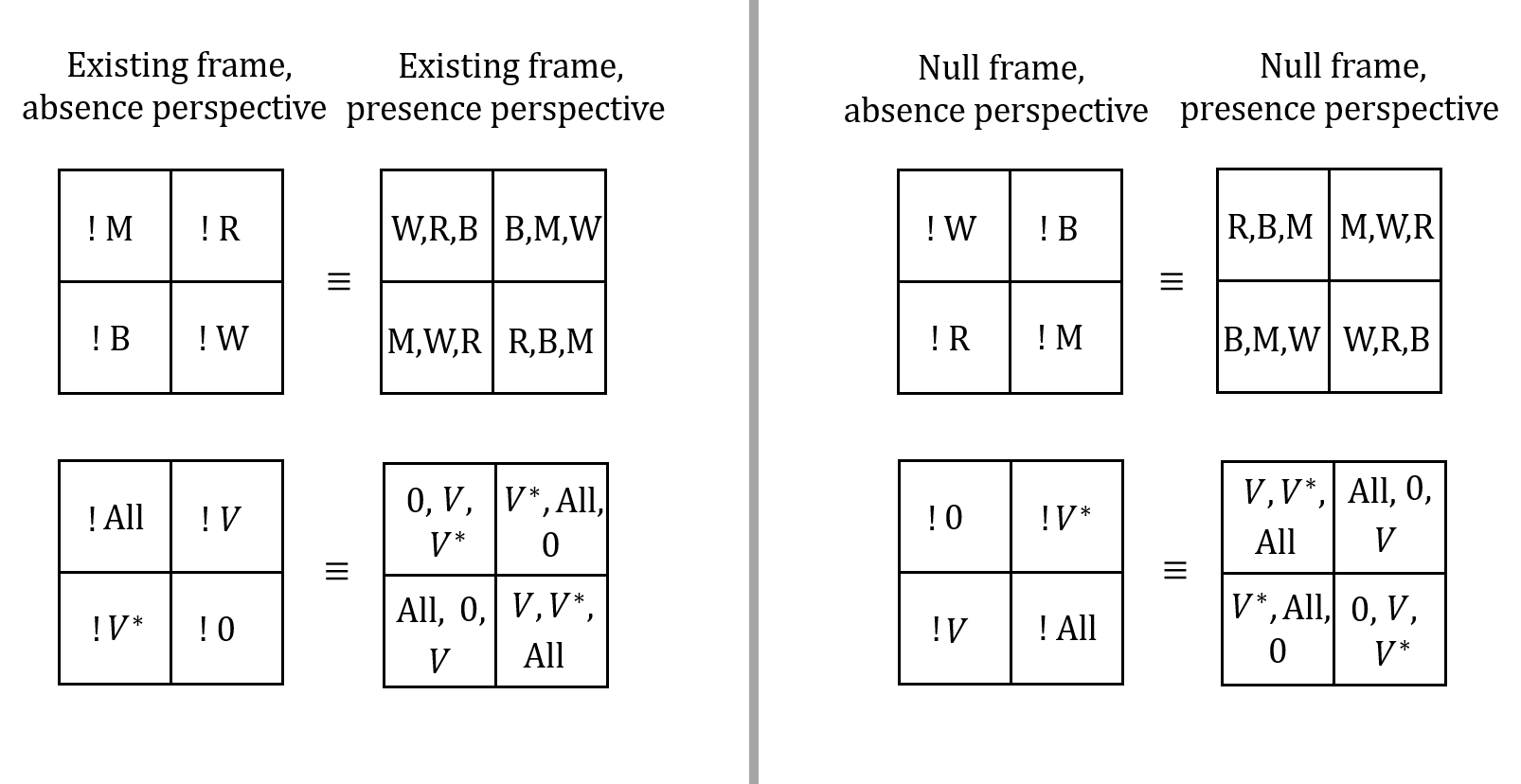}
\caption[Analysis]
{  \label{fig:analysis}}
\end{figure}

The concept of lacking colour denoted by the action `!' is in particular important and requires detailed investigation. Let's give the formal definition. 
\begin{definition}
Let $( \,\,\,\, | \,\,\, )$ be the formal notation for the description of elements under perspectives or frames,  where the left bracket $( \,\,\,\,|$ holds the basis element set $\xi$ that are of interest and the right bracket $| \,\,\,\, ) $ holds the frame or perspective $P$, explicitly $(\xi|P)$.  Define the absence perspective as $0$ and the presence perspective as $1$. 
\end{definition}
The frame and perspective themselves forms a natural basis of irreducible representation of 4-duality group $\mathbb{Z}_2 \times \mathbb{Z}_2$. 
\begin{figure}[H]
\centering
\includegraphics[trim=0cm 0cm 0cm 0cm, clip, scale=0.5]{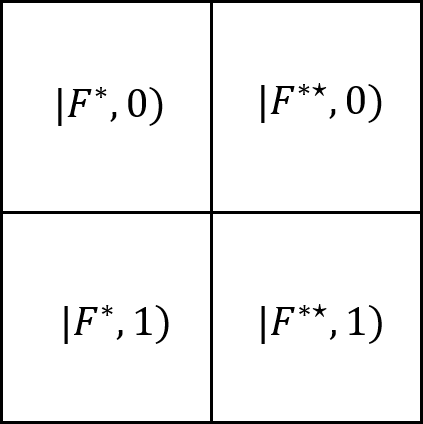}
\caption[Analysis]
{  \label{fig:framdDuality}}
\end{figure}

\begin{definition}
Define in a particular quadrant $Q$ that, the set of elements of lacking $E \in Q$ in the absence perspective $(E|0)$, where $E=\{\text{!}E_1 , \text{!}E_2 ,\cdots, \text{!}E_N \}$; and the set of elements of presence $e \in Q$ in the presence perspective $(e| 1 ) $  where $e = \{e_1 , e_2 , \cdots, e_N \}$.  $E_i$ and $e_j$ are elements in dual set or dual space, which are set or vector space, and $\mathrm{max}\,M, \mathrm{max}\, N = 4$. If there exists more than one $e_i$ such that $e_i \cup e_j = E_k $ ;  or if $E_k \subset e_i $ and $E_k$ is not a subset of all elements in $e$ , then !$E_k$ is pseudo-lacking of $E_k$. Otherwise, !$E_k$ is real-lacking. 
\end{definition}

The concept of pseudo-lacking is introduced because it means it is not really totally lacking. Although the particular color is lacking in the absence perspective, it can be formed or hidden in the elements in the presence perspective. The negation of the two if statements would be real-lacking, as the particular colour cannot be joined by some other elements in the presence perspective, nor it is contained in those elements.

\begin{definition}
For a basis of $\mathbb{Z}_2 \times \mathbb{Z}_2$ represented by the dual diamond lacking representation, the 4-lacking is subdivided into one real-lacking basis and three pseudo-lacking bases and , denoted by $4 = 1 \oplus 3$.  
\end{definition}
This can be easily checked by using the case for existing frame. 
\begin{itemize}
\item For !All under $|F^{*},0 )$, we have $({\mathrm{\text{!}All}}|F^{*} ,0 ) \equiv (0, U, U^* |F^{*} ,1  )$. Yet $\text{All} = U\cup U^* $ thus we can form All in the presence perspective. Hence this is a pseudo-lacking.   
\item For !$V^*$ under $|F^{*},0 )$, we have $({\text{!}}V^* |F^{*} ,0 ) \equiv (\text{All}, 0, V |F^{*} ,0  )$. Yet $V^* \subset \text{All}$, thus $V^*$ is hidden in the presence perspective. This is a pseudo-lacking.
\item For !$V$ under $|F^{*},0 )$, we have $({\mathrm{\text{!}V}}|F^{*} ,0 ) \equiv (V^*, \text{All}, 0 |F^{*} ,0   )$. Yet $V \subset \text{All}$, thus $V$ is hidden in the presence perspective. This is a pseudo-lacking.
\item For $0$,under $|F^{*},0 )$, we have $(0|F^{*} ,0 ) \equiv (V,  V^{*}, \text{All} |F^{*} ,0   )$. These spaces do not contain $0$, but $0$ is a subset or subspace in all elements $V,  V^{*}, \text{All}$. This is a real-lacking.
\end{itemize}
Thus in the diamond representation, we can diagramatically represent as
\begin{figure}[H]
\centering
\includegraphics[trim=0cm 0cm 0cm 0cm, clip, scale=0.5]{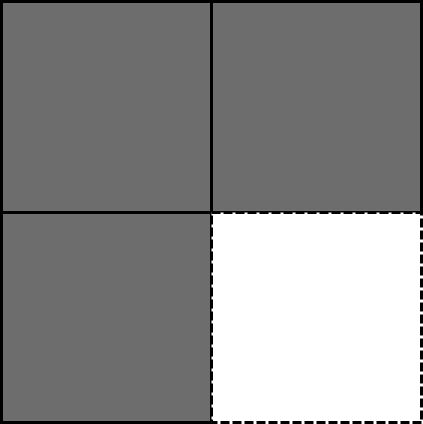}
\caption[4-tableau representation of one real-lacking and 3 pesudo-lacking basis for the diamond representation]
{  \label{fig:1plus3}}
\end{figure}
The white box represents the real-lacking basis while the three dark-grey boxes represent the three pseudo-lacking bases, thus this represents the structure of $4= 1 \oplus 3$. 

Note that this structure is the same as the basis of irreducible representations of SO(4), which is an isomorphic to SU(2)$\times$SU(2). Therefore the diamond representation, which is a dual representation of the box representation, can be used to represent the non-abelian group SO(4).

\section{Representation theory for multi-duality}
Let the basis of $V$ be $|0 \rangle$ and the dual basis of $V^*$ be $|1\rangle$.
The tensor product of vector spaces above has the tensor product states as the basis. For example, $V^* \otimes V \otimes V $ has the basis $|100\rangle$.
In this section, we will study multi-dual symmetry through groups and representations. The dual symmetry here we can naturally refer to the parity symmetry $\mathbb{Z}_2$. The state $|0 \rangle$ and $|1 \rangle$ form the basis of irreducible representation of the duality (parity) group $\mathbb{Z}_2$.

Next we will use two basic theorems from group theory. For two groups $G_1$ and $G_2$ with order $|G_1|$ and $|G_2|$ respectively, the direct product group $G_1 \times G_2$ has group order $|G_1 \times G_2 | = |G_1||G_2 |$. And if $G_1$ and $G_2$ are abelian, then $G_1 \times G_2$ is also abelian. We can apply these two theorems to our parity group $\mathbb{Z}_2 $.

\begin{definition}
Let $\mathbb{Z}_2^N = \mathbb{Z}_2 \times \mathbb{Z}_2 \times \cdots \times \mathbb{Z}_2$ with $N$ be the multiple parity group for the $n$-level duality symmetry group, where the group order $| \mathbb{Z}_2^N| =| \mathbb{Z}_2 |^n = 2^n $. 
\end{definition}
Note that for the multi direct product for $\mathbb{Z}_2$, it is isomorphic to the multiple tensor product of the $\mathbb{Z}_2$, then we can write
\begin{equation}
\mathbb{Z}_2^{N}= \mathbb{Z}_2 \times \mathbb{Z}_2 \times \cdots \times \mathbb{Z}_2 \cong \mathbb{Z}_2 \otimes \mathbb{Z}_2 \otimes \cdots \otimes \mathbb{Z}_2 \,.
\end{equation}
Using the group homomorphism, we have
\begin{equation}
D(\mathbb{Z}_2 \times \mathbb{Z}_2 \times \cdots \times \mathbb{Z}_2 ) = D(\mathbb{Z}_2 \otimes \mathbb{Z}_2 \otimes \cdots \otimes \mathbb{Z}_2) = D(\mathbb{Z}_2) \otimes D(\mathbb{Z}_2) \otimes \cdots \otimes D(\mathbb{Z}_2) \,.
\end{equation} 

Next we will use the following property of abelian groups. For an abelian group $G$, since each group element form the conjugacy class of itself, the number of classes $N_c$ is just the group order $|G|$. Then for our case $N_c$ is just $2^n$. And by group theory the number of classes is equal to the number of irreducible representations $N_{\tau}$, thus the number $N_\tau = 2^n$ for our case. Let $d_i$ be the dimension of the irreducible representation. Then we will use the theorem of group that 
\begin{equation}
\sum_{i=1}^{N_c} d_i^2 = |G| \,.
\end{equation}
But since $N_c = |G| = 2^n$ for our case, then this forces the dimension of each irreducible representation as 1. Therefore, the multiple-duality group of dimension $2^n$ can be decomposed into the direct sum of one-dimensional irreducible representations. Let $\mathrm{g} \in  \mathbb{Z}_2 \otimes \mathbb{Z}_2 \otimes \cdots \otimes \mathbb{Z}_2$ , and $D(\mathrm{g})$ be the matrix representation of the group \footnote{We will use the \emph{non-italic} $\mathrm{g}$ for the group element of  $  Z_2 \otimes Z_2 \otimes \cdots \otimes Z_2$,  while the \emph{italic} $g$ for group elements in each $\mathbb{Z}_2$}.

We will find the representation of the multiple duality group $\mathbb{Z}_2^N$ of general $n \geq 2$  levels. Here we will use the definition by the tensor product $\mathbb{Z}_2 \otimes \mathbb{Z}_2 \otimes \cdots \otimes \mathbb{Z}_2$.

First consider the simplest case, which is the 0-th level with $N=1$, this is just the parity group $\mathbb{Z}_2$. The $\mathbb{Z}_2 $ group only has two elements $\{ I , P \}$ and has two classes. Therefore the representation is reducible to the direct sum of two 1D irreducible representation. Let $g \in \mathbb{Z}_2$, we have
\begin{equation} \label{eq:2sum}
D(g) = \mathcal{A}_1 (g) \oplus \mathcal{A}_2 (g) \,.
\end{equation}
The $\mathcal{A}_1 (g)$ is the trivial irreducible representation, with all characters equal to 1 for all group elements. And we have $\mathcal{A}_2 (I) = 1$ and $\mathcal{A}_2 (P)= -1$. This can be easily checked by the orthogonality theorem in group theory. The basis of reducible representation is $|0\rangle $ and $| 1 \rangle$, which can be written as a basis doublet,
\begin{equation}
| v \rangle = 
\begin{pmatrix}
|0 \rangle \\
|1 \rangle
\end{pmatrix}
\,.
\end{equation}

For any general $n$-levels, the multiple duality group follows the general duality theorem. 
\begin{definition}
Let $|V \rangle$ be the basis of the multiple duality group $\mathbb{Z}_2 \otimes \mathbb{Z}_2 \otimes \cdots \otimes \mathbb{Z}_2$, mathematically
\begin{equation}
|V \rangle = |v_1 v_2 \cdots v_N \rangle = |v_1 \rangle \otimes| v_2  \rangle \otimes \cdots \otimes |v_N \rangle \,.
\end{equation}
The basis is of $N=2^n$ dimension, the whole set of
\begin{equation}
 | \eta_1 \eta_2 \cdots \eta_{N} \rangle = |\eta_1 \rangle \otimes | \eta_2 \rangle \otimes \cdots \otimes |\eta_N \rangle \quad \text{for all}\,\, \eta_j = 0,1
\end{equation}
form the basis of irreducible representation of the multiple duality group. Hence the representation in $n$-levels is the natural basis of the multiple duality group.
\end{definition}
\begin{definition}
The basis $|V \rangle$ transform under the tensor product representation of the parity group $\mathbb{Z}_2$. Let $g_{i_j}$ be the element of the the $j^{\mathrm{th}}$ parity group, then we have
\begin{equation}
|v_1\rangle^\prime \otimes| v_2^\prime \rangle \otimes \cdots \otimes |v_N^\prime \rangle =\Big( D(g_{i_1})\otimes D(g_{i_2}) \otimes \cdots \otimes D(g_{i_N}) \Big) \, ( |v_1 \rangle \otimes| v_2 \rangle \otimes \cdots \otimes |v_N \rangle ) \,.
\end{equation}
It can be written as
\begin{equation}
|V^\prime\rangle = D(\mathrm{g}) |V \rangle
\end{equation}
where
\begin{equation}
D(\mathrm{g}) = D(g_{i_1})\otimes D(g_{i_2}) \otimes \cdots \otimes D(g_{i_N}) \,.
\end{equation}
\end{definition}

\begin{definition}
The multiple duality group $\mathbb{Z}_2 \otimes \mathbb{Z}_2 \otimes \cdots \otimes \mathbb{Z}_2$ of order $N=2^n$ can be decomposed to the direct sum of 1D irreducible representations with each of them having the multiplicity as 1,
\begin{equation}
D(\mathrm{g} \in \mathbb{Z}_2 \otimes \mathbb{Z}_2 \otimes \cdots \otimes \mathbb{Z}_2 ) = A_{1}(\mathrm{g}) \oplus A_2 (\mathrm{g}) \oplus \cdots \oplus A_{2^n} (\mathrm{g}) =\bigoplus_{i=1}^{N_\tau = 2^n} a_i \mathcal{A}_i (\mathrm{g}) \,. 
\end{equation}
for multiplicity $a_1 = a_2 = \cdots a_{2^n} =1$.
\end{definition}

The proof of the above theorems are as follow, 
\begin{equation}
\begin{aligned}
&\quad |v_1\rangle^\prime \otimes| v_2^\prime \rangle \otimes \cdots \otimes |v_N^\prime \rangle \\
&= \Big( D(g_{i_1}) \,|v_1 \rangle \Big) \otimes  \Big( D(g_{i_2})\, |v_1 \rangle \Big) \otimes \cdots \otimes \Big( D(g_{i_N})\, |v_N \rangle \Big) \\
&=\Big(\,D(g_{i_1})\otimes D(g_{i_2}) \otimes \cdots \otimes D(g_{i_N}) \, \Big) \, ( |v_1 \rangle \otimes| v_2 \rangle \otimes \cdots \otimes |v_N \rangle ) \\
&= \Big[ \Big( \, \mathcal{A}_1 (g_{i_1}) \oplus \mathcal{A}_2 (g_{i_1})\Big) \otimes \Big( \mathcal{A}_1 (g_{i_2}) \oplus \mathcal{A}_2  (g_{i_2}) \Big) \otimes  \cdots \otimes \Big( \mathcal{A}_1 (g_{i_N}) \oplus \mathcal{A}_N (g_{i_N}) \Big)\,\Big] \, ( |v_1 \rangle \otimes \cdots \otimes |v_N \rangle ) \\
&= \bigg( \bigoplus_{i_1, i_2 ,\cdots , i_N=1,2} \mathcal{A}_{i_1}(g_{i_1}) \, \mathcal{A}_{i_2} (g_{i_2} ) \cdots \mathcal{A}_{i_N} (g_{i_N}) \bigg) \, ( |v_1 \rangle  \otimes |v_2 \rangle \otimes  \cdots \otimes |v_N \rangle ) \,.
\end{aligned}
\end{equation}
From the second line to the third line we have used the identity of tensor product $(A\otimes B)(u\otimes v) = Au \otimes Bv$. In the forth line we have used (\ref{eq:2sum}) for each $D(g_{i_j})$.  From the forth line to the fifth line, since all the $A_{i_j}$ must be in one dimension, therefore we can apply the distribution rule. Note that the distribution rule for tensor product cannot be generally applied for matrices that are not one dimensional (readers can check that easily). 
We can take the basis as
\begin{equation}
|V \rangle = |v_1 \rangle  \otimes |v_2 \rangle \otimes \cdots \otimes |v_N \rangle = \bigoplus_{\eta_{i_1} , \eta_{i_2} \cdots \eta_{i_N} = 0,1 } | \eta_{i_1} \rangle \otimes  | \eta_{i_2} \rangle \otimes \cdots \otimes  | \eta_{i_N} \rangle \,,
\end{equation}
such that each $\eta_{i_j} = 0,1 $. Then each $ | \eta_{i_1} \rangle \otimes  | \eta_{i_2} \rangle \otimes  | \eta_{i_N} \rangle$ would align with the $A_{i_1}(g_{i_1}) \,A_{i_2} (g_{i_2} ) \cdots A_{i_N} (g_{i_N})$. Then the set of all $| \eta_1 \eta_2 \cdots \eta_N \rangle $ form the basis of irreducible representation of $\mathbb{Z}_2 \otimes \mathbb{Z}_2 \otimes \cdots \otimes \mathbb{Z}_2$.

The consequence of multiplicity for all $A_i (\mathrm{g} )$ is equal to 1 follows directly from the fifth line. This is because there are $N=2^n$ terms for the direct sum therefore we can assign each $A_i$ by
\begin{equation}
A_i (\mathrm{g})= \mathcal{A}_{i_1} ( g_{i_1})\,\mathcal{A}_{i_2} ( g_{i_2}) \cdots \mathcal{A}_{i_N} ( g_{i_N})
\end{equation}
and explicitly we have decomposed $D(\mathrm{g})$  
\begin{equation}
D(\mathrm{g}) = D(g_{i_1})\otimes D(g_{i_2}) \otimes \cdots \otimes D(g_{i_N}) =   \bigoplus_{i_1, i_2 ,\cdots , i_N=1,2} \mathcal{A}_{i_1}(g_{i_1}) \, \mathcal{A}_{i_2} (g_{i_2} ) \cdots \mathcal{A}_{i_N} (g_{i_N})  = \bigoplus_{i=1}^{N_\tau = 2^n} a_i \mathcal{A}_i (\mathrm{g}) 
\end{equation}
that $a_i =1$ for all $i$. Note that each $A_i (\mathrm{g})$ must be either $1$ or $-1$ as it is the products of $1$s and $-1$s , and this comes from the fact that character of the parity group can only be $1$ or $-1$. 

With the basis defined, now we can construct vector. First consider the vector for $\mathbb{Z}_2$, which is a qubit
\begin{equation}
|\psi \rangle = c_0 |0 \rangle + c_1 |1\rangle = c_1 | \pmb{\boldsymbol{\cdot\cdot}} \rangle + c_2 |- \rangle \,,
\end{equation}
where $|c_1|^2 + | c_2 |^2 =1$. This is demanded by the probability of observing the $|0\rangle $ state being $|c_1|^2$ and that of $|c_2|^2$ for the $|1 \rangle$ state. For example, the simplest case would be, having half of the probabilities for getting each state,
\begin{equation}
|\psi \rangle = \frac{1}{\sqrt{2}} ( |0\rangle + | 1 \rangle ) \,.
\end{equation}
In a more formal way to represent the $j^{\rm th}$ state of the parity group we write
\begin{equation}
| \psi_j \rangle =  \sum_{\eta_j = 0,1} c^{(j)}_{\eta_{j}} | \eta_{j} \rangle \,,
\end{equation}
with normalization of
\begin{equation} \label{eq:individualNormalization}
|c^{(j)}_0 |^2 + |c^{(j)}_1 |^2 = 1 \quad \text{for all}\,\,j = 1,2,\cdots, N \,.
\end{equation}
The general state vector with $N$ tensor product of individual vector is given by
\begin{equation}
| \Psi \rangle = | \psi_1 \rangle \otimes | \psi_2 \rangle \otimes \cdots \otimes | \psi_N \rangle = \sum_{\eta_{i_1} , \eta_{i_2}, \cdots , \eta_{i_N} = 0 ,1} c_{\eta_{i_1}}^{(1)} c_{\eta_{i_2}}^{(2)} \cdots c_{\eta_{i_N}}^{(N)} | \eta_{i_1} \rangle \otimes | \eta_{i_2} \rangle \otimes \cdots \otimes | \eta_{i_N} \rangle \,.
\end{equation}
Then we have the tensor component as
\begin{equation}
T_{i_1 \, i_2  \,\cdots \, i_N} =  c_{\eta_{i_1}}^{(1)} c_{\eta_{i_2}}^{(2)} \cdots c_{\eta_{i_N}}^{(N)} \,.
\end{equation}

And we demand the completeness relation by the sum of the probability of each tensor product state be unity, 
\begin{equation} \label{eq:overallNormalization}
\sum_{\eta_{i_1} , \eta_{i_2}, \cdots , \eta_{i_N} = 0 ,1}  |c_{\eta_{i_1}}^{(1)} c_{\eta_{i_2}}^{(2)} \cdots c_{\eta_{i_N}}^{(N)} |^2  =1 \,.
\end{equation}
Then we have to solve (\ref{eq:individualNormalization}) and (\ref{eq:overallNormalization}) simultaneously. The general solution is given simply by
\begin{equation}
c^{(1)}_1 = \pm \sqrt{1- |c_0^{(1)}|^2} \,, \quad  c^{(2)}_1 = \pm \sqrt{1- |c_0^{(2)}|^2} \,, \cdots \, , c^{(N)}_1 = \pm \sqrt{1- |c_0^{(N)}|^2} \,,
\end{equation}
where the set of combinations of all possible sign orientation of $+$ and $-$ form the full solution set, where we can write it as $\{ c^{(1)}_1 ,  c^{(2)}_1 , \cdots c^{(N)}_1  \}$. Since each  $c^{(j)}_1 $ can be $+$ve or $-$ve, then there would be a total of $2^n$ solutions, i.e. the cardinality of the solution set is just $N=2^n$. This solution set satisfies both (\ref{eq:individualNormalization}) and (\ref{eq:overallNormalization}).

The simplest case for the general $n$-level solution would be
\begin{equation}
| \Psi \rangle  = \frac{1}{\sqrt{N}} \sum_{\eta_{i_1} , \eta_{i_2}, \cdots , \eta_{i_N} = 0 ,1}  | \eta_{i_1} \rangle \otimes | \eta_{i_2} \rangle \otimes \cdots \otimes | \eta_{i_N} \rangle  \,,
\end{equation}
so that the probability of getting each state is $1/N$.

\subsubsection{Heterogeneous basis}
The above study has illustrated the tensor product representation of the basis of same representation of $\mathbb{Z}_2$, i.e.
\begin{equation}
\Big( \, D(g) \otimes D(g) \otimes \cdots \otimes D(g)\, \Big)\, ( |v \rangle \otimes | v \rangle \otimes \cdots \otimes |v \rangle  ) \,.
\end{equation}
Now we would like to extend the study to the different representations of the basis of the same vector basis, that means
\begin{equation}
\Big( \, D_1 (g) \otimes D_2 (g) \otimes \cdots \otimes D_N (g)\, \Big)\, ( | v_1 \rangle  \otimes |  v_2 \rangle \otimes \cdots \otimes |v_N \rangle  ) \,,
\end{equation}
such that
\begin{equation}
| v_j \rangle = 
\begin{pmatrix}
|0_j \rangle \\
|1_j \rangle
\end{pmatrix}
\,.
\end{equation}
where each $| v_j \rangle$ is not necessarily same as $| v_k \rangle$. For example for $N= 2^6 = 64$ case, we have a particular basis as
\begin{equation}
| 1_1 0_2 0_3 1_4 1_5 0_6 \rangle\,,
\end{equation} 
which is the basis of $V_1^* \otimes V_2 \otimes V_3 \otimes V_4^* \otimes V_5^* \otimes V_6$. 

\subsection{Binary and diagramatic representation of multi-duality}
For simplicity, we can express all the above concept with the aid of diagrams. And since multi-duality deals with basis of 0s and 1s, it is convenient to use binary representation for the purpose. For two states $|0\rangle$ and $|1\rangle$, consider the following diagramatic approach,
\begin{table}[H]
\begin{center}
\begin{tabular}{l | c c} 
\hline
Decimal representation & 0 & 1  \\
Diagram &
\includegraphics[trim=0cm 0cm 0cm 0cm, clip, scale=0.65]{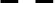} &  
\includegraphics[trim=0cm 0cm 0cm 0cm, clip, scale=0.65]{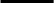} 
\\
Binary representation & 0 & 1 \\
\hline
\end{tabular}
\end{center} 
\end{table} 

Now for second order
\begin{equation}
(V \oplus V^*) \otimes (V \oplus V^*) = (V \otimes V) \oplus (V \otimes V^*) \oplus (V^* \otimes V) \oplus (V^* \otimes V^*) \,,
\end{equation}
which has basis $|00\rangle$, $|01\rangle$, $|10\rangle$ and $|11\rangle$ respectively. Diagramatically,
\begin{table}[H]
\begin{center}
\begin{tabular}{l | c c c c} 
\hline
Decimal representation & 0 & 1 & 2 & 3 \\
Diagram & 
\includegraphics[trim=0cm 0cm 0cm 0cm, clip, scale=0.65]{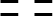} &  
\includegraphics[trim=0cm 0cm 0cm 0cm, clip, scale=0.65]{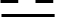} & 
\includegraphics[trim=0cm 0cm 0cm 0cm, clip, scale=0.65]{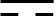} &  
\includegraphics[trim=0cm 0cm 0cm 0cm, clip, scale=0.65]{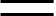} 
\\
Binary representation & (00) & (01) & (10) & (11) \\
\hline
\end{tabular}
\end{center} 
\end{table}

For third order, we have $(V \oplus V^*) \otimes (V \oplus V^*) \otimes (V \oplus V^*) $ and have 8 terms upon expansion. This corresponds to 8 basis, $|000\rangle, |001\rangle, |010\rangle, |011\rangle, |100\rangle, |101\rangle, |110\rangle, |111\rangle$, which can be expressed diagramatically. For convenience we also introduce spectral terms for them.
\begin{table}[H]
\begin{center}
\begin{tabular}{l | c c c c c c c c } 
 \hline
Decimal representation & 0 & 1 & 2 & 3 & 4 & 5 & 6 & 7 \\ 
 Diagram & 
\includegraphics[trim=0cm 0cm 0cm 0cm, clip, scale=0.65]{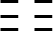} &  
\includegraphics[trim=0cm 0cm 0cm 0cm, clip, scale=0.65]{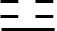} & 
\includegraphics[trim=0cm 0cm 0cm 0cm, clip, scale=0.65]{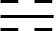} &  
\includegraphics[trim=0cm 0cm 0cm 0cm, clip, scale=0.65]{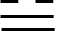} &
\includegraphics[trim=0cm 0cm 0cm 0cm, clip, scale=0.65]{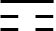} & 
\includegraphics[trim=0cm 0cm 0cm 0cm, clip, scale=0.65]{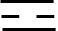} &
\includegraphics[trim=0cm 0cm 0cm 0cm, clip, scale=0.65]{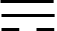} &
\includegraphics[trim=0cm 0cm 0cm 0cm, clip, scale=0.65]{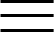} 
  \\ 
Binary representation & (000) & (001) & (010) & (011) & (100) & (101) & (110) & (111) \\ 
Spectural term & K & z & k & d & g & l & x & q \\ 
 \hline
\end{tabular} 
\end{center} 
\end{table}

For forth order, there are 16 basis and it can be formed by the tensor product of two second ordered diagrams, which is shown in figure 19.
\begin{figure}[H]
\centering
\includegraphics[trim=0cm 0cm 0cm 0cm, clip, scale=0.7]{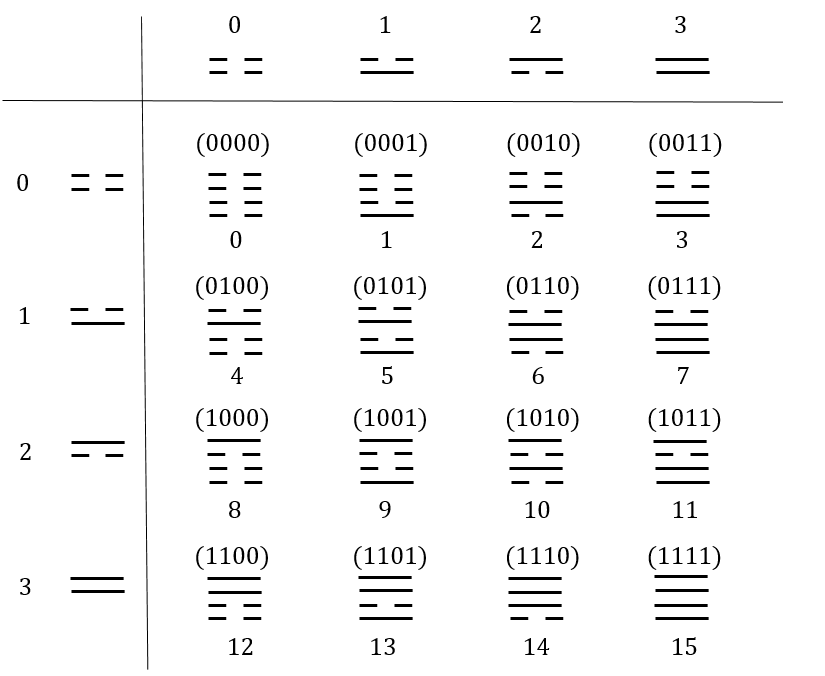}
\caption[64-Gua]
{The 16 basis constructed from the tensor product of two second ordered diagrams. Each number located under the basis diagram is the decimal representation.  \label{fig:16Gua}}
\end{figure}

For higher order, it can be similarly constructed from tensor product of lower ordered diagrams. For example, the 6-th order can be constructed from the tensor product of two third ordered diagrams, which is presented in figure 20. 
\begin{figure}[H]
\centering
\includegraphics[trim=0cm 0cm 0cm 0cm, clip, scale=0.7]{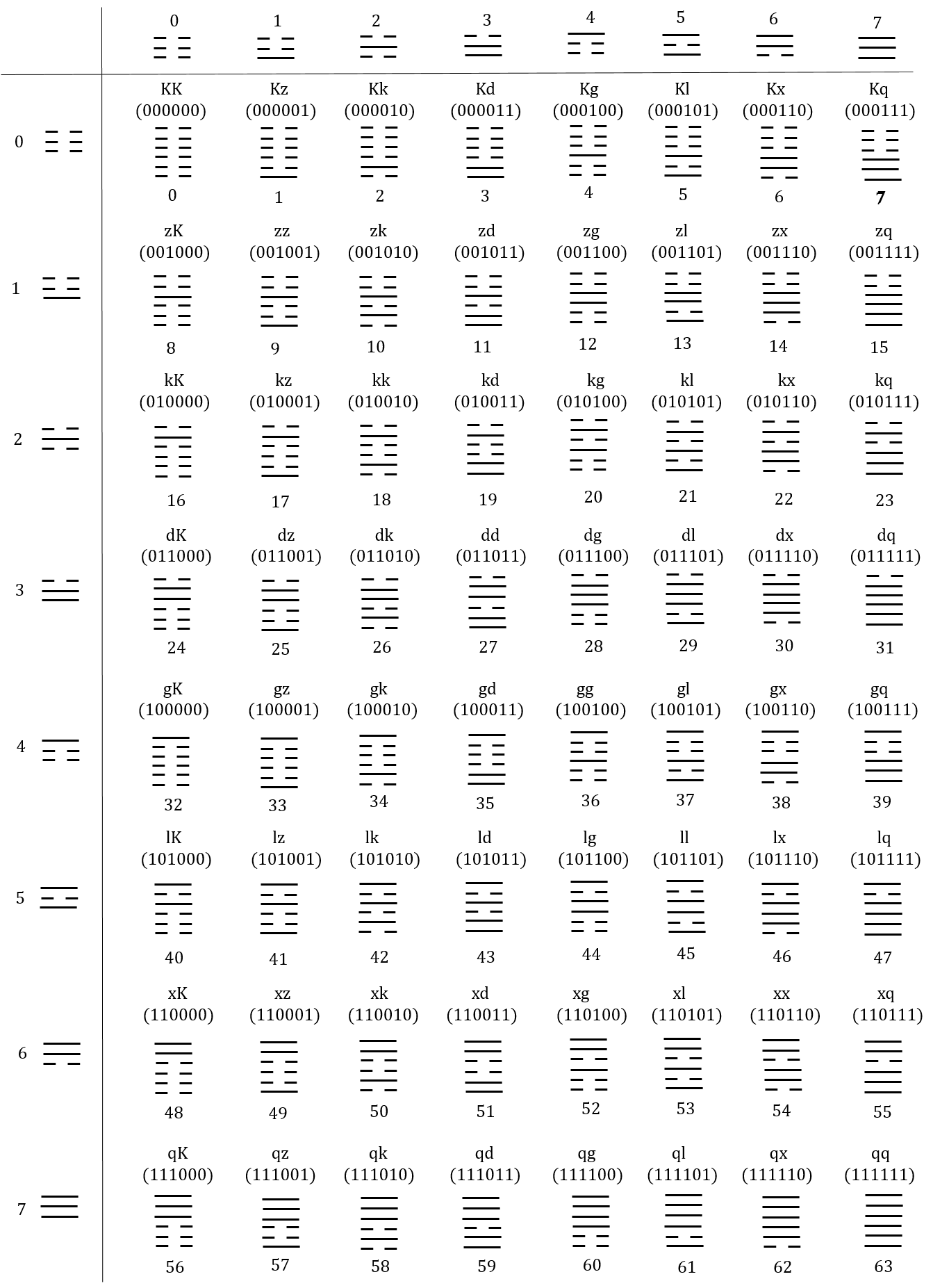}
\caption[64-Gua]
{ The 64 basis constructed from the tensor product of two third ordered diagrams. Each number located under the basis diagram is the decimal representation  \label{fig:64Gua}}
\end{figure}

\subsection{Chiral representation and matrix basis}
We have previously introduced basis representation of the $\mathbb{Z}_2$ and the $\mathbb{Z}_2 \times \mathbb{Z}_2 $ group. Now we will like to see in a more advanced view that the basis are promoted to matrices. In particular we would like to see how this can be related to the context of gamma matrices in fermionic quantum field theory.

Consider that the parity group $\mathbb{Z}_2$ being represented in real general linear space GL(4, $\mathbb{R}$) of 4-dimension that is aroused from Clifford algebra,
\begin{equation}
\{ \gamma^{\mu}, \gamma^{\nu} \} = 2 \eta^{\mu\nu} \pmb{1} \,,
\end{equation}
where $\mu, \nu = 0,1,2,3$ are the Lorentz indices. The identity element is the $4\times 4$ identity matrix and the parity element is the Dirac $\gamma^5$ matrix, where
\begin{equation}
\gamma^5 = i\gamma^0 \gamma^1 \gamma^2 \gamma^3 \,.
\end{equation}
Then the parity group is $\mathbb{Z}_2 = \{ \pmb{1}, \gamma^5 \}$, and we know that $(\gamma^5 )^2 = \pmb{1}$ and is independent of representation (Dirac, Weyl, etc). 

Now recall that in QFT, the global axial (chiral) transformation for a spinor is
\begin{equation}
\psi^\prime (x) = e^{i\theta \gamma^5} \psi(x) \,,
\end{equation} 
where $\theta$ is the global phase (independent of spacetime $x$) of the chiral transformation. And it is easy to show that
\begin{equation} \label{eq:chiraltransformation}
U(\theta) =e^{i\theta \gamma^5} = \cos \theta \,\pmb{1} + i \sin \theta \, \gamma^5 \,.
\end{equation}

Recall that a generic state vector for the duality group can be written as \footnote{Note that the  $|\psi \rangle$ state here has nothing to do with with the spinor field $\psi (x)$ above, readers should be confused by the notation ambiguity. }
\begin{equation}
| \psi \rangle = \cos \theta \,|0\rangle + \sin \theta \, |1 \rangle = \cos \theta \,|--\rangle + \sin \theta \, |- \rangle,
\end{equation}
thus comparing to equation \ref{eq:chiraltransformation}, we can identity the matrix group elements $\pmb{1}, \gamma^5$ of $\mathbb{Z}_2$ as the the basis of the group by
\begin{equation}
|0 \rangle = \pmb{1} \quad \text{and} \quad |1 \rangle = i\gamma^5 \,.
\end{equation} 
Thus the basis can be considered as the matrix group element of $\mathbb{Z}_2$ itself in the chiral representation. 

Therefore, under the basis matrix representation, the tensor product transformation is,
\begin{equation}
U(\theta_1 ) \otimes  U(\theta_2 ) \otimes \cdots \otimes U(\theta_n )=   e^{i\theta_1 \gamma^5} \otimes e^{i\theta_2 \gamma^5} \otimes \cdots \otimes e^{i\theta_n \gamma^5}  \,.
\end{equation}
In full expansion we have
\begin{equation}
 e^{i\theta_1 \gamma^5} \otimes e^{i\theta_2 \gamma^5} \otimes \cdots \otimes e^{i\theta_n \gamma^5} =\bigotimes_{j=1}^n (\cos \theta_j \,\pmb{1} + i \sin \theta_j \, \gamma^5 )\,.
\end{equation}
For example if $n=5$ we can have a particular term like
\begin{equation}
(i^3 \cos \theta_1 \sin \theta_2 \cos \theta_3 \sin\theta_4 \sin\theta_5 )\,\pmb{1} \otimes \gamma^5 \otimes \pmb{1} \otimes \gamma^5 \otimes \gamma^5 \,.
\end{equation}

An important case would be $n=2$, then we have the matrix basis for the 4-duality group $\mathbb{Z}_2 \times \mathbb{Z}_2$, in which
\begin{equation}
\{ \pmb{1}\otimes \pmb{1}, \, \pmb{1}\otimes \gamma^5 , \, \gamma^5 \otimes \pmb{1}, \,\gamma^5 \otimes \gamma^5 \} \mapsto \{ |00\rangle, |01\rangle \, , |10\rangle , |11\rangle \} \,.
\end{equation}
The rank-2 tensor components are
\begin{equation}
T_{ij}  = 
 \begin{pmatrix}
  \cos\theta_1 \cos\theta_2 & i\cos\theta_1 \sin\theta_2  \\
  i\sin\theta_1 \cos\theta_2 & -\sin\theta_1 \sin\theta_2 \\
 \end{pmatrix}
 \,.
\end{equation}
and $\mathrm{det}\,T_{ij} =0$, in which all states are unentangled. 

Finally we would like to define the orthogonality relation of the matrix basis. Recall that we have $\langle 0 |0 \rangle = \langle 1 | 1 \rangle =1$ and $\langle 0 | 1 \rangle = \langle 1 | 0 \rangle =0$. For matrix basis, we can define such by trace. For $g_i , g_j \in \mathbb{Z}_2$,
\begin{equation}
\mathrm{Tr}  \big(\, D(g_i) D(g_j ) \,  \big) = 2 \delta_{ij} \,.
\end{equation}

This follows nicely from that fact that $\mathrm{Tr}\,\gamma^5 = 0$. We can explicitly check that
\begin{equation}
\mathrm{Tr}( \pmb{1} \cdot \pmb{1})=\mathrm{Tr}( \gamma^5 \cdot \gamma^5 ) =2 \quad \text{and} \quad \mathrm{Tr}( \pmb{1} \cdot \gamma^5) = \mathrm{Tr}( \gamma^5 \cdot \pmb{ 1} ) = 0 \,.
\end{equation}

The canonical form for the graded chiral algebra is
\begin{equation}
U(\theta_1) \oplus U(\theta_1) \otimes U(\theta_2) \oplus \cdots \oplus U(\theta_1) \otimes U(\theta_2) \otimes \cdots \otimes U(\theta_n) \,.
\end{equation}

\subsection{Comparison Representation}
In the above studies, we have shown how to represent the $\mathbb{Z}_2^N $ group in duality basis. We can further form duality basis by comparing two diagrams. We will use the notation $(, :: ,)$. Let's $V \times V$ be the comparison vector space of the basis, we have 
\begin{equation}
(,\, ::\, ,)\, : V \times V \rightarrow V \,.
\end{equation}
We compare the two diagrams level by level. If the two levels are of the same state, we assign it as $|0^\prime \rangle$, and $|1^\prime \rangle$ if not. Here we use the primed subscript to indicate this is a new emergent duality basis that is formed by comparison, but for simplicity one can drop such indication as being understood. At the end they are still duality basis. The most fundamental comparison begins from the $n=1$ level,
\begin{equation}
\begin{aligned}
(0 :: 0 ) &= 1 \,, \\
(0 :: 1 ) &= 0 \,, \\
(1 :: 0 ) &= 0 \,, \\
(1 :: 1 ) &= 1 \,.
\end{aligned}
\end{equation}
This can also be referred as the same rule as the normal multiplication for the signs, $-- \rightarrow -$, $-+ \rightarrow +$, $+- \rightarrow -$ and $++ \rightarrow +$. 

For higher $n$, let's take $n=3$ for work out some examples.
\begin{equation}
(010 :: 110 ) = ( 011 ) \,\, , \,\, (111 :: 010 ) = ( 010 ) \,,
\end{equation} 
or diagrammatically
\begin{equation}
(\includegraphics[trim=0cm 0cm 0cm 0cm, clip, scale=0.5]{2.png} :: \includegraphics[trim=0cm 0cm 0cm 0cm, clip, scale=0.5]{6.png} ) = \includegraphics[trim=0cm 0cm 0cm 0cm, clip, scale=0.5]{3.png}  \,\,\,\, , \,\,\,\,
(\includegraphics[trim=0cm 0cm 0cm 0cm, clip, scale=0.5]{7.png} :: \includegraphics[trim=0cm 0cm 0cm 0cm, clip, scale=0.5]{2.png} ) = \includegraphics[trim=0cm 0cm 0cm 0cm, clip, scale=0.5]{2.png}  \,,
\end{equation}
and in decimal places,
\begin{equation}
( 1:: 6) = 3 \,, \,\, (7 :: 2 ) = 2 \,.
\end{equation}
The comparison map can work compositely and satisfies the following axioms. 
\begin{definition}
Let $a,b,c$ be states representation and $(, :: ,)$ be the comparison map, and the composition
\begin{equation}
(, \, :: \,,\, :: \,, \cdots , :: , \, ) \, : V \times V \times \cdots \times V \rightarrow V \,. 
\end{equation}
satisfies the the following axiom
\begin{itemize}
\item $(a::b)::c = a::(b::c)$ \,\,(associativity)
\item $a::b = b::a$ \,\, (commutativity)
\item $a::1 = a$
\item $a::a = 1$
\item $a = a^{-1} $ \,\,(self inverse)
\end{itemize}
\end{definition}
The last axiom directly comes from the second-last axiom. (Note that $1= 1\cdots 1$ and $0= 00 \cdots 0$).
From this definition it follows that
\begin{equation}
b::a ::b =a \,.
\end{equation}
This is because 
\begin{equation}
\begin{aligned}
a::b &= b::a \\
b:: a :: b & = (b::b)::a \\
b:: a :: b &= 1::a \\
b:: a :: b &= a
\end{aligned}
\end{equation}
For example
\begin{equation}
( 101::011:: 000:: 110) = (101::011:: 001) = (101 :: 101 ) = 111\,,
\end{equation}
or diagrammatically
\begin{equation}
(\includegraphics[trim=0cm 0cm 0cm 0cm, clip, scale=0.5]{5.png} :: \includegraphics[trim=0cm 0cm 0cm 0cm, clip, scale=0.5]{3.png} :: \includegraphics[trim=0cm 0cm 0cm 0cm, clip, scale=0.5]{0.png}  :: \includegraphics[trim=0cm 0cm 0cm 0cm, clip, scale=0.5]{6.png}) = (\includegraphics[trim=0cm 0cm 0cm 0cm, clip, scale=0.5]{5.png} :: \includegraphics[trim=0cm 0cm 0cm 0cm, clip, scale=0.5]{3.png} :: \includegraphics[trim=0cm 0cm 0cm 0cm, clip, scale=0.5]{1.png} ) = (\includegraphics[trim=0cm 0cm 0cm 0cm, clip, scale=0.5]{5.png}::\includegraphics[trim=0cm 0cm 0cm 0cm, clip, scale=0.5]{5.png}) =\includegraphics[trim=0cm 0cm 0cm 0cm, clip, scale=0.5]{7.png}
\end{equation}
and in decimal places
\begin{equation}
(5::3 :: 0 :: 6 ) = (5:: 3 :: 1) = (5::5) =1\,.
\end{equation}

There are some general rules that apply to any $n$. First the comparison is 
For $l < n-1$, we have
\begin{equation}
(0 :: l ) = n - l  \,\,\,\, \text{and} \,\,\, ( 2^n :: l ) = l \,.
\end{equation}
We also have
\begin{equation}
( 0 :: 1 :: 2 :: \,, \cdots \,, :: 2^n -1 ) = 2^{n-1} \,, 
\end{equation}
that means of the comparison of all states always return to $(111\cdots 1)$.

\subsection{2-Level, 4-level and General $N=2^n$ Quantum Dual systems}
A dual system with opposite element can be expressed in terms of a qubit,
\begin{equation}
|\psi \rangle =\alpha |u\rangle + \beta |u^* \rangle = \alpha |0\rangle + \beta |1\rangle = \alpha |--\rangle + \beta |-\rangle \,,
\end{equation}
satisfying $|\alpha|^2 + |\beta|^2 =1$ and in particular we are interested in the case for which $\alpha$ and $\beta$ are real. The $|u^* \rangle$ state is the dual of $|u\rangle$, in such $*|u\rangle = |u^*\rangle$ and $*|u^*\rangle = |u\rangle$, with $**=1$ the identity map,  also $\langle u | u^* \rangle =0$. The $\alpha$ values and $\beta$ values are the weight for each of the state respectively. A common form takes the following
\begin{equation}
|\psi \rangle = \cos \frac{\theta}{2} | -- \rangle + \sin \frac{\theta}{2} | - \rangle = \cos \frac{\theta}{2} | 0 \rangle + \sin \frac{\theta}{2} | 1 \rangle \,,
\end{equation}
where $\theta = \omega t$ such that the coefficient of the dual state is oscillatory. The factor of $\frac{1}{2}$ is chosen by convenience so that the probability is at extreme at $k\pi$. Note that we also infer $|-\rangle \equiv |+ \rangle$ as the $|1\rangle$ state and $|--\rangle \equiv |-\rangle $ as the  $|0 \rangle$ state. Suppose the $|--\rangle $ state has energy $E_{--} $ (or $E_0$) and the $|-\rangle$ state has energy $E_{-}$ (or  $E_1) $, then the expectation value of the energy of the system is given by
\begin{equation}
\langle E \rangle = \sum_{j=0,1} |\langle j | \psi \rangle |^2 E_j \,,
\end{equation}
which is
\begin{equation}
\langle E(\theta) \rangle = \cos^2 \frac{\theta}{2} \, E_{-} + \sin^2 \frac{\theta}{2} \, E_{+} \,.
\end{equation}
Thus the oscillatory growing and decreasing of dual energy is described by the probability
\begin{equation}
P_- = \cos^2 \frac{\omega t}{2} \,\,\,\, \text{and} \,\,\,\, P_+ = \sin^2 \frac{\omega t}{2} \,.
\end{equation} 
\begin{figure}[H]
\centering
\includegraphics[trim=0cm 0cm 0cm 0cm, clip, scale=0.6]{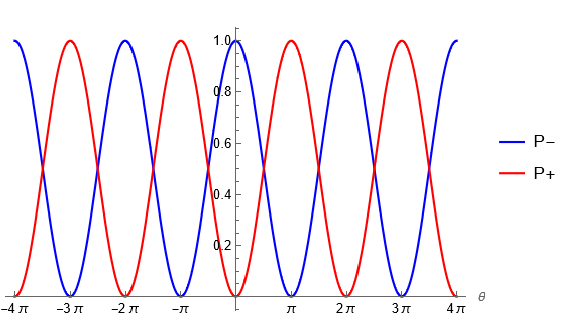}
\caption[Analysis]
{ The probability of $|0\rangle$ and $|1\rangle$ by a given phase. Both probabilities  are the same i.e. $P_+ = P_- = \frac{1}{2}$ at phases $\frac{\pi}{2}, \frac{3\pi}{2}$ , \text{etc}.\label{fig:ProbabilityOscillation}}
\end{figure}
This describes the philosophy of dualism in a mathematical way. The extremes are points at integers of $\pi$, which represent the phases. When the probability of the 0 state rises to its maximum it must fall again where the probability of 1 state grows. When the probability of the 0 state drops to its minimum it must rise again where the probability of 1 state drops. Similarly, the probability of the 1 state rises to its maximum it must fall again where the probability of yin 0 grows. When the probability of the 1 state drops to its minimum it must rise again where the probability of 0 state drops. And 0 and 1 states are in opposition as they do not overlap each other given by the fact that the 0 state and 1 state are orthogonal to each other, $\langle 0 | 1 \rangle = \langle 1 |0 \rangle =0$.

At extreme phase values, this gives the full 0 state or full 1 state with deterministic probability of 1. For example for the full 0 state, at $t=0$
\begin{equation}
|\psi (0) \rangle = | --\rangle\,.
\end{equation}
Suppose now the phase evolve a little such that $t>0$, then we ought to have
\begin{equation}
|\psi (0+\delta) \rangle = (1-\epsilon)| --\rangle + \eta |-\rangle \,,
\end{equation} 
where $\delta, \epsilon\,\text{and} \,\eta$ are small. This means the portion of 0 state is decreasing and the portion of 1 state begins to grows. This is the philosophy that ``at the extreme 0 becomes 1". To be exact, the probability of 0 state becomes less and the probability of 1 state grows a bit, so this gives an illusion that some 0 state converts to 1 state but in fact it is just the probability change. Using small angle approximation, for small $\delta$
\begin{equation}
|\psi (\delta) \rangle = \cos \frac{\delta}{2}|--\rangle + \sin\frac{\delta}{2}|-\rangle \approx \bigg( 1- \frac{\delta^2}{4}\bigg)|--\rangle + \frac{\delta}{2}|-\rangle\,. 
\end{equation}
Therefore we identify $\epsilon = \frac{\delta^2}{4}$ and $\eta=\frac{\delta}{2}$. 

Similarly for the full 1 state,
\begin{equation}
|\psi (\pi) \rangle = | -\rangle\,.
\end{equation}
Now suppose the phase evolve a little bit, let's say before the phase $\pi$, we ought to have
\begin{equation}
|\psi (\pi- \delta) \rangle = \eta|--\rangle + \bigg(1-\frac{\delta^2}{4} \bigg)| -\rangle\,.
\end{equation}
This means the portion of 1 state is decreasing and the portion of 0 state begins to grow. This is the philosophy that ``at the extreme 1 becomes 0''. To be exact, the probability of 1 state becomes less and the probability of 0 state grows a bit, so this gives an illusion that some 1 states converts to 0 state but in fact it is just the probability change. Again, using small angle approximation,

\begin{equation}
\begin{aligned}
|\psi (\pi- \delta) \rangle &= \cos\bigg(\frac{\pi}{2}-\frac{\delta}{2}\bigg)|--\rangle + \sin\bigg(\frac{\pi}{2}-\frac{\delta}{2}\bigg)|-\rangle \\
&=\sin\frac{\delta}{2}|--\rangle + \cos\frac{\delta}{2}|-\rangle \\
&\approx \frac{\delta}{2}|--\rangle + \bigg(1-\frac{\delta^2}{4}\bigg)|-\rangle \,.
\end{aligned}
\end{equation}
Therefore again we identify $\epsilon = \frac{\delta^2}{4}$ and $\eta=\frac{\delta}{2}$.

The first order derivative of the expectation energy is
\begin{equation}
\frac{d\langle E(\theta) \rangle}{d\theta} = \frac{1}{4}\sin\theta ( E_{+} - E_{-} ) \,.
\end{equation}
Thus the expectation energy is at extreme when $\sin \theta =0$ or $ \Delta E = E_{+} - E_{-} = 0$, i.e. $\theta = \pm k\pi$ or $E_{+} = E_{-}$. Whether it is the minimum or maximum can be checked by the sign nature of the second order derivative,
\begin{equation}
\frac{d^2\langle E(\theta) \rangle}{d\theta^2} = \frac{1}{4}\cos\theta ( E_{+} - E_{-} ) \,.
\end{equation}
If $E_{+} > E_{-}$, then minimum occurs at $\theta = 2k \pi$ and maximum occurs at $\theta = \pm(2k +1) \pi$.

Suppose the energy of the $|1\rangle$ state $E_{+} > 0 $ is positive and the  energy if the $|0\rangle$ state $E_{-} <0$ is negative, and their magnitudes are the same $|E_+ | = |E_{-} | = E$, then we have
\begin{equation}
\langle E(t) \rangle = E \cos\, \omega t \,.
\end{equation}
The energy of the dual system oscillate with time.

Next we would like to use the notion of heterogeneous bases for the $\mathbb{Z}_2 \times \mathbb{Z}_2$ group to describe probability relations between the $|0\rangle$ and $|1\rangle$ states along the flow of time $t$. We focuses on the study in one period, $0 \leq \theta < 2\pi$. Define the original $|0\rangle$, $|1\rangle$ states as $|0_1 \rangle = |0\rangle$ and $|1_1 \rangle = | 1\rangle$, and the two dual states of tendency, $|0_2 \rangle = | \downarrow \rangle $ and $|1_2 \rangle = | \uparrow \rangle$. The $| \downarrow \rangle$ state describes the state whenever the probability of $|0\rangle$ or $|1\rangle$ state is decreasing, while the $|\uparrow\rangle$ state describes the state whenever the probability of $|0\rangle$ or $|1\rangle$ state is increasing.  Then we have the heterogeneous basis for the $\mathbb{Z}_2 \times \mathbb{Z}_2$ group as $\{|0_1 0_2 \rangle, |0_1 1_2\rangle, |1_1 0_2 \rangle, |1_1 1_2 \rangle   \}$. 
For $0\leq\theta < \pi$, we have
\begin{equation}
|0 \downarrow \rangle  \,\,\,\,\equiv \,\,\,\, | 1 \uparrow \rangle\,,
\end{equation}
for another half $\pi\leq\theta < 2\pi$ we have
\begin{equation}
|0 \uparrow \rangle \,\,\,\, \equiv  \,\,\,\, | 1 \downarrow \rangle\,.
\end{equation}
The heterogeneous basis of $\mathbb{Z}_2 \times \mathbb{Z}_2 $ can be also interpreted as objects of element with observer frame. One can take $\downarrow$ as the $S_k$ observer and $\uparrow$ as the $S_{k}^\star$ observer, such that we have
\begin{equation}
(0 | S_k ) \,\,\,\, \equiv \,\,\,\, ( 1 | S_k^\star) \quad \quad \text{and} \quad\quad (1 | S_k ) \,\,\,\, \equiv \,\,\,\, ( 0 | S_k^\star) \,.
\end{equation}
So we can apply the operators $I$,$*$, $\star$ and $*\circ \star$ to these objects. We can represent these by a 4-tableau,
\begin{figure}[H]
\centering
\includegraphics[trim=0cm 0cm 0cm 0cm, clip, scale=0.6]{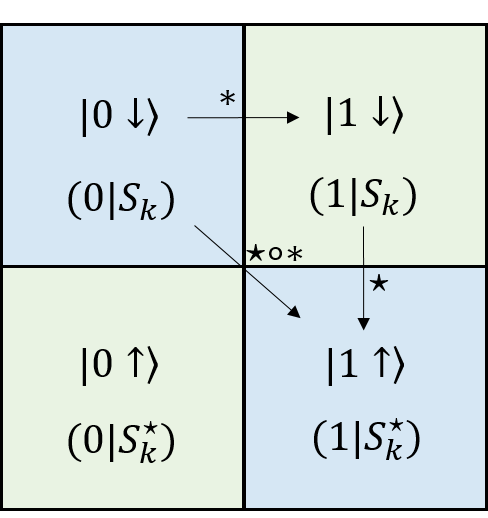}
\caption[dualrelations]
{ The 4-tableau representation of a dual system with heterogeneous basis representation of the 4-duality group. The light green boxes represent the phase regime of $0 \leq \theta < \pi$, the light blue boxes represent the phase regime of $\pi \leq \theta < 2\pi$. Objects with same colour are equipped with equivalent relations. \label{fig:dualityrelation}}
\end{figure}
It is useful to study the entropy of the system. The Shannon entropy is given by
\citep{Shannon1, Shannon2}
\begin{equation}
H=-\sum_{i}p_i \log p_i \,,
\end{equation}
where $p_i$ is the probability of each state. The entropy of the 2-level quantum system is 
\begin{equation}
H(\theta) = -2\cos^2 \frac{\theta}{2} \log  \Big\vert\cos \frac{\theta}{2} \Big\vert -2 \sin^2 \frac{\theta}{2} \log  \Big\vert \sin \frac{\theta}{2} \Big\vert\,.
\end{equation}
Graphically, the entropy is 
\begin{figure}[H]
\centering
\includegraphics[trim=0cm 0cm 0cm 0cm, clip, scale=0.6]{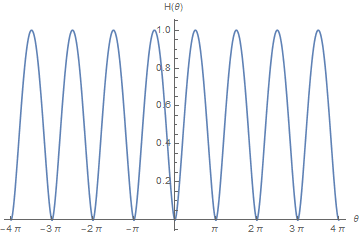}
\caption[Entropy of a 2-level system.]
{ Entropy of a 2-level system. The plot is generated in the range of $-4\pi \leq \theta_1 , \theta_2 \leq 4 \pi$.\label{fig:Entropy}}
\end{figure}
The graph is continuous everywhere and is symmetric. Maximum entropy occurs at $\pm \frac{k\pi}{2}$ and is 1 bit. Minimum entropy is zero at occurs at $\pm k \pi$. 

Next let's study the case of the 4-level system. For heterogeneous basis we have
\begin{equation} \label{eq:4Levelq}
|\psi (\theta_1, \theta_2) \rangle = \cos\frac{\theta_1}{2} \cos\frac{\theta_2}{2} |0_1 0_2 \rangle +\cos\frac{\theta_1}{2}\sin\frac{\theta_1}{2} |0_1 1_2\rangle +\sin\frac{\theta_1}{2}\cos\frac{\theta_2}{2} |1_1 0_2\rangle + \sin\frac{\theta_1}{2}\sin\frac{\theta_1}{2} |1_1 1_2 \rangle \,.  
\end{equation}
The probability tensor is given by the element-wise products of the rank-2 matrices,
\begin{equation}
p_{ij}(\theta_1 ,\theta_2) = T_{ij}(\theta_1) \bullet T_{ij}(\theta_2) \bullet T_{ij}(\theta_1) \bullet T_{ij}(\theta_2) \,.
\end{equation}
Thus the expectation value of the energy is
\begin{equation}
\langle E(\theta_1 , \theta_2 )\rangle = \cos^2\frac{\theta_1}{2} \cos^2 \frac{\theta_2}{2} E_{-_1 - _2} + \cos^2 \frac{\theta_1}{2}\sin^2 \frac{\theta_2}{2} E_{-_1 + _2} + \sin^2 \frac{\theta_1}{2}\cos^2 \frac{\theta_2}{2}E_{+_1 - _1} + \sin^2 \frac{\theta_1}{2}\sin^2 \frac{\theta_2}{2} E_{+_1 + _1} \,.
\end{equation}
Here the random variable $\Theta$ is parametrized by $\theta_1$ and $\theta_2$. We have the entropy as
\begin{equation} \label{eq:Entropy4Level}
\begin{aligned}
H(\Theta) &= -2\Big(\cos\frac{\theta_1}{2}  \cos\frac{\theta_2}{2}\Big)^2 \log \Big\vert \cos\frac{\theta_1}{2} \cos\frac{\theta_2}{2} \Big\vert - 2\Big(\cos\frac{\theta_1}{2} \sin\frac{\theta_2}{2} \Big)^2 \log\Big\vert \cos\frac{\theta_1}{2} \sin\frac{\theta_2}{2} \Big\vert \\
&\quad - 2 \Big(\sin\frac{\theta_1}{2} \cos\frac{\theta_2}{2}\Big)^2 \log \Big\vert \sin\frac{\theta_1}{2} \cos\frac{\theta_2}{2} \Big\vert - 2\Big(\sin\frac{\theta_1}{2} \sin\frac{\theta_2}{2}\Big)^2\log \Big\vert \sin\frac{\theta_1}{2} \sin\frac{\theta_2}{2} \Big\vert \,.
\end{aligned}
\end{equation}
Graphically,
\begin{figure}[H]
\centering
\includegraphics[trim=0cm 0cm 0cm 0cm, clip, scale=0.7]{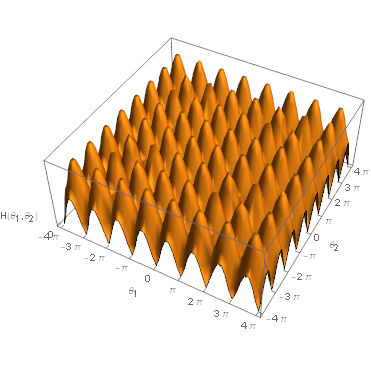}
\caption[Entropy of a 4-level system.]
{ Entropy of a 4-level system. The plot is generated in the range of $-4\pi \leq \theta_1 , \theta_2 \leq 4 \pi$.\label{fig:Entropy}}
\end{figure}
We can see that the entropy function is not continuous everywhere, it is confined in local regions of $(\theta_1 , \theta_2)$ which give discrete comb-liked peaks. The remaining regions would give imaginary entropy values, thus are not well-defined.

Next we would like to find out what what phases $(\theta_1 , \theta_2)$ the entropy is maximum,
\begin{equation}
C= \underset{\Theta}{\mathrm{max}} \, H(\Theta) =  \underset{\theta_1 , \theta_2}{\mathrm{max}} \, H(\theta_1 , \theta_2) \,.
\end{equation}
There are two ways to do so. The first one is to just apply the theorem that an even distribution has maximum entropy. In other words for our case this happens when all the states $|0_1 0_2 \rangle, |0_1 1_2 \rangle, |1_1 0_2 \rangle, |1_1 1_2 \rangle$ have the same probability. This takes place when
\begin{equation}
\begin{aligned}
|\psi \rangle &=\frac{1}{\sqrt{2}} ( |0_1 \rangle + | 1_1 \rangle  ) \, \otimes \, \frac{1}{\sqrt{2}} ( | 0_2 \rangle + | 1_2 \rangle  ) \\
&= \frac{1}{2}|0_1 0_2 \rangle +  \frac{1}{2}|0_1 1_2 \rangle  +\frac{1}{2}|1_1 0_2 \rangle + \frac{1}{2}|1_1 1_2 \rangle \,.
\end{aligned} 
\end{equation} 
Then we have an even probability distribution of
\begin{equation}
p_{00} = p_{01} = p_{10} = p_{11} = \frac{1}{4} \,.
\end{equation}
Obviously this occurs at
\begin{equation}
\cos\frac{\theta_1}{2} =\cos\frac{\theta_2}{2} =\sin\frac{\theta_1}{2} =\sin\frac{\theta_2}{2} = \frac{1}{\sqrt{2}} .
\end{equation}
The general solution $\big( \frac{\pi}{2}+2p\pi, \frac{\pi}{2}+ 2q\pi\big)$ for any positive integers $p$ and $q$.

The maximum entropy is
\begin{equation}
H_{\mathrm{max}} = -\sum_{i=1}^4 \frac{1}{4} \log \frac{1}{4} = \big(\frac{1}{4} \cdot 2 \big) \cdot 4 = 2 \,\, \text{bits}\,.
\end{equation}

For the second way, the local maxima can be found technically by solving simultaneously
\begin{equation}
\frac{\partial H}{\partial\theta_1} =0 \,\, \text{and} \,\,\frac{\partial H}{\partial\theta_2} =0 
\end{equation}
The partial derivatives are very messy, and they are difficult to solve algebraically but computationally with the aid of contour diagram we can read the peak values.

\begin{figure}[H]
\centering
\includegraphics[trim=0cm 0cm 0cm 0cm, clip, scale=0.7]{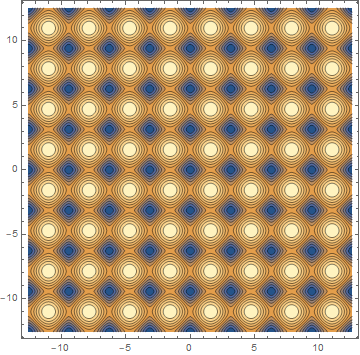}
\caption[Entropy of a 4-level system.]
{Contour plot of the phase angles $(\theta_1 , \theta_2)$ in the range of $-4\pi \leq \theta_1 , \theta_2 \leq 4 \pi$\label{fig:Entropy}}
\end{figure}
On the other hand, the lower limit of the entropy of the system tends to zero, but it cannot be exactly zero. These takes place at $\pm p\pi, \pm q\pi$.  Therefore at the beginning, half-way, and at the end the entropy tends to zero. 

The results make sense, as the phase angles $\theta_i = \omega t_i$, when $\theta=0 ,2\pi$ which are the initial time and final time, things are highly ordered with no uncertainty. But as time goes disorderness increases and reaches maximum.

Now let's reduce the above problem to the condition of same phase. There are two possible meanings for equal phase $\theta_1= \theta_2 $. It can either mean two different angular frequencies but at same time, or same angular frequency but at different times. We will consider the latter case. Suppose now the system we described above is synchronized with the same phase, such that $\theta_1 = \theta_2 = \theta$. Then our original equation (\ref{eq:4Levelq}) reduces to a 2 qubit state,
\begin{equation} \label{eq:4Level}
|\psi (\theta) \rangle = \cos^2\frac{\theta}{2}  |0 0 \rangle +\frac{1}{2}\sin\theta |0 1\rangle +\frac{1}{2}\sin\theta |1 0\rangle + \sin^2\frac{\theta}{2} |1 1 \rangle \,.  
\end{equation}
Then the energy becomes
\begin{equation}
\langle E(\theta) \rangle = \sum_{j=--,-+,+-,++} | \langle j | \psi (\theta) \rangle |^2 E_{j} \,,
\end{equation}
which is
\begin{equation}
\langle E(\theta) \rangle = \cos^4 \frac{\theta}{2}  E_{--} + \frac{1}{4}\sin^2 \theta E_{- + } + \frac{1}{4}\sin^2 \theta  E_{+ - } + \sin^4 \frac{\theta}{2} E_{+ + } \,.
\end{equation}
The probability of each state is plot as
\begin{figure}[H]
\centering
\includegraphics[trim=0cm 0cm 0cm 0cm, clip, scale=0.7]{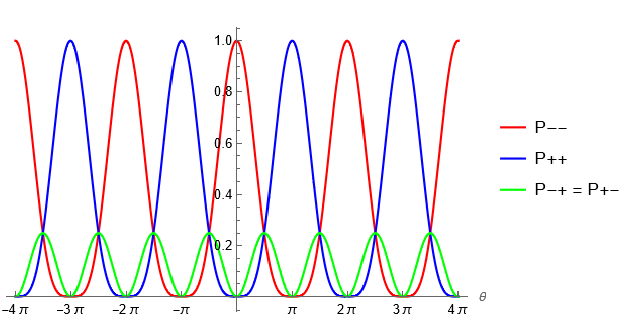}
\caption[Probability ]
{Probability of each state in the range of $-4\pi \leq \theta \leq 4 \pi$. At $\theta = \pm 2k\pi$, $|00\rangle$ is the fully dominating state. At $\theta= \pm (2k+1)\pi$, $|11\rangle$ is the fully dominiating state. At $\theta = \pm\frac{k\pi}{2}$, the system is evenly distributed in all the $|00\rangle , |01 \rangle, |10\rangle , |11\rangle$ states with equal probability of $\frac{1}{4}$.  \label{fig:prob2}}
\end{figure}
We can diagrammatically describe this with the phase evolution of 4-tableau. In particular we are interested in the probability of the states at integer and half-integer multiple of $\pi$. 
\begin{figure}[H]
\includegraphics[trim=0cm 0cm 0cm 0cm, clip, scale=0.42]{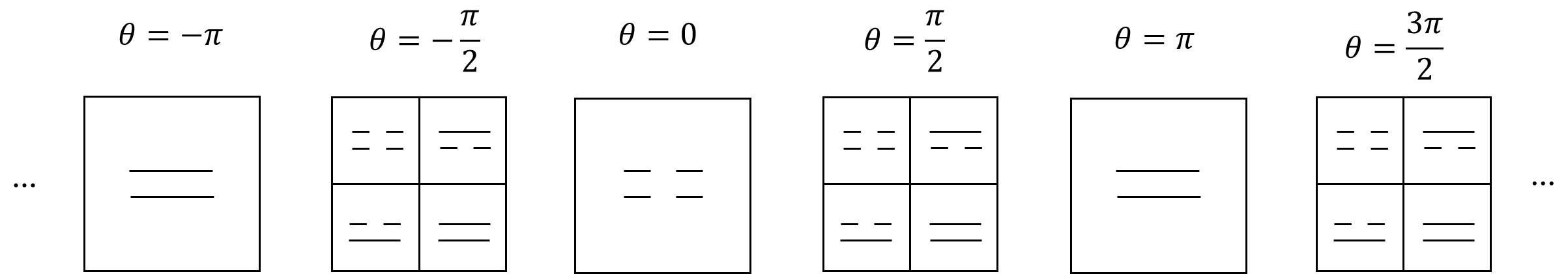}
\caption[Phase evolution of 4-tableau ]
{Phase evolution of 4-tableau.  \label{4dualityoscillation}} 
\end{figure}
We assign the area of outer big square as 1. Then the area of each quarter square as $\frac{1}{4}$. We can map the area of the sub-squares to the probability of each state. Equivalently, we can represent the one big box of full-0 state and one big box of full-1 state in (\ref{fig:4dualityoscillation}) as
\begin{figure}[H]
\centering
\includegraphics[trim=0cm 0cm 0cm 0cm, clip, scale=0.6]{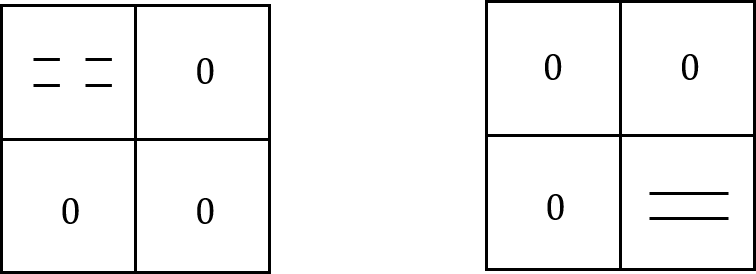}
\caption[ ]
{ \label{fig:4dualityoscillation}} 
\end{figure}
We denote the full-0 state by odd integers $\{ \cdots -1,1,3,5,7 \cdots \}$, the full-1 state by even integers $\{\cdots -2, 0 ,2,4,6 \cdots\}$, and the 4 states with equal probability as half integers $\{\cdots  -\frac{1}{2},0 , \frac{1}{2} , \frac{3}{2} \,\frac{5}{2}\, \cdots  \}$. We say the 4 equally probable states at half integers of $\pi$ \emph{spontaneously} collapse to the full-0 state at odd multiples of $\pi$, and \emph{spontaneously} collapse to the full-1 state at even multiples of $\pi$. The quantum system is said to be momentarily deterministic at $\pm k \pi$ as either the full-0 state or full-1 state is at probability 1.  

The first order derivative of the expectation energy is 
\begin{equation}
\frac{d\langle E(\theta  )\rangle}{d\theta} =\sin\theta \Big(\sin^2 \frac{\theta}{2} E_{++} - \cos^2 \frac{\theta}{2} E_{--} \Big) +\frac{1}{4}\sin 2\theta \,(E_{-+} + E_{+-} )  \,. 
\end{equation}
Therefore the extreme of $\langle E(\theta  )\rangle $ occurs when 
\begin{equation} \label{eq:2sol}
\sin \theta =0 \quad \text{or} \quad   \Big(\sin^2 \frac{\theta}{2} E_{++} - \cos^2 \frac{\theta}{2} E_{--}  \Big)  + \frac{1}{2}\cos\theta ( E_{-+} + E_{+-} ) = 0 \,.
\end{equation}
For the first case again we have $\theta = \pm k \pi$. Thus at the integer multiple of $\pi$, the energy expectation value is at maximum, this correspond to the purely full-0 state of pure full-1 state which is at probability 1. Thus when the system is at its pure state, it has the maximum energy. Since the full-0 state also represents the state of nothing while the full-1 state represent the state of All, then it means when the universe is at the state of purely nothing or purely everything, its energy is the greatest (Note that if we originally off-set by a phase factor of $\pi$, then the $|01\rangle$ and $|10 \rangle$ state would have the maximum energy).  A special attention should be given for the second case (\ref{eq:2sol}), in particular $E_{--} = E_{++} = E_{-+} = E_{+-} = E $ give zero first order derivative regardless of $\theta$. Thus when all energies for the four states are the same, the rate of change of $\langle E (\theta) \rangle$ is always 0 for whatever $\theta $ values. Thus $\langle E(\theta) \rangle$ is constant overall all $\theta$s. It is in fact easy to check that in fact $\langle E(\theta) \rangle = E$ using simple trigonometry. 

Next we find the entropy of the system. Equation \ref{eq:Entropy4Level} is simplified to,
\begin{equation}
H(\theta) = -4\cos^4 \frac{\theta}{2} \log \Big\vert \cos \frac{\theta}{2} \Big\vert - 4 \sin^4 \frac{\theta}{2} \log \Big\vert \sin\frac{\theta}{2} \Big\vert - \sin^2 \theta \log \Big\vert \frac{\sin \theta}{2}  \Big\vert \,.
\end{equation}
The entropy function is plotted,
\begin{figure}[H]
\centering
\includegraphics[trim=0cm 0cm 0cm 0cm, clip, scale=0.7]{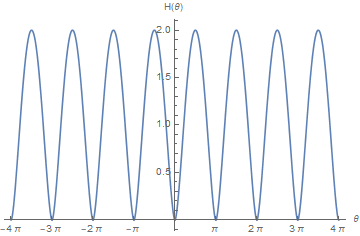}
\caption[Entropy of a 4-level system.]
{$H(\theta)$ plot in the range of $-4\pi \leq \theta \leq 4 \pi$\label{fig:Entropy}}
\end{figure}
 The maximum occurs at $\frac{\pi}{2} \pm k \pi$, which is 2 bits. The roots of $H(\theta)= 0 $ occurs as the limits when $\theta \rightarrow \pm k \pi$.

Our study of $2$-level and $4$-level system can be generalized to $N=2^n$ levels. Since the probability of each state is just the product of sine and cosine, let's write
\begin{equation}
p_{i_1 i_2 \cdots i_N} = \prod_{j=1}^N \mathrm{Trig}^2_{\,\,i_j} ( \theta_j) \,,
\end{equation}
where we define the trigonometry function $\mathrm{Trig}$ by
\begin{equation}
\mathrm{Trig}_{\,\, i_j} ( \theta_j ) =
\begin{cases}
\cos \theta_j & \quad \text{if}\,\, i_j = 0 \\
\sin \theta_j & \quad \text{if}\,\, i_j = 1 
\end{cases}
\,.
\end{equation}

The full quantum state in $2^n$ level is generalized to
\begin{equation} \label{eq:fullquantumstate}
|\psi (\theta_1 , \theta_2 \, \cdots \, \theta_N )\rangle = \sum_{i_1 , i_2 , \cdots , i_N = 0,1} \bigg(\, \prod_{j=1}^N \mathrm{Trig}_{\,\,i_j} ( \theta_j) \,\bigg) \,| \eta_{i_1} \eta_{ i_2 } \cdots \eta_{i_N} \rangle \,.
\end{equation}
The expectation of energy is
\begin{equation}
\langle E(\theta_1 , \theta_2 \, \cdots \, \theta_N )\rangle = \sum_{j\in W} |\langle j | \psi(\theta_1 , \theta_2 \, \cdots \, \theta_N ) \rangle |^2 E_j \,,
\end{equation}
which is
\begin{equation}
\langle E(\theta_1 , \theta_2 \, \cdots \, \theta_N )\rangle = \sum_{i_1 , i_2 , \cdots , i_N = 0,1} \bigg( \,\prod_{j=1}^N \mathrm{Trig}^2_{\,\,i_j} ( \theta_j) \,\bigg) \, E_{i_1 i_2 \cdots i_N} \,,
\end{equation}
where $W$ is the dual set. The entropy is
\begin{equation} \label{eq:EntropyGeneral}
H(\theta_1 , \theta_2 \, \cdots \, \theta_N )= -\sum_{i_1 , i_2 , \cdots , i_N = 0,1}  \bigg[ \, \bigg(\,\prod_{j=1}^N \mathrm{Trig}^2_{\,\,i_j} ( \theta_j) \,\bigg) \,\, \bigg(\, \log \prod_{j=1}^N \ \mathrm{Trig}^2_{\,\,i_j} ( \theta_j) \bigg) \,  \bigg] \,.
\end{equation} 
The entropy is maximized when all the $\theta_j$s are equal such that the probability of each state is even, i.e. equal to $\frac{1}{N}$. 

For synchronized phase, all $\theta_j$s equal to a single $\theta$ variable, and the full state (\ref{eq:fullquantumstate}) is just an unentangled N-qubit in the form of
\begin{equation}
| \psi \rangle = \sum_{j=0}^{2^n} c_j |j \rangle \,.
\end{equation}
The coefficient $c_j$ is in the form of $\cos^p \frac{\theta}{2} \sin^q \frac{\theta}{2}$, where $p$ is the number of 0 and $q$ is the number of 1 in the particular state. Then some of the states have equal probability. For example in $n=3$ case, the 3 states
\begin{equation}
|100 \rangle \,\, , \,\, |010 \rangle \,\, \text{and} \,\, |001 \rangle
\end{equation}
all have the same coefficient $\cos \frac{\theta}{2}\sin^2 \frac{\theta}{2} $, and the same probability $\sin^4 \frac{\theta}{2} \cos^2 \frac{\theta}{2}$. The probability distribution follows the binomial distribution,
\begin{equation}
1 = \Big( \cos^2 \frac{\theta}{2} + \sin^2 \frac{\theta}{2} \Big)^n = \sum_{k=0}^n \frac{n!}{k! (n-k)!} \cos^{2k} \frac{\theta}{2} \sin^{2n - 2k} \frac{\theta}{2}\,, 
\end{equation}
Therefore the number of states have the same probability is given by the binomial coefficient \begin{equation}
\frac{n!}{k! (n-k)!} \,.
\end{equation}
The number of distinct groups of states that have same probabilities is just $n+1$. (For example, in $n=4$, we have binomial coefficients $1,4,6,4,1$ then we have $4+1 =5$ groups). The total number of states is given by the sum of all binomial coefficients, and this is just the identity
\begin{equation}
\sum_{k=0}^n \frac{n!}{k! (n-k)!} = 2^n = N \,,
\end{equation}  
which is just $N = 2^n$ states as expected. 

Consider a large number of levels. When $n \rightarrow$, $N \rightarrow \infty$, we would obtain a normal distribution for the probabilities.

\begin{figure}
        \centering
        \begin{subfigure}[b]{1\textwidth}
                \centering
                \includegraphics[trim=0cm 0cm 0cm 0cm, clip, scale=0.6]{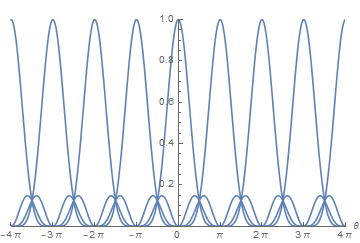}
                \caption{$n= 3$ case, 8 states}
                \label{fig:a}
        \end{subfigure}
        \quad
        \begin{subfigure}[b]{1\textwidth}
                \centering
                \includegraphics[trim=0cm 0cm 0cm 0cm, clip, scale=0.6]{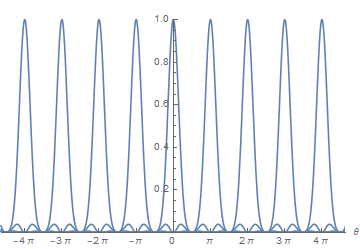}
                \caption{$n= 10$ case, 1024 states}
                \label{fig:b}
        \end{subfigure}
        \quad
        \begin{subfigure}[b]{1\textwidth}
                \centering
                \includegraphics[trim=0cm 0cm 0cm 0cm, clip, scale=0.6]{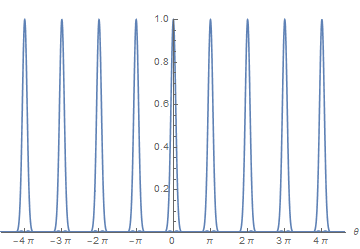}
                \caption{$n=50$ case, $\sim 10^{15}$ states}
                \label{fig:c}
        \end{subfigure}
        \quad
        \begin{subfigure}[b]{1\textwidth}
                \centering
                \includegraphics[trim=0cm 0cm 0cm 0cm, clip, scale=0.6]{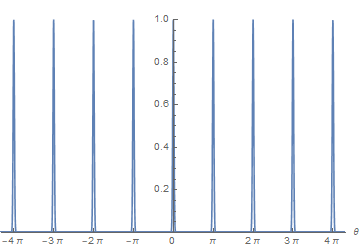}
                \caption{$n=800$ case, $\sim 10^{24}$ states}
                \label{fig:d}
        \end{subfigure}
\caption{ The plots of $p_n$ probability for $n=3,10,50$ and $800$. Note that for each $n$ the probability of the states are plotted with the same colour. These plots are in comparison to the plot \ref{fig:prob2} of $n=2$ case. \label{fig:fourncases}}
\end{figure}

Let's study a few $n$ examples from small $n$ to large $n$. We will plot all the probabilities in a same graph for each $n$. Consider 4 cases $n=3, 10, 50 ,800$, the plots are shown in figure \ref{fig:fourncases}.

We can see that when $n$ grows larger, the contribution of probability from the mixed states of 0 and 1 get lesser, and eventually when $n \rightarrow \infty$ (here $n=800$ is sufficiently large enough ), the $p_{00\cdots 0} = p_0 = \cos^{2n} \frac{\theta}{2} $ and $p_{11\cdots 1} = p_N = \sin^{2n} \frac{\theta}{2} $ fully dominate. The contribution from other individual mixed state is extremely small. Thus basically, when $n \rightarrow\infty$, the system just automatically collapse to the full zero state $|00\cdots 0 \rangle$ at odd multiple of $\pi$ (half cycles) and to full one state $|11\cdots 1 \rangle$ at even multiple of $\pi$ (complete cycles). We call the full zero state $|00\cdots \rangle = |0\rangle$ the beginning state and the full one state $|11\cdots 1 \rangle = |N-1\rangle $ state. Diagrammatically,
\begin{figure}[H]
\centering
\includegraphics[trim=0cm 0cm 0cm 0cm, clip, scale=0.7]{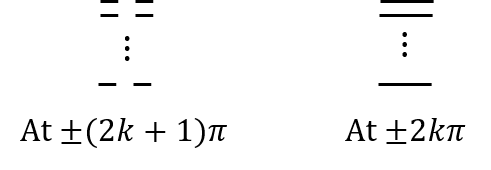}
\caption[]
{\label{fig:diagramdual}}
\end{figure}
Therefore, in fact we can interpret the the state $|\psi_{\infty} \rangle$ as a new dual state $|\infty \rangle$
\begin{equation}
| \bowtie \rangle = \frac{1}{\sqrt{2}} ( |00\cdots 0 \rangle + |11\cdots 1\rangle ) = \frac{1}{\sqrt{2}}( |\bar{0}\rangle + |\bar{1} \rangle  ) = \frac{1}{\sqrt{2}}( |0\rangle + | \infty\rangle  )\,,
\end{equation}
where in the last expression we have used the numerical expression for the state ($0$ is always 0 and $\infty$ is always $\infty$ in any number base ). This is because the $|0\rangle$ and $|\bar{1}\rangle$ dominate over all other states, and hence we can basically think that $| \psi_\infty \rangle$ simply contains two dual state. The $| \bowtie \rangle $ is a EPR pair, which is an entangled state.

Now recall that the zeroth-level state $|\psi_0 \rangle  = | \bigcirc \rangle$ with probability 1. Thus when we go from the beginning to the end $0 \rightarrow \infty$, the originally deterministic $| \bigcirc \rangle$ state now becomes the $| \bowtie \rangle$ state with two dual basis. We denote this as \begin{equation}
0 \rightarrow \infty \,,
\end{equation}  
or in terms of the change in number of states
\begin{equation}
1 \rightarrow 2 \,.
\end{equation}

\begin{figure}[H]
\centering
\includegraphics[trim=0cm 0cm 0cm 0cm, clip, scale=0.7]{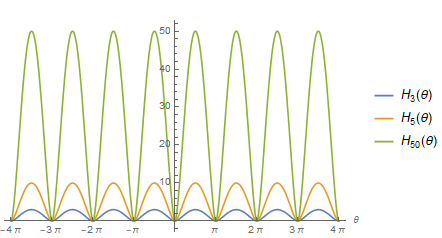}
\caption[]
{\label{fig:Hthetadifferent}}
\end{figure}

The general entropy for synchronized phase is,
\begin{equation}
H_ n (\theta) = -\sum_{k=0}^n \frac{n!}{k! (n-k)!} \Big( \cos^{2k} \frac{\theta}{2} \sin^{2n-2k}  \frac{\theta}{2} \Big) \log \Big( \cos^{2k} \frac{\theta}{2} \sin^{2n-2k} \frac{\theta}{2}  \Big) \,.
\end{equation}
Equivalently,
\begin{equation}
H_ n (\theta) = -2\sum_{k=0}^n \frac{n!}{k! (n-k)!} \Big( \cos^{2k} \frac{\theta}{2} \sin^{2n-2k}  \frac{\theta}{2} \Big) \log \Big\vert \cos^{k} \frac{\theta}{2} \sin^{n-k} \frac{\theta}{2}  \Big\vert \,.
\end{equation}
In general the entropy function $H_ n (\theta)$ for different $n$s have same shapes but just different maximum amplitudes. The local maximum and minimum entropies occur at the same $\pm k \pi$s for all different $n$. The cases for $n=3,10$ and $50$ are illustrated in figure \ref{fig:Hthetadifferent}.

For synchronized phase, if all $E_{i_1 , i_2 \, \cdots \, i_N } = E$ are equal, then it is a degenerate quantum system. The expectation energy is constant independent of phase
\begin{equation}
\langle E(\theta ) \rangle = E \,.
\end{equation}
This is simply because 
\begin{equation}
\langle E(\theta ) \rangle = \sum_{k=0}^n \frac{n!}{k! (n-k)!} \cos^{2k} \frac{\theta}{2} \sin^{2n - 2k} \frac{\theta}{2} \,E = E\cdot 1 = E \,.
\end{equation}

Finally, for the general $n$-qubit case, we have the state vector as
\begin{equation}
|\Psi(\theta_1,\theta_2 ,\cdots \theta_n)\rangle = |\psi_1 (\theta_1) \rangle \otimes |\psi_2 (\theta_2) \rangle \otimes \cdots \otimes |\psi_n (\theta_n) \rangle = \bigotimes_{j=1}^n (\cos\theta_j |0_j\rangle + \sin\theta_j |1_j\rangle  ) \,.
\end{equation}
For each $i_j \in W_j$ where $W_j$ is the jth dual set, the probability of each state is 
\begin{equation}
\begin{aligned}
P_{i_1 i_2\cdots i_n} &= |\langle i_1 i_2\cdots i_n |\Psi(\theta_1,\theta_2 ,\cdots \theta_n)\rangle |^2 \\
&=\bigg\vert \bigg(\bigotimes_{j=1}^n \langle i_j |\bigg)\bigg( \bigotimes_{j=1}^n |\phi_j (\theta_j) \rangle \bigg) \bigg\vert^2 \\
&= |(\langle i_1 |\otimes \langle i_2| \otimes \cdots \otimes\langle i_n| )  (| \phi_1 (\theta_1) \rangle \otimes | \phi_2 (\theta_2) \rangle \otimes \cdots \otimes | \phi_n (\theta_n) \rangle )    |^2 \\
&= | \langle i_1 | \phi_1 (\theta_1)\rangle \otimes  \langle i_2 | \phi_2 (\theta_2)\rangle \otimes \cdots \otimes \langle i_n | \phi_n(\theta_n)\rangle |^2 \\
&= \bigg\vert \bigotimes_{j=1}^{n} \langle i_j | \psi(\theta_j) \rangle \bigg\vert^2 =\bigg\vert \prod_{j=1}^{n} \langle i_j | \psi(\theta_j) \rangle \bigg\vert^2 \\
&= \prod_{j=1}^{n} |\langle i_j | \psi(\theta_j) \rangle |^2 \\
&=\prod_{j=1}^{n} P_{i_j} = P_{i_1}P_{i_2}\cdots P_{i_n}
\end{aligned}
\end{equation}
And the sum of probability is equal to 1,
\begin{equation}
\begin{aligned}
P_{tot} &= \sum_{i_1 , i_2 ,\cdots i_n} P_{i_1 i_2\cdots i_n} \\
&= \sum_{i_1 \in W_1} \sum_{i_2 \in W_2}\cdots  \sum_{i_n \in W_n}\prod_{j=1}^{n} P_{i_j} \\
&=\bigg( \sum_{i_1 \in W_1} P_{i_1}\bigg)\bigg( \sum_{i_2 \in W_2} P_{i_2}\bigg)\cdots \bigg( \sum_{i_n \in W_n} P_{i_n}\bigg)\\
&=\prod_{j=1}^n \sum_{i_j \in W_j} P_{i_j} \\
&=1 
\end{aligned}
\end{equation}
The expectation energy of the $n$-qubit system is given by
\begin{equation}
\begin{aligned}
\langle E(\theta_1 , \theta_2 ,\cdots, \theta_n )\rangle &= \sum_{i_1 \in W_1}\sum_{i_2 \in W_2}\cdots\sum_{i_n \in W_n} |\langle i_1 i_2 \cdots i_n | \Psi (\theta_1 ,\theta_2 , \cdots \theta_n ) \rangle |^2 E_{i_1 i_2 \cdots i_n} \\
&= \sum_{i_1 \in W_1}\sum_{i_2 \in W_2}\cdots\sum_{i_n \in W_n} P_{i_1 i_2 \cdots i_n}(\theta_1 , \theta_2 ,\cdots ,\theta_n) E_{i_1 i_2 \cdots i_n} \\
&= \sum_{i_1 \in W_1 ,\cdots, i_n \in W_n} \prod_{j=1}^n P_{i_j}(\theta_j) E_{i_1 i_2 \cdots i_n}
\end{aligned}
\end{equation}
The entropy for the general case would be
\begin{equation}
\begin{aligned}
H(\theta_1 , \theta_2 ,\cdots\theta_n) &= -\sum_{i_2 \in W_2}\cdots\sum_{i_n \in W_n} P_{i_1 i_2 \cdots i_n}(\theta_1 , \theta_2 ,\cdots\theta_n) \log P_{i_1 i_2 \cdots i_n}(\theta_1 , \theta_2 ,\cdots\theta_n) \\
&= -\sum_{i_2 \in W_2}\cdots\sum_{i_n \in W_n} |\langle i_1 i_2 \cdots i_n | \Psi (\theta_1 ,\theta_2 , \cdots \theta_n ) \rangle |^2 \log |\langle i_1 i_2 \cdots i_n | \Psi (\theta_1 ,\theta_2 , \cdots \theta_n ) \rangle |^2 \\
&= -\sum_{i_1 \in W_1 ,\cdots, i_n \in W_n} \prod_{j=1}^n P_{i_j}(\theta_{i_j}) \log  \prod_{j=1}^n P_{i_j}(\theta_{i_j}) \,.
\end{aligned}
\end{equation}

\subsection{Dual Pairs}
Next, we would introduce a very important concept of quantum states of dual pair. A dual pair is basically a pair of diagrams of which each level is dual to each other. It means if a level is at $|0\rangle$ state then the same level of the dual counterpart is $|1\rangle$, vice versa.  We will give the formal mathematical definition below.

\begin{definition}
Let $|\eta_{i_1} \eta_{ i_2} \cdots \eta_{i_{N}} \rangle $ of a $n$-level with $N=2^n$ be the general state for a diagram, where $\eta_{i_j} = 0 $ or $1$. Define the dual operator $*$ such that $* |0 \rangle = |0^* \rangle = |1\rangle$ and $*|1\rangle = |1^* \rangle = |0\rangle$ which satisfies $*^2 = I$,
\begin{equation}
*|\eta_{i_1} \eta_{ i_2} \cdots \eta_{i_{N}} \rangle  = |\eta_{i_1}^* \eta_{ i_2}^* \cdots \eta_{i_{N}}^* \rangle  \,.
\end{equation} 
Let $k$ be the decimal number recovered from the binary representation and $k^*$ be the dual counter part, then $k$ and $k^*$ is related by
\begin{equation}
k^* = (N-1) - k \,.
\end{equation} \label{eq:dualnumber}
The $(k, k^* )$ is defined as a dual pair, and is arranged for $k < k^*$. The dual pair must contain one odd and one even number. 
\end{definition}

The proof is straight forward. Let $a_j = 0,1 $ be the coefficient and let $a_j^*$ be dual to $a_j$. The duality operator constrains that $a_j + a_j^* =1$. Let
\begin{equation}
k = \sum_{j=1}^n a_j 2^{j-1} \quad \text{and} \quad k^* = \sum_{j=1}^n a_j^* 2^{j-1} \,.
\end{equation}
Then consider $k+k^*$,
\begin{equation}
k +k^* = \sum_{j=1}^n (a_j + a_j^{*} ) \,2^{j-1} = \sum_{j=1}^n 2^{j-1} = \frac{(1)(2^n -1)}{2-1}= 2^n- 1 = N-1 \,, 
\end{equation}
where in the third step we have used the geometric series. It follows that $k^* = (N-1) - k$. Since $N=2^n$ must be even, then $N-1$ must be odd. Thus $k+k^*$ is odd. Using the fact that an odd number is composed of an even number and an odd number, then $k,k^*$ must contain one odd number and one even number. This completes the proof. $\Box$

For example, for $n=6$ level, we have $*|010010 \rangle = |  101101 \rangle$. Then $k =18$ and $k^* = 63-18 = 45 $. The dual pair is $(18,45)$. 

There is a special kind of dual pair which has a common property in all $n$-levels. The difference between $k$ and $k^*$ for this special pair is 3.

\begin{definition}
Let $(k_b , k_b^*)$ be the boundary dual pair. The respective binary representation are $(011 \cdots 110)$ and $(100\cdots001)$; and the respective decimal representation is $\frac{N}{2}-1$ and $\frac{N}{2} +2$ for $N > 2 \,\,\,\,( \text{or} \,\,n>1)$. The two integers in between are called the confined boundary dual pair $\kappa, \kappa^*$. The respective binary representation is $(011\cdots1)$ and $(100\cdots0)$, and the respective decimal representation is $\frac{N}{2}-1$ and $\frac{N}{2}$. The boundary dual is formally defined by $(k_b , k_b^*)$ such that $|k_b^* - k_b | = 3$ and the confined boundary dual is defined by $|\kappa^* - \kappa | = 1$.

\end{definition}
These can be easily shown. First we have
\begin{equation}
k_b =\sum_{j=1}^{n-2} 2^j = 2(1 + 2 +\cdots 2^{n-3}) = \frac{2 (2^{n-1} -1)}{(2-1)} = 2^{n-1} -2 = \frac{N}{2}-2 \,,
\end{equation}
and
\begin{equation}
k_b^* = 1 + 2^{n-1} = \frac{N}{2}+1 \,.
\end{equation}
Then the $\kappa$ in the confined binary dual pair is simply adding $1$ from $k_b^*$, thus $\kappa = k_b + 1 =\frac{N}{2}-1$, and $\kappa^* $ is obtained by subtracting $k_b^*$ from $1$, $\kappa^* = k_b^{*} -1 = \frac{N
}{2}$.  
These can be expressed in terms in a diagramatic way,
\begin{figure}[H]
\centering
\includegraphics[trim=0cm 0cm 0cm 0cm, clip, scale=0.75]{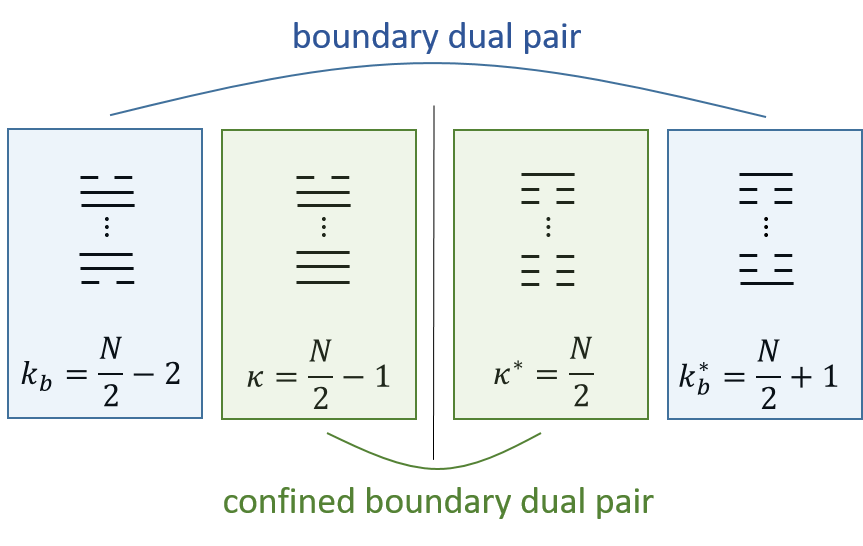}
\caption[Dual pair and boundary dual pair]
{Boundar pair and confined boundary dual pair for general $n$-level. The black middle line is the duality mirror. \label{fig:specialdualpair}}
\end{figure}
Examples for these pairs are shown in table \ref{tab:DualInvariantNumbers}, e.g. for $n=5$ and $n=6$, respectively, we have boundary pairs as $(14,17)$ and $(30,33)$, with boundary confined pair as $(10,21)$ and $(18,45)$. We call it boundary dual pair because the pair appears nearest to the dual mirror, where the dual mirror separates the dual numbers into two halves, the even half and the odd half.

\begin{figure}{H}
\centering
\includegraphics[trim=0cm 0cm 0cm 0cm, clip, scale=0.6]{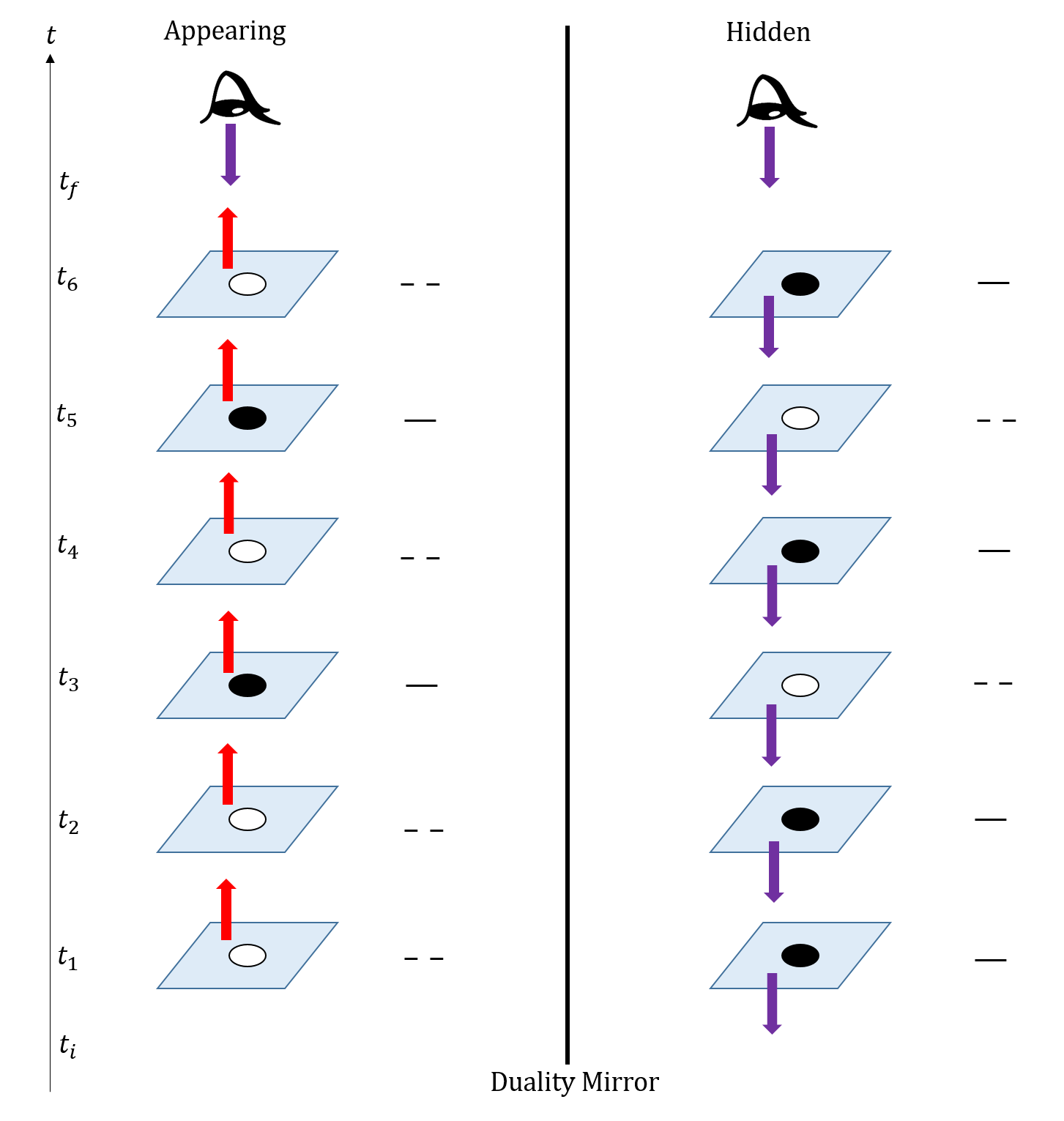}
\caption[Illustration of a dual pair]
{Illustration of a dual pair of $n = 6$ case. \label{fig:dualpair}}
\end{figure}

The dual pair is a very important idea to describe the appearing and hidden states of nature. Suppose we construct a $|\eta_{i_N} \eta_{ i_{N-1}} \cdots \eta_{i_{2}}\eta_{i_{1}} \rangle$ state along a positive flow of time. Each state $| \eta_{i_j} \rangle$ is generated at time $t_j$, where $t_{f} > t_{N} > \cdots > t_1 > t_i$. Consider a surface, for example the surface of the table, then we flip a 2-sided coin which is either head or tail successively in the time interval $t_i \leq t \leq t_f$. This is called the even time interval, in which there is a generation of information in this time interval.  Let's define the head state as $|1\rangle$ and represent it as a black circle, and the tail state as $|0\rangle$ and represent it as a white circle. There are in addition, two more states, which are two perspectives. The direction that the coin shows up to the observer is called the appearing state $|\uparrow\rangle$, while the direction that the coin that is not shown is called the hidden state $|\downarrow\rangle$. The idea is illustrated in \ref{fig:dualpair}.
The two sides of the illustration are equivalent, we have
\begin{equation}
(010100 | \uparrow ) \,\,\equiv \,\, (101011 | \downarrow ) \,,
\end{equation}
where we have the observer frame (or perspective) identified as $S_3 = \uparrow$ and $S_3^\star = \downarrow$. In terms of decimal number we can write
\begin{equation}
(18 | \uparrow)   \,\,\equiv \,\, (45 | \downarrow ) \,. 
\end{equation}
In general we have
\begin{equation}
( \eta_{i_N} \eta_{ i_{N-1}} \cdots \eta_{i_{2}}\eta_{i_{1}}  | S_k ) \,\, \equiv \,\, ( \eta_{i_N}^* \eta_{ i_{N-1}}^* \cdots \eta_{i_{2}}^*\eta_{i_{1}}^*  | S_k^\star ) \,.
\end{equation}
Thus although the diagrams of both sides are not the same in terms of element, when we change from the appearing perspective to the hidden perspective, they are equivalent. 

We can map the two dual states to effective $|\bar{0}\rangle $ and $| \bar{1}\rangle$ states. We add an extra bar so as to distinguish them form the original $|0\rangle$ and $|1\rangle$ states. This means
\begin{equation}
|\eta_{i_N} \eta_{ i_{N-1}} \cdots \eta_{i_{2}}\eta_{i_{1}} \rangle  \rightarrow | \bar{0} \rangle \,\, , \,\, |\eta_{i_N}^* \eta_{ i_{N-1}}^* \cdots \eta_{i_{2}}^*\eta_{i_{1}}^*  \rangle \rightarrow | \bar{1} \rangle
\end{equation}
as $|0^* \rangle  = |1\rangle$.  Therefore in the above example we have
\begin{equation}
|010100 \rangle \rightarrow |\bar{0}\rangle \,\, , \,\, |101011 \rangle \rightarrow | \bar{1}\rangle \,.
\end{equation}
It is interested to study, in particular, the case where $n$ is extremely large. When $n\rightarrow \infty$,
\begin{equation}
(000\cdots 00 | \uparrow ) \equiv (111\cdots 11 |\downarrow ) \,,
\end{equation}
which is, in decimal representation,
\begin{equation}
(0 | \uparrow ) \equiv ( \infty |\downarrow) \,.
\end{equation}
Hence, zero is equivalent to infinity under the appearing-hidden perspective. We interpret as follow, zero in the appearing frame is equivalent to infinity in the hidden frame. Therefore zero can be viewed as everything in the dual frame. 
And of course, we can also have 
\begin{equation}
(0 | \downarrow ) \equiv ( \infty |\uparrow) \,.
\end{equation}
Then zero in the hidden frame is equivalent to infinity in the appearing frame.

The dual pair can be promoted to two dual quantum states. The coefficients can show how much information is appearing and how much information is hidden. The amount of information is described by the probability. For each dual pair in $N = 2^n$ diagrams (we have $2^{n-1}$ dual pairs),
\begin{equation}
| \psi_l \rangle = a_l |l\rangle + a_{l}^* |l^* \rangle= \cos\theta_l |l \rangle + \sin\theta_l | l\rangle^* =\cos\theta_l |l \rangle + \sin\theta_l | 2^n -l -1\rangle \,.
\end{equation} 
The full state is given by
\begin{equation}
|\Psi \rangle  = \sum_{l=0}^{2^{n-1}-1}| \psi_l \rangle =\sum_{l=0}^{2^{n-1}-1} (\, \cos\theta_l |l \rangle + \sin\theta_l | l^* \rangle\,) =\sum_{l=0}^{2^{n-1}-1} (\, \cos\theta_l |l \rangle + \sin\theta_l | 2^n -l -1\rangle \,) \,.
\end{equation}
We can also write it as the sum of paired $| \bar{0}_j \rangle, | \bar{1}_j \rangle$ states,
\begin{equation}
|\Psi \rangle = \sum_{j=0}^{2^{n-1}-1} ( \, \cos \theta_j | \bar{0}_j \rangle + \sin \theta_j | \bar{1} _j \rangle  \,) \,.
\end{equation}
If the appearing state and hidden state evolve under time, we can write it as
\begin{equation}
|\Psi (t) \rangle = \sum_{l=0}^{2^{n-1}-1} (\, \cos\omega_l t |l \rangle + \sin\omega_l t | 2^n -l -1\rangle \,) 
\end{equation}
In the general form, we have
\begin{equation}
|\Psi (t) \rangle = \frac{1}{2} \sum_{p+q = 2^n -1} ( \,a_p (t) |p\rangle + a_q (t) |q \rangle \,) \,.
\end{equation}
The factor of $\frac{1}{2}$ is required due to double counting.

One interesting property arises from the above theory is dual invariance. Suppose the state remains the same regardless of forward time flow or backward time flow, for example $(110011)$ looks the same in either case. Geometrically this is simply left-right invariant by observation, or up-down invariant if diagrammatically. We have
\begin{equation}
(110011 | RL , t>0 ) \,\, \equiv \,\, (110011 | LR , t<0 ) \,.
\end{equation}
Such state can also infer the property of time-reversal symmetry of the state. We will discuss give the formal definition of dual invariant in the next section.

\subsection{Dual Invariant}
We will define dual invariant, which is an invariant under a dual observation frame or perspective. There can be different kinds of dual invariants subjected to the interest of study. Mathematically,
\begin{definition}
Let the observer be situating in a space of dimension $k \geq 3$. Define a dual space $S$ with two elements $S_k$ and $S_k^{\star}$. An object $\xi$ is dual invariant under the two dual observer frames if $\xi$ satisfies
\begin{equation}
( \xi | S_k ) = (\xi | S_k^{\star}) 
\end{equation}
for $S_k\neq S_k^{\star}$ and $S_k \cap S_k^{\star} =\emptyset$. 
\end{definition}
Note for the condition of $k \geq 3$, this is because one at least has to observe perpendicularly to a dual system that is situated in a 2-dimensional plane. The most common ones we will use is the up-down dual observing frame and left-right dual observing frame. If an object is up-down (UD) dual invariant, then the object remains the same regardless of looking from above or looking from below. The two dual frames are $S_{k}^{UD}$ and $S_k^{DU}$.  If an object is left-right dual invariant, then the object remains the same regardless of looking from the left or looking from the right. The two dual frames are $S_{k}^{LR}$ and $S_k^{RL}$. Like-wise, other dual invariants follow similar idea. For example, in the 3rd order case, all $(000)$, $(010)$, $(101)$ and $(111)$ are all dual invariant, this corresponds to the dual invariant decimal number $0,2,5,7$ respectively.

Next we would construct useful diagrams to highlight the important symmetry properties in the multiplication table, they are called the feature diagrams. In these diagrams, only elements in the table which have considerable dual structure that can be form the basis of $\mathbb{Z}_2$ or $\mathbb{Z}_2 \times \mathbb{Z}_2$ will be high-lighted. The feature diagrams are essential to study symmetry patterns of the elements. There are two main parts in the feature diagram. Firstly for even $n$, the numbers are arranged in the a $2^{\frac{n}{2}}\times 2^{\frac{n}{2}}$ grids with each grid follows exactly the same order as in the multiplication table,
\begin{figure}[H]
\centering
\includegraphics[trim=0cm 0cm 0cm 0cm, clip, scale=0.7]{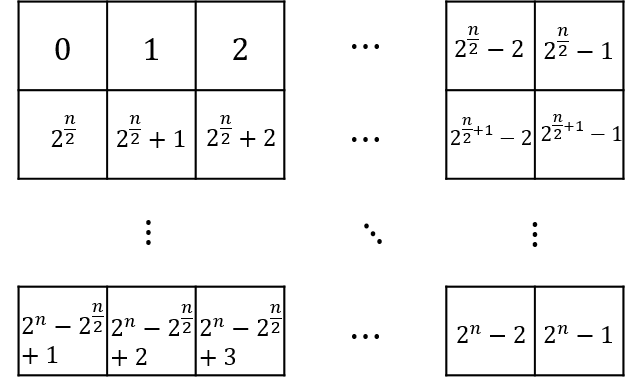}
\caption[Boxes]
{ \label{fig:Boxes}}
\end{figure}

Secondly, coloured symbols represent different dual structures of the diagram. We will define some standard coloured symbols used in the feature diagrams as follow:
\begin{itemize}
\item Blue circle: Represent the diagram(s) that are up-down dual invariant, or equivalently left-right dual invariant for the binary representation $(\eta_{i_1} \eta_{i_2} \cdots \eta_{i_n} | S_k^{LR} ) =(\eta_{i_1} \eta_{i_2} \cdots \eta_{i_n} | S_k^{RL} ) $ where $\eta_i = 0,1$. This corresponds to the condition of $\eta_{i_j} = \eta_{j_{n-j}}$.  The number recovered is defined as dual invariant number (details will be discussed in the next section).
\item Purple square: Represent the diagram(s) that are invariant under the exchange for two blocks of $n/2$-diagram in the multiplication. 
\item Orange hexagon: Represent the diagram(s) where the upper diagram block is dual to the lower diagram block, i.e. the upper block can be formed by acting the dual operator to the lock block, and vice versa.
\item Green triangle: Represent diagram(s) with alternative $0$s and $1$s. There are only two of these for any levels $n >3$, and this is a global property. One of it is the dual of the other.
\item Four colour corners: Represent the diagram(s) of full, half and null states.
\end{itemize}
Basically, the diagrams that fall into Purple square, Orange hexagon, or Green triangle are different basis of the reducible representation of the 2-duality group $\mathbb{Z}_2$
For example, the feature diagram of 6th order is,
\begin{figure}[H]
\centering
\includegraphics[trim=0cm 0cm 0cm 0cm, clip, scale=0.5]{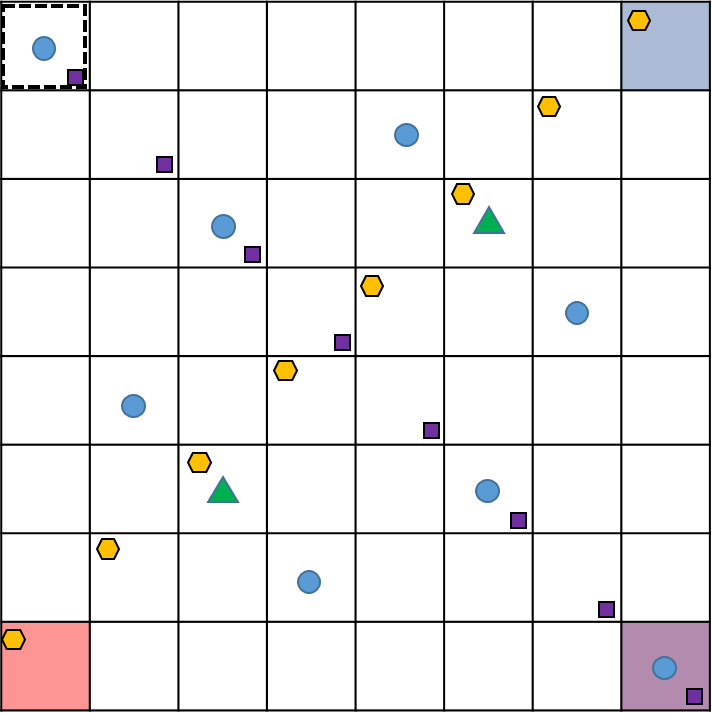}
\caption[Analysis]
{ Feature diagram of 6-th order diagram in correspondence of figure 5. \label{fig:featureDiagram64Gua}}
\end{figure}

\subsection{General properties for general $n$-level}
In this section, we will study the properties of dual invariant number. For each $n$-level, we collect all dual invariant numbers for each $n$-level as a set $D_n$. 

\begin{table}[H] \label{tab:dualinvariantnumber}
\begin{center}
\begin{tabular}{ | l | l | }
\hline
$D_1 $  &  $0,1$   \\
$D_2 $  & $0,3 $    \\ 
$D_3 $  & $0,2,5,7$   \\
$D_4 $  & $0,6,9,15 $   \\
$D_5 $  & $0, 4, 10,14 ,17 ,21, 27 , 31$ \\
$D_6 $  & $0,12, 18, 30, 33,45,51, 63 $ \\
$D_7 $  & $0,8,20,28,34,42,54,62,65,73,85,93,99,107,119,127 $ \\
$D_8 $  & $0,24,36,60,66,90,102,126,129,153,165,189,195,219,231,255 $  \\
\hline
\end{tabular}
\caption[]
{Dual invariant numbers for $n$. \label{tab:DualInvariantNumbers}}
\end{center}
\end{table}
\begin{definition}
Define the set of all dual invariant numbers $D$. It can be considered as the union of all individual $D_n$,
\begin{equation}
D= \bigcup_{i=1}^{\infty } D_i = D_1 \cup D_2 \cup \cdots \cup D_{\infty} \,.
\end{equation} 
Zero is the common element for all of the $D_n$ s,
\begin{equation}
\bigcap_{i=1}^{\infty} D_i = D_1 \cap D_2 \cap \cdots \cap D_{\infty} = \{ 0 \} \,.
\end{equation}
\end{definition}

\begin{figure} \label{fig:dualInvNumTable}
\centering
\includegraphics[trim=0cm 0cm 0cm 0cm, clip, scale=0.7]{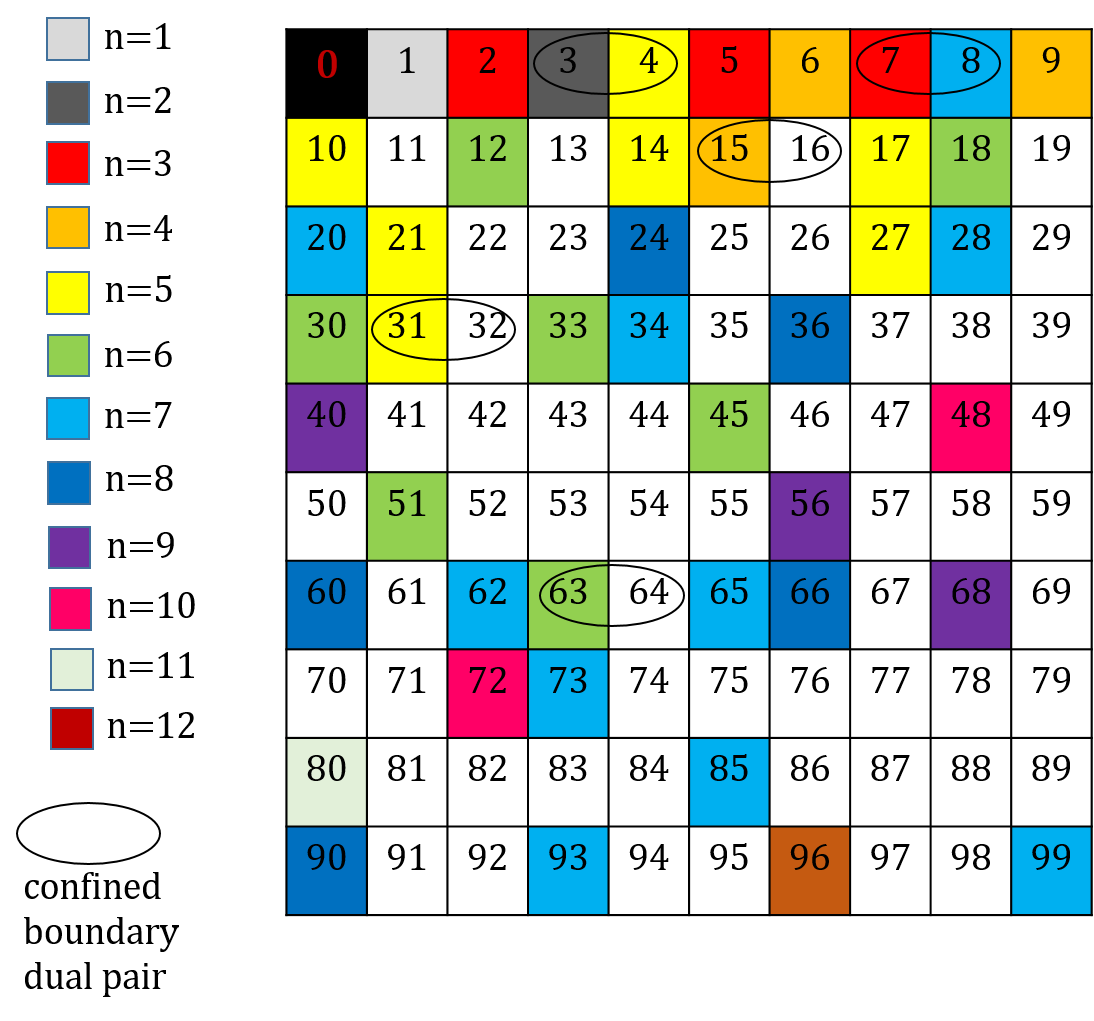}
\caption[Analysis]
{ Dual invariant number $< 100$ from $n=1$ to $n=12$ level. The number $0$ is general to all levels.\label{fig:DualInvNumTable}}
\end{figure}

The dual invariant number $<100$ is shown in figure \ref{fig:DualInvNumTable}. 
It is noted that the distribution pattern of dual invariant number is unclear in figure \ref{fig:DualInvNumTable}. The pattern is only clear in individual $n$-levels when placed in the $n \times n $ grid. 

\begin{definition}
Let $m$ be even and $m>2$. Let the dual invariant number of level $m$ be $d_m$, where $d_m \in D_m$ and is defined by
\begin{equation}
d_m = \sum_{j=0}^{m/2} a_{j+1} ( 2^j + 2^{m-j+1} ) \,,
\end{equation}
with $a_{j+1} = 0 \,\text{or} \,1$.
\end{definition}
This can be easily proved by the following. Consider the general form for a binary number in decimal representation,
\begin{equation}
k(n) = \sum_{j=1}^{n} a_j 2^{j-1} \,\,\text{for}\,\, n \in \mathbb{N} \,.
\end{equation}
We consider the case for even $n$, where $n =m = 2l$. Then the series sum can be explicitly written down as
\[
\begin{aligned}
k(m)&= a_1 + a_2\cdot 2 + a_3 \cdot 2^2 + \cdots + a_{\frac{m}{2}-1} \cdot 2^{\frac{m}{2}-2} + a_\frac{m}{2}\cdot 2^{\frac{m}{2}-1} + a_{\frac{m}{2}}\cdot 2^{\frac{m}{2}-1} + a_{\frac{m}{2}+1} \cdot 2^{\frac{n}{2}} + a_{\frac{n}{2}+2} \cdot 2^{\frac{n}{2}+1} \\
&\quad+ \cdots + a_{m-3} \cdot 2^{m-4} + a_{m-2}\cdot 2^{m-3} + a_{m-1} \cdot 2^{m-2} + a_{m}\cdot 2^{m-1} \\
& = \big( a_1 + a_m \cdot 2^{m-1} \big)+ \big( a_2 \cdot 2 +  a_{m-1} \cdot 2^{m-2} \big) + \big( a_3\cdot 2^2 + a_{m-2} \cdot 2^{m-3} \big) \\ 
& \quad + \cdots + \big( a_{\frac{m}{2}-1}\cdot 2^{\frac{m}{2}-1} + a_{\frac{m}{2}+2} \cdot 2^{\frac{m}{2}+1} \big) + \big( a_{\frac{m}{2}}\cdot 2^{\frac{m}{2}-1} + a_{\frac{m}{2}+1} \cdot 2^{\frac{m}{2}} \big) \,.
\end{aligned}
\]
The dual invariance imposes the condition of 
\begin{equation}
a_j = a_{m+1-j} \,.
\end{equation}
Then we have, for examples
\[
\begin{aligned}
a_1 & = a_m \\
a_2 & = a_{m-1} \\
a_{\frac{m}{2}-1} &= a_{\frac{m}{2}+2} \\
a_{\frac{m}{2} } & = a_{\frac{m}{2}+1} 
\end{aligned} \,.
\]
Thus we can write the series as
\[
d_m = \sum_{j=0}^{m/2} a_{j+1} ( 2^j + 2^{m-j+1} )
\] 
which completes the proof. 

\begin{definition}
The $d_n$ for even $n$ and $n >2$ is divisible by 3.
\end{definition}
This amounts to prove that the term $2^j + 2^{m-j+1}$ in the sum is divisible by 3. We will prove by induction by showing first the propositions $P(m+1, j)$ and $P(m, j+1)$ are true, followed by $P(m+1 , j+1)$ is also true. 
Let $l \in \mathbb{N}$ and suppose
\[
P(l ,j ) =2^j + 2^{2l-j+1} = 3q \quad \text{for some even integers}\,q\,\text{ and max}\,j=\frac{m}{2}. 
\]
For $l=1$, $P(1,j) = 2^j + 2^{3-j}$. And for $j=1$, $P(1,1)=6$ which is divisible by 3 for $q=2$, which is true. Then 
\[
\begin{aligned}
P(l+1 ,j ) &= 2^k + 2^{2(l+1)-j +1} \\
&= 2^j + 4 \cdot 2^{2l - j +1 } \\
&= 2^j +4 (3q- 2^j) \quad \text{(by assumption)}\\
&= 2^j - 2^{l+2} + 6q \\
& = 2^j (1-4) + 6q \\
&= 3 ( 2q -2^j ) \,\,,
\end{aligned}
\]
thus $P(l+1 , j)$ is true. Next
\[
\begin{aligned}
P(l ,j+1 ) &= 2^{j+1} + 2^{2l - {j+1} +1} \\
&= 2^{j+1} + 2^{-1} \cdot 2^{2l -j +1}\\
&= 2^{j+1} + 2^{-1} \cdot (3q - 2^j) \quad \text{(by assumption)} \\
&= 2^{j+1} - 2^{j-1} + \frac{3}{2}q \\
& = 2^{j-1} ( 2^2 -1) + \frac{3}{2}q \\
& = 3 \big( 2^{j+1} + \frac{q}{2}\big) \,\,
\end{aligned}
\]
thus $P(l+1 , j)$ is also true. Finally 
\[
\begin{aligned}
P(l+1 , j+1) & = 2^{j+1} + 2^{2(l+1)-(j+1) +1} \\
&= 2^{j+1} + 2\cdot 2^{2l-j+1} \\
&= 2^{j+1} + 2 (3q  -2^j) \quad \text{(by assumption)} \\
&= 6q
\end{aligned}
\,
\]
thus $P(l+1 , j+1)$ is also true. Hence the proof completes. $\Box$

\begin{definition}
Let positive integers $n >1$ be the level and the number of dual invariant numbers be $d(n)$, where $d(n) =| D_n |$ is the cardinality of the $n^{\rm th}$ level dual invariant number set $D_n$. We have
\begin{equation}
f(n) =
\begin{cases}
 2^{\frac{n}{2}} & \text{if}\,\, n \,\, \text{is even} \\
 2^{\frac{n+1}{2}} & \text{if}\,\, n \,\, \text{is odd} \\ 
\end{cases}
\end{equation}
The initial case is $|D_2 |=2$.
\end{definition}
Consider an dual invariant number number set of an even $n$-level. For the next $n+1$ level which is odd, we can construct new dual invariants by inserting a state in the middle,
\begin{equation}
| \eta_{i_1} \eta_{i_2} \cdots \eta_{i_{\frac{n}{2}}} \eta_{i_{\frac{n}{2}}} \cdots \eta_{i_2} \eta_{i_1} \rangle \rightarrow | \eta_{i_1} \eta_{i_2} \cdots \eta_{i_{\frac{n}{2}}} \eta^{\prime} \eta_{i_{\frac{n}{2}}}  \cdots \eta_{i_2} \eta_{i_1} \rangle \,,
\end{equation}
where again $\eta^\prime  = 0 \,\,\text{or}\,\,1$. Therefore we have $|D_{n+1} | = 2|D_n|$ for $n$ is even. Next to construct new dual invariants from the  $n$ level, we add two states one at the beginning and one with the end,
\begin{equation}
| \eta_{i_1} \eta_{i_2} \cdots \eta_{i_{\frac{n}{2}}} \eta_{i_{\frac{n}{2}}}  \cdots \eta_{i_2} \eta_{i_1} \rangle \rightarrow | \eta_{i}^{\prime\prime} \eta_{i_1} \eta_{i_2} \cdots \eta_{i_{\frac{n}{2}}} \eta^{\prime} \eta_{i_{\frac{n}{2}}}  \cdots \eta_{i_2} \eta_{i_1} \eta^{\prime\prime} \rangle\,,
\end{equation}
where $\eta^{\prime\prime}= 0 \,\,\text{or}\,\,1$. Therefore we have $|D_{n+2} | = 2|D_n|$ for $n$ is even. 
We would like to introduce two set of notations. The center insertion is
\begin{equation}
| \eta_{i_1} \eta_{i_2} \cdots \eta_{i_{\frac{n}{2}}} \,_{\wedge} \, \eta_{i_{\frac{n}{2}}} \cdots \eta_{i_2} \eta_{i_1} \rangle
\end{equation}
and the beginning-ending insertion is
\begin{equation}
| \,_{\wedge} \eta_{i_1} \eta_{i_2} \cdots \eta_{i_{\frac{n}{2}}}  \, \eta_{i_{\frac{n}{2}}} \cdots \eta_{i_2} \eta_{i_1}  \,_{\wedge} \rangle
\end{equation}

Hence, alternative odd and even layers have the same $|D_n |$. Using this way the dual invariant numbers can be constructed recursively. The following diagramatic approach illustrate the process if the state is represented as $n-$ diagram.
\begin{figure}[H]
\centering
\includegraphics[trim=0cm 0cm 0cm 0cm, clip, scale=0.48]{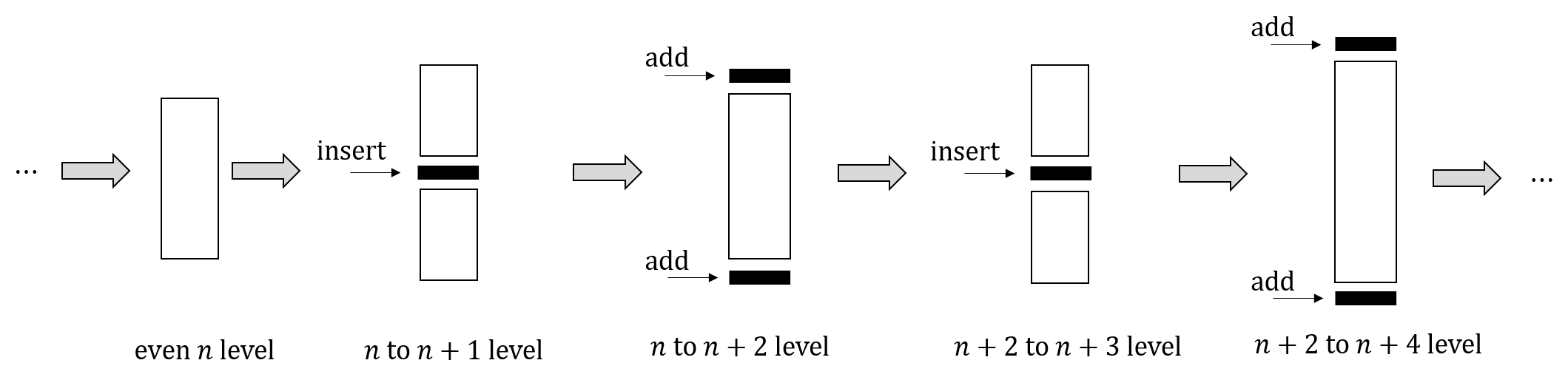}
\caption[]
{\label{fig:insertadd}}
\end{figure}
This allows us to construct \ref{tab:DualInvariantNumbers} and \ref{fig:DualInvNumTable}.

\begin{definition}
Define the ratio of the number of dual invariant number to the number of non-dual invariant as $R(n)$, which is given by
\begin{equation}
R(n) = \frac{2^m}{2^n - 2^m} =
\begin{cases}
 \frac{1}{2^{\frac{n}{2}} -1}   & \text{for} \,\, m = \frac{n}{2} \, \,\text{and}\,\, n \, \text{is even}\\
 \frac{1}{2^{\frac{n-1}{2}} -1} & \text{for} \,\, m = \frac{n+1}{2} \, \,\text{and} \,\, n \, \text{is odd}\\
\end{cases}
\,.
\end{equation}
\end{definition}
Thus we can see that the ratio decreases with increasing $n$, and both even and odd cases converges to $0$ when $n\rightarrow \infty$. It means that as the $n$-level increases, we are getting few dual invariants per $n$-diagram, and eventually they drop to none. 
\begin{figure}[H]
\centering
\includegraphics[trim=0cm 0cm 0cm 0cm, clip, scale=0.8]{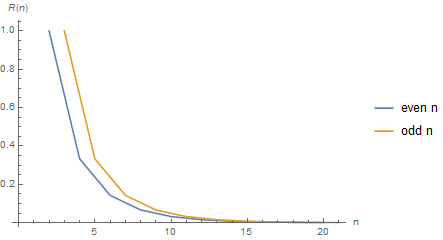}
\caption[]
{The ratio $R(n)$ against $n$.\label{fig:ratio}}
\end{figure}

Next we would like to study the divisibility of weak dual invariant numbers. 
The concept of dual invariant can also be applied to levels of larger group. For $N= 2^n$ of which $n$ is even, we can easily divide the $n-$level diagram into two halves. If the two halves are equal, it is also a dual invariant, however it is a weaker case that the above. In decimal representation,  it takes the form of 
\begin{equation}
| p \rangle | p \rangle \,.
\end{equation}
For example, $|100100 \rangle = |100\rangle  |100 \rangle = |4\rangle |4 \rangle $ is a 3-level dual invariant. The dual invariant we introduced is called the strong dual invariant and is a level-1 dual invariant. 

\begin{definition}
A strong dual invariant is also a weak dual invariant state for even $n$-level(s) .
\end{definition}
\begin{definition}
Let $A$ be the set of strong dual invariant states and $B$ be the set of weak dual invariant states, $B \subseteq A$ for even $n$-levels.
\end{definition}
These two definitions are equivalent. It is note that the converse is not true, a weak dual invariant state cannot be a strong dual invariant state. Only when $n=2$, we have $A=B$. 
\begin{definition}
Any weak dual invariant number is divisible by $N+1$ for even $n$ \,.
\end{definition}
 The proof is easy. First we write the number in decimal representation as
\begin{equation}
x= a_1 2^0 + a_2 2^1 + \cdots a_i 2^{i-1} + a_{i+1} 2^i + \cdots a_{\frac{n}{2}} 2^{\frac{n}{2}-1} +a_{\frac{n}{2}+1}2^{\frac{n}{2}} + \cdots + a_{\frac{n}{2} +j} 2^{\frac{n}{2} +j -1} + a_{\frac{n}{2}+j+1} 2^{\frac{n}{2} +j} +\cdots+ a_n 2^{n-1} \,.
\end{equation}
Due to the condition of weak dual invariant, we must have
\begin{equation}
a_1 = a_{\frac{n}{2}+1}, \,\,\,\, a_2 = a_{\frac{n}{2}+2}\,\,\,\, \cdots \,\,\,\, a_j = a_{\frac{n}{2}+j} \text{and} \,\,\,\, a_n = a_\frac{n}{2} \,.
\end{equation}
Thus we have
\begin{equation}
\begin{aligned}
x &= \sum_{i=0}^{\frac{n}{2}-1} a_{i+1} (2^i + 2^{\frac{n}{2}+i}) \\
&=\sum_{i=0}^{\frac{n}{2}-1} a_{i+1} 2^i (1 + 2^{\frac{n}{2}}) \\
&=(2^{\frac{n}{2}}+1)\sum_{i=0}^{\frac{n}{2}-1} a_{i+1} 2^i \\
&=(N+1)\sum_{i=0}^{\frac{n}{2}-1} a_{i+1} 2^i 
\end{aligned}
\end{equation}
for $n=2k$. Hence, the weak dual invariant number is divisible by $N+1$ for even $n$. It is represented by the magenta boxes along the right diagonal positions in the feature diagram.

\section{Detailed symmetry studies of individual $n$ level}
In this section, we would study the symmetry properties of each $n$-level individually by the concepts we have introduced previously. There are not much new concepts introduced, but instead applying all the previous concepts to give a detailed analysis for each $n$-level. 

\subsection{The n =0 level}
The n=0 level is known as the then null level, this is also called the absolute ground state. We represent it as the a state vector as follow
\begin{equation}
| \Psi \rangle = | \bigcirc \rangle \,.
\end{equation}
This state is completely deterministic and we obtain it by probability equal to 1. The energy of this ground state is given by $E_0$.

\subsection{The n=1 level}
Next we would like to ask how this null state can promote to the dual state, which is the $n=1$ level.
We will model it with spontaneous symmetry breaking. When spontaneous symmetry breaking begins to take place, we call the origin $| \circ \rangle$ state tai chi, which allows the degeneracy of ground states. The process from null to tai chi is  a phase transition, the system changes from a complete deterministic state to a probabilistic dual state.

The process can be model by a potential of scalar field $\phi$ as
\begin{equation}
V(\phi) = \mu^2 \phi^2 + \lambda\phi^4\,.
\end{equation}
for $\mu^2 <0$ and $\lambda >0$. The minimum takes place at
\begin{equation}
v= \pm \sqrt{\frac{-\mu^2}{\lambda}}
\end{equation}
We would promote the field into quantum state, so we have
\begin{equation}
|\Psi \rangle = \frac{1}{\sqrt{2}} ( |-v\rangle + |+v\rangle) \,.
\end{equation}
Then the probability of choosing either the positive or negative vacuum state is $1/2$ and one can infer the $|-v\rangle$ as $|0\rangle$ state and the $|+v\rangle$ as $|1\rangle$.  We have equal probability to choose the either state. Once the either state is chosen, the state becomes classical again. 
The dual state processes complex rotational symmetry, the probability is U(1) invariant, i.e. a local or global transformation
\begin{equation}
| \Psi^\prime \rangle = e^{i\theta}| \Psi \rangle
\end{equation}
leaves the probability invariant.

Next we study the qubit state vector $| \Psi \rangle$. take the form
\begin{equation}
| \psi \rangle = a_0 | 0 \rangle + a_1 | 1 \rangle
\end{equation}
Here first we are interested in the case when $a_0 = b_0$ or $a_0 = -b_0$ (which are $1/\sqrt{2}$ and $-1/\sqrt{2}$) for our study of duality. \begin{equation}
\begin{aligned}
|\psi_{++} \rangle & = \frac{1}{\sqrt{2}} ( |0\rangle + | 1 \rangle ) \\
|\psi_{-+} \rangle & = \frac{1}{\sqrt{2}} ( -|0\rangle + | 1 \rangle ) \\
|\psi_{+-} \rangle & = \frac{1}{\sqrt{2}} ( |0\rangle - | 1 \rangle ) \\
|\psi_{--} \rangle & = \frac{1}{\sqrt{2}} (-|0\rangle - | 1 \rangle ) 
\end{aligned}
\end{equation}
The basis with four elements $\{|\psi_{++} \rangle , |\psi_{-+}\rangle ,  |\psi_{+-} \rangle, |\psi_{--} \rangle  \}$ is isomorphic to $\{ |00 \rangle, |01\rangle, |10\rangle , |11\rangle  \}$. The conserved quantity for such $4-dual$ system is the probability of obtaining each of the state, both are $\frac{1}{2}$.

  The factor of $1/\sqrt{2}$ is not important here for discussion so we just take as $+1, -1$.  Let $U={1 , -1}$ be a dual set and $V={|0 \rangle, | 1 \rangle} $ be another dual set.  We can rewrite the qubit as follow,
\begin{equation}
\psi =  a_0 | 0 \rangle + a_1 | 1 \rangle =  (a_0 \oplus a_1 ) ( | 0 \rangle \oplus  | 1 \rangle) \,,
\end{equation}
where $a_0 , a_1 = \pm 1$ are constant operators. Now define two dual operators $*$ and $\star$, which act on $U$ and $V$ respectively. Note that the two dual operators commute. For example,
\begin{equation}
\begin{aligned}
* | \psi \rangle &= *(a_0 \oplus a_1 ) ( | 0 \rangle \oplus  | 1 \rangle) \\
&=  (*a_0 \oplus *a_1 ) ( | 0 \rangle \oplus  | 1 \rangle) \\
&=  (a_1 \oplus a_0 ) ( | 0 \rangle \oplus  | 1 \rangle) \,,
\end{aligned}
\end{equation}
and 
\begin{equation}
\begin{aligned}
\star | \psi \rangle &= \star\big( (a_0 \oplus a_1 ) ( | 0 \rangle \oplus  | 1 \rangle) \big) \\
&= (a_0 \oplus a_1 )\,\star( | 0 \rangle \oplus  | 1 \rangle) \big) \\
&=  (a_0 \oplus a_1 ) \, ( \star| 0 \rangle \oplus  \star| 1 \rangle) \\
&= (a_0 \oplus a_1 ) ( | 1 \rangle \oplus | 0 \rangle) \,.
\end{aligned}
\end{equation}
Together we have
\begin{equation}
*\star| \psi \rangle = *(a_0 \oplus a_1 ) \star ( | 0 \rangle \oplus  | 1 \rangle) 
=(a_1 \oplus a_0 )( | 1 \rangle \oplus | 0 \rangle)
= a_1 |1 \rangle + a_0 |0 \rangle 
\end{equation}
Thus for the case of $a_0  = 1 , a_1 = -1 $ (or vice versa) which is the dual set $U$, we have 
\begin{equation}
| \psi \rangle = *\star | \psi \rangle \,.
\end{equation}

\subsubsection{Dual Partition evolution}
We would introduce the concept of dual partition evolution for $n=1$ case, this essentially describe the process of a \emph{tai chi} diagram. This allows us to study how the state $|-- \rangle$ and state $|-\rangle $ exchange one another under time evolution. 
\begin{figure}[H]
\centering
\includegraphics[trim=0cm 0cm 0cm 0cm, clip, scale=0.3]{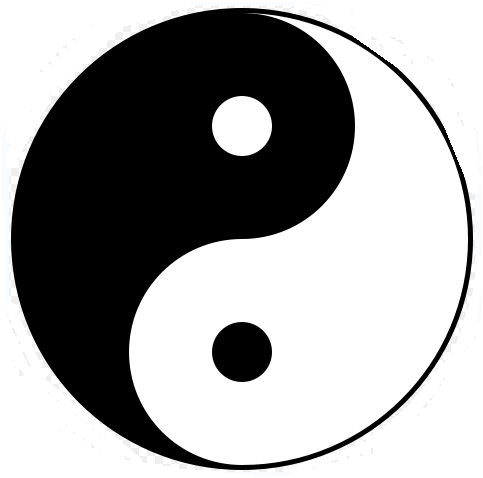}
\caption[A standar tai chi diagram]
{A standard \emph{tai chi} diagram  \label{fig:taichicycle1}}
\end{figure}
We would depict the evolution process in a cycle evolution of the 4-tableau. First consider the initial full set All as $V \cup V^*$ where $V \cap V^* = \emptyset$ such that the full set naturally partitioned into two equal halves $V$ and $V^*$ (represented as $|0 \rangle$ and $|1 \rangle$ with equal probability ). 
\begin{figure}[H]
\centering
\includegraphics[trim=0cm 0cm 0cm 0cm, clip, scale=0.45]{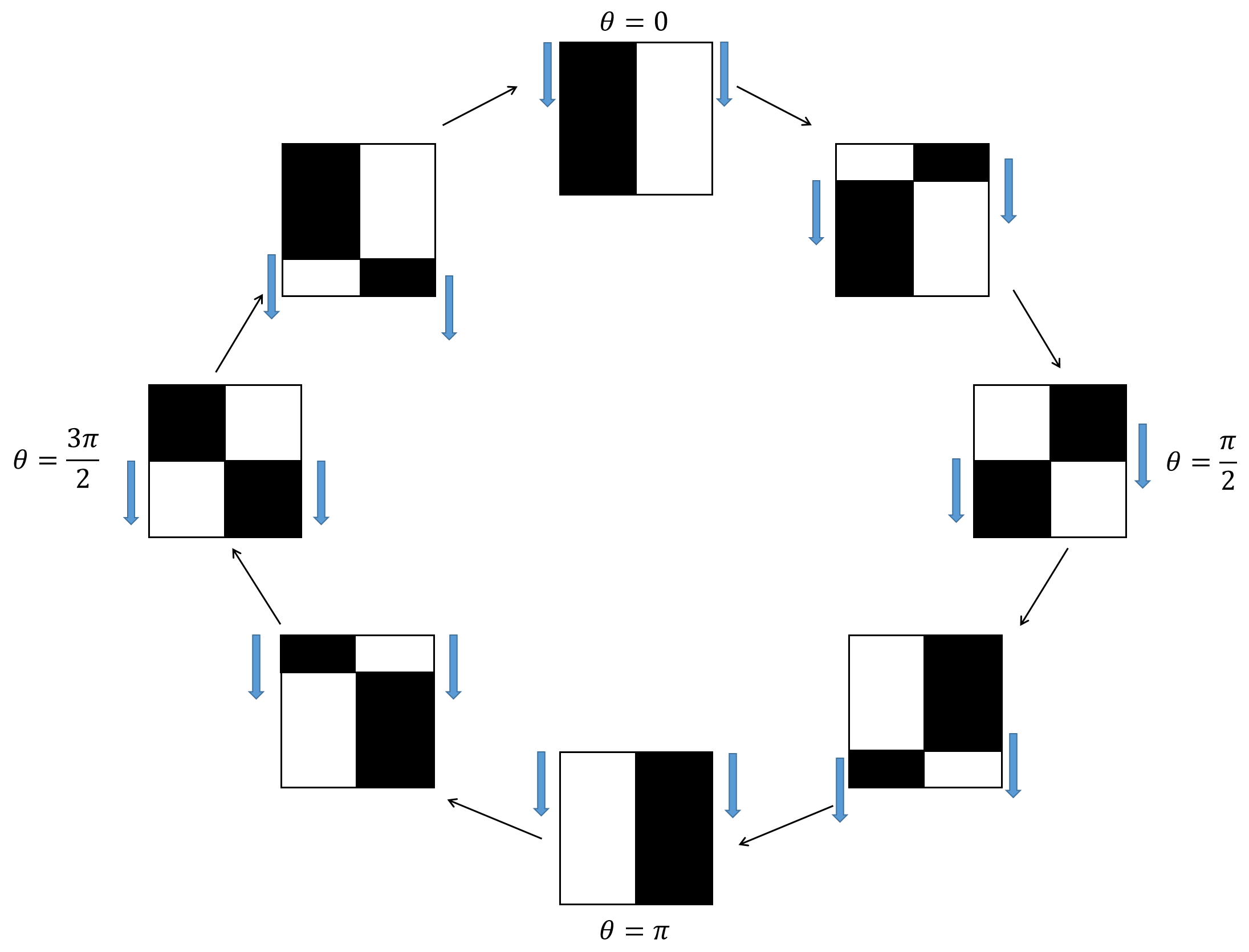}
\caption[A standard tai chi cycle.]
{A standard \emph{tai chi} cycle as the phase evolution of 4-tableau. The $|0\rangle$ is denoted as black and the $|1 \rangle$ is denoted by white. We suppose the flow is moving downwards during the exchange. The phase $\theta$ is related to time by $\theta = \omega t$.  \label{fig:taichicycle1}}
\end{figure}
Let $Q$ be the left partition and $Q^*$ be the dual partition. The general \emph{tai chi } state $|\psi \rangle$ can be expressed as linear superposition of the two states in the two partitions. And we define our observable frame as $S_k$, then
\begin{equation}
( |\psi (t ) \rangle \,\,| S_{k} ) = \Big( c_1 (t) |0 \rangle_{Q} + c_2 (t) |1 \rangle_{Q^*} +  c^*_1 (t) |0^* \rangle_{Q} + c_2^* (t) |1^* \rangle_{Q^*} \,\,\Big\vert S_k \Big) \,.
\end{equation}
The four states satisfy the orthogonality relations,
\begin{equation}
_{P}\langle  i | j \rangle_{P^\prime} = \delta_{ij} \delta_{PP^\prime} \,,
\end{equation}
where $i, j = 0,1$ and $P, P^\prime = Q, Q^{*}$.  Note that one can also infer the state $|0\rangle_Q $ as $|0\bar{0}\rangle$ , $|1\rangle_{Q^*} $ as $|1\bar{1}\rangle$, $|0\rangle_{Q^*}$ as $|0\bar{1}\rangle$ and $|1\rangle_{Q}$ as $|1\bar{0}\rangle$ , which contribute to a 4-dual system emerged from a 2-dual system. The total probability is $ P_1 (t) + P_2 (t) + P_1^* (t) + P_2^* (t) =1$, explicitly
\begin{equation}
 |c_1 (t)|^2 + |c_2 (t)|^2 + |c_1^* (t)|^2 + |c_2^* (t)|^2 =1 \,
\end{equation}
and the probability of each partition is always $\frac{1}{2}$,
\begin{equation}
|c_1 (t)|^2 + |c_1^* (t)|^2 = \frac{1}{2} \quad \text{and} \quad |c_2 (t)|^2 + |c_2^* (t)|^2 = \frac{1}{2} \,. 
\end{equation}
and also we must have
\begin{equation}
|c_1 (t)|^2 + |c_2^* (t)|^2 = \frac{1}{2} \quad \text{and} \quad |c_2 (t)|^2 + |c_1^* (t)|^2 = \frac{1}{2} \,. 
\end{equation}

At $t = 0$, we have
\begin{equation}
c_1 (0 ) = c_2 (0) = \frac{1}{\sqrt{2}} \quad \text{and } \quad c_1^* (0 ) = c_2^* (0) =0 \,,
\end{equation}
thus the probabilities are $P_1=|c_1 (0)|^2 =P_2 =|c_1 (0)|^2 = \frac{1}{2} $ and $P_1^* = |c_1^* (0)|^2 =P_2^* = |c_1^* (0)|^2  =0$. This describe the the initial state, which is the first diagram at $\theta =0$ in \ref{fig:taichicycle1}, which is just the qubit,
\begin{equation} \label{eq:qubittahchi}
|\psi (0 ) \rangle = \Big( \frac{1}{\sqrt{2}} |0 \rangle_{Q} + \frac{1}{\sqrt{2}}  |1 \rangle_{Q^*} \,\,\Big| S_{k} \Big) \,.
\end{equation}

The time evoluting quantum state is
\begin{equation}
|\psi (t) \rangle = \frac{1}{\sqrt2} \bigg( \cos\frac{\omega t}{2} |0\rangle_Q +   \cos\frac{\omega t}{2} |1\rangle_{Q^*} +  \sin\frac{\omega t}{2} |0^*\rangle_Q  + \sin\frac{\omega t}{2} |1^*\rangle_{Q^*}  \bigg)
\end{equation}
We can check that when $\theta = \pi$,
\begin{equation}
|\psi (T/2) \rangle = \frac{1}{\sqrt{2}}  |0^*\rangle_Q +  \frac{1}{\sqrt{2}}  |1^*\rangle_{Q^*} = \frac{1}{\sqrt{2}}  |1\rangle_Q +  \frac{1}{\sqrt{2}}  |0\rangle_{Q^*}
\end{equation}
and when $\theta = \pi/2$,
\begin{equation}
|\psi (T/2) \rangle = \frac{1}{2} \Big( |0\rangle_Q + |1 \rangle_{Q^*} + |1\rangle_Q + |0 \rangle_{Q^*} \Big) \,.
\end{equation}
When we look from the $S_{k}^\star$ dual frame, the phase would be offset by $\pi$, we have
\begin{equation}
( |\psi (\theta ) \rangle \,\,| S_{k} ) =  ( |\psi (\theta - \pi )\rangle \,\,| S_{k}^{\star} ) \,,
\end{equation}
i.e.
\begin{equation}
\begin{aligned}
&\quad \bigg(\frac{1}{\sqrt2} \Big( \cos\frac{\omega t}{2} |0\rangle_Q +   \cos\frac{\omega t}{2} |1\rangle_{Q^*} +  \sin\frac{\omega t}{2} |0^*\rangle_Q  + \sin\frac{\omega t}{2} |1^*\rangle_{Q^*}  \Big) \,\, \bigg\vert S_k  \bigg) \\
&\equiv \bigg( \frac{1}{\sqrt2} \bigg( \sin\frac{\omega t}{2} |0\rangle_Q +   \sin\frac{\omega t}{2} |1\rangle_{Q^*} +  \cos\frac{\omega t}{2} |0^*\rangle_Q  + \cos\frac{\omega t}{2} |1^*\rangle_{Q^*}  \Big) \,\, \bigg\vert S_k^\star \bigg) \\
\end{aligned}
\end{equation}

The \emph{tai chi} diagram and the \ref{fig:taichicycle1} representation have the same topology. We can define the tai-chi process formally by the following:
\begin{definition}
Let $U$ be the full space and $Q$ , $Q^*$ the partition and dual partition space where $U= Q\cup Q^* $, and $U \subset \mathbb{R}^2 $. Let the area function of the partitions be $A(U) = 1$. The exists an isomorphism $f$ between the probability space $P(X)$ with random variable $X=\{ |0\rangle , |1 \rangle\}$ as the state space and the area space $A$, which defines the boxed partition diagram,
\begin{equation}
f : P \rightarrow A \,.
\end{equation} 
\end{definition}
\begin{definition}
The \emph{tai chi} diagram is topological equivalent to the boxed \emph{tai chi} cycle process with the full diagram that has the area of 1 unit. 
\end{definition} 
We can interpret backwards, the \emph{tai chi} cycle process can be deformed to the \emph{tai chi} diagram by homeomorphism.
\begin{figure}[H]
\centering
\includegraphics[trim=0cm 0cm 0cm 0cm, clip, scale=0.6]{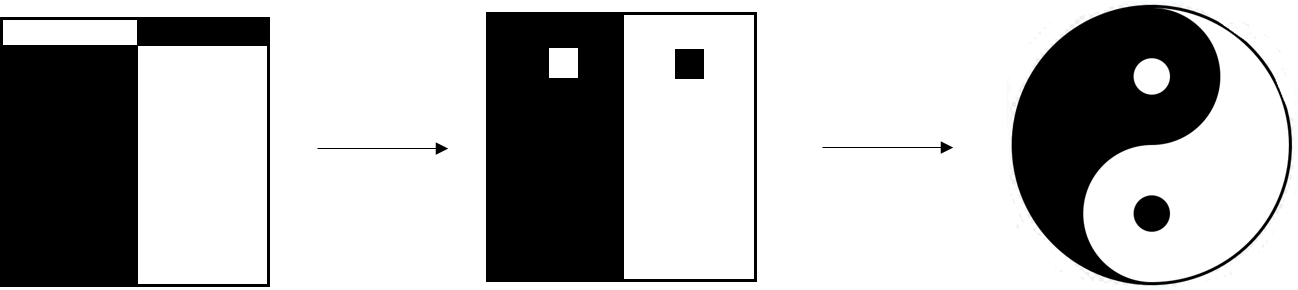}
\caption[TaichiToplogy]
{  Continuous deformation from \emph{tai chi} cycle process to \emph{tai chi } diagram. }
\end{figure}

\subsection{The $n=2$ level }
The basis of $n=2$ level, i.e. $\mathbb{Z}_2 \otimes \mathbb{Z}_2$ 4-duality group is naturally a 4-fundamental tableau. 
\begin{figure}[H]
\centering
\includegraphics[trim=0cm 0cm 0cm 0cm, clip, scale=0.75]{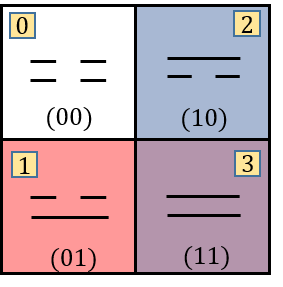}
\caption[Analysis]
{  \label{fig:4xiangtableau}}
\end{figure}
The null state corresponds to $0$, the 01 state corresponds to $V$, the 10 corresponds to $V^*$, and the full 11 state corresponds to All. We can interpret also as either the 01 and 10 state contributes half of the system, and the joining of them represents the fullness All. Therefore, it naturally forms the basis of $\mathbb{Z}_2 \times \mathbb{Z}_2$ 4-duality group and is represented by the 4-fundamental tableau. The basis is $\{ |00\rangle , |01\rangle , |10\rangle , |11\rangle\}$, or in decimal representation $\{ |0\rangle , |1\rangle , |2\rangle , |3\rangle  \}$. 

Now recalling the comparison representation, here we can interpret the 4 states as group elements. The is equivalent to turning the basis as operators. We call it quantization of the 4-basis. We can construct Cayley table as follow. 
\begin{figure}[H]
\centering
\includegraphics[trim=0cm 0cm 0cm 0cm, clip, scale=0.6]{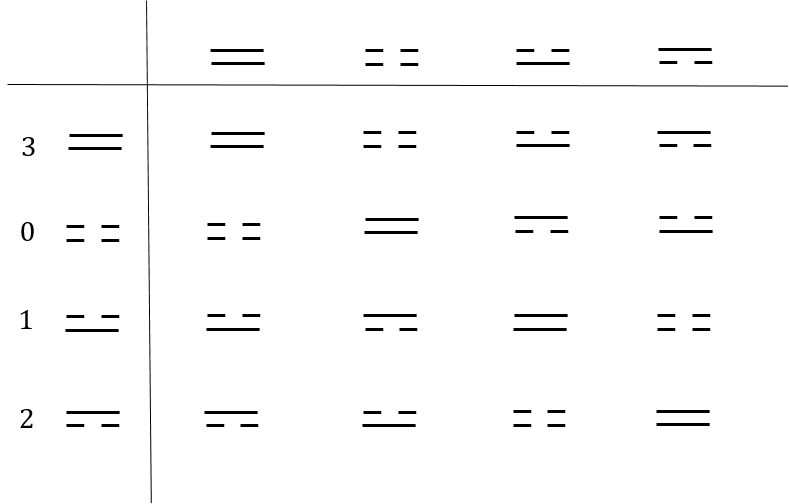}
\caption[Analysis]
{ \label{fig:4CaleyTable}}
\end{figure}
Therefore we write
\begin{equation}
|0\rangle \rightarrow \hat{0} \,\, , \,\, |1\rangle \rightarrow \hat{1}  \,\, , \,\, |2\rangle \rightarrow \hat{2}  \,\,, \,\, |1\rangle \rightarrow \hat{3} \,. 
\end{equation}
Or 
\begin{equation}
|00\rangle \rightarrow \hat{00} \,\, , \,\, |01\rangle \rightarrow \hat{01} \,\, , \,\, |10\rangle \rightarrow \hat{10}  \,\,, \,\, |11\rangle \rightarrow \hat{11} \,. 
\end{equation}
The \ref{fig:4CaleyTable} is isomorphic to the point group $C_{2v} = \{ e, C_2 , \sigma_1 , \sigma_2 \}$, with the element identification as
\begin{equation}
\hat{3} \rightarrow e \,\, , \,\, \hat{0} \rightarrow C_2 \,\, , \,\, \hat{1} \rightarrow \sigma_1 \,\, , \,\, \hat{2} \rightarrow \sigma_2 \,. 
\end{equation}
The feature diagram is 
\begin{figure}[H]
\centering
\includegraphics[trim=0cm 0cm 0cm 0cm, clip, scale=0.75]{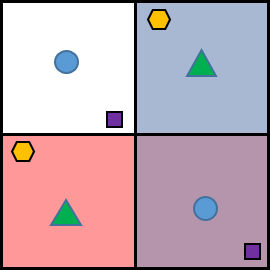}
\caption[Analysis]
{  Feature diagram of 4-yi.\label{fig:featureDiagram4yi}}
\end{figure}

The representation matrix of the 4-duality group is
\begin{equation} \label{eq:AIrreps}
D(\mathbb{Z}_2 \otimes \mathbb{Z}_2)  = 
 \begin{pmatrix}
  A_1 & 0 & 0 & 0 \\
  0 & A_2 & 0 & 0\\
  0  & 0  & A_3 & 0  \\
  0 & 0 & 0 & A_4 \\ 
 \end{pmatrix}
 =
 \begin{pmatrix}
  \mathcal{A}_1 (g) \mathcal{A}_1 (g^\prime) & 0 & 0 & 0 \\
  0 & \mathcal{A}_2 (g) \mathcal{A}_2 (g^\prime) & 0 & 0\\
  0  & 0  & \mathcal{A}_2 (g) \mathcal{A}_1 (g^\prime) & 0  \\
  0 & 0 & 0 & \mathcal{A}_1 (g) \mathcal{A}_2 (g^\prime) \\ 
 \end{pmatrix}
\end{equation}
for $\mathrm{g}=(g, g^\prime) = (I,I) , (P,P) , (P,I) \,\,{\text{and}}\,\,(I,P)$. Explicitly

\begin{equation} \label{eq:resultAIrreps}
\begin{aligned}
D\big( ([0],[0])  \big) &= D\otimes D (I,I) = {\text{diag}}(1,1,1,1) \,, \\
D\big( ([1],[1])  \big) &= D\otimes D (P,P) = {\text{diag}}(1,1,-1,-1) \,,\\
D\big( ([0],[1])  \big) &= D\otimes D (I,P) = {\text{diag}}(1,-1,-1,1) \,,\\
D\big( ([1],[0])  \big) &= D\otimes D (P,I) = {\text{diag}}(1,-1,1,-1) \,.
\end{aligned}
\end{equation}

Next we take the basis as $|00\rangle$ , $|01 \rangle$ , $|10\rangle$ and $|11\rangle$. They correspond to, respectively
\begin{equation}
\begin{aligned}
|00\rangle &= | - - \rangle | - -\rangle & \equiv | \includegraphics[trim=0cm 0cm 0cm 0cm, clip, scale=0.5]{04.png} \rangle \\
|01 \rangle &= |- - \rangle | - \rangle & \equiv | \includegraphics[trim=0cm 0cm 0cm 0cm, clip, scale=0.5]{24.png} \rangle \\
|10 \rangle &= |-  \rangle | - -  \rangle & \equiv | \includegraphics[trim=0cm 0cm 0cm 0cm, clip, scale=0.5]{14.png} \rangle \\
|11 \rangle &= |-  \rangle | -   \rangle & \equiv | \includegraphics[trim=0cm 0cm 0cm 0cm, clip, scale=0.5]{34.png} \rangle \\ \,.
\end{aligned}
\end{equation}

We can further define $A_1 = U_1 , A_2 = U_2$ and $A_3 = W_1 , A_4 = W_2$,
\begin{equation}
\begin{aligned}
&\quad U_1 | {\text{00}} \rangle \oplus U_2 | {\text{11}} \rangle \oplus W_1 | {\text{01}} \rangle \oplus W_2 | {\text{10}} \rangle \\
&= U\big( |{\text{00}} \rangle \oplus  | {\text{11}} \rangle \big) \oplus W\big( |{\text{01}}\rangle \oplus  |{\text{10}} \rangle \big) 
\end{aligned}
\end{equation}
for $U = U_1 \oplus U_2$ and $W= W_1 \oplus W_2$, so we separate them into two large categories, the (CC and DD) one, the fully connected or disconnected case, and the (DC and CD) one which are both halfly connected. Thus one can write
\begin{equation}
D(\mathbb{Z}_2 \otimes \mathbb{Z}_2  )  =
\begin{pmatrix}
  U & 0 \\
  0 & V
 \end{pmatrix}
\end{equation}
which is reducible, and this neatly represents that the 4-duality group as two big categories, with $|00\rangle ,\,|11 \rangle$ being one and $|01\rangle , \,|10\rangle$ another one.  Compactly in dimension notation we can write the tensor product as
\begin{equation}
2 \otimes 2 = 2 \oplus 2 = 1 \oplus 1 \oplus 1 \oplus 1 \,.
\end{equation}

\subsection{Quantum state with embedded 4-duality group}
In this section, we would like to promote the basis of irreps of $\mathbb{Z}_2 \times \mathbb{Z}_2$ to quantum states and study the linear combination of it. The full linear combination of the basis of irreps over the 4-Klein group can be expressed as a rank-2 tensor,
\begin{equation}
|\,\phi (g, g^\prime) \,\rangle = \frac{1}{2} \sum_{i_{1}, i_{2}= 0,1} a_{i_1 i_2 } (g, g^\prime) | \eta_{i_1} \rangle \otimes | \eta_{i_2} \rangle \,,
\end{equation} 
where we sum over repeating indices. The factor of $\frac{1}{2}$ is for normalization. Explicitly we write
\begin{equation}
\begin{aligned}
|\,\phi (g, g^\prime)\,\rangle &= \frac{1}{2} \Big( a_{00} (g, g^\prime) | 00\rangle +  a_{11} (g, g^\prime) |11\rangle +  a_{01}(g, g^\prime) | 01\rangle + a_{10}(g, g^\prime) | 10\rangle  \Big) \\
& =  \frac{1}{2} \Big( a_{00} (g, g^\prime) | -- \rangle |--\rangle +  a_{11} (g, g^\prime) | -\rangle|-\rangle +  a_{01}(g, g^\prime) | -- \rangle|-\rangle + a_{10}(g, g^\prime) | -\rangle |--\rangle  \Big) \,.
\end{aligned}
\end{equation} 
Define the tensor by
\begin{equation}
\mathrm{A}(g , g^\prime) =  \begin{pmatrix}
  a_{00}( g, g^\prime)  & a_{01}( g, g^\prime) \\
  a_{\rm{10}}( g, g^\prime) & a_{\rm{11}}( g, g^\prime) 
 \end{pmatrix}
 = \begin{pmatrix}
   \mathcal{A}_1 (g)  A_1 (g^\prime )  &  \mathcal{A}_1 (g)  \mathcal{A}_2 (g^\prime ) \\
    \mathcal{A}_2 (g) \mathcal{A}_1 (g^\prime ) & \mathcal{A}_2 (g ) \mathcal{A}_2 (g^\prime ) 
 \end{pmatrix} \,, 
\end{equation}
The $\mathcal{A}_1, \mathcal{A}_2$ are 1D irreps we had in \ref{eq:AIrreps}. Using the result we had in \ref{eq:resultAIrreps}, we have
 \begin{equation} \label{eq:4Matrices}
\mathrm{A}(I,I)=  
 \begin{pmatrix}
  1 & 1 \\
  1 &1
\end{pmatrix}
 \,\, , \,\,
\mathrm{A}(P,P)=
 \begin{pmatrix}
  1 & -1 \\
  -1 &1
\end{pmatrix}
\,\, , \,\,
\mathrm{A}(I,P)= 
\begin{pmatrix}
  1 & -1 \\
  1 & -1
 \end{pmatrix}
 \,\, , \,\,
\mathrm{A}(P,I)= 
\begin{pmatrix}
  1 &  1 \\
  -1 & -1
 \end{pmatrix} \,.
 \end{equation} \label{eq:4KleinMatrix}
And since the tensor product of two parity group is just same as the 4-duality group, thus we have written $(g,g^\prime) \in \mathbb{Z}_2 \otimes \mathbb{Z}_2$ as $g\in \mathbb{Z}_2 \times \mathbb{Z}_2$, and therefore from above we find
\begin{equation}
{\rm{det}}\,A(\rm{g}) = 0 \,\,{\text{for all \rm{g}}}\in  \mathbb{Z}_2 \times \mathbb{Z}_2 \,.
\end{equation}
That means the linear combination of the 4 reducible representation basis of the 4-dual set with the tensor components as the transformation of the 4-duality Klein 4 group is untangled. In other words, the doubly parity transformation of a 4-dual set corresponds to 4 untangled basis. 
 
This shows that the full state $\phi$ can be written as two independent tensor product of the same dual duplet. For $g \in \mathbb{Z}_2$
\begin{equation}
|\, \varphi (g)\, \rangle = \frac{1}{\sqrt 2} \big( \mathcal{A}_1 ( g) |{\text{0}} \rangle + \mathcal{A}_2 (g) |  {\text{1}} \rangle \big) .
\end{equation}
Since all $\mathrm{A}_1 , \mathrm{A}_2$ are either $1$ or $-1$, thus it is automatically properly normalized by $\frac{1}{\sqrt{2}}$. And we have
\begin{equation}
|\,\phi(g , g^\prime) \,\rangle = |\,\varphi(g)\,\rangle \otimes |\,\varphi(g^\prime) \,\rangle \,.
\end{equation}  
We can see that we have a probability of $\frac{1}{2}$ of having $| 0 \rangle$ or $| 1 \rangle$ state.  More properties will be discussed later.
 
From the results of \ref{eq:4KleinMatrix} , intuitively we can see that $\mathrm{A}(I,I)$ and $\mathrm{A}(P,P)$ form a dual pair, while $\mathrm{A}(I,P)$ and $\mathrm{A}(P,I)$ for a dual pair, this is because we see that the character (trace) of $\mathrm{A}(I,I)$ and $\mathrm{A}(P,P)$ are the same, while the character of $\mathrm{A}(I,P)$ and $\mathrm{A}(P,I)$ are the same,
\begin{equation}
{\mathrm{Tr}}\, \mathrm{A}(I,I) =  {\mathrm{Tr}}\, \mathrm{A}(P,P)=2 \quad {\text{and}} \quad {\mathrm{Tr}}\, \mathrm{A}(I,P) =  {\mathrm{Tr}}\, \mathrm{A}(P,I)= 0 \,.
\end{equation}
Therefore, one can obtain $\mathrm{A}(P,P)$ from $\mathrm{A}(I,I)$ by some similarity transformation, i.e. $\mathrm{A}(P,P) = U^{-1} \mathrm{A}(I,I) U$ for some matrix $U$ , and likewise $\mathrm{A}(I,P) = V^{-1}\mathrm{A}(P,I) V$ for some matrix $V$. 
In fact, the 4 representation matrices in \ref{eq:4Matrices} form a 4-duality group $\{ \mathrm{A}(I,I) , \mathrm{A}(I,P) , \mathrm{A}(P,I) , \mathrm{A}(P,P)   \}$ under the operation of element-wise matrix multiplication $\bullet$ (known as the Hadamard product). Such operation is Abelian. The identity is $\mathrm{A}(I,I)$, and each element is of its own inverse. For example it is easy to check that
\begin{equation}
\mathrm{A}(I,P) \bullet \mathrm{A}(P,I) = \mathrm{A}(P,P)  \,,
\end{equation} 
\begin{equation}
\mathrm{A}(P,P) \bullet \mathrm{A}(P,I) = \mathrm{A}(I,P) \,,
\end{equation}
\begin{equation}
\mathrm{A}(P,P) \bullet \mathrm{A}(I,P) = \mathrm{A}(P,I) \,,
\end{equation}
\begin{equation}
\mathrm{A}(I,P) \bullet \mathrm{A}(I,P) = \mathrm{A}(I,I) \,,
\end{equation}
etc.  
Therefore $\mathrm{A}$ is actually a function of the group elements of the parity group. The idea can be represented by the following diagram
\begin{figure}[H]
\centering
\includegraphics[trim=0cm 0cm 0cm 0cm, clip, scale=0.6]{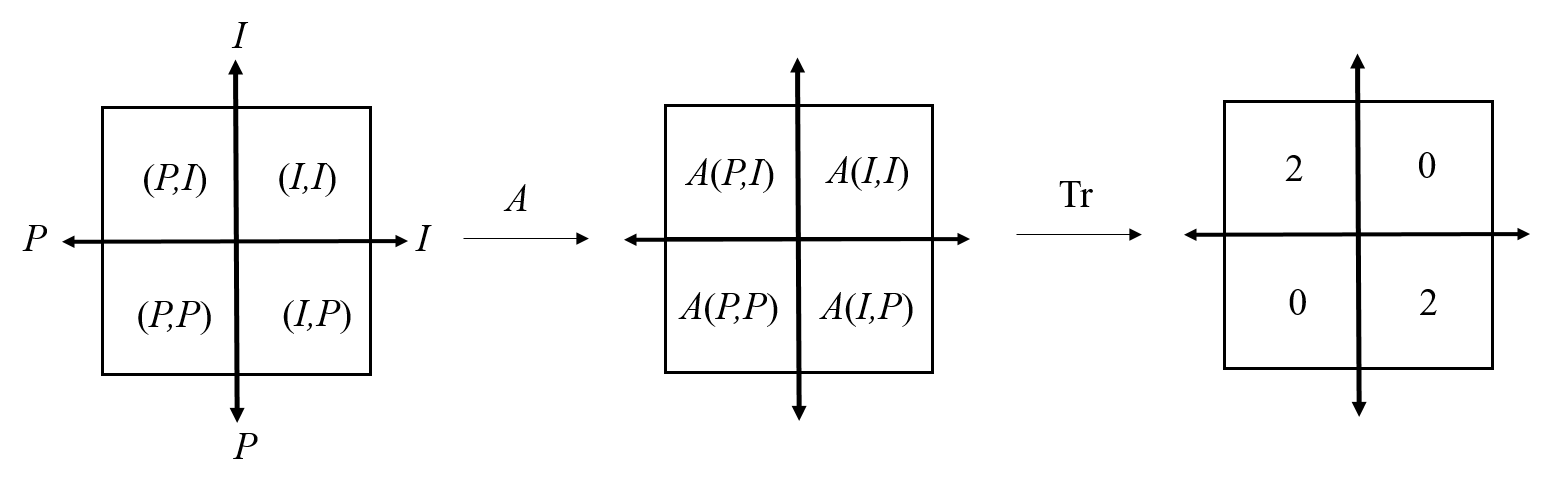}
\caption[Diagramatic representation]
 {Diagramatic representation. \label{fig:CharacterRep}}
\end{figure}

\subsection{4-Duality basis representation by order rearrangement}
In this subsection, we would study a very important construction of the 4-Duality basis by the order permutation of 4-yi of $n=2$ level. This creates a new way of forming basis representation of the 4-duality group, and would extend the originally simple basis representation from $|--\rangle|--\rangle,\, |-\rangle|-\rangle,\,|--\rangle|-\rangle\, \,\text{and}\,\, |-\rangle |--\rangle$. 

Let's first define some clear notations. At the $n=1$ level splitting, we can start with either $-$ or $--$ first, then followed by $--$ and $-$ respectively. The former is said to be left 0 right 1, symbolized by $(--,-)$, while the latter would be left 1 right 0 symbolized by $(-,--)$. Next denote the stacking of a yin state on the left and a yang state on the right as LR and the opposite RL, and this can be done along the upward direction (U) and downward direction (D). Next we assign the Gua's order parameter as I, II, III, IV as usual. The duality mirror lies in the middle to separate I, II and III, IV into two halves. When the stacking carries, if there is no crossing between the two halves we call it is a normal diagram, otherwise a crossing diagram. The crossing reference is labelled by two pairs of order parameter, for example I-III II-IV crossing. This is called the symmetry crossing as one will see that this is symmetric along the dual mirror plane. For each diagram, we collect the possible outcomes as $(ijkl)$, where each of them is the binary number of the outcome. The following illustrates how do these principles work
\begin{figure}[H]
\centering
\includegraphics[trim=0cm 0cm 0cm 0cm, clip, scale=0.5]{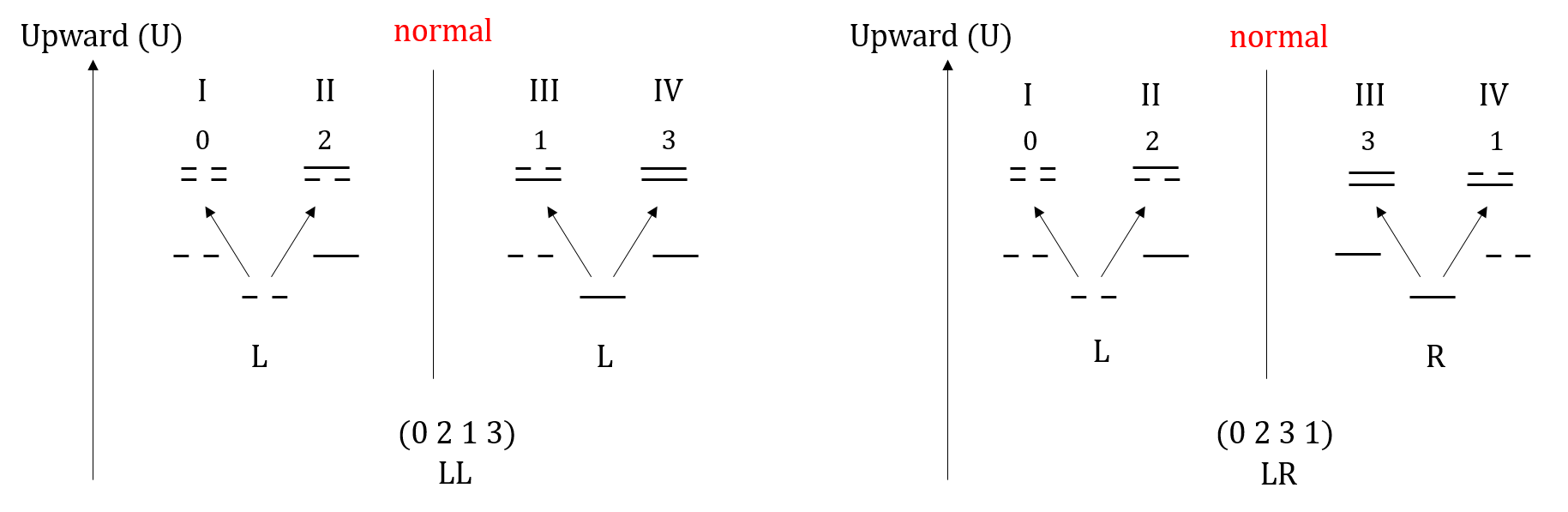}
\caption[Examples for upward direction]
 {Two examples for the splitting processes along the upward direction. Both of them are noraml diagrma. \label{fig:CharacterRep}}
\end{figure}
\begin{figure}[H]
\centering
\includegraphics[trim=0cm 0cm 0cm 0cm, clip, scale=0.5]{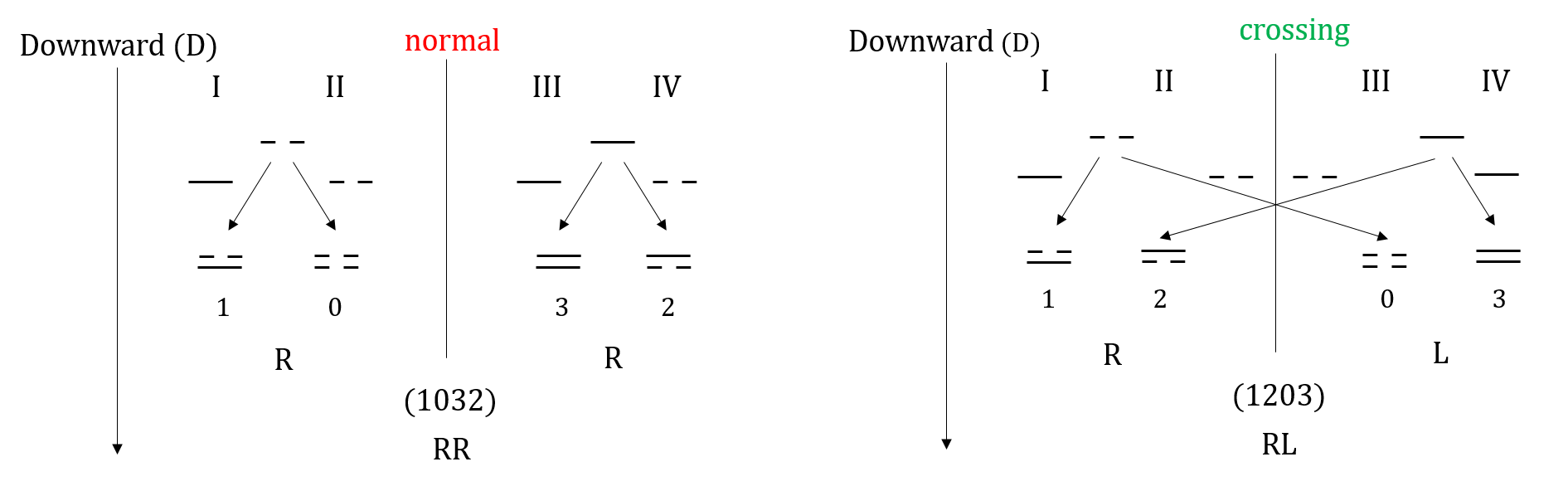}
\caption[Examples for downward direction]
 {Two examples for the splitting processes along the downward direction. The left one is a noraml diagram and the right one is a I-III II-IV crossing diagram. \label{fig:CharacterRep}}
\end{figure}

There would a total of 32 outcomes. The results are tabulated as follow
\begin{table}[H] 
\begin{center}
\begin{tabular}{ c c c c c }
\hline
\multicolumn{5}{l}{Normal diagram: Left 0 Right 1 $(--,-)$} \\
\hline
    & LL       &  RR        & LR       &  RL  \\
 U  & $\color{red}{(0213)}$   &  $\color{red}{(2031)}$    & $\color{red}{(0231)}$   &  $\color{red}{(2013)}$  \\
 D  & $\color{red}{(0123)}$   &  $\color{red}{(1032)}$    & $\color{red}{(0132)}$   &  $\color{red}{(1023)}$  \\
\hline
\multicolumn{5}{l}{Normal diagram: Left 1 Right 0 $(-,--)$} \\
\hline
    & LL       &  RR        & LR       &  RL  \\
 U  & $\color{red}{(1302)}$   &  $\color{red}{(3120)}$    & $\color{red}{(1320)}$   &  $\color{red}{(3102)}$  \\
 D  & $\color{red}{(2301)}$   &  $\color{red}{(3210)}$    & $\color{red}{(2310)}$   &  $\color{red}{(3201)}$ \\
\hline
\multicolumn{5}{l}{Crossing diagram: Left 0 Right 1 $(--,-)$ } \\
\hline
   & LL       &  RR        & LR       &  RL  \\
 U & $\color{blue}{(0303)}$   &  $\color{blue}{(2121)}$    & $\color{ForestGreen}{(0321)}$   &  $\color{ForestGreen}{(2103)}$ \\
 D & $\color{blue}{(0303)}$   &  $\color{blue}{(1212)}$    & $\color{ForestGreen}{(0312)}$   &  $\color{ForestGreen}{(1203)}$ \\
\hline
\multicolumn{5}{l}{Crossing diagram: Left 1 Right 0 $(-,--)$ } \\
\hline
   & LL       &  RR        & LR       &  RL  \\
 U & $\color{blue}{(1212)}$   & $\color{blue}{(3030)}$     & $\color{ForestGreen}{(1230)}$   &  $\color{ForestGreen}{(3012)}$  \\
 D & $\color{blue}{(2121)}$   & $\color{blue}{(3030)}$     & $\color{ForestGreen}{(2130)}$   &  $\color{ForestGreen}{(3021)}$  \\
\hline
\end{tabular}
\end{center}
{\caption{Tabulated results.} \label{tab:result1}}
\end{table}

Hence, the full combination of normal diagrams and crossing diagrams give all the possibilities of the $4! = 24$ elements of permutation group $S_4$ and 8 elements that have two repeated numbers. We call those disconnected diagrams, the reason for calling them `disconnected' with be soon addressed. Now we would represent all the $S_4$ elements diagramatically with categories of closed loops and would map all the red and green $(ijlk)$ the corresponding categories. The result is shown as follow.

\begin{table}[H]
\begin{center}
\begin{tabular}{|c c | c c c c|}
\hline
\includegraphics[trim=0cm 0cm 0cm 0cm, clip, scale=0.5]{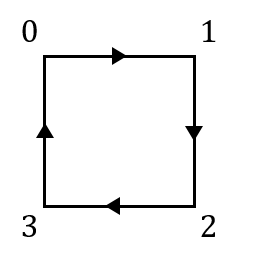} &
\includegraphics[trim=0cm 0cm 0cm 0cm, clip, scale=0.5]{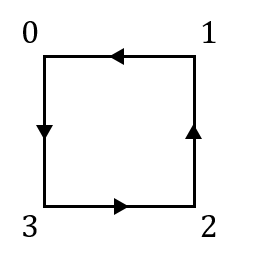} &
\includegraphics[trim=0cm 0cm 0cm 0cm, clip, scale=0.5]{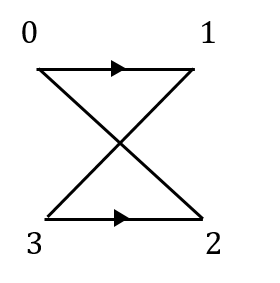} &
\includegraphics[trim=0cm 0cm 0cm 0cm, clip, scale=0.5]{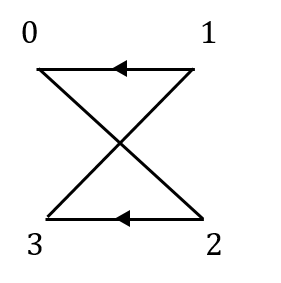} &
\includegraphics[trim=0cm 0cm 0cm 0cm, clip, scale=0.5]{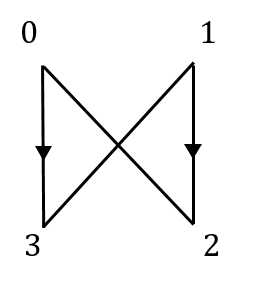} &
\includegraphics[trim=0cm 0cm 0cm 0cm, clip, scale=0.5]{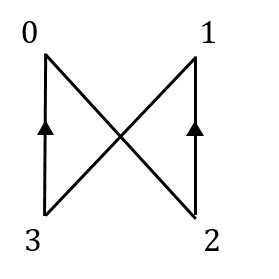} \\
\hline
\color{red}{(0123)} & \color{ForestGreen}{(0321)} & \color{red}{(0132)} & \color{red}{(0231)} & \color{ForestGreen}{(0312)} & \color{red}{(0213)} \\
\color{ForestGreen}{(3012)} & \color{red}{(1032)} & \color{red}{(2013)} & \color{red}{(1023)} & \color{red}{(2031)} & \color{ForestGreen}{(3021)} \\
\color{red}{(2301)} & \color{ForestGreen}{(2103)} & \color{red}{(3201)} & \color{red}{(3102)} & \color{ForestGreen}{(1203)} & \color{red}{(1302)} \\
\color{ForestGreen}{(1230)} & \color{red}{(3210)} & \color{red}{(1320)} & \color{red}{(2310)} & \color{red}{(3120)} & \color{ForestGreen}{(2130)} \\
\hline
\end{tabular}
\end{center}
\caption{ The $S_4$ permutation group represented by diagrams, with red and green values obtained in \ref{tab:result1} mapped onto them. Each successive row is formed by cyclic permutation of the previous one, i.e. $(ijkl) \rightarrow (lijk) \rightarrow$  etc. These operations form a rotation group of $\mathrm{C}_4 = \{I, C_4, C_4^2 , C_4^3 \}$.}
\end{table}

The 8 blue elements with two repeated numbers are not contained in the $S_4$ group representations, as they cannot be drawn as a closed loop. There are redundancies in each of these elements and they do not give rise to $(ijkl)$ for all different $i\neq j \neq k \neq l$. We represent these as two disjoint pieces.
\begin{figure}[H]
\centering
\includegraphics[trim=0cm 0cm 0cm 0cm, clip, scale=0.65]{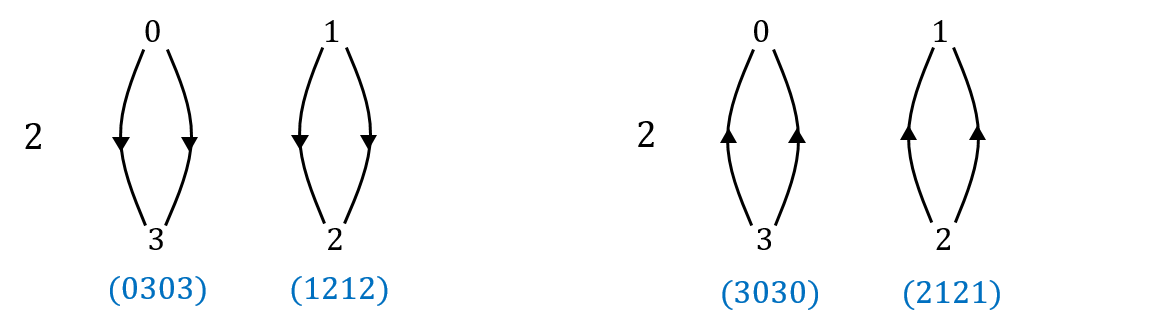}
\caption[Disconnected diagram for symmetric I-III II-IV crossing]
 {Disconnected diagram for symmetric I-III II-IV crossing. There are 2 set of diagrams for each of them.\label{fig:crossing1}}
\end{figure}

There is another choice of crossing , I-IV II-III. This is the asymmetric crossing, as it is not symmetric under the dual mirror plane. This would recover all the 8 green values in \ref{tab:result1}, but different blue values. 
\begin{figure}[H]
\centering
\includegraphics[trim=0cm 0cm 0cm 0cm, clip, scale=0.65]{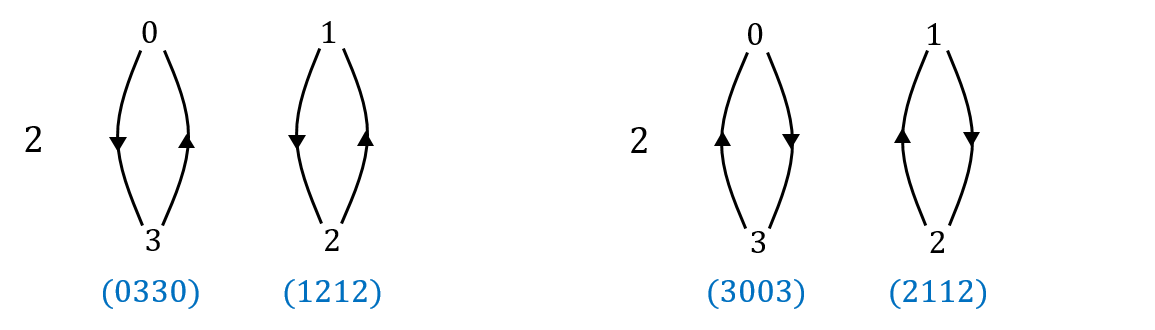}
\caption[Disconnected diagram for symmetric I-III II-IV crossing]
 {Disconnected diagram for symmetric I-IV II-III crossing. There are 2 set of diagrams for each of them.\label{fig:crossing2}}
\end{figure}

Since for the four all orders I, II, III, IV, there are only 3 possible ways for forming two pairs. The first way I-II , \, III-IV generates all 16 normal diagrams, then the remaining two ways are the symmetric I-III, II-IV crossing and the I-IV, II-III asymmetric crossing. Each type of crossing generates another 16 diagrams, in which 8 of them join with the 16  I-II , \, III-IV diagrams that form a representation of $S_4$ group. There are two sets of disconnected diagrams formed by the two different crossing, and the differences are in the direction of the connecting lines (see \ref{fig:crossing1} and \ref{fig:crossing2}). Now we can reorganize the result in \ref{tab:result1} to represent in two sets of 4-tableau due to different crossings as
\begin{figure}[H]
\centering
\includegraphics[trim=0cm 0cm 0cm 0cm, clip, scale=0.7]{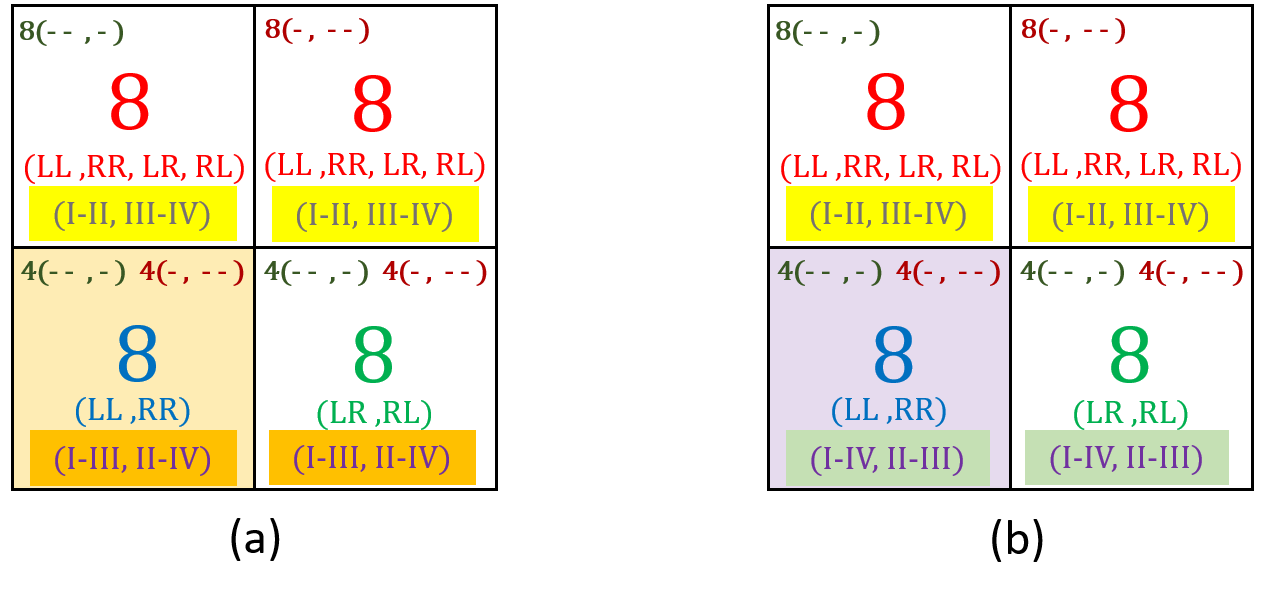}
\caption[Two sets of 4-tableau diagram]
 {Two sets of 4-tableau diagram. Figure (a) is for I-III, II,IV symmetric crossing and figure (b) is for I-IV, II-III asymemtric crossing. The big number 8 in each box means that there are 8 elements of $(ijkl)$ in it. The $8(--,-)$ means that there are 8 elements coming from left 0 right 1 diagrams while $4(--,-)\,4(-,--)$ means the 8 elements come from half of each $(--,-)$ and $(-,--)$. \label{fig:diagramA}}
\end{figure}
Thus for (a) we can mathematically write
\begin{equation}
32_{\rm{sym}} = (24 \oplus 8)_{\rm{sym}}
\end{equation}
and for (b)
\begin{equation}
32_{\rm{asym}} = (24 \oplus 8)_{\rm{asym}} \,.
\end{equation}
where $24_{\rm{sym}} = 24_{\rm{asym}} =24$. Thus we have $32_{\rm{sym}} \cap 32_{\rm{asym}} =  24_{\rm{sym}} = 24_{\rm{asym}}=24 $.   Note that here the `$\oplus$" symbol does not refer to the direct sum but a notation for classifying objects into different categories. In terms of the box representation we can simply write
\begin{equation}
4_{\rm{sym}} =  (3 \oplus 1)_{\rm{sym}}
\end{equation}
and
\begin{equation}
4_{\rm{asym}} = (3 \oplus 1)_{\rm{asym}} \,.
\end{equation}
And we have $4_{\rm{sym}} \cap 4_{\rm{asym}} = 3$.

It is important to note that there are no horizontal disconnect diagrams like $(0101)\,\, (2323)$ and diagonal disconnect diagrams like $(0202)\,\,(1313)$. This is because the former one involve both $|0\rangle$ states or both $|0\rangle$ states at $n=1$ level for the split, which violates our rule. Similarly the latter one involve all 4 $|0 \rangle$ states or all 4 $|1 \rangle$ states for the splitting at $n=2$ level, again this violates our splitting rules. It is always important to bear in mind that the splitting processes must be carried out in emerging one $|0 \rangle$ and $|1 \rangle$ state each time from the original state.  

We can also represent them by
\begin{figure}[H]
\centering
\includegraphics[trim=0cm 0cm 0cm 0cm, clip, scale=0.65]{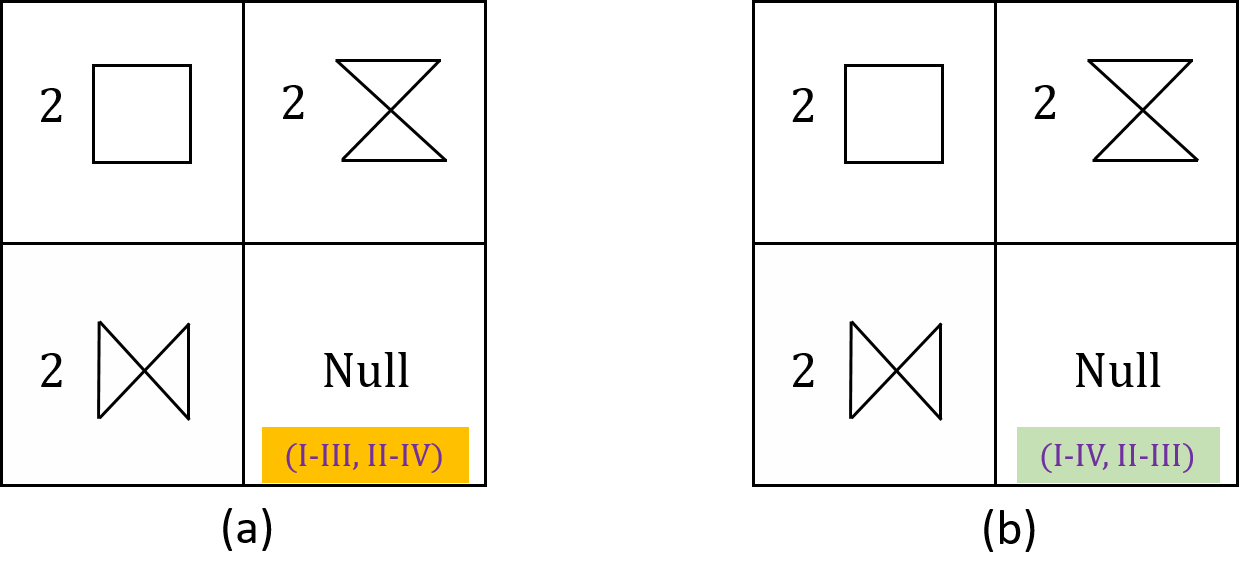}
\caption[Diagramatic representation]
 {Diagramatic representation.\label{fig:dualBox2}}
\end{figure}

There are fruitful information in the arrangement table $\ref{tab:result1}$. It can be represented by two separate 4-tableau, one for normal diagram and the other for crossing diagram. 
\begin{figure}[H]
\centering
\includegraphics[trim=0cm 0cm 0cm 0cm, clip, scale=0.5]{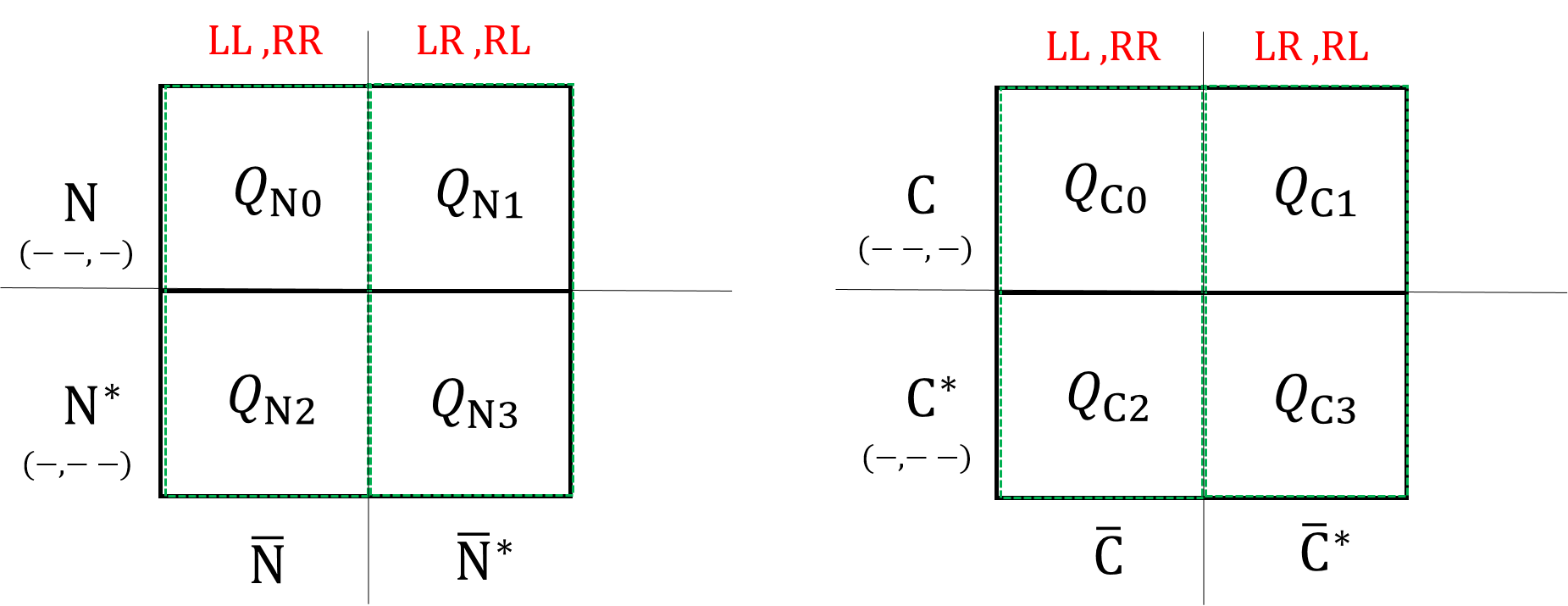}
\caption[QQ]
 {Left standard tableau: Normal diagrams; Right standard tableau: Crossing diagrams. $Q$ with subscripts denote different quadrants.  \label{fig:QQ}}
\end{figure}
For normal diagrams, we have
\begin{equation}
Q_{\mathrm N_0} = \mathrm{N \cap \bar{N}}, \quad Q_{\mathrm N_1} =\mathrm{N \cap \bar{N}^* }, \quad Q_{\mathrm N_2} = \mathrm{N^* \cap \bar{N}}, \quad Q_{\mathrm N_3} = \mathrm{N^* \cap \bar{N}^*} \,.
\end{equation}
We can establish isomorphism by
\begin{equation}
Q_{\mathrm N_0} \mapsto (0_1 0_2) \quad Q_{\mathrm N_1}  \mapsto (0_1 1_1) , \quad Q_{\mathrm N_2} \mapsto (1_1 0_2) , \quad Q_{\mathrm N_3}  \mapsto (1_1 1_2) \,,
\end{equation}
for which unstarred set is represented by 0 and starred set is represented by 1. 

For crossing diagrams, we have
\begin{equation}
Q_{\mathrm C_0} = \mathrm{C \cap \bar{C}}, \quad Q_{\mathrm C_1} =\mathrm{C \cap \bar{C}^* }, \quad Q_{\mathrm C_2} = \mathrm{C^* \cap \bar{C}}, \quad Q_{\mathrm C_3} = \mathrm{C^* \cap \bar{C}^*} \,.
\end{equation}
We can establish isomorphism by
\begin{equation}
Q_{\mathrm C_0} \mapsto (0_1 0_2) \quad Q_{\mathrm C_1}  \mapsto (0_1 1_2) , \quad Q_{\mathrm C_2} \mapsto (1_1 0_2) , \quad Q_{\mathrm C_3}  \mapsto (1_1 1_2) \,.
\end{equation}

Therefore the table establishes an isomorphism of two separate standard 4-tableau, each each of them is isomorphic to the heterogeneous representation of the $\mathbb{Z}_2 \times \mathbb{Z}_2$ group.

Note also that the full sets and empty sets for the normal digram and crossing diagram are given by respectively,
\begin{equation}
\mathrm{X = N \cup N^* = \bar{N}\cup \bar{N}^*  , \quad Y= C \cup C^* = \bar{C}\cup \bar{C}^*   } 
\end{equation}
and
\begin{equation}
\mathrm{ N \cup N^* = \bar{N}\cup \bar{N}^* = \emptyset , \quad C \cup C^* = \bar{C}\cup \bar{C}^* = \emptyset }\,.
\end{equation}

\subsection{ $n=3$ level }
The 3-level case can be studied through the feature diagram, due to the fact that 3 is odd and generally odd levels lack the symmetry property as the even counterpart, there are less interesting properties to study.

The full state is given by
\begin{equation}
|\psi \rangle = \sum_{i=0}^7 a_{i} | i\rangle \,.
\end{equation}
We can write the sum by grouping the terms as 4 dual pairs, in which 2 of them are dual invariants. In binary representation
\begin{equation} \label{eq:31lv1}
\begin{aligned}
|\psi \rangle &= \big[\,( a_{000} |000\rangle + a_{111} |111\rangle  ) +(a_{010} | 101\rangle + a_{101} |101\rangle)\,\big]  \\
&\quad\quad \quad\quad\quad \quad
+ \big[\,(a_{001} |001\rangle + a_{100} | 100\rangle) + (a_{011} |011\rangle + a_{110}|110\rangle)\,\big]
\end{aligned}
\end{equation}
and in decimal representation,
\begin{equation} \label{eq:3l1v2}
|\psi \rangle = \big[\,( a_0 |0\rangle + a_7 |7\rangle  ) +(a_2 | 2\rangle + a_5 |5\rangle)\,\big] + \big[\,(a_1 |1\rangle + a_6 | 6\rangle) + (a_3 |3\rangle + a_4 |4\rangle)\,\big] .
\end{equation}

We can represent this with the feature diagram as $2\times 2\times 2$ hypercube, but what we are more interested is its duality property, thus we will find a way, if possible, to represent it by a 2D 4-duality diagram. This time we can we have two diagrams in a single box instead of 1 as before and each dual pair is treated as one dimension, collectively, in line with the representation theory. The two square brackets in the above equation in \ref{eq:31lv1} or \ref{eq:3l1v2} is showing the classification of two parts: dual invariant and non-dual invariant. The feature diagram and its interpretations are shown as follow.
\begin{figure}[H]
\centering
\includegraphics[trim=0cm 0cm 0cm 0cm, clip, scale=0.55]{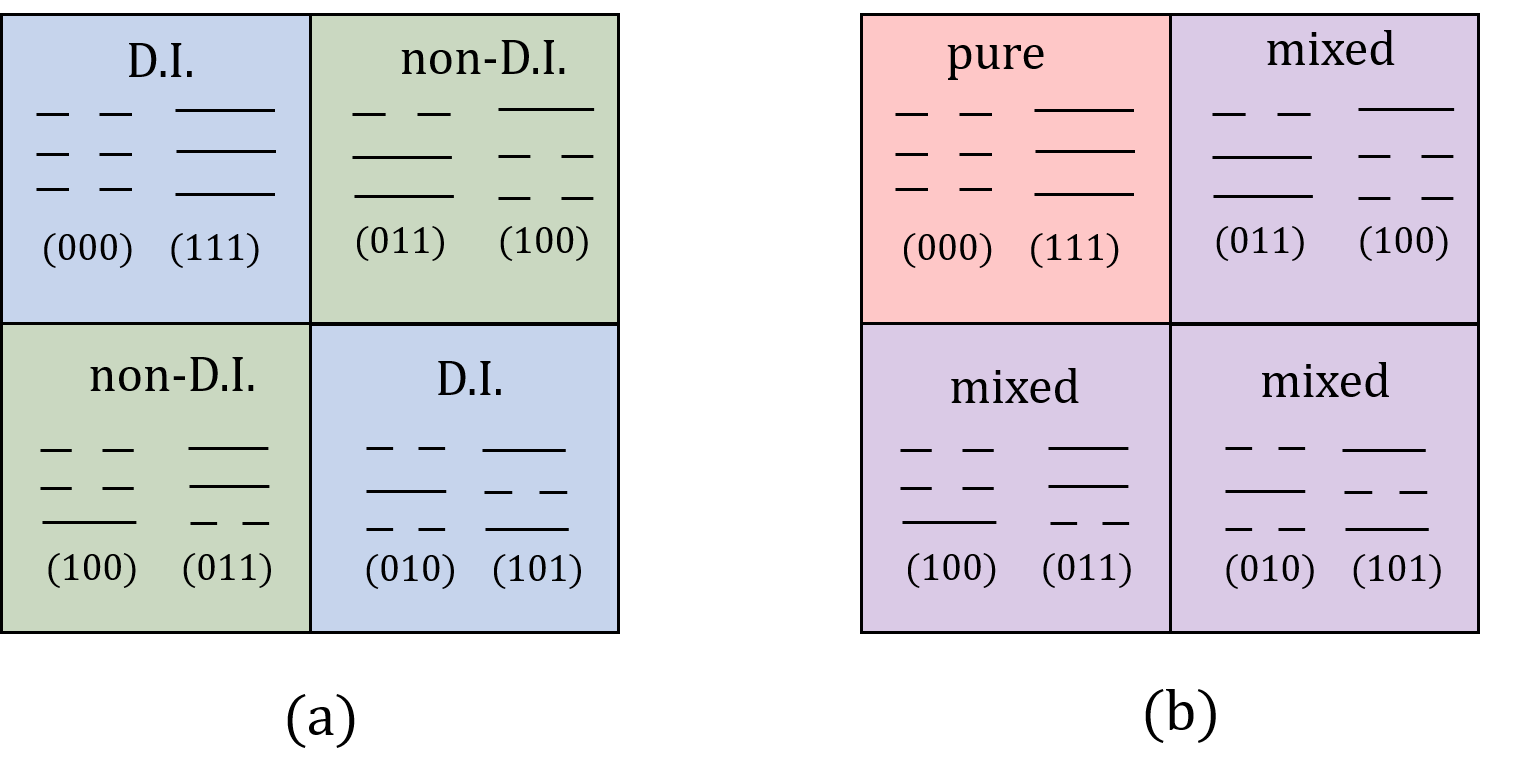}
\caption[4-tableau of 3-level gua.]
{4-tableau of 3-level gua. (a) Dual structure (2+2) of dual invariant states and non-dual invariant states; (b) (1+3) sturcture of pure states and mixed states\label{fig:featureDiagram16Gua}}
\end{figure}
There are two ways of interpretation here. The left one classifies the full state into dual-invariant (indicated by blue and non-dual invariant (indicated by green), giving a $(2\oplus 2)$ structure and hence further a $(1 \oplus 1)$ dual structure. Thus there is the dual classification. Note that dual and non-dual is itself a duality. The right one classifies the full states into pure-states and mixed states, giving a $(1 \oplus 3)$ structure. Pure states are states of which all states are equal in each level, while mixed states are states of which there exist one state that are different.

\subsection{ $n=4$ level}
For $n=4$ level, the full state is given by
\begin{equation}
|\psi \rangle = \sum_{i=0}^{15} a_{i} | i\rangle \,.
\end{equation}
And explicitly we write,
\begin{equation}
|\psi \rangle = \sum_{i_1 , i_2 , i_3 , i_4 = 0, 1} a_{i_1 i_2 i_3 i_4} | \eta_{i_1} \eta_{i_2} \eta_{i_3} \eta_{i_4} \rangle \,.
\end{equation}
with the normalization
\begin{equation}
\sum_{i_1 , i_2 , i_3 , i_4 = 0, 1} |  a_{i_1 i_2 i_3 i_4} |^2 =1 \,.
\end{equation}
Next we would study the feature diagram,
\begin{figure}[H]
\centering
\includegraphics[trim=0cm 0cm 0cm 0cm, clip, scale=0.75]{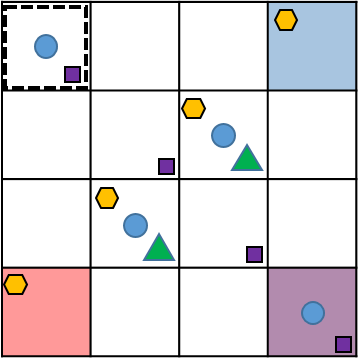}
\caption[Analysis]
{Feature diagram of $16$-Gua. \label{fig:featureDiagram16Gua}}
\end{figure}
We can link up the four spots of dual invariant numbers as a diamond. As we will see later, for higher n-Gua, we can join the spots with different interesting patterns, which can be easily recognized. 

\subsection{The study of $n=6$ level }
For $n=6$ level, the full state is given by
\begin{equation}
|\psi \rangle = \sum_{i=0}^{63} a_{i} | i\rangle \,.
\end{equation}
And explicitly we write,
\begin{equation}
|\psi \rangle = \sum_{i_1 , i_2 , i_3 , i_4 , i_5 , i_6 = 0, 1} a_{i_1 i_2 i_3 i_4 i_5 i_6 }| \eta_{i_1} \eta_{i_2} \eta_{i_3} \eta_{i_4} \eta_{i_5} \eta_{i_6} \rangle \,.
\end{equation}
with the normalization
\begin{equation}
\sum_{i_1 , i_2 , i_3 , i_4 , i_5 , i_6 = 0, 1} |  a_{i_1 i_2 i_3 i_4 i_5 i_6} |^2 =1 \,.
\end{equation}

The feature diagram is 
\begin{figure}[H]
\centering
\includegraphics[trim=0cm 0cm 0cm 0cm, clip, scale=0.5]{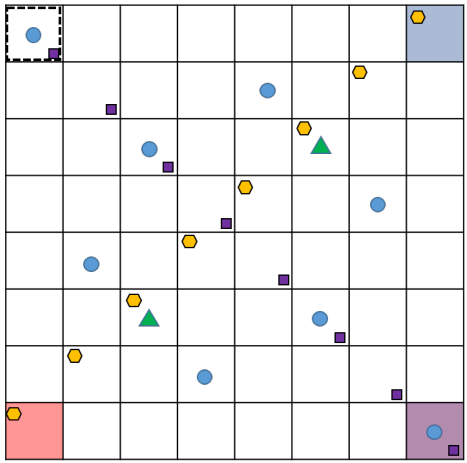}
\caption[Analysis]
{ Feature diagram for $n=6$ level. \label{fig:featureDiagram64Gua}}
\end{figure}
There are two diamonds. Alternatively, we can join it as a hexagon with two external lines.

\subsection{The study of higher n level }
In general for higher even $n$ which can be square rooted, it shows interesting patterns. For example, for $n=8$ case, which is $256$-diagram, which can be represented by a $16\times 16$ feature diagram. 
\begin{figure}[H]
\centering
\includegraphics[trim=0cm 0cm 0cm 0cm, clip, scale=0.4]{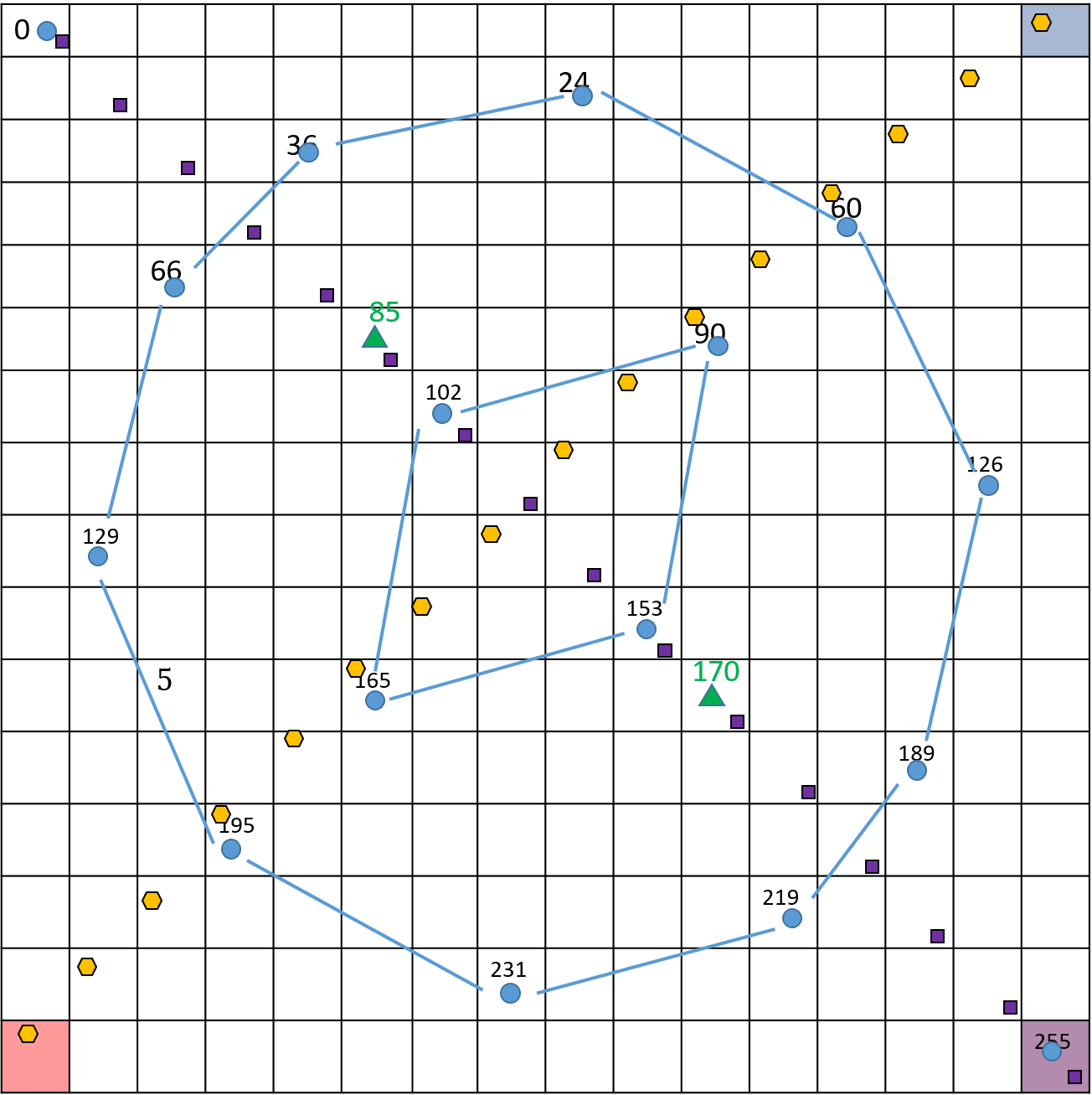}
\caption[Analysis]
{ Feature diagram. \label{fig:featureDiagram64Gua}}
\end{figure}
And for convenience we can just show the spots for dual invariant numbers.
\begin{figure}[H]
\centering
\includegraphics[trim=0cm 0cm 0cm 0cm, clip, scale=0.4]{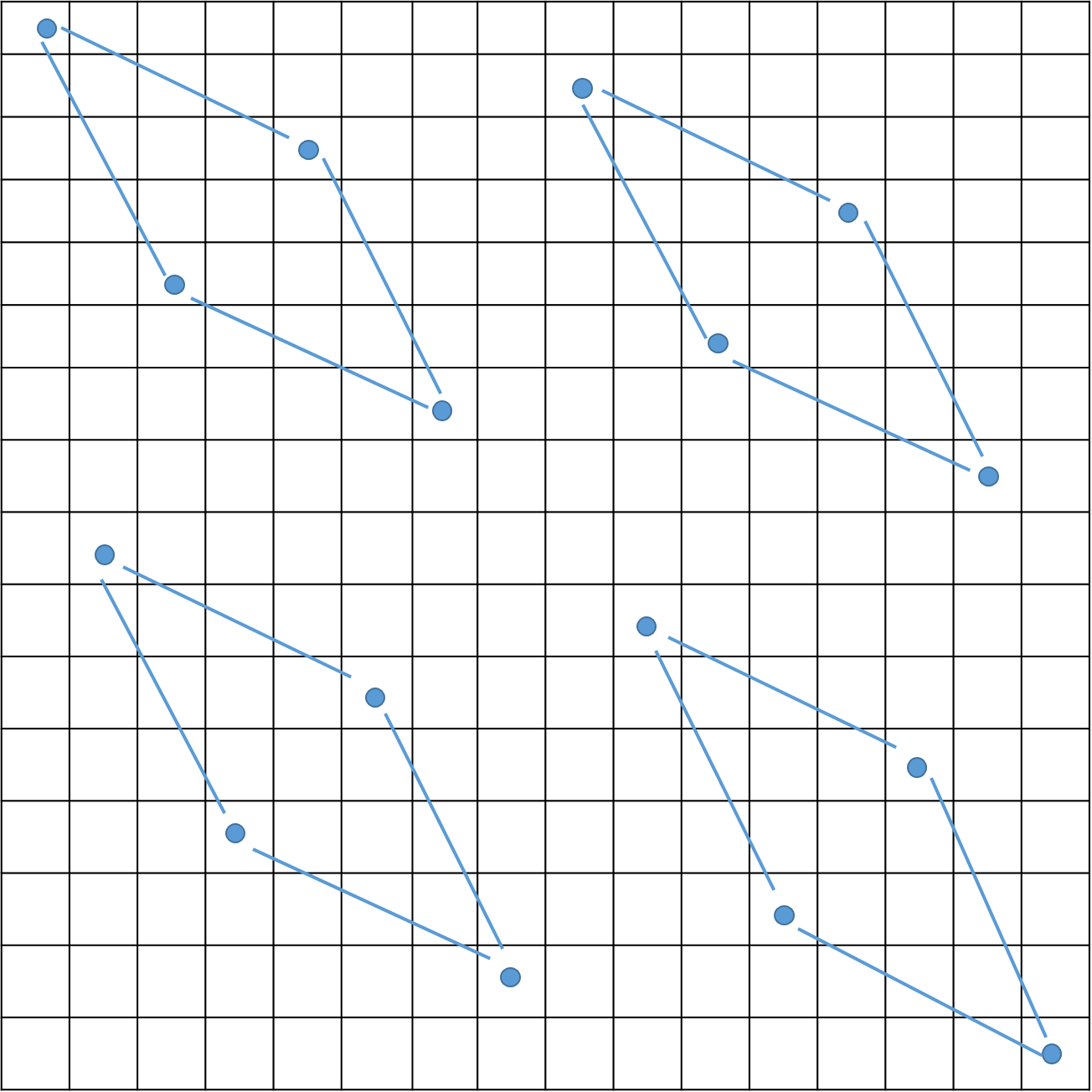}
\caption[Analysis]
{ Feature diagram. \label{fig:featureDiagram64Gua}}
\end{figure}

For example, for $n=10$ case, which is $1024$-diagram, which can be represented by a $32\times 32$ feature diagram. For convenience here we only show the spots for dual invariant number.
\begin{figure}[H]
\centering
\includegraphics[trim=0cm 0cm 0cm 0cm, clip, scale=0.4]{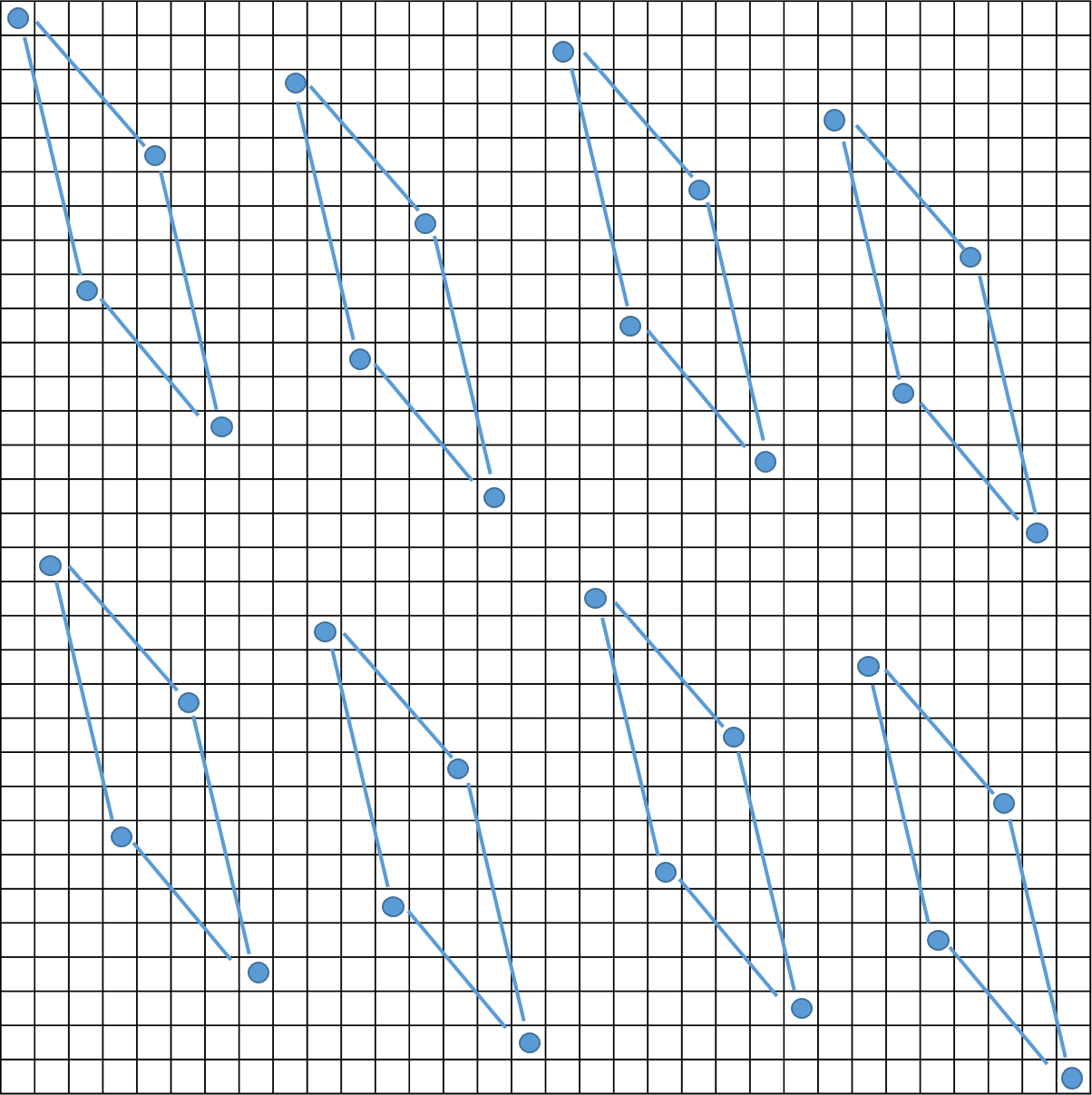}
\caption[Analysis]
{ Feature diagram: diamond pattern \label{fig:featurea}}
\end{figure}

\begin{figure}[H]
\centering
\includegraphics[trim=0cm 0cm 0cm 0cm, clip, scale=0.4]{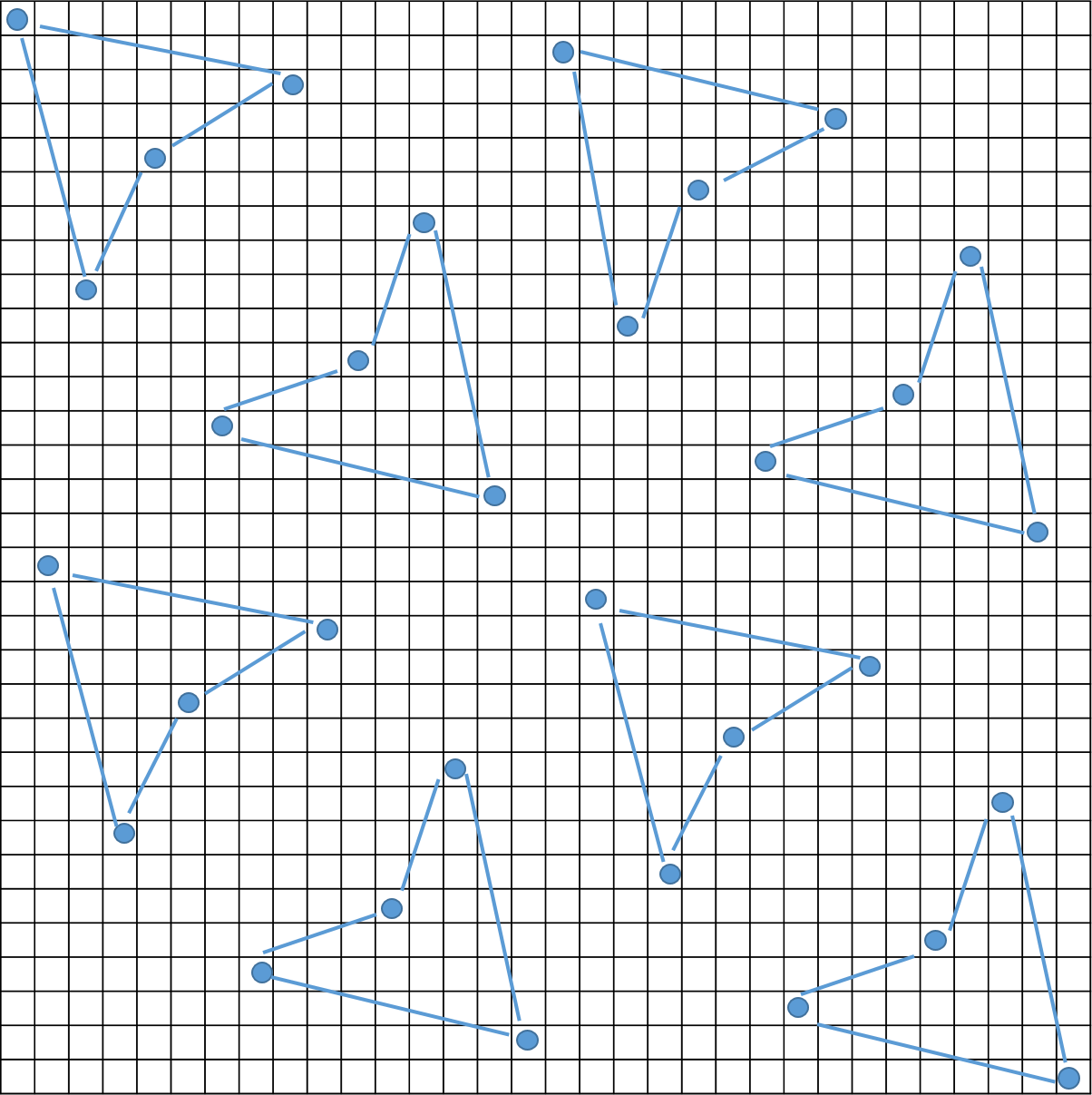}
\caption[Analysis]
{ Feature diagram: second diamond pattern \label{fig:featureb}}
\end{figure}
\begin{figure}[H]
\centering
\includegraphics[trim=0cm 0cm 0cm 0cm, clip, scale=0.4]{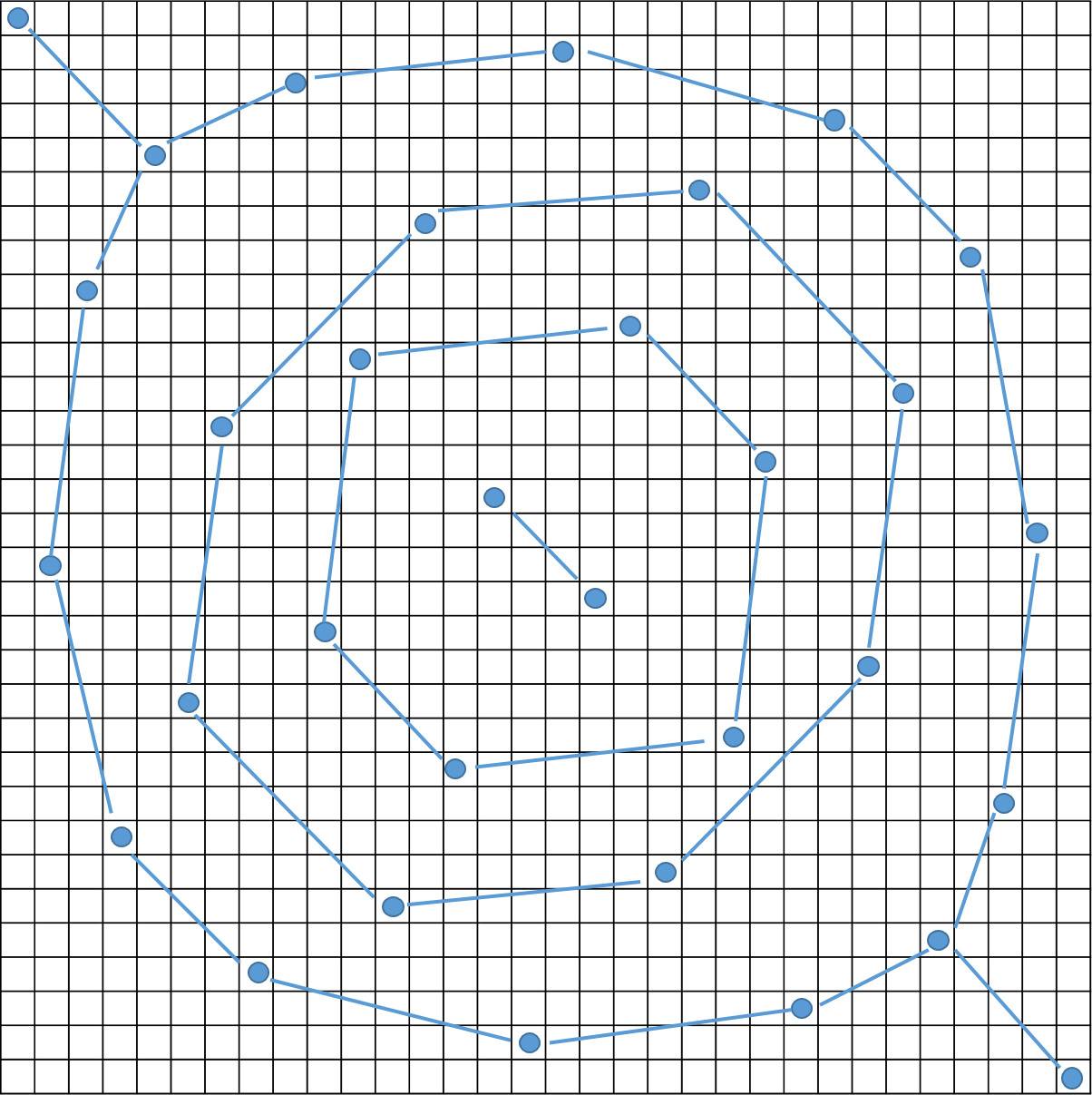}
\caption[Analysis]
{ Feature diagram: polygon pattern \label{fig:featurec}}
\end{figure}
\begin{figure}[H]
\centering
\includegraphics[trim=0cm 0cm 0cm 0cm, clip, scale=0.4]{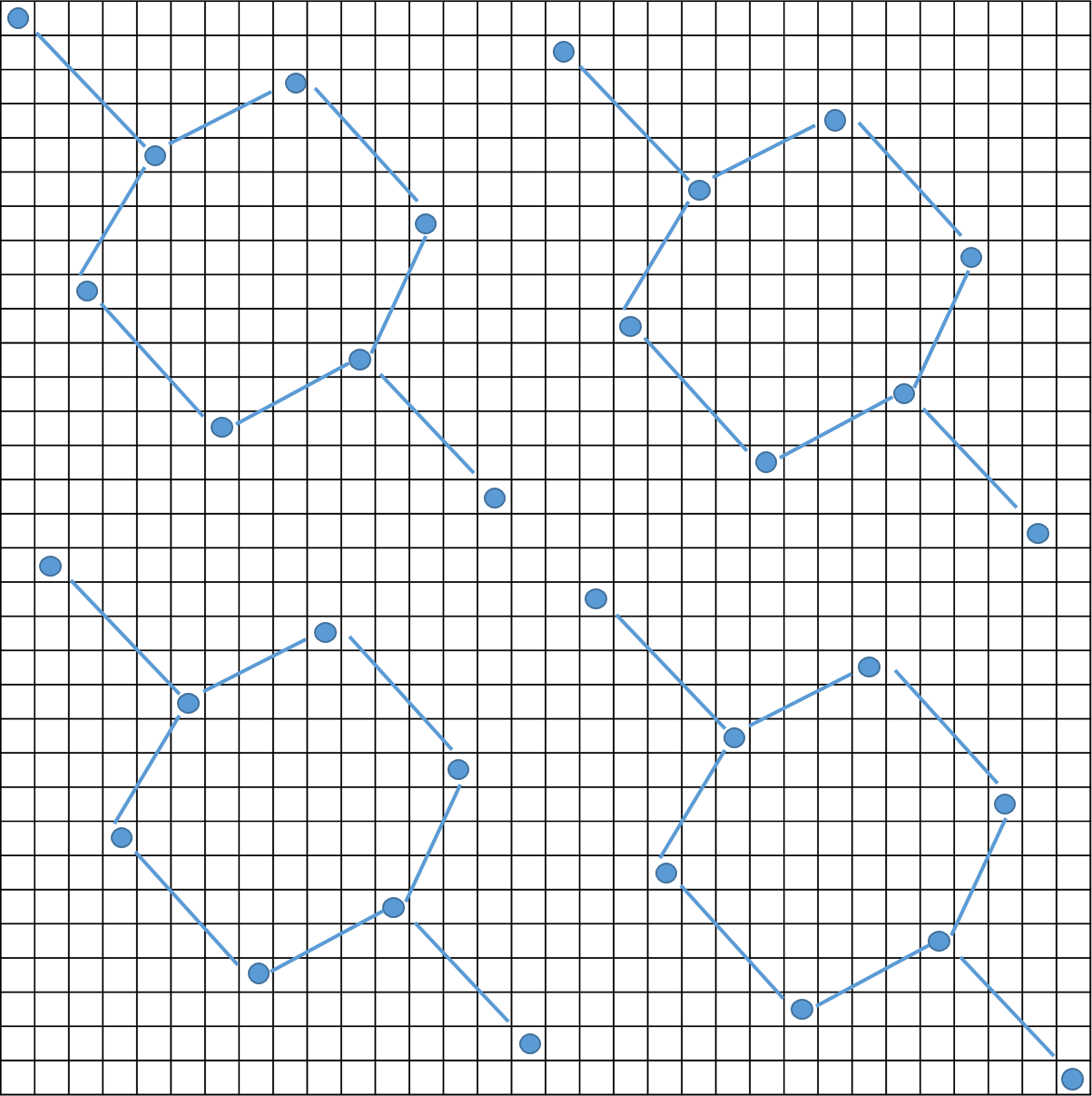}
\caption[Analysis]
{ Feature diagram: hexagon pattern \label{fig:featured}}
\end{figure}

\section{Duality in Rotor Mechanism}
In the section, we would like to introduce duality in a new fashion by rotor mechanism. We will study how duality arises from rotation of circulation of states. 

We will define two disks of states, the external layer and internal layer of states. By rotating the internal or external layer of states, this can general new state with duality structure. 

\subsection{Rotor mechanism in level 2}
Consider an external disk and internal disk, associated with two states in each layer. We rotate the internal disk, keeping the external disk unchanged. The internal rotation introduces a new state which is dual to the original one. 
\begin{figure}[H]
\centering
\includegraphics[trim=0cm 0cm 0cm 0cm, clip, scale=0.6]{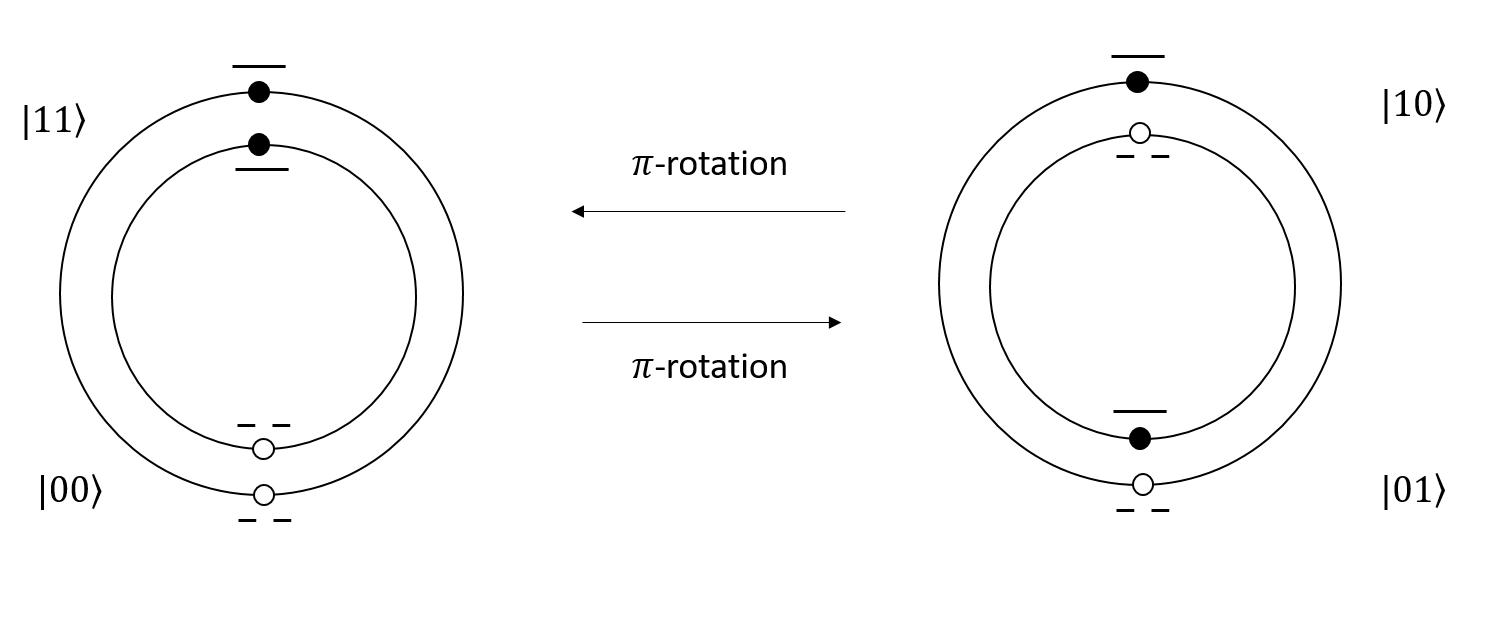}
\caption[Rotor mechanism of 2 level.] {Rotor mechanism of 2 level.\label{fig:rotor1}}
\end{figure}
First notice in the original diagram, we have the entangled state as
\begin{equation}
|\psi \rangle = \frac{1}{\sqrt{2}}( | 00 \rangle + | 11 \rangle )\,.
\end{equation}
After the internal disk rotation for $\pi$ radian, this generates a new entangled state,
\begin{equation}
|\psi^* \rangle = \frac{1}{\sqrt{2}}( | 01\rangle + | 10 \rangle )\,.
\end{equation}
This state is dual to the original state, this is because as shown in chapter 2. Moreover, it can be seen that the first diagram is for $\{ -2n , \cdots -2, 0 ,2 ,\cdots 2n  \}$ rotations and the second diagram is for $\{ -2n+1 , \cdots -1, 1  ,\cdots 2n+1  \}$ rotations, and we know that the set of even numbers and the set of odd numbers are dual to each other. Therefore, in this way, we say $|\psi \rangle = \frac{1}{\sqrt{2}}( | 00 \rangle + | 11 \rangle )$ the even state, while $|\psi^* \rangle = \frac{1}{\sqrt{2}}( | 01\rangle + | 10 \rangle )$ the odd state. The even parity and odd parity can be seen also in the tensor product states. Recalling in equation (\ref{eq:4Level}),
\begin{equation}
|\psi (\theta_1, \theta_2) \rangle = \cos\frac{\theta_1}{2} \cos\frac{\theta_2}{2} |0 0 \rangle +\cos\frac{\theta_1}{2}\sin\frac{\theta_2}{2} |0 1\rangle +\sin\frac{\theta_1}{2}\cos\frac{\theta_2}{2} |1 0\rangle + \sin\frac{\theta_1}{2}\sin\frac{\theta_1}{2} |1 1 \rangle \,.  
\end{equation}
So we see that the $|00\rangle $ and $|11\rangle$ states are associated with even coefficients as $\cos\phi_1 \cos\phi_2$ and $\sin\phi_1 \sin\phi_2$ are even functions; while $|01\rangle $ and $|10\rangle$ states are associated with odd coefficients as $\cos\phi_1 \sin\phi_2$ and $\sin\phi_1 \cos\phi_2$ are odd functions. In addition, $|00\rangle$ and  $|00\rangle$ states are left-right observation invariant, while $|01\rangle$ and  $|10\rangle$ states are not left-right observation invariant. This is an example of, dual invariance and non-dual invariant are itself a duality.
\begin{figure}[H]
\centering
\includegraphics[trim=0cm 0cm 0cm 0cm, clip, scale=0.6]{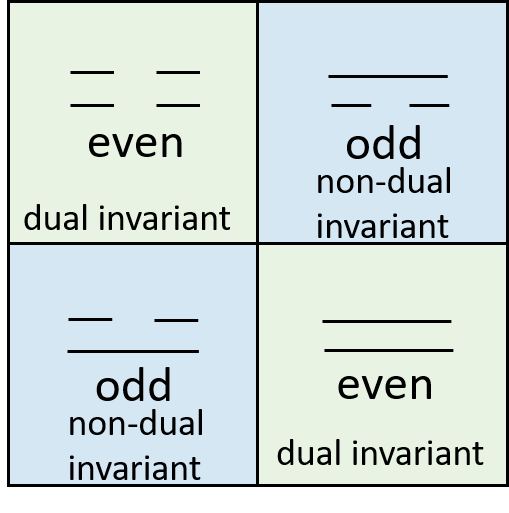}
\caption[] {\label{fig:rotor1}}
\end{figure}

\subsection{Internal and external observation duality }
In this section we investigate the duality between internal and external observer by constructing a two-layer rotation disk. Each layer consists of 4 elements with 0 or 1. The two-layer disk can be viewed from the internal perspective or external perspective, which would give us different decimal numbers respectively. The construction and the operation of the disk is shown below.

\begin{figure}[H]
\centering
\includegraphics[trim=0cm 0cm 0cm 0cm, clip, scale=0.3]{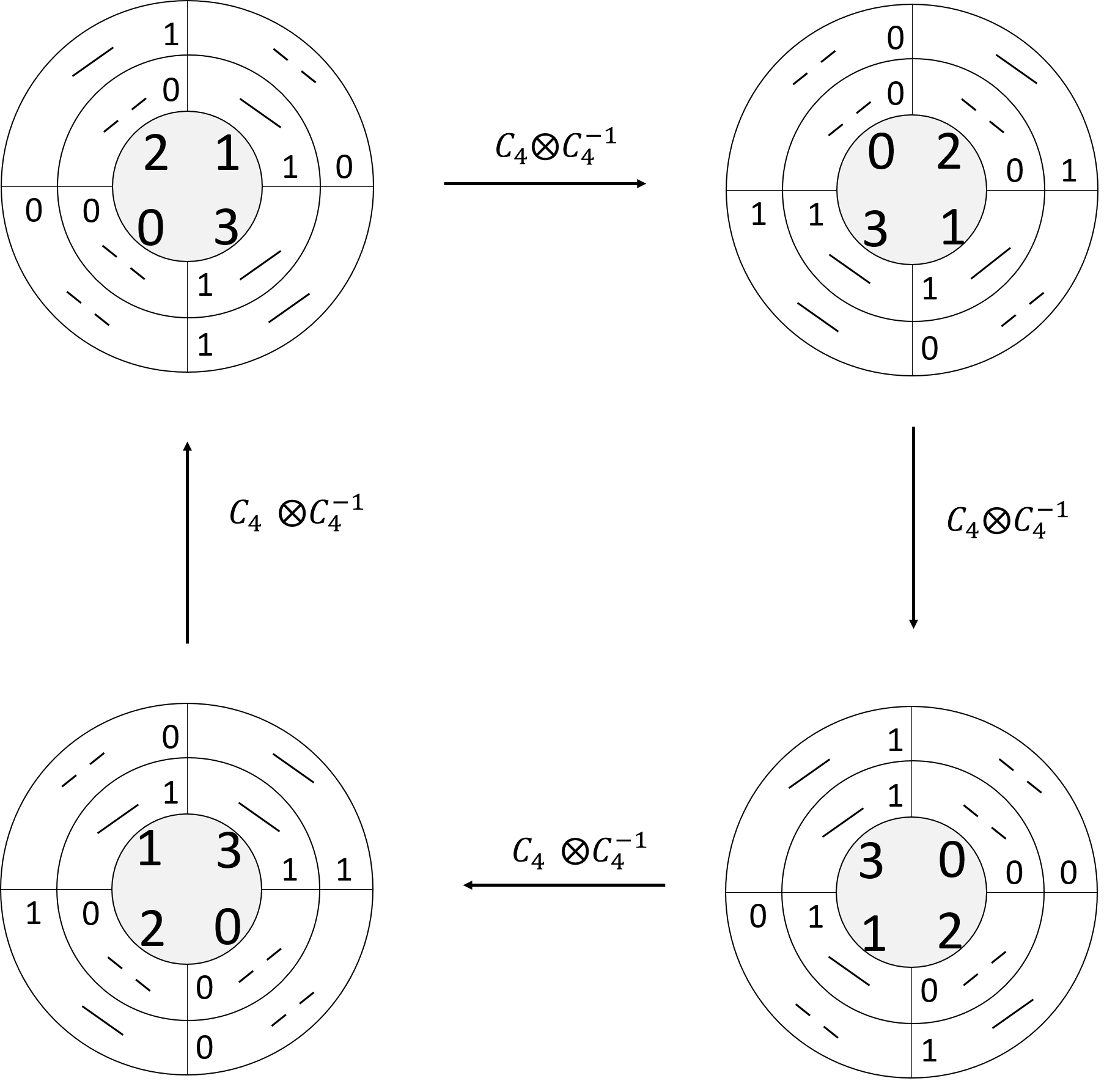}
\caption[Periodic 4-system from external frame]
{Periodic 4-system from internal observer's frame. The outter layer shuffles in the clockwise direction while the internal layer shufffles in the anticlockwise direction.\label{fig:4dualInt}}
\end{figure}
\begin{figure}[H]
\centering
\includegraphics[trim=0cm 0cm 0cm 0cm, clip, scale=0.3]{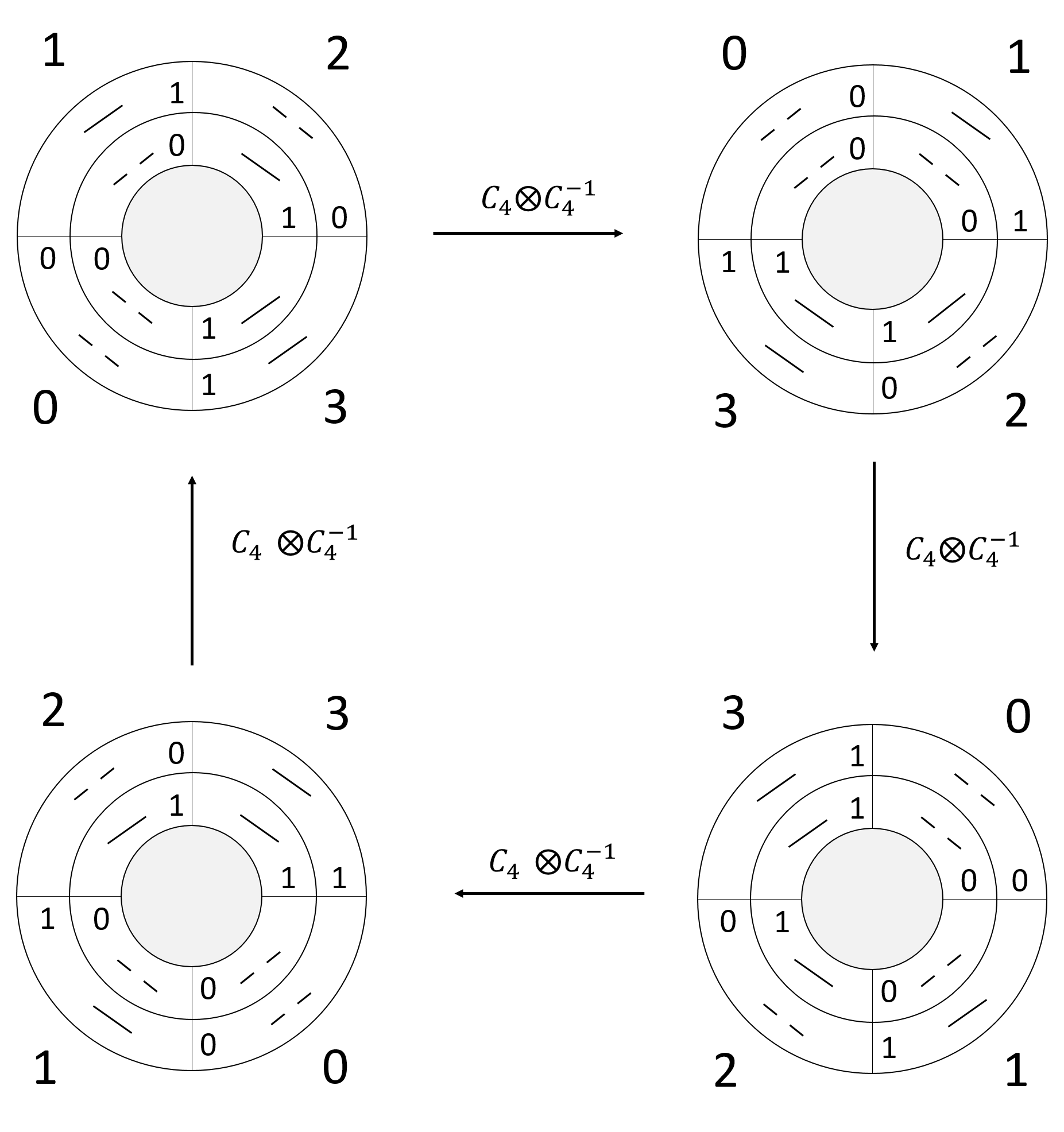}
\caption[Periodic 4-system from external frame]
{Periodic 4-system from external observer's perspetive. The outter layer shuffles in the clockwise direction while the internal layer shufffles in the anticlockwise direction. \label{fig:4dualExt}}
\end{figure}

The maps are given by the cyclic group $\mathbb{Z}_4$. We have the group elements as
\begin{equation}
G= \{ 1 \otimes 1 , C_4 \otimes C_4^{-1}, C_2 \otimes C_2^{-1} , C^3_4 \otimes C_4^{-3}     \}
\end{equation}
The group elements are just isomorphic to $\mathbb{Z}_4$ because we have the tensor product identity of $(A \otimes B)(C\otimes D) = (AC) \otimes (CD)$.

\begin{figure}[H]
\centering
\includegraphics[trim=0cm 0cm 0cm 0cm, clip, scale=0.3]{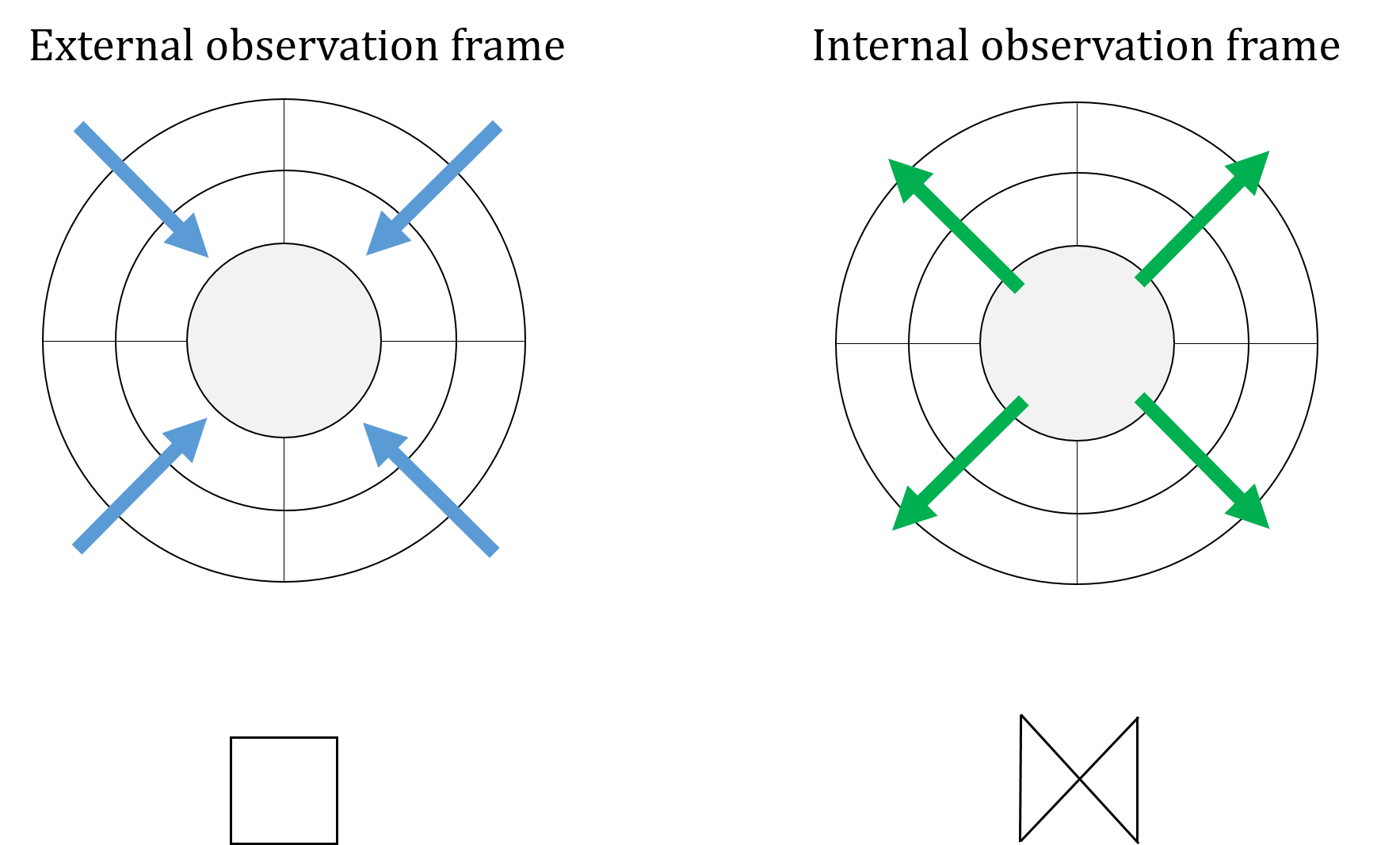}
\caption[Periodic 4-system from external frame]
{Periodic 4-system from external and internal observer's frame \label{fig:IntExtCompare}}
\end{figure}
Therefore we can see that a square is dual to the twisted square, and the hidden meaning is such duality. It is very difficult to see that just from sretch that such two geometrical objects would have such meaning in observation duality. And the difference between these object is that the twisted one contains a node (or intersection) while the plain one does not.

Next let's compute it in a table by unwrapping the periodicity.
\begin{table}[H]
\begin{center}
\begin{tabular}{c|cccc}
\hline
Position        &   1  &  2   & 3    & 4 \\
\hline
 External layer & 0101 & 1010  & 0101 & 1010 \\
 Internal layer & 0011 &  1001 & 1100 & 0110 \\
\hline
\end{tabular}
\end{center}
\end{table}
The position is the label of each quadrant (for both internal and external layer) and must be fixed. 
The property of periodicity also holds for the comparison space. 
\begin{equation}
\begin{aligned}
\text{Position} & \,\,1 : \,\,0101 :: 0011 = 1001 \\
\text{Position} & \,\,2 : \,\, 1010 :: 1001 = 1100 \\
\text{Position} & \,\,3 : \,\,0101 :: 1100 = 0110 \\
\text{Position} & \,\,4 : \,\, 1010 :: 0110 = 0011 \\
\end{aligned}
\end{equation}

We have two dual invariants after the $::$ computation, 1001 and its dual 0110 ; and two non-dual invariants, 1100 and its dual 0011. And thus we have 1,3 dual and 2,4 dual respectively in terms of the position. Note that before, for internal layer, we have 1,3 identical, 2,4 identical; for external layer, we have non-dual invariants 1,3 dual and dual invariants 2,4 dual. The comparison computation $::$ retains the dual structure followed by the external layer, but the role of dual invariant and non-dual invariant has swapped.

\subsection{Rotor mechanism in level 4}
Next we promote to study the duality properties of rotor mechanism in level 4. There are more complications than the case in level 2. Consider each layer with four states $|00\rangle$, $|01\rangle$, $|10\rangle$ and $|11\rangle$, in decimal place which are $|0\rangle$, $|1\rangle$, $|2\rangle$ and $|3\rangle$. First consider that we rotate the internal disk in the anti-clockwise direction, while keeping the external disk constant. Here first we employ the internal observation perspective. We can see that in even number of rotation of $\pi/4$, the structure of duality is remained.

\begin{figure}[H]
\centering
\includegraphics[trim=0cm 0cm 0cm 0cm, clip, scale=0.6]{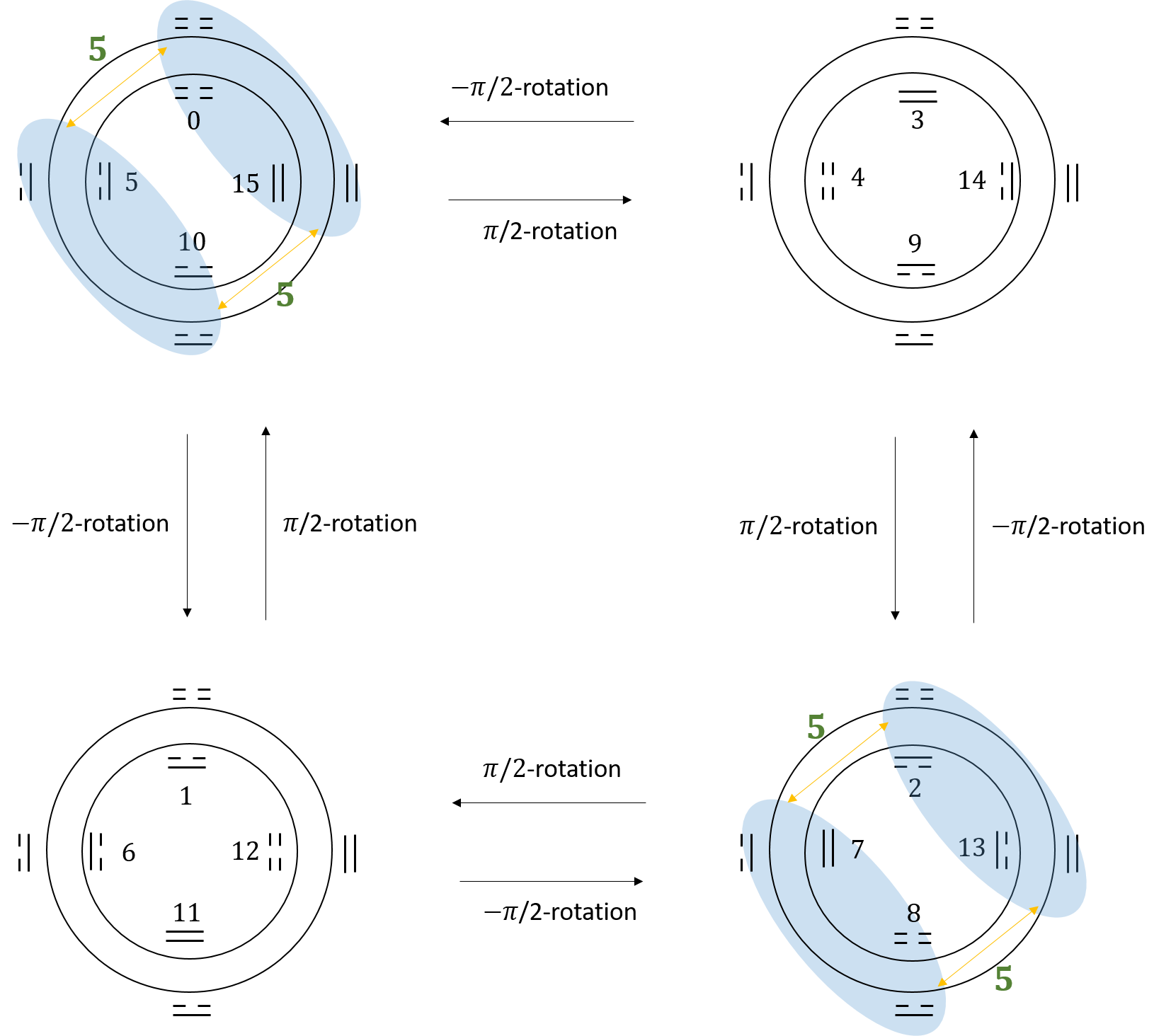}
\caption[Rotor mechanism in level 4]
{Rotor mechanism in level 4. The blue ovals represent the pairs of dual diagrams \label{fig:rotor2}}
\end{figure}
For the unrotated diagram, we have the state
\begin{equation}
| \psi_0 \rangle = \frac{1}{2}|00\rangle\otimes|00\rangle + \frac{1}{2} |01\rangle \otimes |01\rangle + \frac{1}{2}|10\rangle\otimes |10\rangle + \frac{1}{2} |11\rangle \otimes |11\rangle  
\end{equation}
and similar for the others.

In the original diagram (diagram of the upper-left corner), we have two sets of dual diagrams, $\{5,10 \}$ and  $\{0,15 \}$. The numerical difference of the two sets is 5. When the internal layer is rotated by first even number of $\pi/2$, such duality set structure is remained. We have $\{7,8\}$ and $\{2,13 \}$ two dual sets and the numerial difference of the two dual sets is also 5. Therefore $n=0$, $n=2$ even $n$ (where $n$ is the number of $\pi/4$ phase rotation of the internal disk), there is an isomorphism of the two structures. However, for $n=1$, $n=3$ odd $n$, there are no dual sets. 

The following table illustrates the result of the rotor mechanism in level 4.
\begin{figure}[H]
\centering
\includegraphics[trim=0cm 0.5cm 0cm 0.38cm, clip, scale=0.6]{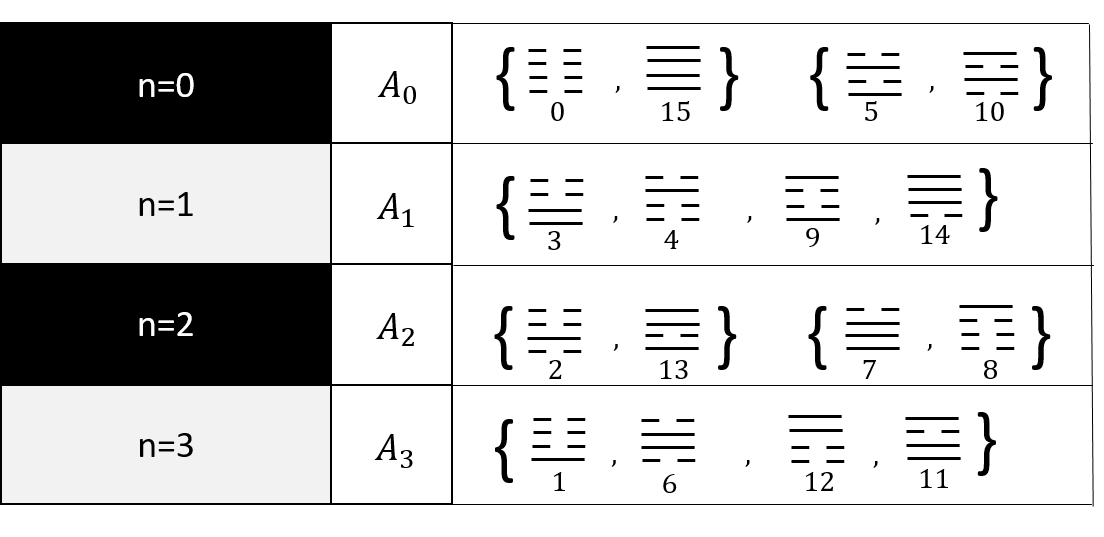}
\caption[]
{ \label{fig:table1}}
\end{figure}
The full space can be decomposed into four partitions due to four rotations, mathematically we write,
\begin{equation}
16 = 4 \oplus 4 \oplus 4 \oplus 4 \,.
\end{equation}
The collection of elements of each rotor diagram defines a vector space $A_i$, where we have four of them. For $A_0$ and $A_2$, the space can be decomposed into two dual spaces, which is
\begin{equation}
4 = 2 \oplus 2 = 1 \oplus \bar{1} \oplus 1 \oplus \bar{1}\,.
\end{equation} 
For $A_1$ and $A_3$, there do not exist any decomposition into dual spaces, but 
four distinctive unrelated spaces. Mathematically we write,
\begin{equation}
4 = 1 \oplus 1 \oplus 1 \oplus 1 \,.
\end{equation}
One also finds that for each partition, the sum of all elements is always a constant,
\begin{equation}
\sum_{a\in A_i} a = 30 \,.
\end{equation}
Next we want to show that the total (net) change in numerical diagram is zero. The proof is very easy. First define the change
\begin{equation}
\Delta a_{i,j} = a_{A_{i+1 ,j}} - a_{A_i ,j} \,,
\end{equation}
where $j$ is the $j$-th element of the $i$-th partition Now it follows that
\begin{equation}
\sum_j \Delta a_{i,j} = \sum_{a_j\in A_{i+1}}a_j - \sum_{a_j\in A_i}a_j = 15-15=0 \,.
\end{equation}

Next we would like to study the conservation constants among the rotor diagrams. In general, we see that properties within the same partition will remain the same. We see that for the even partition (even number of $\pi/2$)
\begin{equation}
a_{A_{0}\,1} + a_{A_{0}\,3} = 0 +10 = a_{A_{2}\,1} + a_{A_{2}\,3} = 2+8 =10 \,.
\end{equation}
\begin{equation}
a_{A_{0}\,2} + a_{A_{0}\,4} =5 +15 = a_{A_{2}\,2} + a_{A_{2}\,4} = 7 +13 =20 \,.
\end{equation}
Also we see that for the odd partition (odd number of $\pi/2$)
\begin{equation}
a_{A_{1}\,1} + a_{A_{1}\,3} = 3 +9 = a_{A_{3}\,1} + a_{A_{3}\,3} = 1+11 =12 \,.
\end{equation}
\begin{equation}
a_{A_{1}\,2} + a_{A_{1}\,4} = 4 +14 = a_{A_{3}\,2} + a_{A_{3}\,4} = 6+12 =18 \,.
\end{equation}
And we also see that
\begin{equation}
a_{A_{1}\,1} + a_{A_{1}\,2} = 3 +4 = a_{A_{3}\,1} + a_{A_{3}\,2} = 1+6 =7 \,.
\end{equation}
\begin{equation}
a_{A_{1}\,3} + a_{A_{1}\,4} = 9 +14 = a_{A_{3}\,3} + a_{A_{3}\,4} = 11+12 =23 \,.
\end{equation}
Finally we will study the idea of sub-duality. We have already seen that the even $n$ and odd $n$ are dual partition as shown in table (\ref{fig:table1}). Now we will show that in fact the $n=0$ partition is dual to the $n=2$ partition, and $n=1$ partition is dual to the $n=3$ partition. 

For the $n=1$ partition, the sum of difference is $(+3)+(-1)+(-1)+(-1)=0$. For the $n=3$ partition, the sum of difference is $(-3)+(+1)+(+1)+(+1)=0$. Thus we have
\begin{equation}
\sum_j \Delta a_{1,j} = -\sum_j \Delta a_{3,j} \,.
\end{equation}
Therefore $n=1$ partition is dual to the $n=3$ partition. And the difference in the odd partition is asymmetric.

For the $n=0$ partition, the sum of difference is trivial $0+0+0+0=0$. For the $n=2$ partition, the sum of difference is $(+2)+(+2)+(-2)+(-2)=0$, which is symmetric. 

\begin{figure}[H]
\centering
\includegraphics[trim=0cm 0cm 0cm 0cm, clip, scale=0.6]{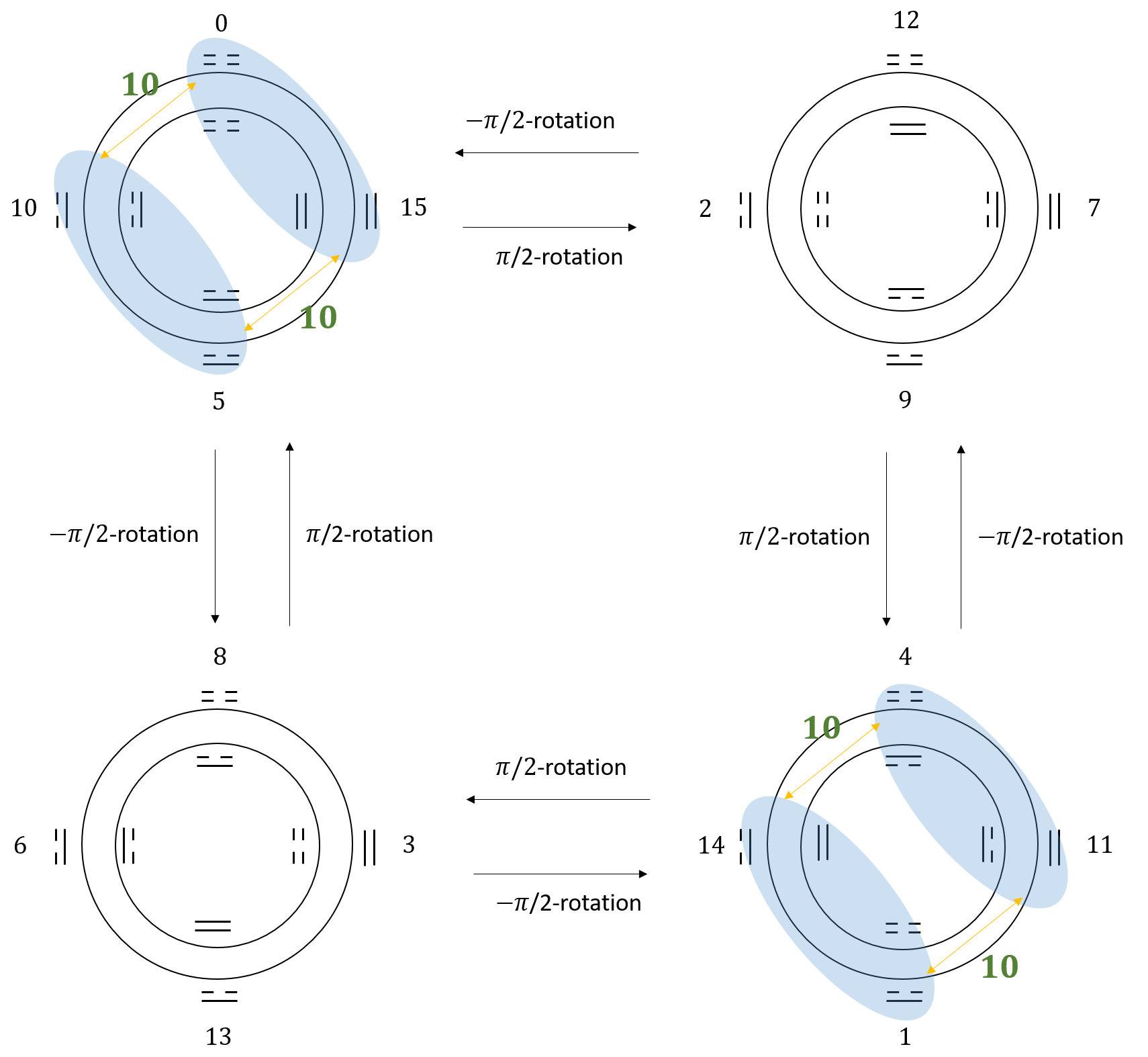}
\caption[]
{ \label{fig:rotor3}}
\end{figure}

\begin{figure}[H]
\centering
\includegraphics[trim=0cm 0cm 0cm 0cm, clip, scale=0.6]{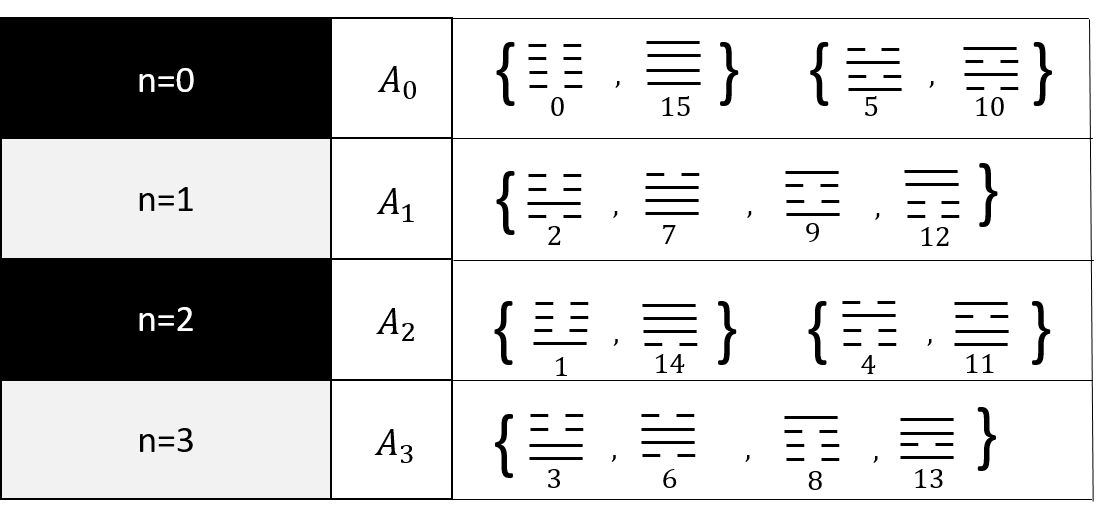}
\caption[]
{ \label{fig:table2}}
\end{figure}

Now we consider the external perspective. We have the following result in \ref{fig:rotor3} and \ref{fig:table2}. The properties from the internal perspective are basically inherited to the external perspective, but just the numerical values are different. For example, for the even partition $a_{A_{0}\,1} + a_{A_{0}\,3} = 0 +5 = a_{A_{2}\,1} + a_{A_{2}\,3} = 4+1 =5 $ ; $a_{A_{0}\,2} + a_{A_{0}\,4} =10 +15 = a_{A_{2}\,2} + a_{A_{2}\,4} = 14 +11 =25 .$. For the odd partition, $a_{A_{1}\,1} + a_{A_{1}\,3} = 12 +9 = a_{A_{3}\,1} + a_{A_{3}\,3} = 8+13 =21 .$ ; $a_{A_{1}\,2} + a_{A_{1}\,4} = 2 +7 = a_{A_{3}\,2} + a_{A_{3}\,4} = 6+3 =9 \,.$ Also, $a_{A_{1}\,1} + a_{A_{1}\,2} = 12 +2 = a_{A_{3}\,1} + a_{A_{3}\,2} = 8+6 =14$ and $a_{A_{1}\,3} + a_{A_{1}\,4} = 9 +7 = a_{A_{3}\,3} + a_{A_{3}\,4} = 13+3 =26 \,.$ For the $n=1$ partition, the sum of difference is $(+12)+(-8)+(+4)+(-8)=0$. For the $n=3$ partition, the sum of difference is $(-12)+(+8)+(-4)+(+8)=0$. Thus all the properties are preserved.

Now we consider anti-clockwise rotation of the external layer, while keeping the internal layer constant. We have the following result. 
\begin{figure}[H]
\centering
\includegraphics[trim=0cm 0cm 0cm 0cm, clip, scale=0.6]{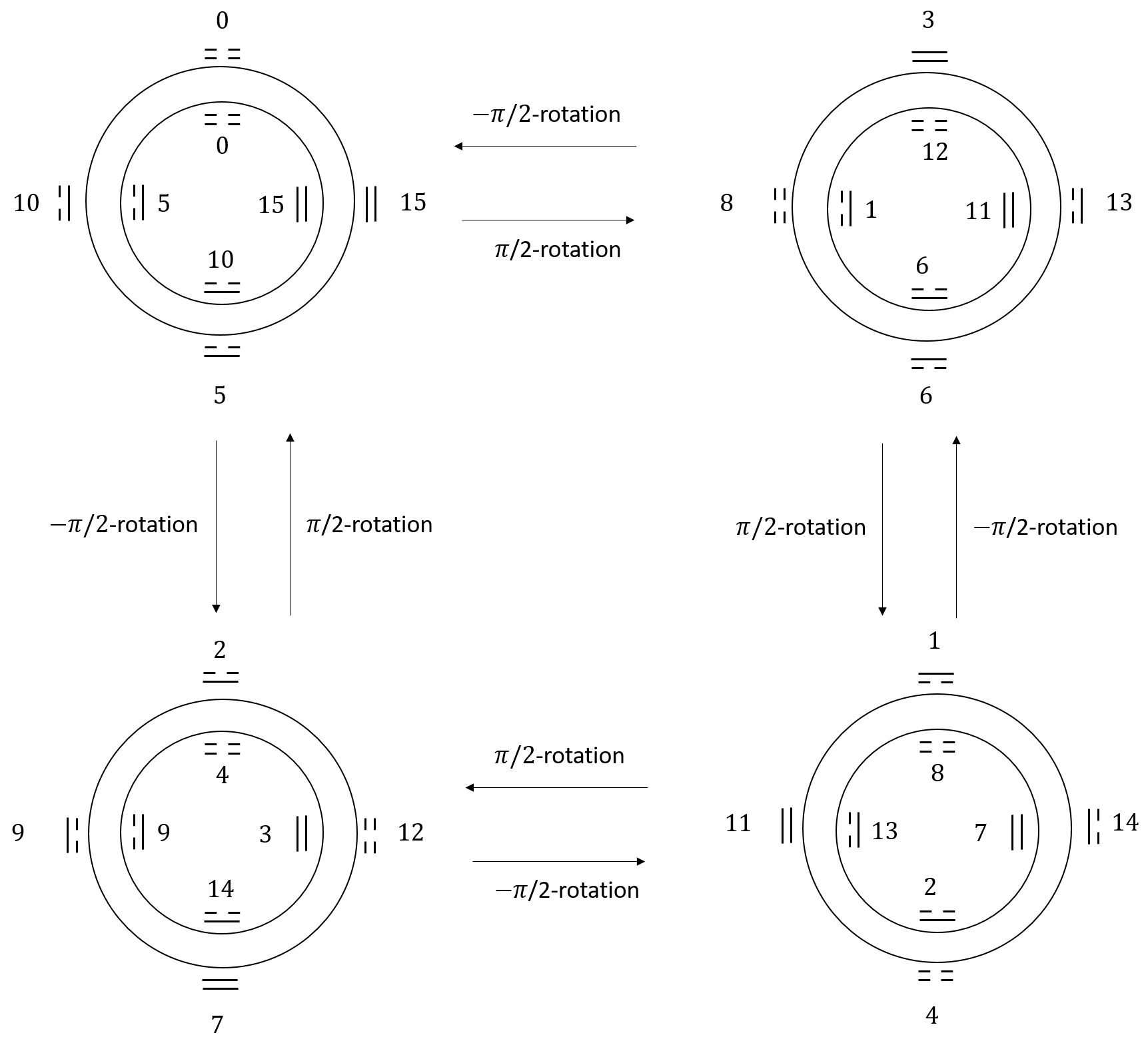}
\caption[]
{ \label{fig:rotor4}}
\end{figure}
Then we can establish mathematical relationships between the case of internal disk rotation and external disk rotation. Denote the diagram function, which is dependent on the rotation angle $\theta$ as $\mathcal{D}(\theta)$ being considered under the internal rotation perspective or the external rotation perspective. For example, the upper left diagram in figure \ref{fig:rotor2} is denoted as $(\mathcal{D}(0)| \text{internal rotation})$; the lower left diagram of figure \ref{fig:rotor2} is denoted as $(\mathcal{D}(-\pi/2)| \text{internal rotation})$ and so on. The same notion goes for the external rotation case in figure \ref{fig:rotor3}. We find the following dual equivalent relationships:
\begin{equation}
(\mathcal{D}(\pi/2)| \text{internal rotation}) \equiv (\mathcal{D}(-\pi/2)| \text{external rotation})
\end{equation}
with an offset of phase of $+\pi/2$ in the $(\mathcal{D}(-\pi/2)| \text{external rotation})$ diagram. Similarly,
\begin{equation}
(\mathcal{D}(-\pi/2)| \text{internal rotation}) \equiv (\mathcal{D}(\pi/2)| \text{external rotation})
\end{equation}
with an offset of phase of $-\pi/2$ in the $(\mathcal{D}(\pi/2)| \text{external rotation})$ diagram.
This is simply noted as
\begin{equation}
(\text{anti-clockwise}| \text{internal rotation}) \equiv (\text{clockwise}| \text{external rotation})
\end{equation}
and
\begin{equation}
(\text{clockwise}| \text{internal rotation}) \equiv (\text{anti-clockwise}| \text{external rotation})
\end{equation}
The above relationship just satisfies
\begin{equation}
(u | S_k) \equiv (u^* | S_k^\star) \quad \text{and} \quad  (u^* | S_k) \equiv (u | S_k^\star) 
\end{equation}
by identifying $u (\theta) = \pi/2$, $u^* (\theta) = -\pi/2$, $S_k = \text{internal rotation}$ and $S_k^\star = \text{external rotation}$.

The meaning of offset explicitly refers to the following. Let's compare the diagram for internal rotation and external rotation.
\begin{figure}[H]
\centering
\includegraphics[trim=0cm 0cm 0cm 0cm, clip, scale=0.6]{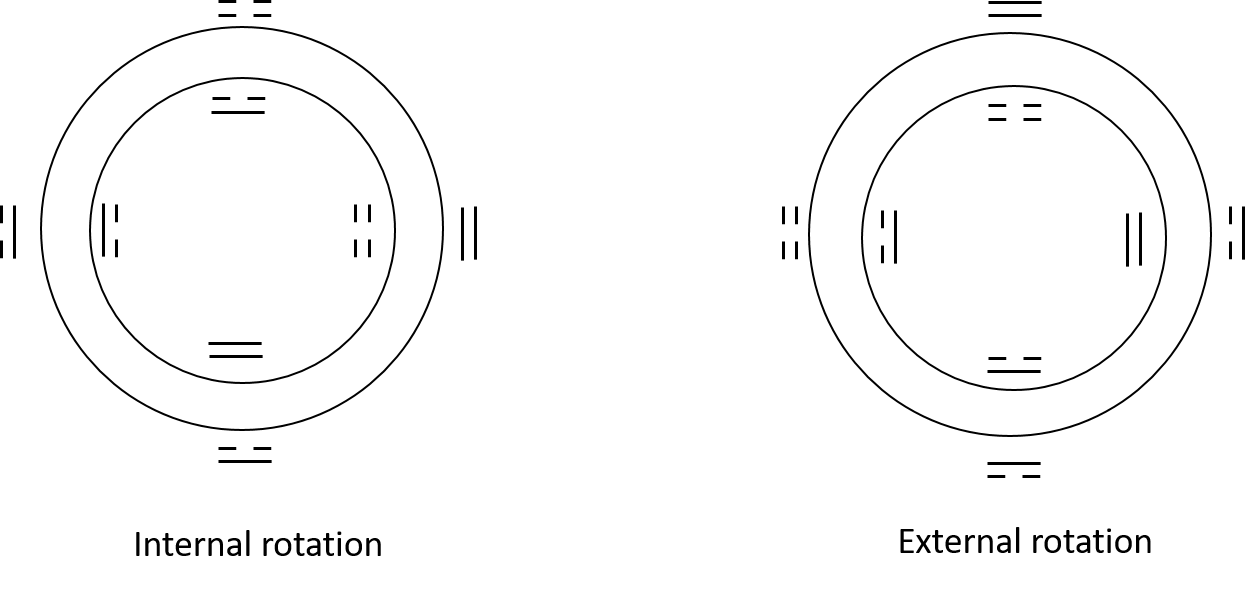}
\caption[]
{ \label{fig:rotor5a}}
\end{figure}
We can see that for the external rotation, although the diagram is the same, it is rotated by $+\pi/2$. And similarly
\begin{figure}[H]
\centering
\includegraphics[trim=0cm 0cm 0cm 0cm, clip, scale=0.6]{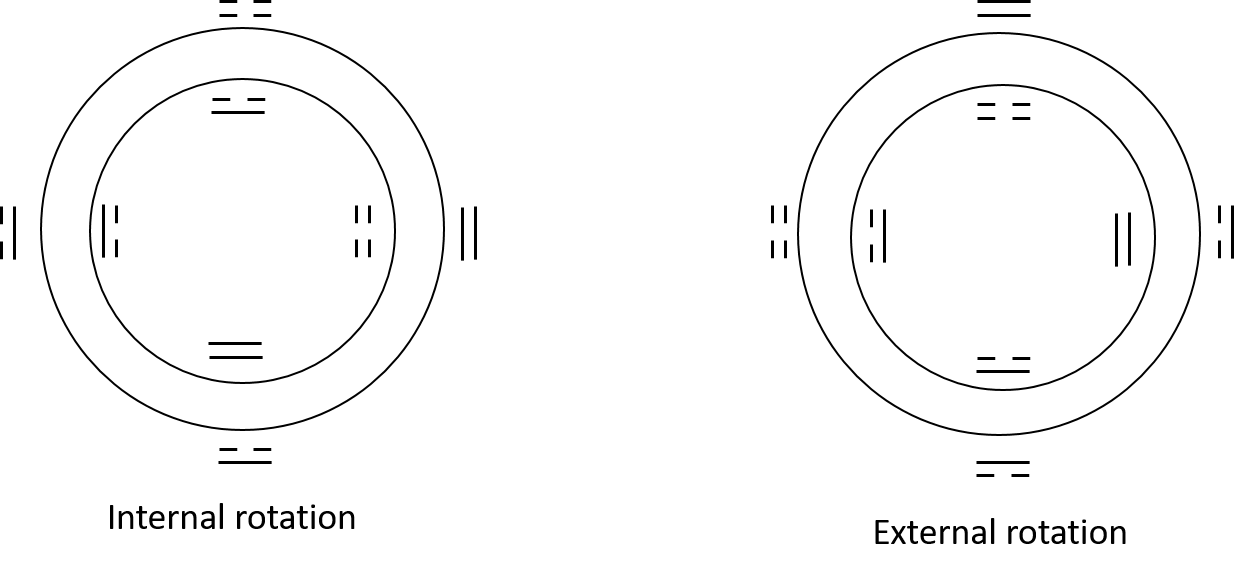}
\caption[]
{ \label{fig:rotor5b}}
\end{figure}
for which the external diagram is rotated by $-\pi/2$. 

Finally we have the dual invariant diagram for $\theta = \pi$ and $\theta =-\pi$. We can see that
\begin{equation}
(\mathcal{D}(\pi) | \text{internal rotation}) \equiv (\mathcal{D}(\pi) | \text{external rotation})
\end{equation}
and
\begin{equation}
(\mathcal{D}(-\pi) | \text{internal rotation}) \equiv (\mathcal{D}(-\pi) | \text{external rotation}) \,,
\end{equation}
with an offset of phase of $\pi$ (or  $-\pi$) in the external diagram. 
\begin{figure}[H]
\centering
\includegraphics[trim=0cm 0cm 0cm 0cm, clip, scale=0.6]{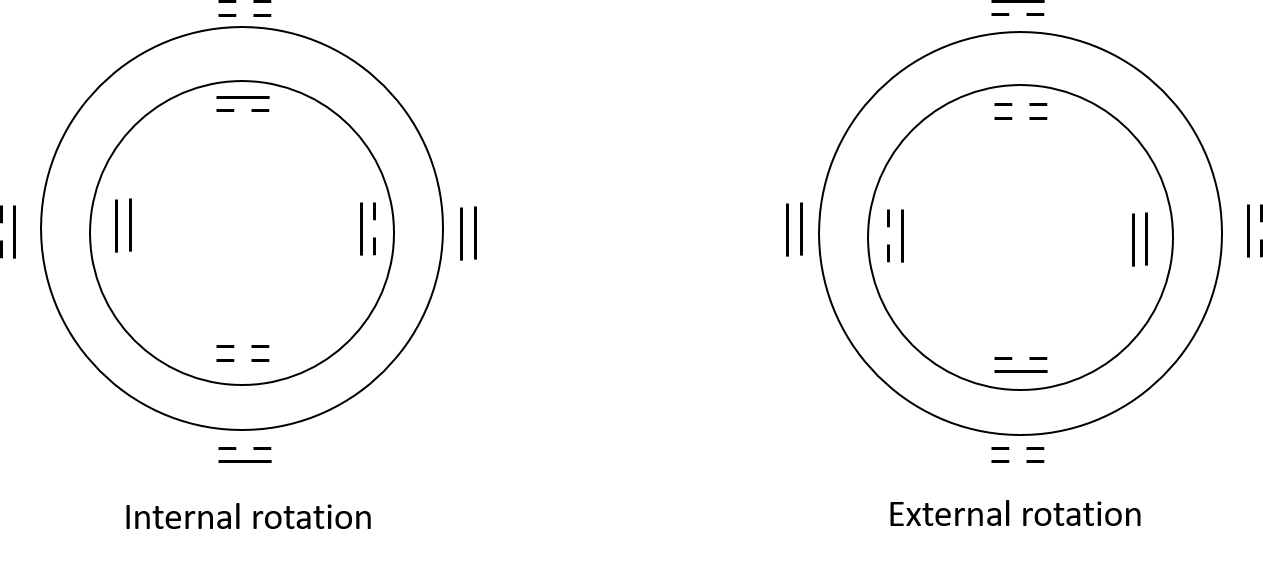}
\caption[]
{ \label{fig:rotor5c}}
\end{figure}

\section{Complementary Theorem in Duality}
In this section, we will investigate a very important theorem in duality-The Complementary Theorem. The theorem states a very deep principle: when a system is dual invariant in a specific way, it has to be non-dual invariant in another way, vice versa. 

The theorem will be proved as the follows. Consider a duality of filled (full) and unfilled (empty) states. The full state is denoted as $|0\rangle$ and the empty state is denoted as $|1\rangle$. Then we have the dual element set as $\{\text{empty},\text{full}\} = \{|0\rangle , |1\rangle  \}$. A qubit constructed will be $|\psi (\theta)\rangle = \cos\theta|0\rangle +\sin\theta |1\rangle$. Now consider the dual observer perspective. Consider $S_k = \text{matter}$ and  $S_k^\star = \text{empty space}$. Then we have the dual observer set as $\{\text{matter},  \text{empty space} \} = \{S_k , S_k^\star   \}$. Now first consider a fully filled system, a glass of fully filled water as an example. This is the full state $\psi(\pi/2) = |1\rangle$ with probability equal to 1. We can say it is full of matter, or in another way we can say it is empty in empty space. They are equivalent statements. Mathematically we write $( |1\rangle | S_k) \equiv (|0\rangle | S_k^\star )$, or simply $( 1 | S_k) \equiv (0 | S_k^\star)$. Next consider its dual counter part, an empty glass. This is the empty state $\psi(0) = |0\rangle$ with probability equal to 1.  We say it is empty in matter, but equivalently we can say it is full (filled) of empty space. Mathematically, we have $(0 | S_k) \equiv ( 1|S_k^\star) $. Next consider a halfly filled glass. This is the state of $|\psi(\pi/4)\rangle\equiv |\frac{1}{2}\rangle = \frac{1}{\sqrt{2}}(|0\rangle + |1\rangle)$ We can say it is halfly filled with matter or equivalently halfly filled with empty space. Mathematically, $(\frac{1}{2} |S_k) \equiv (\frac{1}{2} |S_k^\star) $. It's dual is also a halfly filled glass, but the position of matter and empty space is interchanged. It has a state of  $|\psi(\pi/4)\rangle \equiv |\frac{1}{2}\rangle = \frac{1}{\sqrt{2}}(|1\rangle + |0\rangle) =\frac{1}{\sqrt{2}}(|0\rangle + |1	\rangle) $. Mathematically $(\frac{1}{2} |S_k^\star) \equiv (\frac{1}{2} |S_k) $. Diagramatically, we have the following:
\begin{figure}[H]
\centering
\includegraphics[trim=0cm 0cm 0cm 0cm, clip, scale=0.6]{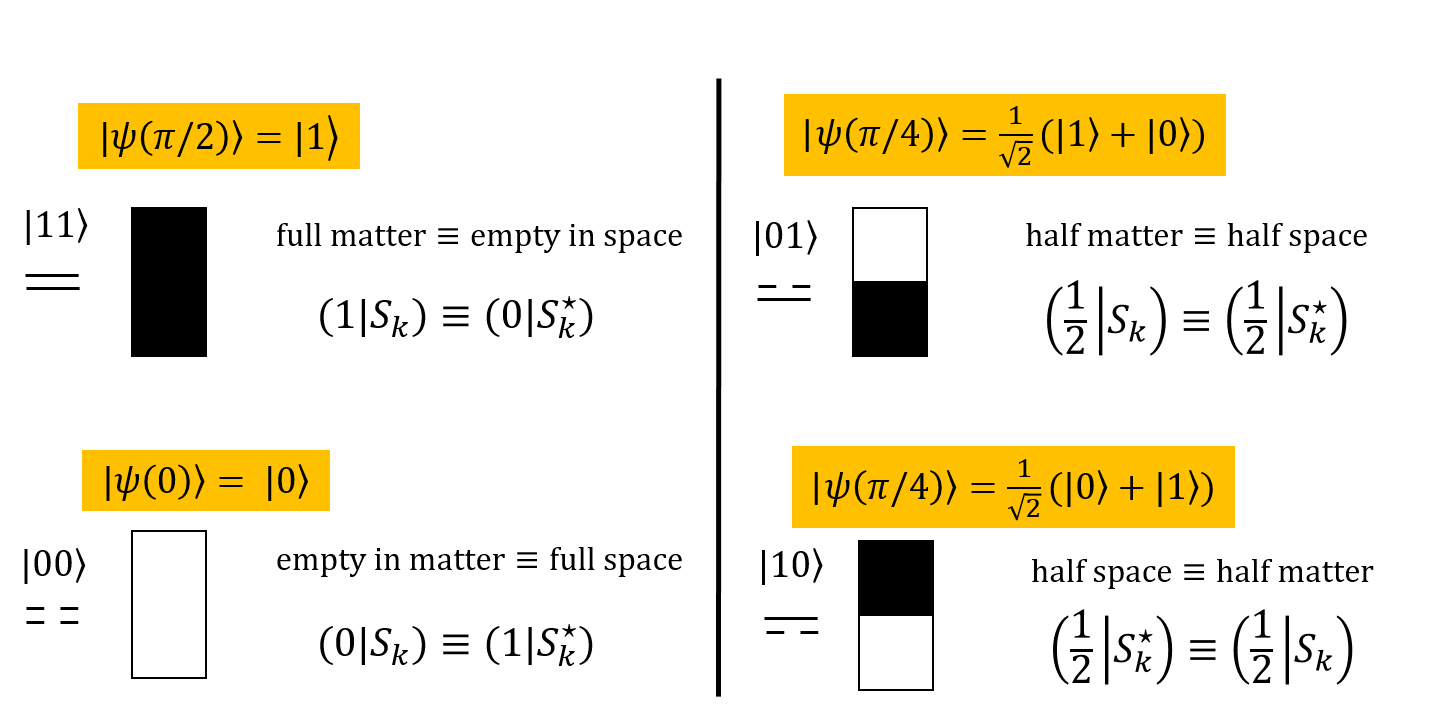}
\caption[ ]
{ \label{fig:water}}
\end{figure}
We can see that for dual invariant representation states $|11\rangle$ and  $|00\rangle$, we have non-dual relationships of $( 1 | S_k) \equiv (0 | S_k^\star)$ and $(0 | S_k) \equiv ( 1|S_k^\star) $ respectively. Or we can think in terms of the states, both $|\psi(\pi/2)\rangle=|1\rangle$ and $|\psi(0)\rangle=|0\rangle$ are non-dual invarinat states  ; while for non-dual representation states $|01\rangle$ and  $|10\rangle$, we have dual relationships of $(\frac{1}{2} |S_k) \equiv (\frac{1}{2} |S_k^\star) $ (or $(\frac{1}{2} |S_k^\star) \equiv (\frac{1}{2} |S_k) $ ). Or we can think in terms of the states that $*|\psi (\pi/4)\rangle = |\psi (\pi/4)\rangle$. In a summary, we can put the results in a 4-tableau,
\begin{figure}[H]
\centering
\includegraphics[trim=0cm 0cm 0cm 0cm, clip, scale=0.75]{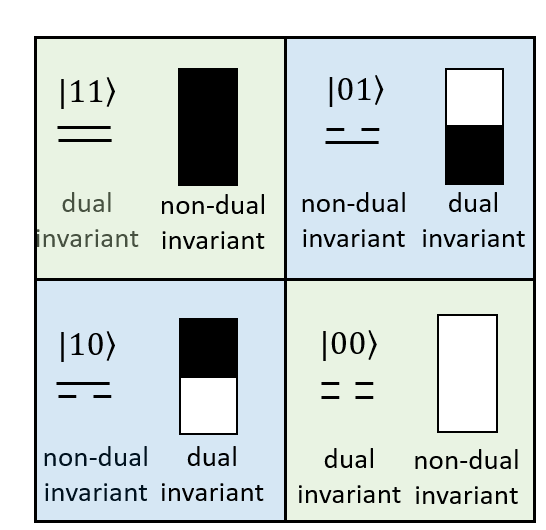}
\caption[ ]
{ \label{fig:water}}
\end{figure}
This marks the complementary theorem of duality. A dual invariant state adopts a non-dual relationship establishment and a non-dual invariant state adopts a dual relationship establishment. 

\subsubsection*{Implication}
From the above analysis, in certain extent, it implies the incompleteness of information carried by each state. When we are in the $|11\rangle $ state with exact probability equal to 1, we basically know everything about the matter realm but know nothing about the empty space realm. Vice versa, when we are situated in the $|00\rangle$ state with exact probability equal to 1, we know everything about the empty space realm but know nothing about the matter world.
When we are in the $|10\rangle$ state or $|01\rangle$ state, we only know at most know half of each realm, therefore the information obtained is still incomplete. We adopt the view that the full state, which is the linear combination of four states contain the complete information of the universe,
\begin{equation}
\psi = a_{00} |00\rangle+a_{11} |11\rangle +a_{01} |01\rangle+a_{10} |10\rangle \,.
\end{equation}  

The complementary theorem will be best demonstrated through the following identities. Upon direct computation, it is found that
\begin{equation}\label{eq:eqb}
\begin{aligned}
|0\rangle \oplus |1\rangle &= |0\rangle \otimes |0\rangle + |1\rangle \otimes |1\rangle \\
|1\rangle \oplus |0\rangle &= |0\rangle \otimes |1\rangle + |1\rangle \otimes |0\rangle \,.
\end{aligned}
\end{equation}
The first line is dual to the second line. From the left hand side, it is clearly that it is is non-left,right invariant. However, for the right hand side, it is $0\leftrightarrow 1$ dual invariant. Therefore, we have non-dual invariant in one side and dual invariant in another, but the two expressions are equivalent. Next we have
\begin{equation} \label{eq:eqc}
\begin{aligned}
|0\rangle \oplus |0\rangle &= |0\rangle \otimes |0\rangle + |1\rangle \otimes |0\rangle \\
|1\rangle \oplus |1\rangle &= |1\rangle \otimes |1\rangle + |0\rangle \otimes |1\rangle \,.
\end{aligned}
\end{equation}
Again, the first line is dual to the second line. From the left hand side, it is clearly that it is left, right dual invariant. However, for the right hand side, it is non $0\leftrightarrow 1$ dual invariant. Therefore, we have dual invariant in one side and non-dual invariant in another, but the two expressions are equivalent. Therefore, from \ref{eq:eqb} and \ref{eq:eqc}, we have demonstrated the complementary theorem in duality. 

We can further demonstrate the complementary theorem of duality in the following illustration:
\begin{figure}[H]
\centering
\includegraphics[trim=0cm 0cm 0cm 0cm, clip, scale=0.5]{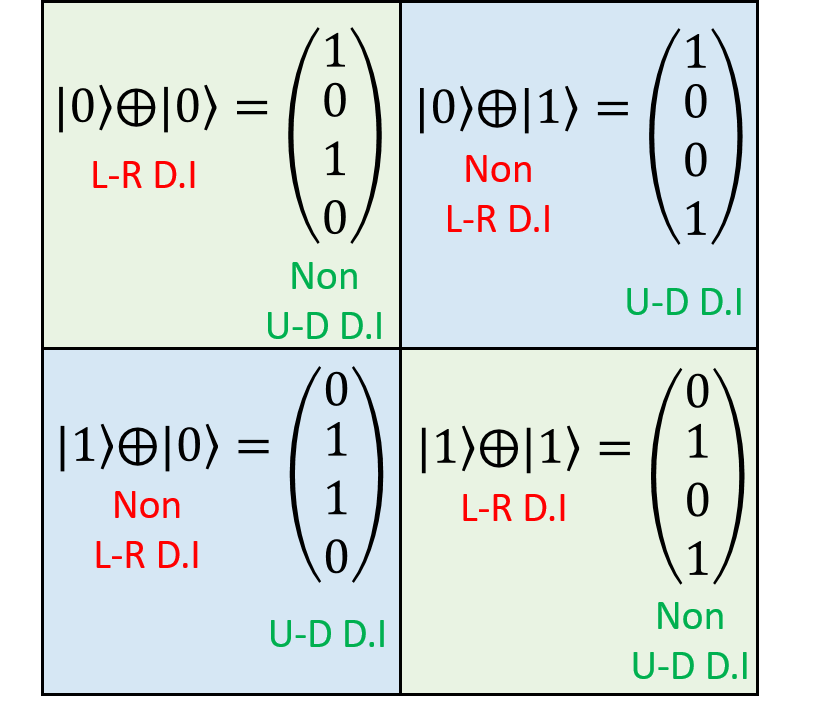}
\caption[ ]
{ \label{fig:water}}
\end{figure}
When we express the direct sum of the vectors explicitly in column vectors, we can see the following property. For example, in the upper left box, $|0\rangle \oplus |0\rangle$ is dual invariant under left and right observation perspective, but the explicit column vector is  not invariant under up and down observation perspective. The same idea goes for the lower right box, which is dual to the upper left box. Then in the upper right box, $|0\rangle \oplus |1\rangle$ is not dual invariant under left and right observation perspective, but the explicit column vector is dual invariant under up and down observation perspective. The same idea goes for the lower left box, which is dual to the upper right box. 

And we have the identity
\begin{equation}
|0\rangle \oplus |1\rangle + |1\rangle \oplus |0\rangle = |0\rangle \oplus |0\rangle + |1\rangle \oplus |1\rangle = \sum_{i,j=0,1}|i\rangle \otimes |j\rangle \,.
\end{equation}
The full dual invariant state, for which here is $(1\,\, 1\,\, 1\,\, 1)^T$ can be always decomposed into a sum of two non-dual invariant states or sum of two dual invariant states.

\section{Quantum field theory with duality symmetry}
In this chapter, we would like apply the idea of duality to study Lagrangian that process the duality $\mathbb{Z}_2$ symmetry. Consider a particle that is described by the scalar field $\phi_+$ field, and its dual particle being described by the $(\phi_+)^* = \phi_-$ field. Next we want to construct an interaction term that preserve $\mathbb{Z}_2 $ symmetry, i.e. the Lagrangian is invariant under the transformation of fields under duality,
\begin{equation} \label{eq:dualTransform}
\phi_+ \rightarrow \phi_- \quad\text{and}\quad \phi_- \rightarrow \phi_+ \,.
\end{equation} 
This means a field transformation as
\begin{equation} \label{eq:transform}
\begin{pmatrix}
\phi_{-} \\
\phi_{+}
\end{pmatrix}
=
\pmb{\mathrm{M}}
\begin{pmatrix}
\phi_{+} \\
\phi_{-}
\end{pmatrix} \,,
\end{equation}
where
\begin{equation}
\pmb{\mathrm{M}}
=
\begin{pmatrix}
0 & 1 \\
1 & 0 
\end{pmatrix}
\end{equation}
is the matrix representation of the duality group $\mathbb{Z}_2$. A Lagrangian with dual symmetry can always be written as two separations of dual terms,
\begin{equation}
\mathcal{L} = \,\,^{*}\mathcal{L} + \mathcal{L^*} \,,
\end{equation}
where $* ^{*}\mathcal{L}=\mathcal{L^*}$ and $*\mathcal{L^*} =\,\,^{*}\mathcal{L} $. The simplest dual invariant Lagrangian for scalar field is
\begin{equation}
\mathcal{L}= \frac{1}{2}\partial_{\mu} \phi_+ \partial^{\mu} \phi_+  + \frac{1}{2}\partial_{\mu} \phi_- \partial^{\mu} \phi_- -g \phi_+ \phi_- \,,
\end{equation}
where $g$ is the constant coupling. Upon symmetrization, we can write the above as
\begin{equation} \label{eq:La1}
\mathcal{L}= \frac{1}{2}\partial_{\mu} \phi_+ \partial^{\mu} \phi_+  + \frac{1}{2}\partial_{\mu} \phi_- \partial^{\mu} \phi_- - \frac{g}{2} (\phi_+ \phi_- + \phi_- \phi_+ )  \,,
\end{equation}
which is invariant under the $\mathbb{Z}_2$ transformation in \ref{eq:transform}. We can identify the two dual Lagrangians as,
\begin{equation}
^{*}\mathcal{L}= \frac{1}{2}\partial_{\mu} \phi_+ \partial^{\mu} \phi_+ -\frac{g}{2} \phi_+ \phi_- 
\end{equation}
and
\begin{equation}
\mathcal{L^*} = \frac{1}{2}\partial_{\mu} \phi_- \partial^{\mu} \phi_- -\frac{g}{2} \phi_- \phi_+ \,.
\end{equation}
The equation of motions for the $\phi^+$ field and the $\phi_-$ field are,
\begin{equation}
\Box \phi_+ = -g\phi_-  \quad \text{and} \quad \Box \phi_- = -g\phi_+ \,.
\end{equation}
Therefore, the dynamics of  $\phi_+$ is sourced by $\phi_-$ and the dynamics of $\phi_-$ is sourced by $\phi_+$.
It is noted that we cannot add heterogeneous mass terms to the above Lagrangian as this violates $\mathbb{Z}_2$ duality symmetry. Consider,
\begin{equation}
\mathcal{L}= \frac{1}{2}\partial_{\mu} \phi_+ \partial^{\mu} \phi_+  + \frac{1}{2}\partial_{\mu} \phi_- \partial^{\mu} \phi_-  +\frac{1}{2}m_+^2 \phi_+ \phi_+ + \frac{1}{2}m_-^2 \phi_- \phi_-     - \frac{g}{2} (\phi_+ \phi_- + \phi_- \phi_+ )  \,,
\end{equation}
Under the duality field transformation, we obtain
\begin{equation}
\mathcal{L}^\prime = \frac{1}{2}\partial_{\mu} \phi_- \partial^{\mu} \phi_-  + \frac{1}{2}\partial_{\mu} \phi_+ \partial_{\mu} \phi_+  +\frac{1}{2}m_+^2 \phi_- \phi_- + \frac{1}{2}m_-^2 \phi_+ \phi_+     - \frac{g}{2} (\phi_- \phi_+ + \phi_+ \phi_- )  \,,
\end{equation}
which is not equal to $\mathcal{L}$. Thus the mass terms break the $\mathbb{Z}_2$ invariance. The dual symmetry is only preserved if $m_+ = m_-$. Therefore, for a dual field theory, both fields have to be of same mass or massless.

Another example for $\mathbb{Z}_2$ invariant Lagrangian is
\begin{equation} \label{eq:La2}
\mathcal{L}= \frac{1}{2}\partial_{\mu} \phi_+ \partial^{\mu} \phi_+  + \frac{1}{2}\partial_{\mu} \phi_- \partial^{\mu} \phi_- - \lambda (\phi_+ \phi_+ + \phi_- \phi_- )  \,,
\end{equation}

In fact, the Lagrangian that possess $\mathbb{Z}_2$ symmetry is not unique. The combination of the Lagrangians in \ref{eq:La1} and \ref{eq:La2} is also $\mathbb{Z}_2$ invariant,
\begin{equation}
\begin{aligned}
\mathcal{L} &= \frac{1}{2}\partial_{\mu} \phi_+ \partial^{\mu} \phi_+  + \frac{1}{2}\partial_{\mu} \phi_- \partial^{\mu} \phi_- - g(\phi_+ \phi_+ + \phi_- \phi_- + \phi_+ \phi_- + \phi_- \phi_+) \\
&= \frac{1}{2}\partial_{\mu} \phi_+ \partial^{\mu} \phi_+  + \frac{1}{2}\partial_{\mu} \phi_- \partial^{\mu} \phi_- - g(\phi_+ + \phi_-)^2 \,.
\end{aligned}
\end{equation}
The interaction term is invariant under $\mathbb{Z}_2 \times \mathbb{Z}_2$ double duality transformation. We can construct representation matrices in $\mathbb{R}^4$ for our purpose. The four elements are
\begin{equation}
\begin{aligned}
& I = D([0],[0])=\mathbb{I}\otimes\mathbb{I}= \begin{pmatrix}
1 & 0 & 0 & 0\\
0 & 1 & 0 & 0 \\
0 & 0 & 1 & 0 \\
0 & 0 & 0 & 1 
\end{pmatrix} \,\,,\,\,
a = D([0],[1])=\mathbb{I}\otimes\pmb{\mathrm{M}}= \begin{pmatrix}
0 & 0 & 1 & 0\\
0 & 0 & 0 & 1 \\
1 & 0 & 0 & 0 \\
0 & 1 & 0 & 0 
\end{pmatrix} \,\,,\,\, \\
& b = D([1],[0])=\pmb{\mathrm{M}}\otimes\mathbb{I}= \begin{pmatrix}
0 & 1 & 0 & 0\\
1 & 0 & 0 & 0 \\
0 & 0 & 0 & 1 \\
0 & 0 & 1 & 0 
\end{pmatrix} \,\,,\,\,
c=D([1],[1])=\pmb{\mathrm{M}}\otimes\pmb{\mathrm{M}}= \begin{pmatrix}
0 & 0 & 0 & 1\\
0 & 0 & 1 & 0 \\
0 & 1 & 0 & 0 \\
1 & 0 & 0 & 0 
\end{pmatrix} \,\,.\,\,
\end{aligned}
\end{equation}
Note that the identity element $\mathbb{I}\otimes\mathbb{I}$ has all positive eigenvalue of $+1$, while the remaining elements $\pmb{\mathrm{M}}\otimes\mathbb{I}, \mathbb{I}\otimes\pmb{\mathrm{M}} \,\,\text{and}\,\,\pmb{\mathrm{M}}\otimes\pmb{\mathrm{M}}$ would give two positive eigenvalues of $+1$ and two negative eigenvalues of $-1$. Upon diagonalization, this will recover the result in \ref{eq:resultAIrreps}. These $4\times 4$ representation matrices act on the basis $\{ |00\rangle,|01\rangle , |10\rangle, |11\rangle    \}$, which is $\{\phi_+ \phi_+ , \phi_+ \phi_- , \phi_- \phi_+ , \phi_- \phi_- \}$.

 For each term in the interaction, we can carry out further analysis. First notice that both $\phi_+ \phi_+$ and $\phi_- \phi_-$ are observer dual invariant, i.e. looking from left makes no difference to looking at right. However, this is not the case for $\phi_+ \phi_-$ and $\phi_- \phi_+$, which are non-dual invariant under observer. We notice that
\begin{equation}
(\phi_\pm \phi_\pm | S_3 ) = (\phi_\pm \phi_\pm | S_3^\star ) \,.
\end{equation}
On the contrary,
\begin{equation}
(\phi_\pm \phi_\mp | S_3 ) = (\phi_\mp \phi_\pm | S_3^\star ) 
\end{equation}
Therefore we introduce two partitions $B$ and $Q$
\begin{equation}
B= \{ \phi_+\phi_+ ,\phi_-\phi_-\, | \,(\phi_\pm \phi_\pm | S_3 ) = (\phi_\pm \phi_\pm | S_3^\star )  \}
\end{equation}
\begin{equation}
Q= \{ \phi_+\phi_- ,\phi_-\phi_+\, | \, (\phi_\pm \phi_\mp | S_3 ) \neq (\phi_\pm \phi_\mp | S_3^\star )   \}
\end{equation}
Therefore, $B$ and $Q$ are dual partitions.

In general, the following Lagrangian is dual invariant
\begin{equation}
\mathcal{L}=\frac{1}{2}\partial_{\mu} \phi_+ \partial^{\mu} \phi_+  + \frac{1}{2}\partial_{\mu} \phi^- \partial^{\mu} \phi^- - g(\phi_+ + \phi_-)^n \,. 
\end{equation}
The interaction term $g(\phi_+ + \phi_-)^n $ is invariant under $\mathbb{Z}_2 \times \cdots \times    \mathbb{Z}_2 $ $(\mathrm{or}\, \mathbb{Z}_2 \otimes \cdots \otimes    \mathbb{Z}_2 )$ multi-duality transformation.

Let $U$ be $\{+\}$ and its dual $U^*$ be $\{-\}$ and the complete set be $W= U \cup U^* = \{+,-\}$.  Since the interaction term can be written as the following
\begin{equation}
(\phi_+ + \phi_-)^n  = \sum_{i_1 , i_2 ,\cdots , i_n \in W   } \phi_{i_1} \phi_{i_2}\cdots \phi_{i_N }= \sum_{i_1 , i_2 ,\cdots , i_n \in W   } \prod_{k=1}^n \phi_{i_k} \,,
\end{equation}
Therefore the Lagrangian can be written as
\begin{equation} \label{eq:LL}
\mathcal{L}=\frac{1}{2}\partial_{\mu} \phi_+ \partial^{\mu} \phi_+  + \frac{1}{2}\partial_{\mu} \phi_-\partial^{\mu} \phi_- - g  \sum_{i_1 , i_2 ,\cdots , i_n \in W   } \prod_{k=1}^n \phi_{i_k} \,.
\end{equation}
It is noted that such interaction term form the basis of representation of homogeneous 

In addition, for $n$ is even, the following four Lagrangians are also dual invariant,
\begin{equation}
\mathcal{L}=\frac{1}{2}\partial_{\mu} \phi_+ \partial^{\mu} \phi_+  + \frac{1}{2}\partial_{\mu} \phi_- \partial^{\mu} \phi_- - g(\pm \phi_+ \pm \phi_-  )^{2k} \,,
\end{equation}
and
\begin{equation} \label{eq:secondcase}
\mathcal{L}=\frac{1}{2}\partial_{\mu} \phi_+ \partial^{\mu} \phi_+  + \frac{1}{2}\partial_{\mu} \phi_- \partial^{\mu} \phi_- - g(\pm \phi_+ \mp \phi_-  )^{2k} \,,
\end{equation}
Note that for the second case in \ref{eq:secondcase}, if $n$ is odd, we have anti-dual invariant symmetry, i.e. the Lagrangian in invariant under $\phi_+ \rightarrow -\phi_-$ and $\phi_- \rightarrow -\phi_+$. For $\ref{eq:secondcase}$, the interaction can be separated further into the positive part and the negative part. With some simple algebra, it can be shown that the positive part is where the number of $+$ and the number of $-$ are even, and the negative part is where the number of $+$ and the number of $-$ are odd. For example,
\begin{equation}
\begin{aligned}
(\phi_+ - \phi_-)^{2k} &= \sum_{\substack{i_1 ,i_2 \cdots, i_n \in W \\ \#+, \#- = \mathrm{even}}} \phi_{i_1} \cdots \phi_{i_{}2k} - \sum_{\substack{i_1 ,i_2 \cdots, i_n \in W \\ \#+, \#- = \mathrm{odd}}} \phi_{i_1} \cdots \phi_{i_{k}} \\
&= \sum_{\substack{i_1 ,i_2 \cdots, i_n \in W \\ \#+, \#- = \mathrm{even}}} \prod_{l=1}^{2k} \phi_{i_l} - \sum_{\substack{i_1 ,i_2 \cdots, i_n \in W \\ \#+, \#- = \mathrm{odd}}} \prod_{l=1}^{2k} \phi_{i_l}  \,.
\end{aligned}
\end{equation}
The positive part and negative part, respectively, for each can be further split into two dual partitions with $2^{n-1}$ elements. Let $K_+$ be the partition and its dual $K_+^*$ for the positive partition, and the complete positive partition is given by $K^{(+)} =K_+ \cup K_+^* $ and $K_+ \cap K_+^* = \emptyset $; while $K_-$ be the partition and its dual $K_-^*$ for the negative partition, and the complete negative partition is given by $K^{(-)} =K_- \cup K_-^* $  and $K_- \cap K_-^* = \emptyset $. It is important to note that the positive positive and negative partition are not dual to each other. For simplicity denote $I=\{i_1 ,i_2 \cdots i_n \}$, then we have
\begin{equation}
\sum_{\substack{I \in W \\ \#+, \#- = \mathrm{even}}}\prod_{l=1}^{2k} \phi_{i_l}  = \sum_{I \in K_+}\prod_{l=1}^{2k} \phi_{i_l} + \sum_{I \in K_+^*}\prod_{l=1}^{2k} \phi_{i_l} \,,
\end{equation}
and
\begin{equation}
\sum_{\substack{I\in W \\ \#+, \#- = \mathrm{odd}}}\prod_{l=1}^{2k} \phi_{i_l}  = \sum_{I \in K_-}\prod_{l=1}^{2k} \phi_{i_l} + \sum_{I \in K_-^*}\prod_{l=1}^{2k} \phi_{i_l} \,,
\end{equation}
This can be checked, for example $n =4$,
\begin{equation}
\begin{aligned}
(\phi_+ - \phi_-)^{4} &= \phi_{+}\phi_{+}\phi_{+}\phi_{+} - \phi_{+}\phi_{+}\phi_{+}\phi_{-} - \phi_{+}\phi_{+}\phi_{-}\phi_{+} + \phi_{+}\phi_{-}\phi_{+}\phi_{-} \\
& -\phi_{+}\phi_{-}\phi_{+}\phi_{+} + \phi_{+}\phi_{-}\phi_{+}\phi_{-} + \phi_{+}\phi_{-}\phi_{-}\phi_{+} - \phi_{+}\phi_{-}\phi_{-}\phi_{-} \\
& -\phi_{-}\phi_{+}\phi_{+}\phi_{+} + \phi_{-}\phi_{+}\phi_{+}\phi_{-} + \phi_{-}\phi_{+}\phi_{-}\phi_{+} - \phi_{-}\phi_{+}\phi_{-}\phi_{-} \\
& + \phi_{-}\phi_{-}\phi_{+}\phi_{+} - \phi_{-}\phi_{-}\phi_{+}\phi_{-} -\phi_{-}\phi_{-}\phi_{-}\phi_{+} + \phi_{-}\phi_{-}\phi_{-}\phi_{-}
\end{aligned}
\end{equation}
There are $2^{4-1} =8$ positive terms and 8 negative terms respectively. For the positive partition, we have 
\begin{equation}
K_+ = \{ \phi_{+}\phi_{+}\phi_{+}\phi_{+}, \phi_{+}\phi_{+}\phi_{-}\phi_{-} , \phi_{+}\phi_{-}\phi_{+}\phi_{-} , \phi_{+}\phi_{-}\phi_{-}\phi_{+}  \}
\end{equation} 
and
\begin{equation}
K_+^* = \{ \phi_{-}\phi_{-}\phi_{-}\phi_{-},\, \phi_{-}\phi_{-}\phi_{+}\phi_{+} ,\, \phi_{-}\phi_{+}\phi_{-}\phi_{+} ,\, \phi_{-}\phi_{+}\phi_{+}\phi_{-}  \}
\end{equation}
We can see that $K_+$ and $K_+^*$ are dual to each other in which $K_+^*  = * K_+$. For the negative partition, we have
\begin{equation}
K_- = \{ -\phi_{+}\phi_{+}\phi_{+}\phi_{-} ,\, -\phi_{+}\phi_{+}\phi_{-}\phi_{+},\, -\phi_{+}\phi_{-}\phi_{+}\phi_{+} ,\, -\phi_{+}\phi_{-}\phi_{-}\phi_{-}   \}
\end{equation}
and
\begin{equation}
K_-^* = \{ -\phi_{-}\phi_{-}\phi_{-}\phi_{+} ,\, -\phi_{-}\phi_{-}\phi_{+}\phi_{-},\, -\phi_{-}\phi_{+}\phi_{-}\phi_{-} ,\, -\phi_{-}\phi_{+}\phi_{+}\phi_{+}   \}
\end{equation}
We can see that $K_-$ and $K_-^*$ are dual to each other in which $K_-^*  = * K_-$. Therefore the full interaction term can be written as 
\begin{equation}
K^{(+)} \cup K^{(-)} = (K_+ \cup K_+^* ) \cup  (-K_- \cup -K_-^* ) \,.
\end{equation}

Now let's reconsider for the $n= \mathrm{odd}$ case. The Lagrangian is given by
\begin{equation} \label{eq:Laodd}
\mathcal{L}=\frac{1}{2}\partial_{\mu} \phi_+ \partial^{\mu} \phi_+  + \frac{1}{2}\partial_{\mu} \phi_- \partial^{\mu} \phi_- - g(\pm \phi_+ \pm \phi_-  )^{2k-1} \,,
\end{equation}
The interaction term $(\phi_+ - \phi_-)^{2k-1}$, upon expansion can be separated into positive partition and negative partition that is dual to each other. Let $P^{(+)}$ be the positive partition and $P^{(-)}$ be the negative partition, we ill have $*P^{(+)} = P^{(-)}$. Mathematically

\begin{equation}
(\phi_+ - \phi_-)^{2k-1} = \sum_{I \in W} \prod_{l=1}^{2k-1} \phi_{i_l}   = \sum_{I \in P_+}\prod_{l=1}^{2k-1} \phi_{i_l} - \sum_{I \in P_-}\prod_{l=1}^{2k-1} \phi_{i_l} \,,
\end{equation} 
where
\begin{equation}
\bigg(\sum_{I \in P_+}\prod_{l=1}^{2k-1} \phi_{i_l} \bigg)^* = \sum_{I \notin P_+}\prod_{l=1}^{2k-1} \phi_{i_l}^* =\sum_{I \in P_-}\prod_{l=1}^{2k-1} \phi_{i_l}
\end{equation}
For example, this can be verified by $n=3$ case. Consider
\begin{equation} \label{eq:3order}
\begin{aligned}
(\phi_+ - \phi_-)^3 &= \phi_{+}\phi_{+}\phi_{+} - \phi_{+}\phi_{-}\phi_{+} - \phi_{-}\phi_{+}\phi_{+} + \phi_{-}\phi_{-}\phi_{+} \\
& -\phi_{+}\phi_{+}\phi_{-} + \phi_{+}\phi_{-}\phi_{-}+\phi_{-}\phi_{+}\phi_{-} - \phi_{-}\phi_{-}\phi_{-} \,.
\end{aligned}
\end{equation}
We have for the positive partition,
\begin{equation}
P^{(+)} = \{ \phi_{+}\phi_{+}\phi_{+},\,\phi_{+}\phi_{-}\phi_{-} ,\,\phi_{-}\phi_{+}\phi_{-} ,\, \phi_{-}\phi_{-}\phi_{+} \} \,,
\end{equation}
and the negative partition,
\begin{equation}
P^{(-)} = \{ -\phi_{-}\phi_{-}\phi_{-},\,-\phi_{-}\phi_{+}\phi_{+} ,\,-\phi_{+}\phi_{-}\phi_{+} ,\, -\phi_{+}\phi_{+}\phi_{-} \} \,.
\end{equation}
We can clearly see that $P^{(-)} = * P^{(+)}$ with $**=1$, where the matrix representation of $*$ is $\pmb{\mathrm{M}}(*)$
\begin{equation}
\pmb{\mathrm{M}}(*)
=
\begin{pmatrix}
0 & -1 \\
-1 & 0 
\end{pmatrix}
\end{equation}
In conclusion, when $n=2k$ is even, $K^{(+)}$ and $K^{(-)}$ are non-dual; while when $n=2k-1$ is odd, $P^{(+)}$ and $P^{(-)}$ are dual to each other. Notice that dual and non-dual are a duality itself, this is reflects by the odd and even power respectively. Define $J_{\mathrm{odd}} = \{ (\phi_+ - \phi_-)^{2k-1} | \forall k \in \mathbb{N} \}$ and $J_{\mathrm{even}} = \{ (\phi_+ - \phi_-)^{2k} | \forall k \in \mathbb{N} \}$. They form a duality. This is because, consider the operator $*= (\phi_+ + \phi_-)$, 
\begin{equation}
\begin{aligned}
*J_{\mathrm{odd}} &= (\phi_+ - \phi_-)J_{\mathrm{odd}} =(\phi_+ - \phi_-) \{ (\phi_+ + \phi_-)^{2k-1} | \forall k \in \mathbb{N} \} \\
&= \{ (\phi_+ - \phi_-)^{2k} | \forall k \in \mathbb{N} \} \\
&=J_{\mathrm{even}}
\end{aligned}
\end{equation}
Then
\begin{equation}
\begin{aligned}
**J_{\mathrm{odd}} &= *J_{\mathrm{even}} =(\phi_+ - \phi_-)J_{\mathrm{even}}\\
&=(\phi_+ - \phi_-)\{ (\phi_+ - \phi_-)^{2k} | \forall k \in \mathbb{N} \}\\
&=\{ (\phi_+ - \phi_-)^{2k+1} | \forall k \in \mathbb{N} \}\\
&=J_{\mathrm{odd}}
\end{aligned}
\end{equation}
Therefore $**=1$ is an identity map. Since there exists an bijective map between odd and even numbers, and also that $J_{\mathrm{odd}} \cap J_{\mathrm{even}} = \emptyset$, therefore $J_{\mathrm{odd}}$ and $J_{\mathrm{even}}$ is dual to each other. Thus the Lagrangians in \ref{eq:secondcase} and $\ref{eq:Laodd}$ are dual to each other.

Next we would like to construct another possible classification for the terms in \ref{eq:3order}. Notice that the following 4 terms are observational dual invariant, which remains the same regardless the direction looking at it.
\begin{equation}
(\phi_{+}\phi_{+}\phi_{+},\,-\phi_{-}\phi_{-}\phi_{-},\,-\phi_{+}\phi_{-}\phi_{+},\,\phi_{-}\phi_{+}\phi_{-} | S_3 ) = (\phi_{+}\phi_{+}\phi_{+},\,-\phi_{-}\phi_{-}\phi_{-},\,-\phi_{+}\phi_{-}\phi_{+},\,\phi_{-}\phi_{+}\phi_{-} | S_3^\star )
\end{equation}
Thus we have the $B$ partition as
\begin{equation}
B= \{ \phi_{+}\phi_{+}\phi_{+},\,-\phi_{-}\phi_{-}\phi_{-},\,-\phi_{+}\phi_{-}\phi_{+},\,\phi_{-}\phi_{+}\phi_{-}   \} \,.
\end{equation}
The remaining $Q$ partition which does not have observation dual invariance contains all the remaining terms
\begin{equation}
Q=\{ -\phi_{+}\phi_{+}\phi_{-}, \phi_{-}\phi_{-}\phi_{+}, -\phi_{-}\phi_{+}\phi_{+} , \phi_{+}\phi_{-}\phi_{-} \}
\end{equation}
Mathematically, we can define a parity operator $\hat{P}=(-1)*$ acting on the third therm or first term of the product, denoting $\hat{P}_3 = (-1)*_3$ or $\hat{P}_1=(-1)*_1$ respectively. We can see that
\begin{equation}
\begin{aligned}
\hat{P}_3 B &= \{  \phi_{+}\phi_{+}[(-1)*\phi_{+}],\,-\phi_{-}\phi_{-}[(-1)*\phi_{-}],\,-\phi_{+}\phi_{-}[(-1)*\phi_{+}],\,\phi_{-}\phi_{+}[(-1)*\phi_{-}]   \} \\
&=\{-\phi_{+}\phi_{+}\phi_{-}, \phi_{-}\phi_{-}\phi_{+}, -\phi_{-}\phi_{+}\phi_{+} , \phi_{+}\phi_{-}\phi_{-} \} \\
&=Q \,.
\end{aligned}
\end{equation}
One can see that $\hat{P}_3^2 = ((-1)*_3)^2 =(-1)(-1)*_3 *_3 =1$ which is the identity. 
Next, we also see that
\begin{equation}
\begin{aligned}
\hat{P}_1 B &= \{ [(-1)*\phi_{+}]\phi_{+}\phi_{+},\,-[(-1)*\phi_{-}]\phi_{-}\phi_{-},\,-[(-1)*\phi_{+}]\phi_{-}\phi_{+},\,[(-1)*\phi_{-}]\phi_{+}\phi_{-}   \} \\
&=\{ -\phi_{-}\phi_{+}\phi_{+},\,\phi_{+}\phi_{-}\phi_{-},\,\phi_{-}\phi_{-}\phi_{+},\,-\phi_{+}\phi_{+}\phi_{-}   \} \\
&=Q \,.
\end{aligned}
\end{equation}
And we have $\hat{P}_1^2 =1$. Therefore, $B$ and $Q$ are dual to each other under such parity maps. We interpret as, $B$ is a dual invariant while $Q$ is a non-dual invariant under observation, such that $Q$ is dual to $B$.

The equation of motion for the $n-$th order interaction term is the following
\begin{equation}
\Box\phi_+ = n(\phi_+ - \phi_-)^{n-1} \quad\text{and}\quad \Box\phi_- = - n(\phi_+ - \phi_-)^{n-1} \,.
\end{equation}
Together we have
\begin{equation}
\Box(\phi_+ + \phi_- ) = 0 \,.
\end{equation}

To cover all the diagrams of different order for $g^k$, we demand a full theory as follow
\begin{equation}
\begin{aligned}
\mathcal{L} &= \frac{1}{2}\partial_{\mu}\phi_+ \partial^{\mu}\phi_+  + \frac{1}{2}\partial_{\mu}\phi_- \partial^{\mu}\phi_- -\sum_{k=2}^n g_k (\phi_+ + 
\phi_-)^k \\
&=\frac{1}{2}\partial_{\mu}\phi_+ \partial^{\mu}\phi_+  + \frac{1}{2}\partial_{\mu}\phi_- \partial^{\mu}\phi_- -\sum_{k=2}^n \sum_{i_1 , i_2 ,\cdots , i_k \in W   } g_k \prod_{l=1}^k \phi_{i_l} \,.
\end{aligned}
\end{equation}
The the Lagrangian is invariant under the global degraded symmetry group of
\begin{equation} \label{eq:degraded}
\mathbb{Z}_2 \oplus (\mathbb{Z}_2 \otimes  \mathbb{Z}_2 ) \oplus (\mathbb{Z}_2 \otimes  \mathbb{Z}_2 \oplus \mathbb{Z}_2 ) \otimes \cdots = \bigoplus_{k=1}^n \bigotimes_{l=1}^k \mathbb{Z}_2 \,.
\end{equation}

Finally we would like to study the case of heterogeneous field case. We have the following Lagrangian,
\begin{equation}
\mathcal{L} =\frac{1}{2}\sum_{a=+,-}\sum_{i=1}^n \partial_{\mu}\phi_{ai}\partial^{\mu}\phi_{ai} -g \prod_{i=1}^n (\pm \phi_{+i}\pm \phi_{-i}) \,,
\end{equation}
and
\begin{equation} 
\mathcal{L} =\frac{1}{2}\sum_{a=+,-}\sum_{i=1}^n \partial_{\mu}\phi_{ai}\partial^{\mu}\phi_{ai} -g \prod_{i=1}^n (\pm \phi_{+i}\mp \phi_{-i}) \,,
\end{equation}
Using $n=2$ as an example, and the interaction term as  $(\phi_{+i}- \phi_{-i})^2$, we have four scalar fields in the Lagrangian
\begin{equation}
\begin{aligned}
\mathcal{L}&=\frac{1}{2}\partial_{\mu}\phi_{+1}\partial^{\mu}\phi_{+1} + \frac{1}{2}\partial_{\mu}\phi_{-1}\partial^{\mu}\phi_{-1} + \frac{1}{2}\partial_{\mu}\phi_{+2}\partial^{\mu}\phi_{+2}+ \frac{1}{2}\partial_{\mu}\phi_{-2}\partial^{\mu}\phi_{-2} \\
&-g(\phi_{+1} \phi_{+2} - \phi_{-1} \phi_{-2} - \phi_{+1} \phi_{-2} + \phi_{1-} \phi_{+2} )
\end{aligned}
\end{equation}
The full Lagrangian for all orders will be, take the (+,+) for example,
\begin{equation}
\mathcal{L} =\frac{1}{2}\sum_{a=+,-}\sum_{i=1}^n \partial_{\mu}\phi_{ai}\partial^{\mu}\phi_{ai} -g\sum_{k=1}^n\prod_{i=1}^k ( \phi_{+i} + \phi_{-i}) \,. 
\end{equation}

\section{Dual Invariant Topological Field Theory}
The previous section investigates dual invariant interaction. In this chapter, we would like to give a formal study of dual invariant action. Let's suppose we define the observation duality as left and right observation perspective, $L$ and $R$. We want to construct an action which remains the same whenever you look from the left or look from the right, i.e. mathematically
\begin{equation}
(S| L ) \equiv (S|R) \,,
\end{equation}
such that $S=S^*$.

The scalar field action takes the generic form 
of \begin{equation}
S = \frac{1}{2}\int d^4 x \,\partial_{\mu} \phi \eta^{\mu\nu}  \partial_\nu \phi  
\end{equation}
is not dual invariant. First of all, $\partial_{\mu}\phi \partial^\mu \phi$, when looking on the left is $-\phi\partial^\mu \phi\partial_{\mu}$, which is an operator and not a scalar. If we carry out integration by parts,
\begin{equation}
S = \frac{1}{2}\int d^4 x (-\phi \Box \phi)
\end{equation}
It is tempted to think that $\phi \Box \phi$ is dual invariant, however bare in mind that the $\Box$ operator when looking from the right, it is $-\Box$ instead of $\Box$ due to the minus signs in the $\Box$ operators. Also the negative sign in front makes the action impossible to be dual invariant. Next if we include a mass term of $-\frac{1}{2}m^2 \phi^2$, this is not dual invariant due to the negative sign. Similar reason goes to the action of gauge field,
\begin{equation}
S =-\frac{1}{4}\int d^4 x \,F_{\mu\nu} F^{\mu\nu} = \frac{1}{2} \int d^4 x A^{\mu}(\Box \eta_{\mu\nu} - \partial_{\mu}\partial_{\nu})A^\nu = \frac{1}{2}\int d^4 x \, (\pmb{\mathrm{E}}^2 - \pmb{\mathrm{B}}^2 )  \,,
\end{equation}
which is non-dual invariant due to the minus sign and the $\Box$ operator. Or we can see that the term $\pmb{\mathrm{E}}^2 - \pmb{\mathrm{B}}^2 $ is non-dual invariant due to the minus sign as in the form of $a-b$. If we add a mass term $\frac{1}{2}m^2 A_{\mu}A^{\mu}$, although we have a positive sign, the $A_{\mu}A^{\mu}$ would induce a negative sign when we look from 
the right.

The Dirac action is also not dual invariant. The term $i\bar{\psi}\gamma^\mu\partial_{\mu}\psi $ has a structure of row times matrix times column, but when we look form the right we have something column times matrix times a row, which is not possible. For the mass term $-m\bar{\psi}\psi$, it has a minus sign it is immediately non-dual invariant. 

In other words, we come into a conclusion that the standard model action is not dual invariant. One of the major problems is due to the negative signs in the metric, so the flat Minkowski metric tensor is the problematic cause of non-dual invariance. This shows that the intrinsic structure of spacetime itself is naturally non-dual invariant. 

Generally speaking, a dual invariant action in the left and right observation perspective must be positive, and it should possess $ada$ form if it contains derivatives. It should also be independent of the metric tensor. 

The most possible candidate so far we know is the Chern-Simon's action. We will show that in fact it is dual invariant. The Chern-Simons action is given by
\begin{equation}
S_{\mathrm{CS}} = \int_M d^3 x \,\epsilon^{\mu\nu\rho} A_{\mu} \partial_{\nu} A_{\rho} \,,
\end{equation}
where $M$ is the topological manifold. We can check that it is left, right observation dual invariant by expanding the terms in the action explicitly,
\begin{equation}
\begin{aligned}
S_{\mathrm{CS}} &= \int_M d^3 x \,\Big(( \epsilon^{012} A_0 \partial_{1} A_2 + \epsilon^{210}  A_2 \partial_{1} A_0   ) + (\epsilon^{120} A_1 \partial_{2} A_0 + \epsilon^{021}  A_0 \partial_{2} A_1 ) \\
&\quad\quad \quad \quad \quad \quad \quad \quad \quad  +(\epsilon^{201} A_2 \partial_{0} A_1 + \epsilon^{102}  A_1 \partial_{0} A_2 ) \Big) \\
&=\int_M d^3 x \,\Big(( A_0 \partial_{1} A_2 -  A_2 \partial_{1} A_0   ) + ( A_1 \partial_{2} A_0 -  A_0 \partial_{2} A_1 ) \\
&\quad\quad \quad \quad \quad \quad \quad \quad \quad  +( A_2 \partial_{0} A_1  -  A_1 \partial_{0} A_2 ) \Big)
\end{aligned}
\end{equation} 
So we have terms of $A_a \partial_b A_c - A_c \partial_b A_a$, which is dual invariant. Therefore the Chern-Simons action is dual invariant, $(S_{\mathrm{CS}} |L) \equiv (S_{\mathrm{CS}}|R) $.

The Chern-Simons action is topological, which is independent of the metric, so it perfectly fits our necessary criteria. The dual invariant nature of the Chern-Simons action echoes the idea in section 2.1.6 (General Analysis and Topological Duality) that, if a system is dual invariant, its corresponding geometry is topological. 

The most general Chern-Simons action takes the following form,
\begin{equation}
S_{\mathrm{CS}} = \int_M d^3 x \,\kappa\epsilon^{\mu\nu\rho} A_{\mu} \partial_{\nu} A_{\rho} \,,
\end{equation}
where $\kappa$ is some constant which can be positive or negative. It is noted that if $\kappa$ is negative, let's say $-1$, the Chern-Simons action becomes,
\begin{equation}
\begin{aligned}
S_{\mathrm{CS}} &= \int_M d^3 x \,\Big((A_2 \partial_{1} A_0 - A_0 \partial_{1} A_2    ) + ( A_0 \partial_{2} A_1 - A_1 \partial_{2} A_0   ) \\
&\quad\quad \quad \quad \quad \quad \quad \quad \quad  +( A_1 \partial_{0} A_2 - A_2 \partial_{0} A_1 )  \Big) \,.
\end{aligned} 
\end{equation}
Hence now we have  terms of $A_c \partial_b A_a - A_a \partial_b A_c$, which is another dual invariant. Therefore no matter $\kappa$ is positive or negative, the Chern-Simons action is still dual invariant. Not that we cannot add a mass term $\frac{1}{2}m^2 A_{\mu}A^{\mu}$ nor add an external source term coupling to to the gauge field $A_{\mu} J^\mu$ as both of these terms depend on the metric, this will break the dual invariant symmetry. In other words, in order to preserve dual invariance, the gauge field particle must be massless. In order to gain mass, there must be involving some new mass acquisition mechanisms.  

The dual invariance of the Chern-Simons action can be proved by integration by parts. Consider
\begin{equation}
\begin{aligned}
(S_{\mathrm{CS}} |L) &= \int_M d^3 x \, \epsilon^{\mu\nu \rho} A_{\mu} \partial_{\nu} A_{\rho} \\
&= \mathrm{B.C.} - \int_M d^3 x \,\epsilon^{\mu\nu\rho} A_{\rho} \partial_\nu A_{\mu} \\
&= -\int_M d^3 x \,(-1)^3 \epsilon^{\rho\nu\mu} A_{\rho} \partial_\nu A_{\mu} \\
&=\int_M d^3 x \,\epsilon^{\rho\nu\mu} A_{\rho} \partial_\nu A_{\mu} \\
&= (S_{\mathrm{CS}} |R) \,,
\end{aligned}
\end{equation}
where we have dropped the boundary term. We can see that $A_{\mu}\partial_{\nu} A_{\rho} \rightarrow -A_{\rho} \partial_\nu A_{\mu}$ and $\epsilon^{\mu\nu\rho} \rightarrow -\epsilon^{\rho\nu\mu}$, so when we exchange perspectives of the two terms from $L$ to $R$, each term will induce a negative sign and the two negative signs become positive again. Thus we prove that the Chern-Simons action is dual invariant under the left-right perspective.

For non-Abelian case, where the gauge field are matrices and suppose $A_{\mu} = A_{\mu}^a t^a$ where $t^a$ are the SU$(N)$ generators the dual invariance still holds,
\begin{equation}
S_{\mathrm{CS}} = \int_M d^3 x \, \mathrm{Tr}(\kappa\epsilon^{\mu\nu\rho} A_{\mu} \partial_{\nu} A_{\rho}) = \int_M d^3 x \,\kappa\epsilon^{\mu\nu\rho} A^a_{\mu} \partial_{\nu} A^a_{\rho}
\end{equation}

If the gauge field is some generic matrix, for example Berry connection for the degenerate case, we have
\begin{equation}
A_{\mu mn} = \langle   \psi_m | \partial_{\mu}  \psi_n \rangle \,.
\end{equation}
Then the Chern-Simons action is 
\begin{equation}
\begin{aligned}
S_{\mathrm{CS}} = \int_M d^3 x \, \kappa\sum_{m}\sum_n \epsilon_{\mu\nu\rho} A^{\mu}_{ mn} \partial^\nu A^{\rho}_{ nm} \,,
\end{aligned} 
\end{equation}
which is obviously dual invariant. Explicitly, this is
\begin{equation}
\begin{aligned}
S_{\mathrm{CS}} &= \int_M d^3 x \, \kappa\sum_{m}\sum_n \epsilon_{\mu\nu\rho}\langle   \psi_m | \partial^{\mu}  \psi_n \rangle \partial^{\nu}  \langle   \psi_n | \partial^{\mu}  \psi_m \rangle \\
&= \int_M d^3 x \, \kappa\sum_{m}\sum_n \epsilon_{\mu\nu\rho}\langle   \psi_m | \partial^{\mu}  \psi_n \rangle (\langle \partial^\nu \psi | \partial^{\rho} \psi_n \rangle +\langle \psi_n | \partial^\nu \partial^\rho \psi_m \rangle )\\
&= \int_M d^3 x \, \kappa\sum_{m}\sum_n \epsilon_{\mu\nu\rho}\langle   \psi_m | \partial^{\mu}  \psi_n \rangle \langle \partial^\nu \psi | \partial^{\rho} \psi_n \rangle \,.
\end{aligned}
\end{equation}
The last term vanishes because $\partial_\nu \partial_\rho$ is symmetric.

The most generic Chern-Simons Three form endowed in SU(N) Lie algebra takes the following action:
\begin{equation}
S_{\mathrm{CS}} = \kappa \int_M  \mathrm{Tr} (A \wedge dA + \frac{2}{3} A\wedge A \wedge A  ) \,.
\end{equation}
The $A$ is gauge-field 1-form, $dA$ is the exterior on $A$ which is a two-form. Explicitly we have,
\begin{equation}
\begin{aligned}
S_{\mathrm{CS}} = \kappa \int_M  \mathrm{Tr} 
\Big((A^a_\mu t^a dx^\mu)\wedge \partial_\rho A_\nu^b t^b dx^\rho \wedge dx^\nu + \frac{2}{3}A_\mu^a t^a A_\nu^b t^b A_{\rho}^c t^c dx^\mu \wedge dx^\nu \wedge dx^\rho \Big)
\end{aligned}
\end{equation}
By relabelling indices and grouping terms, we have 
\begin{equation} \label{eq:CS1}
S_{\mathrm{CS}} = \kappa \int_M  \mathrm{Tr} 
\Big( A_\mu^a \partial_\nu A_\rho^b t^a t^b dx^\mu \wedge dx^\nu \wedge dx^\rho + \frac{2}{3} A_\mu^a A_\nu^b A_\rho^c t^a t^b t^c dx^\mu \wedge dx^\nu \wedge dx^\rho \Big) \,.\end{equation} 
Next using using the the standard form of wedge product
\begin{equation}
d^n x = \frac{1}{n!} \epsilon_{\mu_1 \mu_2 \cdots \mu_n} dx^{\mu_1}\wedge dx^{\mu_2} \wedge\cdots \wedge dx^{\mu_n} \,.
\end{equation}
Then we use the fact that 
\begin{equation}
\epsilon^{\mu_1 \mu_2 \cdots \mu_n}\epsilon_{\mu_1 \mu_2 \cdots \mu_n} = n! \,.
\end{equation}
Therefore
\begin{equation}
\epsilon^{\mu_1 \mu_2 \cdots \mu_n} d^n x =dx^{\mu_1}\wedge dx^{\mu_2} \wedge\cdots \wedge dx^{\mu_n} \,.  
\end{equation}
Therefore, \ref{eq:CS1} simplifies into
\begin{equation}
S_{\mathrm{CS}} = \kappa \int_M  d^3 x\, \epsilon^{\mu\nu\rho} \mathrm{Tr} \Big(
 A_\mu^a \partial_\nu A_\rho^b t^a t^b   + \frac{2}{3} A_\mu^a A_\nu^b A_\rho^c t^a t^b t^c  \Big)
\end{equation}
Next, we use the trace identity,
\begin{equation}
\mathrm{Tr}(t^a t^b) = \frac{1}{2}\delta^{ab} \quad \text{and}\quad \mathrm{Tr}(t^a t^b t^c) = \frac{1}{4}(d^{abc}+if^{abc}).
\end{equation}
Therefore finally we get,
\begin{equation}
S_{\mathrm{CS}} = \frac{\kappa}{2} \int_M  d^3 x\, \epsilon^{\mu\nu\rho}  \Big(
 A_\mu^a \partial_\nu A_\rho^a + \frac{1}{3}if^{abc}   A_\mu^a A_\nu^b A_\rho^c \Big) \,.
\end{equation}
where $d^{abc}$ does not contribute as it is a totally symmetric tensor. For the first term we have already shown that it is dual invariant, the remain task to show is the $AAA$ term is dual invariant. First notice that
\begin{equation}
\frac{1}{3}i\epsilon^{\mu\nu\rho}f^{abc}A_\mu^a A_\nu^b A_\rho^c =\frac{1}{3}i(-\epsilon^{\rho\nu\mu})(-f^{cba}) A_\rho^c A_\nu^b A_\mu^a =\frac{1}{3}i\epsilon^{\rho\nu\mu} f^{cba} A_\rho^c A_\nu^b A_\mu^a
\end{equation}
Thus the last term is left-right dual invariant. 

The dual invariant action is not unique to the Chern-Simons 3-form. It can involve more derivatives, for example $adada$, which is also a left-right dual invariant. The follow action is dual invariant:
\begin{equation}
S=\int_M d^5 x \, \epsilon^{\mu\nu\rho\alpha\beta} A_{\mu} \partial_{\nu} A_{\rho} \partial_{\alpha} A_{\beta} \,.
\end{equation} 
The sum has a total of $5!=120$ terms. If we analyse it in detail, we can write the action explicitly as follow:
\begin{equation}
\begin{aligned}
S &= \int_M d^5 x \, \Big( \sum_{\mu\nu\alpha\beta} \epsilon^{\mu\nu 0 \alpha \beta} A_{\mu} \partial_{\nu} A_{0} \partial_{\alpha} A_{\beta} + \sum_{\mu\nu\alpha\beta} \epsilon^{\mu\nu 1 \alpha \beta} A_{\mu} \partial_{\nu} A_{1} \partial_{\alpha} A_{\beta}  + \sum_{\mu\nu\alpha\beta} \epsilon^{\mu\nu 2 \alpha \beta} A_{\mu} \partial_{\nu} A_{2} \partial_{\alpha} A_{\beta}\\
& \quad\quad\quad \quad\quad\quad +\sum_{\mu\nu\alpha\beta} \epsilon^{\mu\nu 3 \alpha \beta} A_{\mu} \partial_{\nu} A_{3} \partial_{\alpha} A_{\beta} + \sum_{\mu\nu\alpha\beta} \epsilon^{\mu\nu 4 \alpha \beta} A_{\mu} \partial_{\nu} A_{4} \partial_{\alpha} A_{\beta} \Big) \,.
\end{aligned}
\end{equation}
Take the first term as an example,
\begin{equation}
\sum_{\mu\nu\alpha\beta} \epsilon^{\mu\nu 0 \alpha \beta} A_{\mu} \partial_{\nu} A_{0} \partial_{\alpha} A_{\beta} \,.
\end{equation}
There will be $5! /5 = 24$ terms, corresponding to $24/2 = 12 $ pairs. We will list our the 12 pairs as an example for the $\rho = 0$ case, it will be similar for the remaining $\rho = 1 ,2 ,3 ,4$ case. 
\begin{table}[H]
\begin{center}
\begin{tabular}{ c | c  }
\hline
\hline
$L-$perspective & $R-$perspective	\\
\hline
\hline	
$\epsilon^{12034}$ & $\epsilon^{43021}$  \\
$\epsilon^{21034}$ & $\epsilon^{43012}$  \\
$\epsilon^{12043}$ & $\epsilon^{34021}$  \\
$\epsilon^{21043}$ & $\epsilon^{34012}$  \\
\hline
$\epsilon^{13024}$ & $\epsilon^{42031}$  \\
$\epsilon^{31024}$ & $\epsilon^{42013}$  \\
$\epsilon^{13042}$ & $\epsilon^{24031}$  \\
$\epsilon^{31042}$ & $\epsilon^{24013}$  \\
\hline  
$\epsilon^{14023}$ & $\epsilon^{32041}$  \\
$\epsilon^{41023}$ & $\epsilon^{32014}$  \\
$\epsilon^{14032}$ & $\epsilon^{23041}$  \\
$\epsilon^{41032}$ & $\epsilon^{23014}$  \\
\hline  
\end{tabular}
\end{center}
\end{table}
Since noting that
\begin{equation}
\epsilon^{\mu\nu\rho \alpha \beta} = (-1)^4 (-1)^3 (-1)^2 (-1)^1 \epsilon^{\beta\alpha\rho\nu\mu} = \epsilon^{\beta\alpha\rho\nu\mu}\,,
\end{equation}
so the terms in left perspective and right perspective are also +1. For example, we will have
\begin{equation}
\epsilon^{12034} A_{1} \partial_2 A_0 \partial_3 A_4 + \epsilon^{43021}A_{4} \partial_3 A_0 \partial_2 A_1 = A_{1} \partial_2 A_0 \partial_3 A_4 + A_{4} \partial_3 A_0 \partial_2 A_1 
\end{equation}
which is obviously an left-right dual invariant. For example,
\begin{equation}
\epsilon^{21034} A_{2} \partial_1 A_0 \partial_3 A_4 + \epsilon^{43012}A_{4} \partial_3 A_0 \partial_2 A_1 = -A_{2} \partial_1 A_0 \partial_3 A_4 - A_{4} \partial_3 A_0 \partial_1 A_2 
\end{equation}
is also dual invariant.

We can in fact prove that this action is dual invariant using integration by parts. 
\begin{equation}
\begin{aligned}
(S|L) &=\int_M d^5 x \, \epsilon^{\mu\nu\rho\alpha\beta} A_{\mu} \partial_{\nu} A_{\rho} \partial_{\alpha} A_{\beta} \\
&= \mathrm{B.C.} - \int_M d^5 x \,\epsilon^{\mu\nu\rho \alpha\beta} A_\beta\partial_\alpha (A_\mu \partial_\nu A_\rho) \\
&= -\int_M d^5 x \,\epsilon^{\mu\nu\rho \alpha\beta} A_\beta (\partial_\alpha A_\mu \partial_\nu A_\rho + A_\mu \partial_\alpha \partial_\nu A_\rho ) \\
&= -\int_M d^5 x \,\epsilon^{\mu\nu\rho \alpha\beta} A_\beta \partial_\alpha A_\mu \partial_\nu A_\rho \\
&=-\int_M d^5 x \, (-1)^4 \epsilon^{\beta\mu\nu\rho \alpha} A_\beta \partial_\alpha A_\mu \partial_\nu A_\rho \\
&= -\int_M d^5 x \, (-1)^4 \epsilon^{\beta\rho\nu\mu \alpha} A_\beta \partial_\alpha A_\rho \partial_\nu A_\mu \\
&= -\int_M d^5 x \, (-1)^4 (-1)^3 \epsilon^{\beta\alpha\rho\nu\mu } A_\beta \partial_\alpha A_\rho \partial_\nu A_\mu \\
&=\int_M d^5 x \,  \epsilon^{\beta\alpha\rho\nu\mu } A_\beta \partial_\alpha A_\rho \partial_\nu A_\mu \\
&= (S|R) \,,
\end{aligned}
\end{equation}
where we have dropped the boundary term.

Finally, we would like to investigate the  action with chiral anomaly, which is given by the following:
\begin{equation}
S_{A} =  \frac{\alpha}{4\pi}\int_M d^4 x \, \theta \epsilon^{\mu\nu\alpha\beta}F_{\mu\nu} F_{\alpha\beta} = \frac{\alpha\theta}{\pi} \int_M d^4 x \,  \pmb{\mathrm{E}} \cdot \pmb{\mathrm{B}} \,,
\end{equation}
where $\alpha$ is the fine-structure constant, $\theta \in [0,2\pi]$ is a constant. This action must be dual-invariant because 
\begin{equation}
(\pmb{\mathrm{E}} \cdot \pmb{\mathrm{B}} | L ) \equiv (\pmb{\mathrm{B}} \cdot\pmb{\mathrm{E}} |R ) \,,
\end{equation}
where $\pmb{\mathrm{E}} \cdot \pmb{\mathrm{B}}= \pmb{\mathrm{B}} \cdot\pmb{\mathrm{E}} $. At the first sight, the term $\epsilon^{\mu\nu\alpha\beta}F_{\mu\nu} F_{\alpha\beta}$ is non-dual invariant, but in fact we can prove that it is a dual invariant using divergence theorem. First, we rewrite the action as follow:
\begin{equation}
\begin{aligned}
S_{A} &=  \frac{\alpha}{4\pi}\int_M d^4 x \, \theta\epsilon^{\mu\nu\alpha\beta}F_{\mu\nu} F_{\alpha\beta} \\
&= \frac{\alpha\theta}{4\pi}\int_M d^4 x \,\epsilon^{\mu\nu\alpha\beta}  (\partial_{\mu}A_{\nu} - \partial_{\nu} A_\mu)(\partial_{\rho} A_{\sigma} -\partial_\sigma A_\rho ) \\
&= \frac{\alpha\theta}{4\pi}\int_M d^4 x \, \epsilon^{\mu\nu\alpha\beta} (2 \partial_{\mu}A_{\nu} )(2 \partial_{\rho} A_\sigma ) \\
&= \frac{\alpha \theta}{\pi} \int_M d^4 x \, \partial_{\mu}(\epsilon^{\mu\nu\rho\sigma}A_\nu \partial_{\rho} A_\sigma ) \\
&= \frac{\alpha \theta}{\pi} \int_{\partial M} d^3 x \, n_{\mu} \epsilon^{\mu\nu\rho\sigma} A_{\nu} \partial_{\rho} A_{\sigma} \\
&= \frac{\alpha \theta}{\pi} \int_{\partial M} d^3 x \, (n_0 \epsilon^{0\nu\rho\sigma} A_\nu \partial_\rho A_\sigma + n_1 \epsilon^{1\nu\rho\sigma} A_\nu \partial_\rho A_\sigma +n_2 \epsilon^{2\nu\rho\sigma} A_\nu \partial_\rho A_\sigma + n_3 \epsilon^{3\nu\rho\sigma} A_\nu \partial_\rho A_\sigma ) \,,
\end{aligned}
\end{equation}  
where $n_{\mu}$ is the unit vector normal to the 3D volume surface. Let's investigate the first term. We have for the left perspective,
\begin{equation}
\begin{aligned}
&\quad\frac{\alpha \theta}{\pi} \int_{\partial M} d^3 x \,\Big( ( n_0 \epsilon^{0123} A_1 \partial_2 A_3 + n_0 \epsilon^{0321} A_3 \partial_2 A_1) + (n_0 \epsilon^{0312} A_3 \partial_1 A_2 + n_0 \epsilon^{0213} A_2 \partial_1 A_3) + \\
&\quad\quad\quad\quad\quad\quad\quad\quad (n_0 \epsilon^{0231} A_2 \partial_3 A_1 + n_0 \epsilon^{0132} A_1 \partial_3 A_2) \Big)\\
&=\frac{\alpha \theta}{\pi} \int_{\partial M} d^3 x \,\Big( ( n_0  A_1 \partial_2 A_3 - n_0  A_3 \partial_2 A_1) + (n_0  A_3 \partial_1 A_2 - n_0  A_2 \partial_1 A_3) + \\
&\quad\quad\quad\quad\quad\quad\quad\quad (n_0  A_2 \partial_3 A_1 - n_0  A_1 \partial_3 A_2) \Big) \\
&= \frac{\alpha \theta}{\pi} \int_{\partial M} d^3 x \, n_0 \Big( (  A_1 \partial_2 A_3 -  A_3 \partial_2 A_1) + ( A_3 \partial_1 A_2 -  A_2 \partial_1 A_3) + \\
&\quad\quad\quad\quad\quad\quad\quad\quad ( A_2 \partial_3 A_1 -  A_1 \partial_3 A_2) \Big) \\
&= \frac{\alpha \theta}{\pi} \int_{\partial M} d^3 x \,\Big( (  A_1 \partial_2 A_3 -  A_3 \partial_2 A_1) + ( A_3 \partial_1 A_2 -  A_2 \partial_1 A_3) \\
&\quad\quad\quad\quad\quad\quad\quad\quad ( A_2 \partial_3 A_1 -  A_1 \partial_3 A_2) \Big) n_0 \,, \\
\end{aligned}
\end{equation}
which is the same as looking from the right perspective. The same idea goes to the second, third and fourth term. Therefore, the chiral anomaly action is left-right dual invariant, and it is topological.

\section{New exotic matter state-Dualiton}
In this section, we will study the the duality wave oscillation, its promotion to duality field, and how is second quantization give rise to the quasi particle which we call the dualiton. Dualiton is a new quasi particle which arises from the theory of duality, and is regarded as the quantization of the states with $\mathbb{Z}_2 \times \mathbb{Z}_2$ symmetry. 

First, we will use the mathematics of 4-box tableaux we constructed in section 3. Let $U=\{ u \}$ and $U^*=\{ u^* \}$. We have the following
\begin{figure}[H] 
\centering
\includegraphics[trim=0cm 0cm 0cm 0cm, clip, scale=0.6]{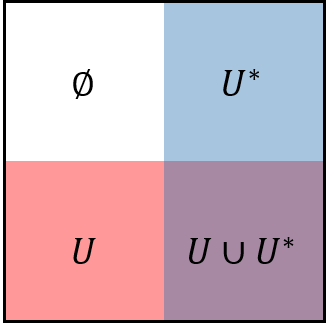}
\caption[]
{\label{fig:4tab}}
\end{figure}
The fullness is considered as $U\cup U^*$. By taking the negation,
\begin{equation}
*(U\cup U^*) = *U \cap *U^* = *U \cap U = U^* \cap U= \emptyset \,, 
\end{equation}
where in the second step we have used the de-Morgan's theorem. For example, let $u =\text{creation}$, $u^* = \text{annihilation}$, the fullness is creation or annihilation. Not creation or annihilation is nor creation and nor annihilation, which we will prove here is indeed the empty set. Here we have also excluded the middle of $\{0\}$ as we do not include it in our definition, so the exclusion of middle is automatically satisfied. 

First we would like investigate some their properties. We have respectively the full set and the null set as
\begin{equation}
W = U \cup U^*  \quad \text{and} \quad \emptyset = U \cap U^* \,.
\end{equation}
Explicitly,
\begin{equation}
W = \{ u , u^*  \}   \quad \text{and} \quad  \emptyset = \{ \} \,.
\end{equation}
The respective cardinality is $|W| =2$ and $|\emptyset| =0$. 
The full set and the empty set are dual to each other,
\begin{equation} \label{eq:dualsets}
\emptyset = *W  \quad \text{and} \quad W= *\emptyset \,,
\end{equation}
which is shown by the de-Morgan's law. In a detailed look, this can be manifested as
\begin{equation}
(1)\emptyset = (*)W  \quad \text{and} \quad (1)W= (*)\emptyset \,,
\end{equation}
where $1$ is the identity operator element while $*$ is the dual operator element. We can take $u=1$ and $u^* = *$ and the abstract perspective $S_k = W$ and $S_k^\star = \emptyset $, then \ref{eq:dualsets} means to be the standard form,
\begin{equation}
( 1| \emptyset  ) \equiv (*|W )\quad \text{and} \quad ( 1|W  ) \equiv (*|\emptyset ) \,.
\end{equation}
In a compact form, this is just the standard relation we had before,
\begin{equation} \label{eq:compactform}
(1|0) \equiv (0|1) \quad \text{and} \quad (1|1) \equiv (0|0) \,.
\end{equation}

Similarly, for the half sets, 
\begin{equation}
U= \{ u\}  \quad \text{and} \quad U^* = \{ u^*\} \,.
\end{equation}
The two half sets are dual to each other,
\begin{equation}
U^* = *U \quad \text{and} \quad U = *U^* \,.
\end{equation}
Explicitly this means
\begin{equation}
( 1| U^*  ) \equiv (*|U )\quad \text{and} \quad ( 1|U  ) \equiv (*|U^* ) \,.
\end{equation}
and the compact form shares the one in \ref{eq:compactform}. Element-wise, figure \ref{fig:4tab} can be represented as follows,
\begin{figure}[H]
\centering
\includegraphics[trim=0cm 0cm 0cm 0cm, clip, scale=0.8]{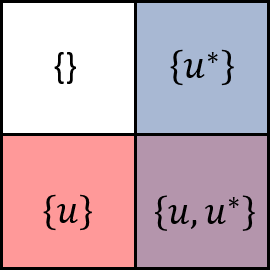}
\caption[]
{\label{fig:4elements}}
\end{figure}

\subsection{The canonical 4-duality structure}
Next, we define the 4-canonical structure as follows ,
\begin{equation}
\psi = \sum_{i=0,1}^\oplus \sum_{j=0,1}^\oplus (\hat{P}_i \oplus \hat{P}_j) (|u\rangle \oplus |u^*\rangle ) \,,
\end{equation}
where $\hat{P}_0 = \hat{e}$ and  $\hat{P}_1 = \star\hat{e}$. Explicitly, the sum is
\begin{equation} \label{eq:can}
\psi = (\hat{e}|u\rangle\oplus \hat{e}|u^*\rangle ) \oplus (\hat{e}|u\rangle \oplus \star \hat{e}|u^*\rangle) \oplus (\star\hat{e}|u\rangle \oplus e|u^*\rangle) \oplus (\star \hat{e} |u\rangle \oplus \star \hat{e}|u^*\rangle)\,.
\end{equation}
First consider the first statement of negation: ``nor created and nor annihilated". In terms of state, we have $|u\rangle = |\text{created}\rangle$ and its dual state $|u^*\rangle = |\text{annihilated}\rangle$. The negation nor operator is defined as $\hat{P}_1 = \star\hat{e} = \hat{-}$. So the statement of ``nor created and nor annihilated" is mathematically represented as
\begin{equation}
\begin{aligned}
&\quad\star\hat{e}( |\text{created}\rangle \oplus |\text{annihilated}\rangle   )\\ 
 &=(\star\hat{e}\oplus\star\hat{e})(|\text{created}\rangle \oplus |\text{annihilated}\rangle ) \\
 &=  \star\hat{e} |\text{created}\rangle \oplus \star\hat{e} |\text{annihilated}\rangle \\
 &=\hat{-}|\text{created}\rangle \oplus \hat{-} |\text{annihilated}\rangle\,.
\end{aligned}
\end{equation}
However, there also exists a state with the dual of it. The normal operator is defined as $\hat{P}_0 = \star\hat{e}\star\hat{e} = \hat{e}= \hat{+}$,
\begin{equation}
\begin{aligned}
&\quad \star\hat{e}\big(\star\hat{e}( |\text{created}\rangle \oplus |\text{annihilated}\rangle   )\big) \\
&=  (\star\hat{e}\oplus\star\hat{e})(\star\hat{e}\oplus\star\hat{e})(|\text{created}\rangle \oplus |\text{annihilated}\rangle ) \\
&=\hat{e} |\text{created}\rangle \oplus\hat{e} |\text{annihilated}\rangle \\
&=\hat{+}|\text{created}\rangle \oplus \hat{+} |\text{annihilated}\rangle \,. \\
\end{aligned}
\end{equation}
This is the statement of ``being created and being annihilated", which is the negation statement of ``nor created and nor annihilated". 

Finally we have two more states, ``being created and nor annihilated" and its negation ``nor created and being annihilated", which are mathematically, respectively,
\begin{equation}
\hat{+}|\text{created}\rangle \oplus  \hat{-}|\text{annihilated}\rangle \quad \text{and} \quad \hat{-}|\text{created}\rangle \oplus  \hat{+}|\text{annihilated}\rangle
\end{equation}
Therefore, the full state consists of two dual pairs as follow,
\begin{equation}
\begin{aligned}
\psi &= (\hat{+}|\text{created}\rangle \oplus \hat{+} |\text{annihilated}\rangle) \oplus (\hat{-}|\text{created}\rangle \oplus \hat{-} |\text{annihilated}\rangle) \\
& \quad\oplus (\hat{+}|\text{created}\rangle \oplus \hat{-} |\text{annihilated}\rangle) \oplus (\hat{-}|\text{created}\rangle \oplus \hat{+} |\text{annihilated}\rangle) \\
\end{aligned}
\end{equation}
Or neatly we write symbolically as a canonical 4-duality structure,
\begin{equation} \label{eq:neat}
\psi= (\hat{+}|u\rangle \oplus \hat{+} |u^*\rangle) \oplus (\hat{-}|u\rangle \oplus \hat{-} |u^*\rangle)\oplus (\hat{+}|u\rangle \oplus \hat{-} |u^*\rangle) \oplus (\hat{-}|u\rangle \oplus \hat{+} |u^*\rangle) ,
\end{equation}
In terms of 4-tableau, we have
\begin{figure}[H] 
\centering
\includegraphics[trim=0cm 0cm 0cm 0cm, clip, scale=0.6]{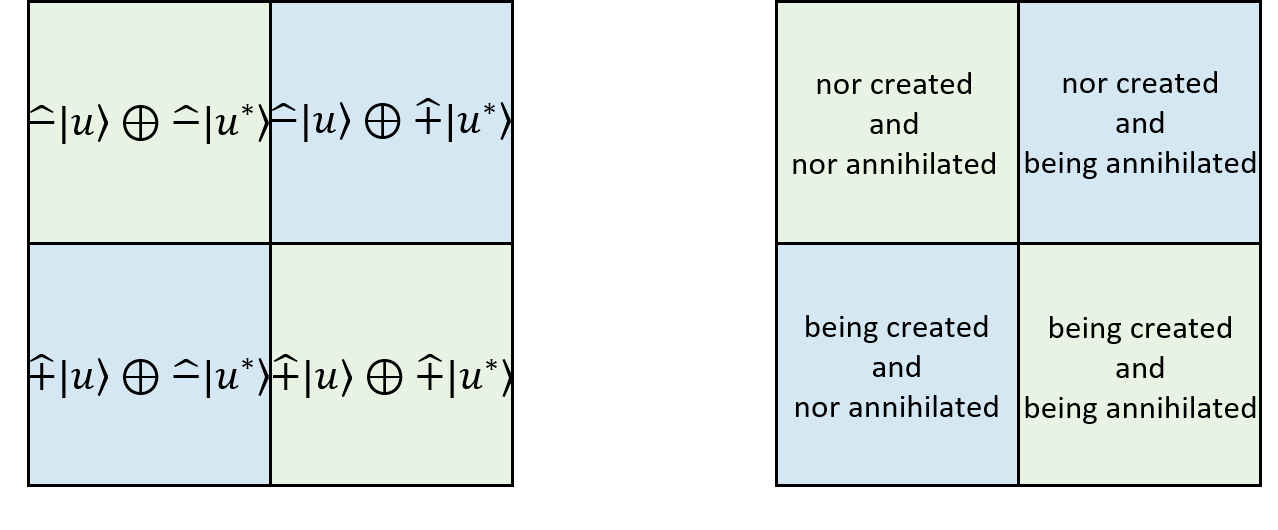}
\caption[]
{\label{fig:4rep}}
\end{figure}

Therefore, $\hat{-}|u\rangle \oplus \hat{-}|u^*\rangle$ state corresponds to the Null state, which is one of the four states of the canonical structure. On the other hand, its dual is the illustration of the $\hat{+}|u\rangle \oplus \hat{+}|u^*\rangle$ state. The $(\hat{+}|\text{created}\rangle \oplus \hat{+} |\text{annihilated}\rangle)$ state means that appearance is created and annihilated,  Then $(\hat{-}|\text{created}\rangle \oplus \hat{-} |\text{annihilated}\rangle)$ is that state of no appearance. 
We remain to have two duals states left, these are the half states. 

Now, we evaluate the operations. First we can see that
\begin{equation}
\hat{+} |i \rangle = |i\rangle \quad \text{and} \quad \hat{-} |i \rangle = \bar{\pmb{0}}_i = \bar{\pmb{0}} \,.
\end{equation}
Here, $\bar{\pmb{0}}$ is defined as a empty vector which does not contain any entries, it is a hollow column vector with zero dimension, which we call the vacuum state,
\begin{equation}
\bar{\pmb{0}} = 
\begin{pmatrix}
&\\
&\\
\end{pmatrix} \,,
\end{equation}
so that for all $|v\rangle \in {V}$, $\bar{\pmb{0}}\oplus |v\rangle = |v\rangle\oplus \bar{\pmb{0}} $ and $\bar{\pmb{0}}\oplus \bar{\pmb{0}} = \bar{\pmb{0}} $.

This is because $\hat{+}$ is the being-operator, this gives the state itself; while $\hat{-}$ is the nor-operator, the action on the state means the state does not exist, so this gives the empty vector.  Therefore, equation \ref{eq:neat} is evaluated to be
\begin{equation}
\begin{aligned}
\psi&= (|u\rangle \oplus  |u^*\rangle) \oplus (\bar{\pmb{0}}_{u} \oplus \bar{\pmb{0}}_{u^*}) \oplus (|u\rangle \oplus \bar{\pmb{0}}_{u^*}) \oplus (\bar{\pmb{0}}_{u} \oplus |u^*\rangle) \\
&=(|u\rangle \oplus  |u^*\rangle) \oplus \bar{\pmb{0}}  \oplus |u\rangle \oplus    |u^*\rangle) \,.
\end{aligned}
\end{equation}

This gives the result of figure \ref{fig:tablele}. Therefore in terms of direct sum vector space representation, this is figure \ref{fig:tablele},
\begin{figure}[H]
\centering
\includegraphics[trim=0cm 0cm 0cm 0cm, clip, scale=0.8]{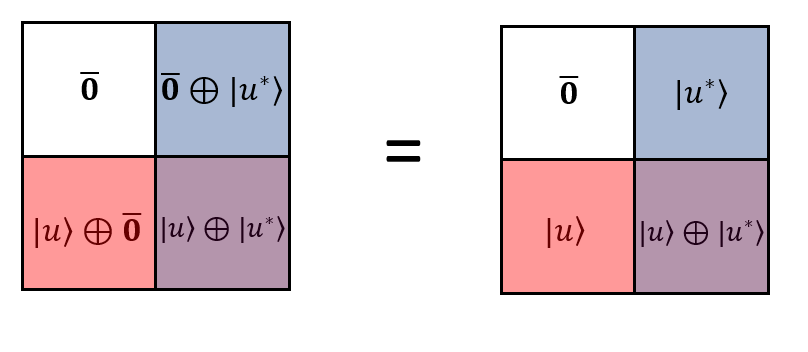}
\caption[]
{} \label{fig:tab8}
\end{figure}
Therefore we have the following results
\begin{equation}
\begin{aligned}
&\hat{-}|\text{created}\rangle \oplus \hat{-} |\text{annihilated}\rangle =\bar{\pmb{0}} \,. \\
&\hat{+}|\text{created}\rangle \oplus \hat{+} |\text{annihilated}\rangle =|\text{created}\rangle \oplus |\text{annihilated}\rangle \,. \\
&\hat{+}|\text{created}\rangle \oplus \hat{-} |\text{annihilated}\rangle =|\text{created}\rangle \,. \\
&\hat{-}|\text{created}\rangle \oplus \hat{+} |\text{annihilated}\rangle =|\text{annihilated}\rangle \,.
\end{aligned}
\end{equation}
Explicitly in terms of diagram,
\begin{figure}[H]
\centering
\includegraphics[trim=0cm 0cm 0cm 0cm, clip, scale=0.6]{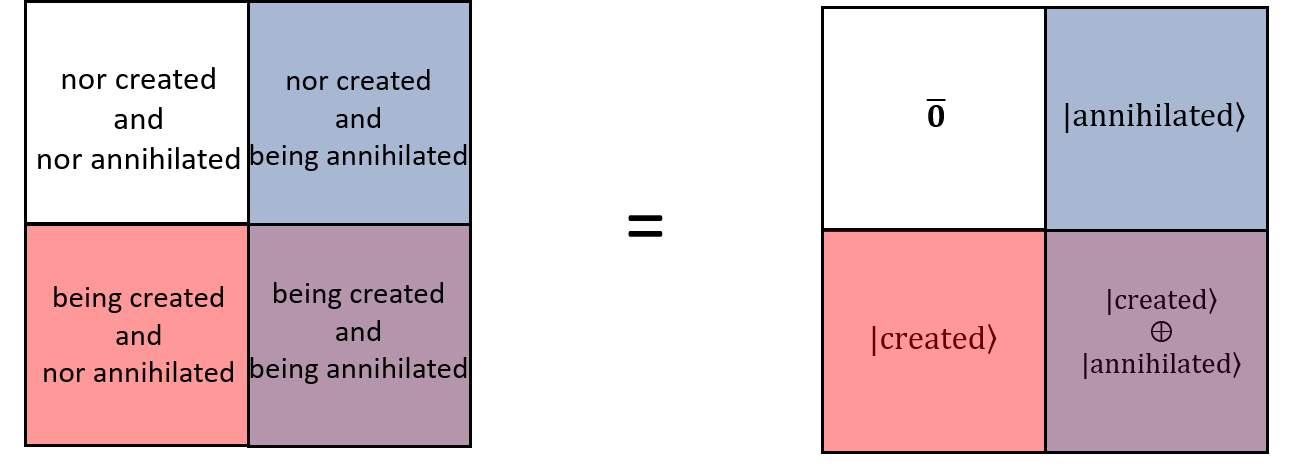}
\caption[]
{} \label{fig:finaltab}
\end{figure}

In terms of set, we have \ref{fig:4elements}. We find the structure of the set is identical to the case of direct-sum space. Therefore one has the isomorphism,
\begin{figure}[H]
\centering
\includegraphics[trim=0cm 0cm 0cm 0cm, clip, scale=0.8]{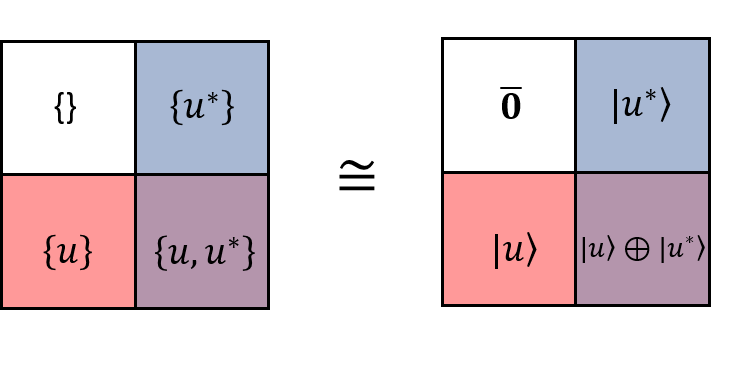}
\caption[]
{} \label{fig:isomorphism}
\end{figure}
The operation isomorphism for the two cases hold as follow,
\begin{equation}
\begin{aligned}
& U\cup \emptyset = \emptyset \cup U = U \quad,\quad |v\rangle\oplus \bar{\pmb{0}}= \bar{\pmb{0}}\oplus |v\rangle  = |v\rangle \,. \\
&U^*\cup \emptyset = \emptyset \cup U^* = U^* \quad,\quad  |v^*\rangle\oplus \bar{\pmb{0}} = \bar{\pmb{0}}\oplus |v^*\rangle  =  |v^*\rangle \,.
\end{aligned}
\end{equation}
In addition, the power set of the full set $W$ is the discrete topology of $W$,
\begin{equation}
\tau = \{ \emptyset , \{u\} , \{u^* \} , \{u, u^* \}\} \,.
\end{equation}
This is isomorphic to the binary number set of 
\begin{equation}
\tau=\{00, 10 ,01, 11 \} \,.
\end{equation}
This topology set is dual invariant, i.e. it is the same no matter you look from the left perspective or right perspective, i.e.
\begin{equation}
(\tau |L ) \equiv (\tau |R) \,.
\end{equation}
Therefore, we arrive at a very important theorem that topology implies dual invariance.

In conclusion, the ``nor created and nor annihilated" state means empty space (set) mathematically. Since the empty set does not contain any elements, this implies such object has no appearance. The dual perspective is the view of the full set (space). It contains both the appearance of creation and the appearance of annihilation. 

\subsubsection{Cardinality representation}
Finally, we would like to work on the cardinality representation of the above representation. If there exists a $u$ or $|u^\star\rangle$, we label as 1; while if it is $\pmb{0}$ then we have 0. In other words we get, 
\begin{figure}[H]
\centering
\includegraphics[trim=0cm 0cm 0cm 0cm, clip, scale=0.8]{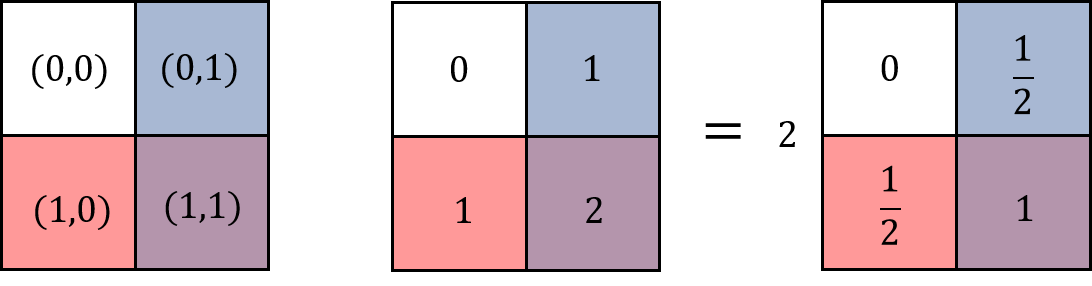}
\caption[]
{} \label{fig:cardin}
\end{figure}
In the first diagram, we can actually see that the cardinality representation is the representation of the $\mathbb{Z}_2 \times \mathbb{Z}_2$ group. In the right diagram, we calculate the cardinality of the set for each quadrant. In the last diagram, the $1/2$ representation denotes the red and blue quadrant are half quadrants , while the 0 representation is the empty quadrant and the 1 representation is the full quadrant. 
Next, we would like to construct the cardinality representation matrix by directing summing the cardinality elements of the state,
\begin{figure}[H]
\centering
\includegraphics[trim=0cm 0cm 0cm 0cm, clip, scale=0.8]{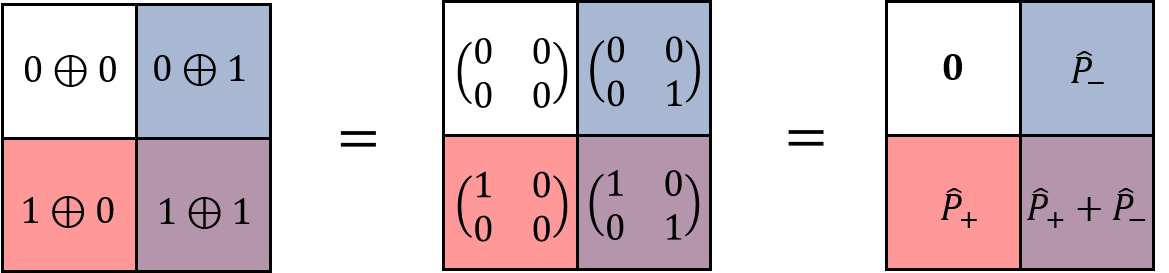}
\caption[]
{} \label{fig:cardmatrix}
\end{figure}
Now define the dual doublet as the full state
\begin{equation}
\phi = \begin{pmatrix}
|u\rangle \\
|u^*\rangle
\end{pmatrix}
=|u\rangle  \oplus |u^*\rangle \,.
\end{equation}
Each representation matrix can be defined by
\begin{equation}
\pmb{0} =\begin{pmatrix}
0 & 0 \\
0 & 0
\end{pmatrix} \quad , \quad
\pmb{1} =\begin{pmatrix}
1 & 0 \\
0 & 1
\end{pmatrix} \quad , \quad
\hat{P}_+ =\begin{pmatrix}
1 & 0 \\
0 & 0
\end{pmatrix} \quad , \quad
\hat{P}_- =\begin{pmatrix}
0 & 0 \\
0 & 1
\end{pmatrix}\,.
\end{equation}
where
$\hat{P}_+$  and $\hat{P}_-$ are projection matrices constructed by the elements of the cardinality, which satisfies
\begin{equation}
\hat{P}_+ + \hat{P}_-  =\pmb{1} \quad , \quad \hat{P}_+  \hat{P}_- = \hat{P}_-  \hat{P}_+ =\pmb{0} \,.
\end{equation}
Define also
\begin{equation}
\underline{\pmb{0}} =\begin{pmatrix}
0  \\
0 
\end{pmatrix} = 0\oplus 0 \quad , \quad
\phi_+  =\begin{pmatrix}
|u\rangle \\
0 
\end{pmatrix} = |u\rangle \oplus 0  \quad , \quad
\phi_-  =\begin{pmatrix}
0 \\
|u^*\rangle 
\end{pmatrix} = 0\oplus |u^*\rangle   \,.
\end{equation}
Then we have
\begin{equation} \label{eq:4eq}
\pmb{1}\phi=\phi \quad,\quad \pmb{0}\phi=\underline{\pmb{0}}  \quad,\quad \hat{P}_+ \phi = \phi_+ \quad , \quad \hat{P}_- \phi = \phi_- \,.
\end{equation}
Also we have
\begin{equation}
\hat{P}_+ \phi_- = \underline{\pmb{0}} \quad , \quad \hat{P}_- \phi_+ = \underline{\pmb{0}} \,.
\end{equation}
Since $*\phi_+ = \phi_-$ and $*\phi_- = \phi_+$ ; $*\phi = \underline{\pmb{0}}$ and $*\underline{\pmb{0}} = \phi$, it follows from equation \ref{eq:4eq} that for operators, we have
\begin{equation}
*\hat{P}_+ = \hat{P}_- \,\,,\,\, *\hat{P}_- = \hat{P}_+ \quad ; \quad *\pmb{1} = \pmb{0} \,\,,\,\, *\pmb{0} = \pmb{1} \,.
\end{equation}
We term the second dual equivalence as \emph{associated duality} given in \ref{eq:dual10}, for which such duality is not given by normal matrix operation that we introduced before. In addition, for $\hat{P}_+$ and $\hat{P}_-$, the dual operator can be regarded as off-diagonal transpose $S$, such that $\hat{P}_+ = (\hat{P}_+)^S $. We can also check that the dual operators are orthogonal.
\begin{equation}
\hat{P}_+ \cdot \hat{P}_-  =\mathrm{Tr}(\hat{P}_+^T \hat{P}_-  ) = 0 \quad , \quad \hat{P}_- \cdot \hat{P}_+ = \mathrm{Tr}(\hat{P}_-^T \hat{P}_+  ) = 0 \,,
\end{equation}
and
\begin{equation}
\pmb{0} \cdot \pmb{1}  =\mathrm{Tr}(\pmb{0}^T \pmb{1}  ) = 0 \quad , \quad \pmb{1} \cdot \pmb{0} = \mathrm{Tr}(\pmb{1}^T \pmb{0}  ) = 0 \,.
\end{equation}
The cardinality is given by the norm of the state vector,
\begin{equation}
\begin{aligned}
&\text{card}(\underline{\pmb{0}}) = \langle \phi | \phi\rangle =\mathrm{Tr}(0\oplus 0)  =0 \,, \\
&\text{card}(\phi_+) = \langle \phi_+ | \phi_+ \rangle =\mathrm{Tr}(1\oplus 0)  =1 \,,\\
&\text{card}(\phi_-) = \langle \phi_- | \phi_- \rangle =\mathrm{Tr}(0\oplus 1)  =1 \,,\\
&\text{card}(\phi) = \langle \phi | \phi \rangle =\mathrm{Tr}(1\oplus 1)  =2 \,.
\end{aligned}
\end{equation}
There is also one important property, note that
\begin{equation}
(\hat{P}_+ + \hat{P}_- )^2 = \hat{P}_+^2 + \hat{P}_-^2 \,,
\end{equation}
which takes the form of $(a+b)^2 = a^2 + b^2 $, this is prohibited when $a,b$ are real numbers (except zero), but allowed when $a,b$ are matrices. Similarly, we have
\begin{equation}
(\hat{P}_+ + \hat{P}_- )^2 =  (\hat{P}_+ - \hat{P}_- )^2  
\end{equation}
and this only holds for the matrix case.

\subsection{The Duality Wave}
With all the studies above, we now can proceed to study the duality wave given by the $\mathbb{Z}_2 \times \mathbb{Z}_2 $ representation.
For convenience take 
\begin{equation}
|u\rangle =|0\rangle = |\text{creation}\rangle=\begin{pmatrix}
1 \\ 
0
\end{pmatrix}
\quad , \quad
|u^*\rangle = |1\rangle = |\text{annihilation}\rangle =\begin{pmatrix}
0 \\ 
1
\end{pmatrix} \,.
\end{equation}
In physical sense, the duality system, can also infer as system with two spins which are dual to each other, the up-spin and down spin. By isomorphism, we take an ansatz that we can construct the quantum states as follow,
\begin{equation} \label{eq:canonicalduality}
\bar{\pmb{0}} \rightarrow |0\oplus 0\rangle \quad,\quad |u\rangle \oplus  \bar{\pmb{0}} \rightarrow |1 \oplus 0\rangle  \quad,\quad \bar{\pmb{0}}\oplus |u^*\rangle \rightarrow |0 \oplus 1\rangle\quad,\quad |u\rangle \oplus |u^*\rangle \rightarrow |1 \oplus 1\rangle \,.
\end{equation} 
where $|a\oplus b\rangle = |a\rangle \oplus |b\rangle$ for $a,b = 0,1$. The general form takes the following
\begin{equation}
\hat{P}|0\rangle \oplus \hat{Q}|1\rangle = |a\rangle \oplus |b\rangle \,,
\end{equation}
where $\hat{P} , \hat{Q} $ are $\hat{+}$ or $\hat{-}$.

 In the original representation
\begin{equation}
\hat{-} = \pmb{0} = \begin{pmatrix}
0 & 0 \\
0 & 0
\end{pmatrix} \quad , \quad
\hat{+} = \pmb{1} = \begin{pmatrix}
1 & 0 \\
0 & 1
\end{pmatrix}
\end{equation}
where $\hat{-}$ and $\hat{+}$ are dual to each other as we have stated above. Then we have the following computation,
\begin{equation}
\hat{-}|0\rangle \oplus \hat{-}|1\rangle = \begin{pmatrix}
0\\
0\\
0\\
0
\end{pmatrix} \quad , \quad
\hat{+}|0\rangle \oplus \hat{+}|1\rangle = \begin{pmatrix}
0\\
1\\
1\\
0
\end{pmatrix} \,,
\end{equation}
\begin{equation}
\hat{+}|0\rangle \oplus \hat{-}|1\rangle = \begin{pmatrix}
0\\
1\\
0\\
0
\end{pmatrix} \quad ,\quad
\hat{-}|0\rangle \oplus \hat{+}|1\rangle = \begin{pmatrix}
0\\
0\\
1\\
0
\end{pmatrix} \,,
\end{equation}
We can see that $*(\hat{-}|0\rangle \oplus \hat{-}|1\rangle) = \pmb{M}(\hat{-}|0\rangle \oplus \hat{-}|1\rangle) =\hat{+}|0\rangle \oplus \hat{+}|1\rangle) $. Therefore $\hat{-}|0\rangle \oplus \hat{-}|1\rangle)$ and $\hat{+}|0\rangle \oplus \hat{+}|1\rangle)$ are dual and orthogonal to each other. Also $\hat{-}|0\rangle \oplus \hat{-}|1\rangle $ and $\hat{+}|0\rangle \oplus \hat{+}|1\rangle $ are of associated duality, and they are orthogonal. 

Now we would like to use another matrix representation of $\hat{-}$ and $\hat{+}$, so that we can obtain the canonical duality in equation \ref{eq:canonicalduality}. We take 
\begin{equation} \label{eq:rep}
\hat{-}  = \begin{pmatrix}
1 & 1 \\
0 & 0
\end{pmatrix} \quad , \quad
\hat{+}  = \begin{pmatrix}
0 & 0 \\
1 & 1
\end{pmatrix} \,.
\end{equation}
Clearly that $*\hat{-} = \pmb{M}\hat{+}$ and $*\hat{+} = \pmb{M}\hat{-}$.
\begin{equation}
\begin{pmatrix}
0 & 1 \\
1 & 0
\end{pmatrix}
\begin{pmatrix}
1 &1 \\
0 &0
\end{pmatrix} =
\begin{pmatrix}
0&0 \\
1 &1
\end{pmatrix} \quad \text{and} \quad
\begin{pmatrix}
0 & 1 \\
1 & 0
\end{pmatrix}
\begin{pmatrix}
0 &0 \\
1 &1
\end{pmatrix} =
\begin{pmatrix}
1&1 \\
0 &0
\end{pmatrix} \,.
\end{equation}
Using this canonical representation,
\begin{equation}
\hat{-}|0\rangle \oplus \hat{-}|1\rangle = \begin{pmatrix}
1\\
0\\
1\\
0
\end{pmatrix} =|0\rangle\oplus|0\rangle \quad , \quad
\hat{+}|0\rangle \oplus \hat{+}|1\rangle = \begin{pmatrix}
0\\
1\\
0\\
1
\end{pmatrix} = |1\rangle\oplus|1\rangle  \,,
\end{equation}
\begin{equation}
\hat{+}|0\rangle \oplus \hat{-}|1\rangle = \begin{pmatrix}
0\\
1\\
1\\
0
\end{pmatrix} =|0\rangle\oplus|1\rangle\quad ,\quad
\hat{-}|0\rangle \oplus \hat{+}|1\rangle = \begin{pmatrix}
1\\
0\\
0\\
1
\end{pmatrix} =|1\rangle\oplus|0\rangle \,,
\end{equation}
Now we can see that $*(\hat{-}|0\rangle \oplus \hat{-}|1\rangle)=\pmb{M}(\hat{-}|0\rangle \oplus \hat{-}|1\rangle)= (\hat{+}|0\rangle \oplus \hat{+}|1\rangle)$ is dual and orthogonal to each other $\langle 0 \oplus 0|1 \oplus 1 \rangle = \langle 0 | 1 \rangle + \langle 0 | 1 \rangle =0 $, while $\hat{+}|0\rangle \oplus \hat{-}|1\rangle$ and $\hat{-}|0\rangle \oplus \hat{+}|1\rangle$ are associated dual and orthogonal  $\langle 0 \oplus 1|1 \oplus 0 \rangle = \langle 0 | 1 \rangle + \langle 1 |0 \rangle =0 $ .

Diagramatically,
\begin{figure}[H]
\centering
\includegraphics[trim=0cm 0cm 0cm 0cm, clip, scale=0.6]{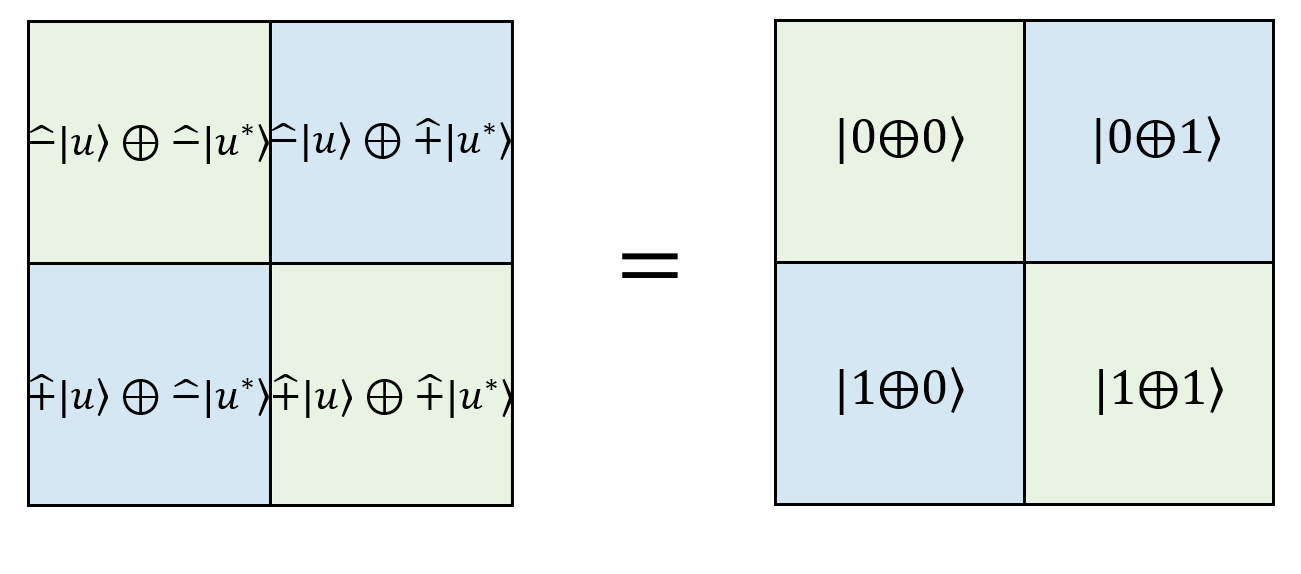}
\caption[]
{} \label{fig:4rep3}
\end{figure}

The states $\{|0\oplus0 \rangle , |0\oplus1 \rangle , |1\oplus0 \rangle , |1\oplus1 \rangle \}$ form the basis of $\mathbb{Z}_2 \times \mathbb{Z}_2$. The transformation of each of the basis is given by
\begin{figure}[H]
\centering
\includegraphics[trim=0cm 0cm 0cm 0cm, clip, scale=0.6]{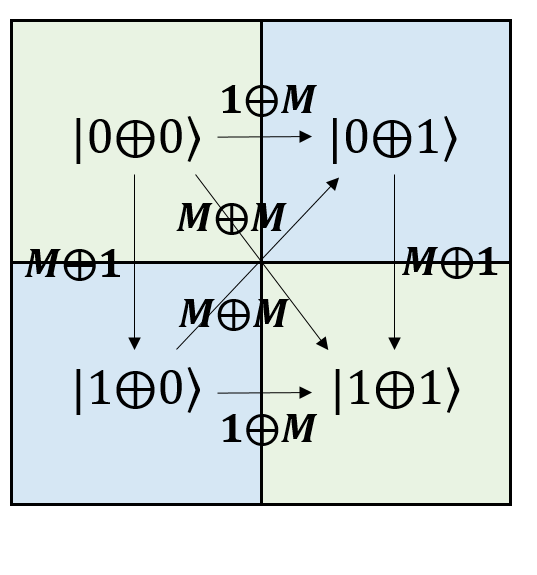}
\caption[]
{} \label{fig:operators}
\end{figure}
Explicitly, the all six transformation reads
\begin{equation}
\begin{aligned}
& \pmb{1}\oplus\pmb{M}|0\oplus 0\rangle=(\pmb{1}\oplus\pmb{M})(|0\rangle \oplus 0\rangle ) = |0\rangle \oplus |1\rangle \,, \\
&\pmb{M}\oplus\pmb{1}|0\oplus 1\rangle=(\pmb{M}\oplus\pmb{1})(|0\rangle \oplus 1\rangle ) = |1\rangle \oplus |1\rangle \,, \\
&\pmb{M}\oplus\pmb{M}|0\oplus 0\rangle=(\pmb{M}\oplus\pmb{M})(|0\rangle \oplus 0\rangle ) = |1\rangle \oplus |1\rangle \,, \\
&\pmb{M}\oplus\pmb{1}|0\oplus 0\rangle=(\pmb{M}\oplus\pmb{1})(|0\rangle \oplus 0\rangle ) = |1\rangle \oplus |0\rangle \,, \\
& \pmb{1}\oplus\pmb{M}|1\oplus 0\rangle=(\pmb{1}\oplus\pmb{M})(|1\rangle \oplus 0\rangle ) = |1\rangle \oplus |1\rangle \,, \\
&\pmb{M}\oplus\pmb{M}|1\oplus 0\rangle=(\pmb{M}\oplus\pmb{M})(|1\rangle \oplus 0\rangle ) = |0\rangle \oplus |1\rangle \,.\\
\end{aligned}
\end{equation}
It is noted that
\begin{equation}
\pmb{M}\oplus\pmb{M} = (\pmb{1}\oplus\pmb{M})(\pmb{M}\oplus\pmb{1}) = (\pmb{M}\oplus\pmb{1})(\pmb{1}\oplus\pmb{M}) \,.
\end{equation}
We have the Klein-4 group as
\begin{equation}
\mathbb{Z}_2 \times \mathbb{Z}_2 = \{\pmb{1}\oplus\pmb{1} , \pmb{1}\oplus\pmb{M} , \pmb{M}\oplus\pmb{1}  , \pmb{M}\oplus\pmb{M}  \} \,.
\end{equation}

The identity operator is given by the completeness relation,
\begin{equation}
\pmb{1} =|00\rangle \langle 00| + |11\rangle \langle 11| + |01\rangle \langle 01 | + |10\rangle \langle 01 | \,,
\end{equation}
and the duality operator is  given by
\begin{equation}
\pmb{M} = |00\rangle\langle 11| + |11\rangle \langle 00| + |01\rangle\langle 10| + |10\rangle \langle 01 | \,.
\end{equation}
Now we define an qubit operator,
\begin{equation}
\hat{\Psi}(\theta) = \cos \theta (\hat{-} \oplus \hat{-}) + \sin \theta (\hat{+} \oplus \hat{+})
\,.
\end{equation}
where the phase dependent qubit is
\begin{equation}
|\Psi(\theta)\rangle = \hat{\Psi}(\theta)|u\oplus u^*\rangle \,.
\end{equation}
The qubit operator acts on the full state $|u\oplus u^*\rangle$,
\begin{equation}
\begin{aligned}
\hat{\Psi}(\theta)|u\oplus u^*\rangle &=\cos\theta(\hat{-}|u\rangle \oplus \hat{-}|u^*\rangle) + \sin\theta (\hat{+}|u\rangle \oplus \hat{+}|u^*\rangle)\\
&=\cos\theta(\hat{-}|0\rangle \oplus \hat{-}|1\rangle) + \sin\theta (\hat{+}|0\rangle \oplus \hat{+}|1\rangle)\\
&=\cos\theta|0 \oplus 0\rangle +\sin\theta|1\oplus1\rangle 
\end{aligned} \,,
\end{equation}
 Thus we write
\begin{equation}
|\Psi(\theta)\rangle = \cos\theta |\text{nothingness} \rangle + \sin\theta|\text{All} \rangle \,,
\end{equation}
where $|\text{nothingness} \rangle = *|\text{All} \rangle $ and $*|\text{nothingness} \rangle = |\text{All} \rangle $. For the probabilities, we have
\begin{equation}
P_{\text{nothingness}} (\theta)= \langle 0 \oplus 0 | \Psi(\theta) \rangle= \cos^2\theta \,,
\end{equation}
and
\begin{equation}
P_{\text{All}}(\theta) = \langle 1 \oplus 1 | \Psi(\theta)\rangle = \sin^2\theta \,.
\end{equation}
Therefore, the probability of nothingness and All transforms one-another. In particular we have
\begin{equation}
|\Psi(0)\rangle =|\text{nothingness}\rangle \,,
\end{equation}
with probability equal to 1 for the nothingness state. And
\begin{equation}
|\Psi(\pi/2)\rangle = |\text{All}\rangle \,.
\end{equation}
with probability equal to 1 for the All state. The two states are orthogonal,
\begin{equation}
\langle\Psi(0) | \Psi(\pi/2)\rangle = \langle\text{nothingness}|\text{All} \rangle = 0 \,.
\end{equation}
There is a special state when $\theta = \pi/2$, then we have the analogous Bell state (which is a entangled EPR pair)
\begin{equation}
|\Psi(\pi/4)\rangle = \frac{1}{\sqrt{2}}(|0\oplus 0\rangle + |1\oplus 1\rangle)
\end{equation}
which is equal to
\begin{equation}
|\Psi(\pi/4)\rangle =\frac{1}{\sqrt{2}}(|\text{nothingness}\rangle + |\text{All}\rangle)
\end{equation}

This is same as the famous case of  Schrodinger's cat in physics, where the cat is regarded as nor alive nor dead, or both alive or dead, unless measurement is taken. Therefore, the ultimate truth here is a qubit, which is indeterministic unless an observation is made, where observation means measurement. 

Next we will study some properties of the qubit. Define the orthogonal state by differentiating the state with respect to $\theta$, then we have 
\begin{equation}
\langle \Psi(\theta) | \frac{d}{d\theta}| \Psi (\theta) \rangle = 0\,.
\end{equation}
Consider the state where we first take the derivative and the the dual operator,
\begin{equation}
|\psi^\prime (\theta) \rangle = *\frac{d}{d\theta}| \Psi (\theta) \rangle \,.
\end{equation}
This new state will be probability invariant as compared to the original state.

Finally, we are particularly interested in local phase for which $\theta$ is dependent on space and time and is linear of the spacetime variables. In other words, we are interested in the qubit wave operator as
\begin{equation}
\hat{\Psi}(x,y,z,t) = \cos (\omega t - k_x x - k_y y - k_z z) (\hat{+} \oplus \hat{+}) + \sin (\omega t - k_x x - k_y y - k_z z) (\hat{-} \oplus \hat{-})
\end{equation}
The qubit wave operator acts on the full state $|u\oplus u^*\rangle$,
\begin{equation} \label{eq:oscillation}
\begin{aligned}
\hat{\Psi}|u\oplus u^*\rangle &= \cos (\omega t - k_x x - k_y y - k_z z) (\hat{-}|u\rangle \oplus \hat{-}|u^*\rangle)+\sin (\omega t - k_x x - k_y y - k_z z) (\hat{+}|u\rangle \oplus \hat{+}|u^*\rangle)\\
&=\cos (\omega t - k_x x - k_y y - k_z z)|0 \oplus 0\rangle +\sin(\omega t - k_x x - k_y y - k_z z)|1 \oplus 1\rangle \,.
\end{aligned}
\end{equation}
The qubit operator satisfies the wave equation,
\begin{equation}
\Box \hat{\Psi} |u\oplus u^*\rangle = \bigg(\nabla^2 \hat{\Psi} - \frac{1}{c^2}\frac{\partial^2}{\partial t^2}\hat{\Psi}\bigg)|u\oplus u^*\rangle = 0
\end{equation}
where $c$ is the speed of the qubit wave.

Next for the dual qubit wave, we have
\begin{equation}
\hat{\Psi^\star}(\theta) = \cos \theta (\hat{+} \oplus \hat{-}) + \sin \theta (\hat{-} \oplus \hat{+})
\end{equation}

By acting on the full state $|u\otimes u^* \rangle$,
\begin{equation}
\begin{aligned}
\hat{\Psi}^\star (\theta)|u\oplus u^*\rangle &=\cos\theta(\hat{+} |u\rangle \oplus \hat{-} |u^*\rangle) + \sin\theta (\hat{-} |u\rangle \oplus\hat{+}_3 |u^*\rangle)\\
&=\cos\theta(\hat{+}|0\rangle \oplus \hat{-} |1\rangle) + \sin\theta (\hat{-} |0\rangle \oplus \hat{+} |1\rangle)\\
&=\cos\theta|1\oplus 0\rangle +\sin\theta|0 \oplus 1\rangle 
\end{aligned} \,,
\end{equation}
We have 
\begin{equation}
\hat{\Psi}^\star (\theta) =  \cos\theta|\text{creation}\rangle +\sin\theta|\text{annihilation}\rangle
\end{equation}
where $|\text{creation}\rangle = *|\text{annihilation}\rangle$ and $*|\text{creation}\rangle = |\text{annihilation}\rangle$.
\begin{equation}
P_{\text{creation}} (\theta)= \langle 1 \oplus 0 | \Psi^\star(\theta) \rangle= \cos^2\theta \,,
\end{equation}
and
\begin{equation}
P_{\text{annihilation}}(\theta) = \langle 0\oplus 1 | \Psi^\star (\theta)\rangle = \sin^2\theta \,.
\end{equation}
In phase equal to zero,  we have
\begin{equation}
|\Psi^\star(0)\rangle = |\text{creation}\rangle  
\end{equation}
with probability equal to 1 for creation state. And
\begin{equation}
|\Psi^\star (\pi/2)\rangle = |\text{annihilation}\rangle  
\end{equation}
with probability equal to 1 for the annihilation state. The two states are orthogonal,
\begin{equation}
\langle\Psi^\star(0) | \Psi^\star(\pi/2)\rangle = \langle\text{creation}|\text{annihilation} \rangle = 0 \,.
\end{equation}
There is a special state when $\theta = \pi/2$, then we have the entangled EPR state as
\begin{equation}
|\Psi(\pi/4)\rangle = \frac{1}{\sqrt{2}}(|1 \oplus 0\rangle + |0 \oplus 1\rangle)
\end{equation}
which is equal to
\begin{equation}
|\Psi(\pi/4)\rangle =\frac{1}{\sqrt{2}}(|\text{creation}\rangle + |\text{annihilation}\rangle)
\end{equation}

Then the dual wave qubit operator is
\begin{equation}
\hat{\Psi}^\star (x,y,z,t) = \cos (\omega t - k_x x - k_y y - k_z z) (\hat{+} \oplus \hat{-} ) + \sin (\omega t - k_x x - k_y y - k_z z) (\hat{-} \oplus \hat{+} ) \,.
\end{equation}
By acting on the full state
\begin{equation} \label{eq:oscillate}
\begin{aligned}
\hat{\Psi}^\star |u\oplus u^*\rangle &= \cos (\omega t - k_x x - k_y y - k_z z) (\hat{+} |u\rangle \oplus \hat{-} |u^*\rangle)+\sin (\omega t - k_x x - k_y y - k_z z) (\hat{-} |u\rangle \oplus \hat{+} |u^*\rangle)\\
&=\cos (\omega t - k_x x - k_y y - k_z z)|1\oplus 0\rangle +\sin(\omega t - k_x x - k_y y - k_z z)|0 \oplus 1\rangle \,.
\end{aligned}
\end{equation}
The dual qubit also satisfies the wave equation,
\begin{equation}
\Box \hat{\Psi}^\star |u\oplus u^*\rangle = \bigg(\nabla^2 \hat{\Psi}^\star - \frac{1}{c^2}\frac{\partial^2}{\partial t^2}\hat{\Psi}^\star\bigg)|u\oplus u^*\rangle = 0 \,.
\end{equation}
The pair of states are orthogonal to each other
\begin{equation}
\langle \Psi(\theta)| \Psi^\star (\theta) \rangle = 0 \,.
\end{equation}
Note that the linear combination of the two states is also the solution of the wave equation,
\begin{equation}
\phi = A\Psi + B\Psi^\star \,.
\end{equation}
Therefore, when the dual states oscillate, it generates a wave propagation. For example, in equation \ref{eq:oscillate}, the periodic cycle of creation and annihilation is a wave. Since mind creates and annihilates, this shows that mind is a wave. And for equation \ref{eq:oscillation}, when the state of nothingness and All oscillate, it creates a wave. So the periodic transformation of nothingness and All will produce a wave signal, that maybe detected in experimental sense.

Next we would like to investigate what happens when the operators act on the dual of $|u\oplus u^*\rangle$, i.e. $*|u\oplus u^*\rangle =|u^* \oplus u\rangle $. Using the representation in \ref{eq:rep} we find that the following holds,
\begin{equation}
\hat{P}*|0\rangle \oplus \hat{Q}*|1\rangle = \hat{P}|1\rangle \oplus \hat{Q}|0\rangle  \,.
\end{equation}
Explicitly, we can check that
\begin{equation} \label{eq:resultss}
\begin{aligned}
\hat{-}*|0\rangle \oplus \hat{-}*|1\rangle &=\hat{-}|1\rangle \oplus \hat{-}|0\rangle = |0\rangle \oplus |0\rangle =\hat{-}|0\rangle \oplus \hat{-}|1\rangle \,,\\
\hat{+}*|0\rangle \oplus \hat{+}*|1\rangle &=\hat{+}|1\rangle \oplus \hat{+}|0\rangle = |1\rangle \oplus |1\rangle=\hat{+}|0\rangle \oplus \hat{+}|1\rangle  \,, \\
\hat{+}*|0\rangle \oplus \hat{-}*|1\rangle &=\hat{+}|1\rangle \oplus \hat{-}|0\rangle = |0\rangle \oplus |1\rangle =\hat{+}|0\rangle \oplus \hat{-}|1\rangle \,, \\
\hat{-}*|0\rangle \oplus \hat{+}*|1\rangle &=\hat{-}|1\rangle \oplus \hat{+}|0\rangle = |1\rangle \oplus |0\rangle=\hat{-}|0\rangle \oplus \hat{+}|1\rangle  \,.
\end{aligned}
\end{equation}
Therefore, we have
\begin{equation}
\hat{-}* = \hat{-} \quad , \quad \hat{+}* = \hat{+} \,.
\end{equation} \label{eq:twoeq}
We can explicitly check that
\begin{equation}
\begin{pmatrix}
1 & 1 \\
0 & 0
\end{pmatrix}
\begin{pmatrix}
0 & 1\\
1 & 0
\end{pmatrix} =
\begin{pmatrix}
1 & 1\\
0 & 0
\end{pmatrix} \quad , \quad
\begin{pmatrix}
0 & 0\\
1 & 1
\end{pmatrix}
\begin{pmatrix}
0 & 1\\
1 & 0
\end{pmatrix} =
\begin{pmatrix}
0 & 0\\
1 & 1
\end{pmatrix}  \,.
\end{equation}
Now we would like to give a more detailed study in the $\pmb{M}$ dual matrix operator. First notice that the action of $*=\pmb{M}$ swap the position of rows as a vertical reflection. For example, for a general matrix
\begin{equation}
A=\begin{pmatrix}
a & b\\
c & d
\end{pmatrix}
\quad \text{then} \quad
*A = \pmb{M}A = A^* =
\begin{pmatrix}
 c & d \\
 a & b
\end{pmatrix} \,.
\end{equation}
If $A$ acts on $\pmb{M}$, the action is to swap the position of the columns as a horizontal reflection,
\begin{equation}
A* =  A\pmb{M} = \,^*A = 
\begin{pmatrix}
 b & a \\
 d & c
\end{pmatrix} \,.
\end{equation}
Also we have
\begin{equation}
** A = A** = A \,.
\end{equation}
We have the following identity, recalling $T$ is the transpose action and $R$ is the off-transpose action, we have
\begin{equation}
A^{(TR)} = A^{(RT)} = \pmb{M}A\pmb{M}^{-1} = \pmb{M}A\pmb{M} = *A*^{-1}  =*A* \,,
\end{equation}
as $\pmb{M}^{-1} = \pmb{M}$. It follows that $\mathrm{Tr} (A^{(TR)}) = \mathrm{Tr}A$ and $\mathrm{det} (A^{(TR)}) = \mathrm{det}A$.
According to equations \ref{eq:twoeq},
\begin{equation}
*\hat{-}* = \hat{+} \quad ,\quad *\hat{+}* = \hat{-} \,.
\end{equation}
And as $*=*^{-1}$, thus $\hat{+}$ and $\hat{-}$ are related by similarity transform
\begin{equation}
*\hat{-}*^{-1} = \hat{+} \quad ,\quad *\hat{+}*^{-1} = \hat{-} \,.
\end{equation}
Next we would like to show that the identity
\begin{equation}
(*A*)^{-1} = *A^{-1}* \,.
\end{equation}
This is because $(*A*)^{-1} =*^{-1} (*A)^{-1} =  *A^{-1}*$.

Based on the result of \ref{eq:resultss}, if follows that $|u^* \oplus u\rangle $ also satisfies the wave equation,
\begin{equation}
\Box \hat{\Psi} |u^* \oplus u\rangle = \bigg(\nabla^2 \hat{\Psi} - \frac{1}{c^2}\frac{\partial^2}{\partial t^2}\hat{\Psi}\bigg)|u^* \oplus u\rangle = 0
\end{equation}
and
\begin{equation}
\Box \hat{\Psi}^\star |u^* \oplus u\rangle = \bigg(\nabla^2 \hat{\Psi} - \frac{1}{c^2}\frac{\partial^2}{\partial t^2}\hat{\Psi}^\star \bigg)|u^* \oplus u\rangle = 0
\end{equation}

Finally we will investigate the entropy of the system. The entropy for the both systems $|\Psi\rangle$ and $|\Psi^\star\rangle$ are 
\begin{equation}
H(\theta) = -2\cos^2 \theta\log |\cos\theta|  -2\sin^2 \theta\log |\sin\theta| \,.
\end{equation}
When the probability is equal the entropy is maximized, thus the EPR state is the equilibrium state, which is the most stable state. The system has 0.693 bits. When it is either at its designated state with probability equal to 1, the entropy is zero, which is minimum and hence unstable.

\subsection{Canonical Quantization}
In this section, we will dive into the core part of the whole subject, the quantization of duality wave. Upon quantization, this will give the quanta of the duality field. The procedure is called second quantization, which is originated from quantum field theory. Second quantization turns a wave into a particle quanta. 

First for the $\hat{\Phi}(\theta)$ we define the conjugated momentum operator by
\begin{equation}
\Pi = \frac{h}{i}\frac{\partial}{\partial\theta} \Psi\,.
\end{equation}
Therefore the conjugated momentum of $\Psi$ is 
\begin{equation}
\Pi(\theta) =\frac{h}{i} \Big(-\sin\theta(\hat{-}\oplus\hat{-}) + \cos\theta(\hat{+} \oplus \hat{+})\Big) \,. 
\end{equation}
Now we compute the commutator $[\hat{\Psi}(\theta), \hat{\Pi}(\theta)]$,
\begin{equation}
\begin{aligned}
\left[\Psi(\theta), \Pi(\theta) \right] &= (-ih)[ \cos\theta(\hat{-}\oplus\hat{-})+\sin\theta(\hat{+} \oplus \hat{+}) \,\,  ,\,\,-\sin\theta(\hat{-}\oplus\hat{-}) + \cos\theta(\hat{+} \oplus \hat{+})  ] \\
&=-ih\Big(-\cos\theta\sin\theta [(\hat{-}\oplus\hat{-}), (\hat{-}\oplus\hat{-}) ] -\sin^2\theta [(\hat{+}\oplus\hat{+}), (\hat{-}\oplus\hat{-}) ]  \\
&\,\,\,\,\,\,+\cos^2 \theta [(\hat{-}\oplus\hat{-}), (\hat{+}\oplus\hat{+}) ] +\sin\theta \cos\theta [(\hat{+}\oplus\hat{+}), (\hat{+ }\oplus\hat{+}) ] \Big)\\
&= -ih [(\hat{-} \oplus \hat{-} ) , (\hat{+} \oplus \hat{+} )  ]  \\
&= -ih \Big((\hat{-} \oplus \hat{-} )(\hat{+} \oplus \hat{+})  -(\hat{+} \oplus \hat{+} )(\hat{-} \oplus \hat{-} )\Big) \\
&=-ih (\hat{-}\hat{+} \oplus \hat{-}\hat{+} -\hat{+}\hat{-} \oplus \hat{+}\hat{-}) \\
\end{aligned}
\end{equation}
Now we take an ansatz for the replacement of $\hat{-},\hat{+}$ matrix operators by bosonic creation and annihilation operators $\hat{a},\hat{a}^{\dagger}$ with commutation relation of $[\hat{a},\hat{a}^{\dagger}]=1$
\begin{equation}
\hat{-} \rightarrow \hat{a}^\dagger \quad,\quad \hat{+} \rightarrow \hat{a}
\end{equation}
such that the quantum fields for duality are
\begin{equation}
\hat{\Psi}(\theta) = \cos\theta(\hat{a}^\dagger\oplus\hat{a}^\dagger)+\sin\theta(\hat{a} \oplus \hat{a}) 
\end{equation}
and
\begin{equation}
\hat{\Pi}(\theta) =\frac{h}{i} \Big(-\sin\theta(\hat{a}^\dagger\oplus\hat{a}^\dagger) + \cos\theta(\hat{a} \oplus \hat{a}) \Big)  \,. 
\end{equation}
Then
\begin{equation}
\begin{aligned}
\left[\hat{\Psi(\theta)}, \hat{\Pi(\theta)} \right] &= -ih( \hat{a}^\dagger\hat{a} \oplus  \hat{a}^\dagger\hat{a} - \hat{a}\hat{a}^\dagger \oplus \hat{a}\hat{a}^\dagger )\\
&= -ih \bigg(
\begin{pmatrix}
\hat{a}^\dagger \hat{a} & 0\\
0 &   \hat{a}^\dagger \hat{a}
\end{pmatrix} -
\begin{pmatrix}
\hat{a} \hat{a}^\dagger & 0\\
0 &  \hat{a} \hat{a}^\dagger
\end{pmatrix}
\bigg) \\
&=-ih\begin{pmatrix}
-[\hat{a},\hat{a}^\dagger] & 0\\
0 & -[\hat{a},\hat{a}^\dagger] 
\end{pmatrix} \\
& = -ih\begin{pmatrix}
-1 & 0 \\
0 & -1
\end{pmatrix} \\
&= ih\pmb{1} \,.
\end{aligned}
\end{equation}
In the meantime, we have $\left[\hat{\Psi}(\theta) , \hat{\Psi}(\theta) \right] = 0$ and $\left[\hat{\Pi}(\theta) , \hat{\Pi}(\theta)\right]$. Therefore we have the canonical quantization condition as
\begin{equation}
\left[\hat{\Psi}(\theta), \hat{\Pi}(\theta) \right]=ih\pmb{1} \quad,\quad \left[\hat{\Psi}(\theta) , \hat{\Psi}(\theta)\right]=\left[\hat{\Pi}(\theta) , \hat{\Pi}(\theta)\right] =0\,.
\end{equation}
Next for the $\hat{\Phi}^\star (\theta)$ we define the conjugated momentum operator by
\begin{equation}
\hat{\Pi}^\star(\theta) = \frac{h}{i}\frac{\partial}{\partial\theta} \hat{\Psi}^\star \,.
\end{equation}
Therefore the conjugated momentum of $\Psi$ is 
\begin{equation}
\Pi^\star (\theta) =\frac{h}{i} \Big(-\sin\theta(\hat{+}\oplus\hat{-}) + \cos\theta(\hat{-} \oplus \hat{+})\Big) \,. 
\end{equation}
Now we compute the commutator $[\hat{\Psi}(\theta), \hat{\Pi}(\theta)]$,
\begin{equation}
\begin{aligned}
\left[\Psi^\star(\theta), \Pi^\star (\theta) \right] &= (-ih)\left[ \cos\theta(\hat{+}\oplus\hat{-})+\sin\theta(\hat{-} \oplus \hat{+}) \,\,  ,\,\,-\sin\theta(\hat{+}\oplus\hat{-}) + \cos\theta(\hat{-} \oplus \hat{+})  \right] \\
&=-ih\Big(-\cos\theta\sin\theta \left[(\hat{+}\oplus\hat{-}), (\hat{+}\oplus\hat{-}) \right] -\sin^2\theta \left[(\hat{-}\oplus\hat{+}), (\hat{+}\oplus\hat{-}) \right] \\
&\,\,\,\,\,\,+\cos^2 \theta \left[(\hat{+}\oplus\hat{-}), (\hat{-}\oplus\hat{+}) \right] +\sin\theta \cos\theta \left[(\hat{-}\oplus\hat{+}), (\hat{- }\oplus\hat{+}) \right] \Big)\\
&= -ih \left[(\hat{+} \oplus \hat{-} ) , (\hat{-} \oplus \hat{+} )  \right]  \\
&= -ih \Big((\hat{+} \oplus \hat{-} )(\hat{-} \oplus \hat{+})  -(\hat{-} \oplus \hat{+} )(\hat{+} \oplus \hat{-} )\Big) \\
&=-ih (\hat{+}\hat{-} \oplus \hat{-}\hat{+} -\hat{-}\hat{+} \oplus \hat{+}\hat{-}) \\
\end{aligned}
\end{equation}
Then we have 
\begin{equation}
\begin{aligned}
\left[\hat{\Psi}^\star(\theta), \hat{\Pi}^\star (\theta) \right] &= -ih(\hat{a}\hat{a}^\dagger \oplus \hat{a}^\dagger \hat{a} - \hat{a}^\dagger \hat{a} \oplus \hat{a}\hat{a}^\dagger) \\
&=-ih \bigg(\begin{pmatrix}
\hat{a} \hat{a}^\dagger & 0\\
0 &   \hat{a}^\dagger \hat{a}
\end{pmatrix} -
\begin{pmatrix}
\hat{a}^\dagger \hat{a} & 0\\
0 &  \hat{a} \hat{a}^\dagger
\end{pmatrix}
\bigg) \\
&=-ih\begin{pmatrix}
[\hat{a},\hat{a}^\dagger] & 0\\
0 &-[\hat{a},\hat{a}^\dagger] 
\end{pmatrix} \\
& = ih\begin{pmatrix}
-1 & 0 \\
0 & 1
\end{pmatrix} \\
&= ihM  \,,
\end{aligned}
\end{equation}
where $M$ is another representation matrix of the duality operator with $M^2 = \pmb{1}$. In the meantime, we have $[\hat{\Psi}^\star (\theta) , \hat{\Psi}^\star (\theta)] = 0$ and $[\hat{\Pi}^\star (\theta) , \hat{\Pi}^\star (\theta)] =0$. Therefore we have the canonical quantization condition as
\begin{equation}
[\hat{\Psi}^\star (\theta), \hat{\Pi}^\star (\theta) ]=ihM \quad,\quad [\hat{\Psi}^\star(\theta) , \hat{\Psi}^\star(\theta)]=[\hat{\Pi}^\star(\theta) , \hat{\Pi}^\star(\theta)] =0\,.
\end{equation}
Notice that the cross terms are not zero, this is given by the fact that the states are not independent of each other. For example
\begin{equation}
[\hat{\Psi}(\theta) , \hat{\Psi}^\star(\theta) ] = \begin{pmatrix}
-\cos 2\theta &0 \\
0 & 0
\end{pmatrix} \,.
\end{equation}
\subsubsection{Hamiltonian}
The canonical momentum operator was defined by $\hat{\Pi} = \frac{1}{i}\frac{\partial}{\partial\theta}$, the $h$ is introduced such that we will get the same form of commutation relation as quantum mechanics does. From now on, to calculate the canonical Hamiltonian, let's define the dimensionless conjugated momentum, still denoted as $\hat{\Pi}$,
\begin{equation}
\hat{\Pi} = \frac{1}{i}\frac{\partial}{\partial\theta}\hat{\Psi} \quad \text{and} \quad  \hat{\Pi}^\star = \frac{1}{i}\frac{\partial}{\partial\theta}\hat{\Psi}^\star
\end{equation}
We ansatz that the Hamiltonian is defined as
\begin{equation}
\hat{H}=\bigg(\frac{1}{2}\hat{\Pi}^2 + \frac{1}{2}\hat{\Psi}^2 \bigg) \hbar\omega \quad \text{and} \quad 
\hat{H}^\star=\bigg(\frac{1}{2}\hat{\Pi}^{\star 2} + \frac{1}{2}\hat{\Psi}^{\star 2} \bigg)\hbar\omega \,.
\end{equation}
Now first we evaluate $\hat{H}(\theta)$.
\begin{equation}
\begin{aligned}
&\quad\hat{\Pi}^2 (\theta)\\
 &= \frac{1}{2i^2}\Big( -\sin\theta(\hat{-}\oplus\hat{-}) + \cos\theta(\hat{+}\oplus\hat{+}) \Big)\Big( -\sin\theta(\hat{-}\oplus\hat{-}) + \cos\theta(\hat{+}\oplus\hat{+}) \Big) \\
&= -\frac{1}{2}\Big(\sin^2\theta (\hat{-}\oplus\hat{-}  )(\hat{-}\oplus\hat{-}  ) -\sin\theta\cos\theta(\hat{-}\oplus\hat{-}  )(\hat{+}\oplus\hat{+}  ) \\
&\quad\quad-\cos\theta\sin\theta(\hat{+}\oplus\hat{+}  )(\hat{-}\oplus\hat{-}  ) + \cos^2\theta(\hat{+}\oplus\hat{+}  )(\hat{+}\oplus\hat{+}  ) \Big)\\
&=-\frac{1}{2}\Big(\sin^2\theta(\hat{-}\hat{-}\oplus\hat{-}\hat{-}) -\sin\theta\cos\theta(\hat{-}\hat{+}\oplus\hat{-}\hat{+})-\sin\theta\cos\theta(\hat{+}\hat{-}\oplus\hat{+}\hat{-}) +\cos^2 \theta (\hat{+}\hat{+}\oplus \hat{+}\hat{+})\Big) \\
&=-\frac{1}{2}\bigg[(\sin^2 \theta\begin{pmatrix}
\hat{-}\hat{-} & 0 \\
0 & \hat{-}\hat{-}
\end{pmatrix}
-\sin\theta\cos\theta\begin{pmatrix}
\hat{-}\hat{+} & 0 \\
0 & \hat{-}\hat{+}
\end{pmatrix}
-\sin\theta\cos\theta\begin{pmatrix}
\hat{+}\hat{-} & 0 \\
0 & \hat{+}\hat{-}
\end{pmatrix}
+\cos^2 \theta \begin{pmatrix}
\hat{+}\hat{+} & 0 \\
0 & \hat{+}\hat{+}
\end{pmatrix} \bigg]\\
&= -\frac{1}{2}\begin{pmatrix}
\sin^2 \theta \hat{-}\hat{-} + \cos^2\theta \hat{+}\hat{+} -\sin\theta\cos\theta(\hat{-}\hat{+}+\hat{+}\hat{-}) & 0 \\
0& \sin^2 \theta \hat{-}\hat{-} + \cos^2\theta\hat{+}\hat{+} -\sin\theta\cos\theta(\hat{-}\hat{+}+\hat{+}\hat{-}) 
\end{pmatrix}\\
&= -\frac{1}{2}\Big(\sin^2 \theta \hat{-}\hat{-} + \cos^2\theta\hat{+}\hat{+} -\sin\theta\cos\theta(\hat{-}\hat{+}+\hat{+}\hat{-}) \Big) \begin{pmatrix}
1 & 0 \\
0& 1
\end{pmatrix} \\
&= -\frac{1}{2}\Big( \sin^2 \theta \hat{a}^\dagger\hat{a}^\dagger + \cos^2\theta\hat{a}\hat{a} -\sin\theta\cos\theta(\hat{a}\hat{a}^\dagger+\hat{a}^\dagger\hat{a}) \Big)\pmb{1}\\
&= -\frac{1}{2}\Big( \sin^2 \theta \hat{a}^\dagger\hat{a}^\dagger + \cos^2\theta\hat{a}\hat{a} \Big) +\frac{1}{2}(2\sin\theta\cos\theta)\bigg(\hat{a}^\dagger\hat{a}+\frac{1}{2}\bigg)\pmb{1}\\
&=-\frac{1}{2}\Big( \sin^2 \theta \hat{a}^\dagger\hat{a}^\dagger + \cos^2\theta\hat{a}\hat{a} \Big) +\sin\theta\cos\theta\bigg(\hat{a}^\dagger\hat{a}+\frac{1}{2}\bigg)\pmb{1} \,. 
\end{aligned}
\end{equation}
Then 
\begin{equation}
\begin{aligned}
&\quad\hat{\Psi}^2 (\theta)\\
 &= \frac{1}{2}\Big( \cos\theta(\hat{-}\oplus\hat{-}) + \sin\theta(\hat{+}\oplus\hat{+}) \Big)\Big( \cos\theta(\hat{-}\oplus\hat{-}) + \sin\theta(\hat{+}\oplus\hat{+}) \Big) \\
&= \frac{1}{2}\Big(\cos^2\theta (\hat{-}\oplus\hat{-}  )(\hat{-}\oplus\hat{-}  ) +\cos\theta\sin\theta(\hat{-}\oplus\hat{-}  )(\hat{+}\oplus\hat{+}  ) \\
&\quad\quad+\sin\theta\cos\theta(\hat{+}\oplus\hat{+}  )(\hat{-}\oplus\hat{-}  ) + \sin^2\theta(\hat{+}\oplus\hat{+}  )(\hat{+}\oplus\hat{+}  ) \Big)\\
&=\frac{1}{2}\Big(\cos^2\theta(\hat{-}\hat{-}\oplus\hat{-}\hat{-}) +\cos\theta\sin\theta(\hat{-}\hat{+}\oplus\hat{-}\hat{+})+\cos\theta\sin\theta(\hat{+}\hat{-}\oplus\hat{+}\hat{-}) +\sin^2 \theta (\hat{+}\hat{+}\oplus \hat{+}\hat{+})\Big) \\
&=\frac{1}{2}\bigg[(\cos^2 \theta\begin{pmatrix}
\hat{-}\hat{-} & 0 \\
0 & \hat{-}\hat{-}
\end{pmatrix}
+\cos\theta\sin\theta\begin{pmatrix}
\hat{-}\hat{+} & 0 \\
0 & \hat{-}\hat{+}
\end{pmatrix}
+\cos\theta\sin\theta\begin{pmatrix}
\hat{+}\hat{-} & 0 \\
0 & \hat{+}\hat{-}
\end{pmatrix}
+\sin^2 \theta \begin{pmatrix}
\hat{+}\hat{+} & 0 \\
0 & \hat{+}\hat{+}
\end{pmatrix} \bigg]\\
&= \frac{1}{2}\begin{pmatrix}
\cos^2 \theta \hat{-}\hat{-} + \sin^2\theta \hat{+}\hat{+} +\cos\theta\sin\theta(\hat{-}\hat{+}+\hat{+}\hat{-}) & 0 \\
0& \cos^2 \theta \hat{-}\hat{-} + \sin^2\theta\hat{+}\hat{+} +\cos\theta\sin\theta(\hat{-}\hat{+}+\hat{+}\hat{-}) 
\end{pmatrix}\\
&= \frac{1}{2}\Big(\cos^2 \theta \hat{-}\hat{-} + \sin^2\theta\hat{+}\hat{+} +\cos\theta\sin\theta(\hat{-}\hat{+}+\hat{+}\hat{-}) \Big) \begin{pmatrix}
1 & 0 \\
0& 1
\end{pmatrix} \\
&= \frac{1}{2}\Big( \cos^2 \theta \hat{a}^\dagger\hat{a}^\dagger + \sin^2\theta\hat{a}\hat{a} +\cos\theta\sin\theta(\hat{a}\hat{a}^\dagger+\hat{a}^\dagger\hat{a}) \Big)\pmb{1}\\
&= \frac{1}{2}\Big( \cos^2 \theta \hat{a}^\dagger\hat{a}^\dagger + \sin^2\theta\hat{a}\hat{a} \Big)\pmb{1}  +\frac{1}{2}(2\cos\theta\sin\theta)\bigg(\hat{a}^\dagger\hat{a}+\frac{1}{2}\bigg)\pmb{1}\\
&=\frac{1}{2}\Big( \cos^2 \theta \hat{a}^\dagger\hat{a}^\dagger + \sin^2\theta\hat{a}\hat{a} \Big) \pmb{1} +\sin\theta\cos\theta\bigg(\hat{a}^\dagger\hat{a}+\frac{1}{2}\bigg)\pmb{1} \,. 
\end{aligned}
\end{equation}
Therefore, finally we obtain the Hamiltonian as follow
\begin{equation} \label{eq:hamiltonian}
\hat{H}(\theta)=\bigg(\frac{1}{2}\hat{\Pi}^2 (\theta) + \frac{1}{2}\hat{\Psi}^2 (\theta)\bigg)\hbar\omega = \frac{1}{2}\cos 2\theta (\hat{a}^\dagger\hat{a}^\dagger -\hat{a}\hat{a})\hbar\omega\pmb{1}  + \sin 2\theta \bigg(\hat{a}^\dagger\hat{a}+\frac{1}{2}\bigg)\hbar\omega \pmb{1} 
\end{equation} 
The expectation energy is hence given by
\begin{equation}
\begin{aligned}
\langle E\rangle= \langle n | \hat{H}(\theta) |n \rangle &= \frac{1}{2}  \cos 2\theta\langle n |(\hat{a}^\dagger\hat{a}^\dagger -\hat{a}\hat{a})|n\rangle \hbar\omega \, \pmb{1}+ \sin 2\theta \langle n |\bigg(\hat{a}^\dagger\hat{a}+\frac{1}{2}\bigg) |n\rangle \hbar\omega \pmb{1} \\
&=\sin 2\theta \bigg(n+\frac{1}{2} \bigg) \hbar\omega\pmb{1}\\
&= H\sin 2\theta \,\pmb{1}\,,
\end{aligned}
\end{equation}
where
\begin{equation}
H = \bigg(n+\frac{1}{2}\bigg)\hbar\omega.
\end{equation}
This is because $\langle n |\hat{a}^\dagger \hat{a}^\dagger | n\rangle =\langle n |\hat{a} \hat{a}| n\rangle =0$ and $\langle n |\hat{a}^\dagger \hat{a}| n\rangle =n$. Thus this gives the oscillation of expectation of energy which depends upon the phase, which defines the quanta of duality wave, the dualiton. The vacuum expectation energy of the dualiton is given by
\begin{equation} \label{eq:groundstateenergy}
\langle 0 | \hat{H}(\theta) |0 \rangle = \frac{1}{2} \sin 2\theta \hbar\omega\,\pmb{1}= \sin\theta\cos\theta \hbar\omega \,\pmb{1}\,.
\end{equation}
Now let's evaluate the vacuum expectation value for different important phases. When $\theta = 0$ or $\frac{\pi}{2}$, we have 
\begin{equation}
E_0 =\langle 0 | \hat{H}(\theta) |0 \rangle = \pmb{0} \,.
\end{equation}
Recalling that $\Psi(\theta)=\cos\theta |0\oplus 0\rangle + \sin\theta|1\oplus1 \rangle =\cos\theta|\text{nothingness} \rangle + \sin\theta|\text{All} \rangle$ , this means when the dual wave function is at a particular state, then it has zero vacuum expectation energy value. On the other hand, when $\theta = \pi/4$,
\begin{equation}
E_0 = \langle 0 | \hat{H}(\theta) |0 \rangle = \frac{1}{2}\hbar\omega \,\pmb{1} =\begin{pmatrix}
\frac{1}{2}\hbar\omega & 0 \\
0 & \frac{1}{2}\hbar\omega 
\end{pmatrix} \,.
\end{equation}
This means that when $\Psi(\frac{\pi}{4})=\frac{1}{\sqrt{2}} |0\oplus 0\rangle + \frac{1}{\sqrt{2}}|1\oplus1 \rangle =\frac{1}{\sqrt{2}}|\text{nothingness} \rangle + \frac{1}{\sqrt{2}}|\text{All} \rangle$, the state is in linear superposition of the two states has its greatest entropy, it has the maximized energy. This shows that when the system is in its determined state of $|0\oplus 0\rangle$ or $|1\oplus 1\rangle$ it has lower ground state energy than the undetermined  superpositioned state of  $\Psi=\frac{1}{\sqrt{2}} |0\oplus 0\rangle + \frac{1}{\sqrt{2}}|1\oplus1 \rangle$, thus is considered to be more stable. We say the superpositioned state is spontaneously collapsed into one of the state of full probability, this statement is equivalent to spontaneous symmetry breaking when a state is completely determined.

Next we evaluate $\hat{H}^\star (\theta)$.
\begin{equation}
\begin{aligned}
&\quad\hat{\Pi}^{\star 2} (\theta)\\
 &= \frac{1}{2i^2}\Big( -\sin\theta(\hat{+}\oplus\hat{-}) + \cos\theta(\hat{-}\oplus\hat{+}) \Big)\Big( -\sin\theta(\hat{+}\oplus\hat{-}) + \cos\theta(\hat{-}\oplus\hat{+}) \Big) \\
&= -\frac{1}{2}\Big(\sin^2\theta (\hat{+}\oplus\hat{-}  )(\hat{+}\oplus\hat{-}  ) -\sin\theta\cos\theta(\hat{+}\oplus\hat{-}  )(\hat{-}\oplus\hat{+}  ) \\
&\quad\quad-\cos\theta\sin\theta(\hat{-}\oplus\hat{+}  )(\hat{+}\oplus\hat{-}  ) + \cos^2\theta(\hat{-}\oplus\hat{+}  )(\hat{-}\oplus\hat{+}  ) \Big)\\
&=-\frac{1}{2}\Big(\sin^2\theta(\hat{+}\hat{+}\oplus\hat{-}\hat{-}) -\sin\theta\cos\theta(\hat{+}\hat{-}\oplus\hat{-}\hat{+})-\sin\theta\cos\theta(\hat{-}\hat{+}\oplus\hat{+}\hat{-}) +\cos^2 \theta (\hat{-}\hat{-}\oplus \hat{+}\hat{+})\Big) \\
&=-\frac{1}{2}\bigg[(\sin^2 \theta\begin{pmatrix}
\hat{+}\hat{+} & 0 \\
0 & \hat{-}\hat{-}
\end{pmatrix}
-\sin\theta\cos\theta\begin{pmatrix}
\hat{+}\hat{-} & 0 \\
0 & \hat{-}\hat{+}
\end{pmatrix}
-\sin\theta\cos\theta\begin{pmatrix}
\hat{-}\hat{+} & 0 \\
0 & \hat{+}\hat{-}
\end{pmatrix}
+\cos^2 \theta \begin{pmatrix}
\hat{-}\hat{-} & 0 \\
0 & \hat{+}\hat{+}
\end{pmatrix} \bigg]\\
&= -\frac{1}{2}\begin{pmatrix}
\sin^2 \theta \hat{+}\hat{+} + \cos^2\theta \hat{-}\hat{-} -\sin\theta\cos\theta(\hat{+}\hat{-}+\hat{-}\hat{+}) & 0 \\
0& \sin^2 \theta \hat{-}\hat{-} + \cos^2\theta\hat{+}\hat{+} -\sin\theta\cos\theta(\hat{-}\hat{+}+\hat{+}\hat{-}) 
\end{pmatrix}\\
&= -\frac{1}{2}\begin{pmatrix}
\sin^2 \theta \hat{a}\hat{a} + \cos^2\theta \hat{a}^\dagger\hat{a}^\dagger -\sin\theta\cos\theta(\hat{a}\hat{a}^\dagger+\hat{a}^\dagger\hat{a}) & 0 \\
0& \sin^2 \theta \hat{a}^\dagger\hat{a}^\dagger + \cos^2\theta\hat{a}\hat{a} -\sin\theta\cos\theta(\hat{a}^\dagger\hat{a}+\hat{a}\hat{a}^\dagger) 
\end{pmatrix}\\
&=-\frac{1}{2}\begin{pmatrix}
\sin^2 \theta \hat{a}\hat{a} + \cos^2\theta \hat{a}^\dagger\hat{a}^\dagger -\sin 2\theta\Big(\hat{a}^\dagger\hat{a} + \frac{1}{2}\Big) & 0 \\
0& \sin^2 \theta \hat{a}^\dagger\hat{a}^\dagger + \cos^2\theta\hat{a}\hat{a}-\sin 2\theta\Big(\hat{a}^\dagger\hat{a} + \frac{1}{2}\Big) 
\end{pmatrix}\,.
\end{aligned}
\end{equation}
Then 
\begin{equation}
\begin{aligned}
&\quad\hat{\Psi}^{\star 2} (\theta)\\
 &= \frac{1}{2}\Big( \cos\theta(\hat{+}\oplus\hat{-}) + \sin\theta(\hat{-}\oplus\hat{+}) \Big)\Big( \cos\theta(\hat{+}\oplus\hat{-}) + \sin\theta(\hat{-}\oplus\hat{+}) \Big) \\
&= \frac{1}{2}\Big(\cos^2\theta (\hat{+}\oplus\hat{-}  )(\hat{+}\oplus\hat{-}  ) +\cos\theta\sin\theta(\hat{+}\oplus\hat{-}  )(\hat{-}\oplus\hat{+}  ) \\
&\quad\quad+\sin\theta\cos\theta(\hat{-}\oplus\hat{+}  )(\hat{+}\oplus\hat{-}  ) + \sin^2\theta(\hat{-}\oplus\hat{+}  )(\hat{-}\oplus\hat{+}  ) \Big)\\
&=\frac{1}{2}\Big(\cos^2\theta(\hat{+}\hat{+}\oplus\hat{-}\hat{-}) +\cos\theta\sin\theta(\hat{+}\hat{-}\oplus\hat{-}\hat{+})+\cos\theta\sin\theta(\hat{-}\hat{+}\oplus\hat{+}\hat{-}) +\sin^2 \theta (\hat{-}\hat{-}\oplus \hat{+}\hat{+})\Big) \\
&=\frac{1}{2}\bigg[(\cos^2 \theta\begin{pmatrix}
\hat{+}\hat{+} & 0 \\
0 & \hat{-}\hat{-}
\end{pmatrix}
+\cos\theta\sin\theta\begin{pmatrix}
\hat{+}\hat{-} & 0 \\
0 & \hat{-}\hat{+}
\end{pmatrix}
+\cos\theta\sin\theta\begin{pmatrix}
\hat{-}\hat{+} & 0 \\
0 & \hat{+}\hat{-}
\end{pmatrix}
+\sin^2 \theta \begin{pmatrix}
\hat{-}\hat{-} & 0 \\
0 & \hat{+}\hat{+}
\end{pmatrix} \bigg]\\
&= \frac{1}{2}\begin{pmatrix}
\cos^2 \theta \hat{+}\hat{+} + \sin^2\theta \hat{-}\hat{-} +\cos\theta\sin\theta(\hat{+}\hat{-}+\hat{-}\hat{+}) & 0 \\
0& \cos^2 \theta \hat{-}\hat{-} + \sin^2\theta\hat{+}\hat{+} +\cos\theta\sin\theta(\hat{-}\hat{+}+\hat{+}\hat{-}) 
\end{pmatrix}\\
&= \frac{1}{2}\begin{pmatrix}
\cos^2 \theta \hat{a}\hat{a} + \sin^2\theta \hat{a}^\dagger\hat{a}^\dagger +\cos\theta\sin\theta(\hat{a}\hat{a}^\dagger+\hat{a}^\dagger \hat{a}) & 0 \\
0& \cos^2 \theta \hat{a}^\dagger\hat{a}^\dagger + \sin^2\theta\hat{a}\hat{a} +\cos\theta\sin\theta(\hat{a}^\dagger\hat{a}+\hat{a}\hat{a}^\dagger) 
\end{pmatrix}\\
&=\frac{1}{2}\begin{pmatrix}
\cos^2 \theta \hat{a}\hat{a} + \sin^2\theta \hat{a}^\dagger\hat{a}^\dagger +\sin 2\theta\Big(\hat{a}^\dagger\hat{a} + \frac{1}{2}\Big) & 0 \\
0& \cos^2 \theta \hat{a}^\dagger\hat{a}^\dagger + \sin^2\theta\hat{a}\hat{a}+\sin 2\theta\Big(\hat{a}^\dagger\hat{a} + \frac{1}{2}\Big) 
\end{pmatrix} \,.
\end{aligned}
\end{equation}
Therefore, finally we obtain the Hamiltonian as follow
\begin{equation}
\begin{aligned}
&\quad\hat{H}^\star (\theta) =\bigg(\frac{1}{2}\hat{\Pi}^{\star 2} (\theta) + \frac{1}{2}\hat{\Psi}^{\star 2} (\theta)\bigg)\hbar\omega \\
&= \begin{pmatrix}
\frac{1}{2}\cos 2\theta (\hat{a}\hat{a} - \hat{a}^\dagger \hat{a}^\dagger) + \sin 2\theta \Big(\hat{a}^\dagger\hat{a} + \frac{1}{2}\Big) & 0 \\
0&\frac{1}{2}\cos 2\theta (\hat{a}^\dagger \hat{a}^\dagger -\hat{a}\hat{a}) +\sin 2\theta \Big(\hat{a}^\dagger\hat{a} + \frac{1}{2}\Big) 
\end{pmatrix} \hbar\omega \\
&= \frac{1}{2}\cos 2\theta (\hat{a}^\dagger\hat{a}^\dagger -\hat{a}\hat{a})\hbar\omega M  + \sin 2\theta \bigg(\hat{a}^\dagger\hat{a}+\frac{1}{2}\bigg)\hbar\omega \pmb{1} \,.
\end{aligned}
\end{equation}

The expectation energy is hence given by
\begin{equation}
\begin{aligned}
\langle E^\star \rangle= \langle n | \hat{H}^\star (\theta) |n \rangle &= \frac{1}{2} \cos 2\theta\langle n |(\hat{a}^\dagger\hat{a}^\dagger -\hat{a}\hat{a}) |n\rangle h\omega M+ \sin 2\theta \langle n |\bigg(\hat{a}^\dagger\hat{a}+\frac{1}{2}\bigg) |n\rangle h\omega \pmb{1} \\
&=\sin 2\theta \bigg(n+\frac{1}{2} \bigg) \hbar\omega\pmb{1}\\
&= H\sin 2\theta\,\pmb{1} \,,
\end{aligned}
\end{equation}
Therefore the expectation energy of $\Psi^* (\theta)$ is same as the case of $\Psi (\theta)$. That means
\begin{equation}
\langle E(\theta) \rangle = \langle E^\star (\theta) \rangle =H\sin 2\theta \pmb{1} \,.
\end{equation}
Hence the phase analysis remains the same as the $\Psi (\theta)$ case.
\begin{figure}[H]
\centering
\includegraphics[trim=5cm 0cm 3cm 0cm, clip, scale=0.5]{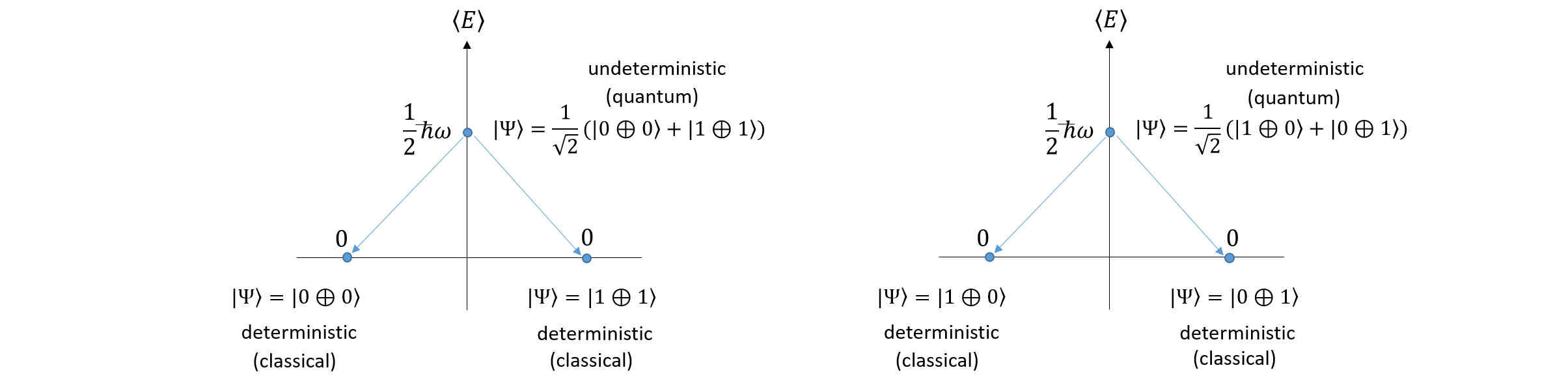}
\caption[]
{For the left figure, we compute the ground state $n=0$ for phase $\theta =\pi/4 , 0 ,\pi/2 $, corresponding to $|\Psi (\pi/4)\rangle =\frac{1}{\sqrt{2}}(|0\oplus 0 \rangle +|1\oplus 1 \rangle ),|\Psi (0)\rangle =|0\oplus 0 \rangle \,\, \text{and} \,\, |\Psi (\pi/2)\rangle=|1\oplus 1 \rangle $ respectively. The $|\Psi (\pi/4)\rangle$ is a quantum state with probability of each state of $1/2$, it has the average ground state energy of $\frac{1}{2}\hbar\omega$, while $|\Psi (0)\rangle$ and $|\Psi (\pi/2)\rangle$ are classical states, with probability equal to 1 with average ground energy equal to zero.  The idea is same for the right figure. \label{fig:energy}}
\end{figure}

In the Hamiltonian, we have the extra term of $\hat{a}^\dagger\hat{a}^\dagger-\hat{a}\hat{a} $ in addition to the standard Hamiltonian of simple harmonic oscillator. We would like to investigate its meaning here. First consider
\begin{equation}
\begin{aligned}
&\quad(\hat{a}^\dagger + \hat{a})(\hat{a}^\dagger - \hat{a}) \\
&= \hat{a}^\dagger\hat{a}^\dagger-\hat{a}^\dagger\hat{a} + \hat{a}\hat{a}^\dagger-\hat{a}\hat{a}\\
&=\hat{a}^\dagger\hat{a}^\dagger-\hat{a}\hat{a} + [\hat{a},\hat{a}^\dagger  ]\\
&= \hat{a}^\dagger\hat{a}^\dagger-\hat{a}\hat{a} +1
\end{aligned}
\end{equation}
Therefore
\begin{equation}
\hat{a}^\dagger\hat{a}^\dagger -\hat{a}\hat{a} = (\hat{a}^\dagger + \hat{a})(\hat{a}^\dagger - \hat{a}) -1 \,.
\end{equation}
But recall that in quantum mechanics
\begin{equation}
\hat{x} =\sqrt{\frac{\hbar}{2m\omega}}(\hat{a}^\dagger + \hat{a}) \quad \text{and} \quad \hat{p} = i\sqrt{\frac{\hbar m\omega}{2} }(\hat{a}^\dagger  -\hat{a}) \,.
\end{equation}
It follows that
\begin{equation}
\hat{a}^\dagger\hat{a}^\dagger -\hat{a}\hat{a} = \sqrt{\frac{2m\omega}{\hbar}} \hat{x}\cdot\frac{1}{i}\sqrt{\frac{2}{\hbar m\omega}}\hat{p} -1
=\frac{2}{i\hbar}\hat{x}\hat{p}-1\,.
\end{equation}
Therefore
\begin{equation} \label{eq:phaseh}
\begin{aligned}
\hat{H}(\theta) &=\cos 2\theta \bigg(\frac{1}{ih}\hat{x}\hat{p} -\frac{1}{2}\bigg) \hbar\omega \pmb{1} +\sin 2\theta\bigg(\hat{a}^\dagger \hat{a} + \frac{1}{2}\bigg)\hbar \omega \pmb{1} \\
&=\bigg( \frac{1}{i\hbar}\cos 2\theta \,\hat{x}\hat{p} +\sin 2\theta \hat{a}^\dagger \hat{a} \bigg) \hbar\omega\,\pmb{1} +\frac{1}{2}(\sin 2\theta-\cos 2\theta) \hbar\omega\,\pmb{1}\\
&= ( \sin 2\theta \hat{n}- i\cos 2\theta \hat{\theta} )\hbar\omega\,\pmb{1}+\frac{1}{2}(\sin 2\theta-\cos 2\theta) \hbar\omega\,\pmb{1}\,,
\end{aligned}
\end{equation}
where
\begin{equation}
\hat{\theta} = \frac{\hat{x}\hat{p}}{\hbar} = \hat{x}\frac{\partial}{\partial x}
\end{equation}
is the phase operator.

However, instead we can have
\begin{equation}
\begin{aligned}
&\quad(\hat{a}^\dagger - \hat{a})(\hat{a}^\dagger + \hat{a}) \\
&= \hat{a}^\dagger\hat{a}^\dagger+\hat{a}^\dagger\hat{a} - \hat{a}\hat{a}^\dagger-\hat{a}\hat{a}\\
&=\hat{a}^\dagger\hat{a}^\dagger-\hat{a}\hat{a} + [\hat{a}^\dagger , \hat{a}]\\
&= \hat{a}^\dagger\hat{a}^\dagger-\hat{a}\hat{a} -1
\end{aligned}
\end{equation}
Therefore
\begin{equation}
\hat{a}^\dagger\hat{a}^\dagger -\hat{a}\hat{a} = (\hat{a}^\dagger - \hat{a})(\hat{a}^\dagger + \hat{a}) +1 = \frac{2}{i\hbar} \hat{p}\hat{x}+1\,.
\end{equation}
Then
\begin{equation}
\begin{aligned}
\hat{H}(\theta) &=\cos 2\theta \bigg(\frac{1}{ih}\hat{p}\hat{x} +\frac{1}{2}\bigg) \hbar\omega \pmb{1} +\sin 2\theta\bigg(\hat{a}^\dagger \hat{a} + \frac{1}{2}\bigg)\hbar \omega \pmb{1} \\
&= \bigg(\frac{1}{i\hbar} \cos 2\theta \,\hat{p}\hat{x} +\sin 2\theta \hat{a}^\dagger \hat{a} \bigg)\hbar\omega\,\pmb{1} +\frac{1}{2}(\sin 2\theta+\cos 2\theta) \hbar\omega\,\pmb{1} \\
&= (\sin 2\theta \hat{n}- i\cos 2\theta \hat{\theta}^* )\hbar\omega \,\pmb{1}+\frac{1}{2}(\sin 2\theta+\cos 2\theta)  \hbar\omega\,\pmb{1}\,,
\end{aligned}
\end{equation}
where
\begin{equation}
\hat{\theta}^* = \frac{\hat{p}\hat{x}}{\hbar} = \frac{\partial}{\partial x}\hat{x} 
\end{equation}
is the  dual phase operator.

Next we evaluate the phase expectation value. Consider
\begin{equation}
\langle n|\hat{a}^\dagger\hat{a}^\dagger -\hat{a}\hat{a}|n\rangle = 0 =\langle n|\frac{2}{i\hbar}\hat{x}\hat{p}-1 |n\rangle = \frac{2}{i\hbar} \langle n | \hat{x}\hat{p}|n\rangle -\langle n | n\rangle =\frac{2}{i\hbar} \langle n | \hat{x}\hat{p}|n\rangle -1 \\ 
\end{equation}
Therefore we have
\begin{equation}
\langle n |\hat{\theta} |n\rangle = \frac{i}{2} \,.
\end{equation}
On the other hand
\begin{equation}
\langle n|\hat{a}^\dagger\hat{a}^\dagger -\hat{a}\hat{a}|n\rangle = 0 =\langle n|\frac{2}{i\hbar}\hat{p}\hat{x}+1 |n\rangle = \frac{2}{i\hbar} \langle n | \hat{x}\hat{p}|n\rangle +\langle n | n\rangle =\frac{2}{i\hbar} \langle n | \hat{x}\hat{p}|n\rangle +1 \\ 
\end{equation}
Therefore we have
\begin{equation}
\langle n | \hat{\theta}^* |n\rangle  = \frac{-i}{2}\,. 
\end{equation}
Then we have
\begin{equation}
\langle n | \hat{\theta} + \hat{\theta}^* |n\rangle = 0 \,. 
\end{equation}
Thus we have the dual invariant operator as
\begin{equation}
\hat{\theta} + \hat{\theta}^*  =\hat{x}\frac{\partial}{\partial x} + \frac{\partial}{\partial x}\hat{x}\,.
\end{equation}
Therefore, we can see that $\hat{a}^\dagger\hat{a}^\dagger -\hat{a}\hat{a} $ relates to the dimensionless phase term, contrasting to then occupation number term $\hat{n}$.

From equation \ref{eq:phaseh} , in particular at phase $\theta = \pi /8$, we have equal portion of $\hat{n}$ and $\hat{\theta}$,
\begin{equation}
\hat{H}\Big(\frac{\pi}{8}\Big) = \frac{1}{\sqrt{2}} (\hat{n} - i\hat{\theta}) \hbar\omega \,\pmb{1}\,.
\end{equation}
We can see that at different phase, the Hamiltonian is dominated by different behaviour. At $\theta =  \pi/4$, it is energy dominated,
\begin{equation}
H\Big(\frac{\pi}{4}\Big) = \bigg(\hat{a}^\dagger\hat{a}+\frac{1}{2}\bigg)\hbar\omega \pmb{1} \,.
\end{equation}
At $\theta =0$, it is phase dominated,
\begin{equation}
H(0)=\frac{1}{2} (\hat{a}^\dagger\hat{a}^\dagger -\hat{a}\hat{a})\hbar\omega\pmb{1} = \bigg(- i\hat{\theta} -\frac{1}{2} \bigg) \hbar\omega \pmb{1} \,.
\end{equation}

\subsection{Partition function}
In this section, we would study the partition function of the duality wave system. First, the phase dependent partition function is given by
\begin{equation}
Z(\theta) =\mathrm{Tr} \Big[ \exp{\Big(-\beta\hat{H}(\theta)}\Big) \Big] \,.
\end{equation}
For the $\hat{\Psi}(\theta)$ field case, according to the Hamiltonian in \ref{eq:hamiltonian}, it follows that
\begin{equation}
\begin{aligned}
Z(\theta) &= \mathrm{Tr} \Big[ \exp \Big( -\frac{\beta}{2}\cos 2\theta (\hat{a}^\dagger\hat{a}^\dagger -\hat{a}\hat{a})\hbar\omega\pmb{1}  -\beta\sin 2\theta \Big(\hat{a}^\dagger\hat{a}+\frac{1}{2}\Big)\hbar\omega \pmb{1}  \Big) \Big] \\
&=\sum_{n} \langle n| \exp \Big(-\frac{\beta}{2}\cos 2\theta (\hat{a}^\dagger\hat{a}^\dagger -\hat{a}\hat{a})\hbar\omega\pmb{1}  -\beta\sin 2\theta \bigg(\hat{a}^\dagger\hat{a}+\frac{1}{2}\bigg)\hbar\omega \pmb{1}  \Big) |n\rangle \\
&= \sum_{n}\exp\Big( -\sin 2\theta \Big( n +\frac{1}{2} \Big)\beta\hbar\omega \pmb{1} \Big)
\end{aligned}
\end{equation}
The expectation energy is 
\begin{equation}
\begin{aligned}
&\quad\langle E(\theta) \rangle = -\frac{\partial}{\partial\beta}\ln Z(\theta) \\
&= -\frac{\partial}{\partial\beta}\ln \Big( \sum_{n}\exp\Big( -\sin 2\theta \Big( n +\frac{1}{2} \Big)\beta\hbar\omega \pmb{1}\Big)\Big)\\
&= \frac{\sum_n \sin 2\theta \big( n+\frac{1}{2}\big)\hbar\omega\exp\Big( -\sin 2\theta \Big( n +\frac{1}{2} \Big)\beta\hbar\omega \pmb{1}\Big) }{\sum_n \exp\Big( -\sin 2\theta \Big( n +\frac{1}{2} \Big)\beta\hbar\omega \pmb{1}\Big) } \\
&=\frac{\exp\big( -\frac{\sin 2\theta}{2} \beta \hbar \omega \pmb{1}\big) \big[\sum_n n\hbar\omega \sin 2\theta \exp\big( -\sin 2\theta  n \beta\hbar\omega \pmb{1}\big) +\sum_{n}\frac{1}{2}\hbar\omega \sin 2\theta \exp\big( -\sin 2\theta\beta n\hbar\omega \pmb{1} \big)\big] }{ \exp\big( -\frac{\sin 2\theta}{2}\beta \hbar \omega \pmb{1} \big)\sum_{n}  \exp\big( -\sin 2\theta  n \beta\hbar\omega \pmb{1}\big)} \\
&=\frac{\sum_n n\hbar\omega \sin 2\theta \exp\big( -\sin 2\theta  n \beta\hbar\omega \pmb{1}\big)}{\sum_n    \exp\big( -\sin 2\theta  n \beta\hbar\omega \pmb{1}\big)} + \frac{1}{2}\sin 2\theta \hbar\omega \pmb{1} \\
&=\hbar\omega \sin 2\theta \frac{\sum_n n \exp\big( -\sin 2\theta  n \beta\hbar\omega \pmb{1}\big)}{\sum_n    \exp\big( -\sin 2\theta  n \beta\hbar\omega \pmb{1}\big)} + \frac{1}{2}\sin 2\theta \hbar\omega \pmb{1} \,.
\end{aligned}
\end{equation}
Then let $X=\exp\big(-\sin 2\theta \hbar\omega \pmb{1}\big) $ , and using the identity,
\begin{equation}
\sum_{n=0}^\infty X^n = \frac{\pmb{1}}{\pmb{1} - X} \,.
\end{equation}
It follows that
\begin{equation}
\sum_{n=0}^\infty   \exp\big( -\sin 2\theta  n \beta\hbar\omega \pmb{1}\big) = \frac{\pmb{1}}{\pmb{1}-\exp\big(-\sin 2\theta \hbar\omega \pmb{1}\big)}\,.
\end{equation}
Then using the identity of 
\begin{equation}
\begin{aligned}
 \sum_{n=0}^\infty \frac{d}{dX}X^n &= \frac{\pmb{1}} {(\pmb{1}-X)^2} \\
 \lim_{n\rightarrow \infty} (X+2X^2 + 3X^3 +\cdots + nX^n )& = \frac{X}{(\pmb{1} - X)^2} \,.
\end{aligned} 
\end{equation}
Therefore,
\begin{equation}
\sum_{n=0}^\infty n \exp\big( -\sin 2\theta  n \beta\hbar\omega \pmb{1}\big) =  \frac{\exp\big( -\sin 2\theta  n \beta\hbar\omega \pmb{1}\big)}{\pmb{1} -\exp\big( -\sin 2\theta  n \beta\hbar\omega \pmb{1}\big)} \,.
\end{equation}
Next
\begin{equation}
\lim_{N\rightarrow\infty}\frac{\sum^N_{n=0} nX^n}{\sum_{n=0}^N X^n} = \frac{X}{1-X} \,.
\end{equation}
Then
\begin{equation}
\lim_{N\rightarrow\infty}\frac{\sum_{n=0}^N n \exp\big( -\sin 2\theta  n \beta\hbar\omega \pmb{1}\big)}{\sum_{n=0}^N    \exp\big( -\sin 2\theta  n \beta\hbar\omega \pmb{1}\big)} = \frac{ \exp\big( -\sin 2\theta  n \beta\hbar\omega \pmb{1}\big)}{\pmb{1}- \exp\big( -\sin 2\theta  n \beta\hbar\omega \pmb{1}\big)} = \frac{\pmb{1}}{ \exp\big( \sin 2\theta  n \beta\hbar\omega \pmb{1}\big)-\pmb{1}}\,.
\end{equation}
Hence finally we obtain
\begin{equation}
\langle E(\theta)\rangle =\frac{\hbar\omega \sin 2\theta \pmb{1}}{\exp\big( \sin 2\theta  n \beta\hbar\omega \pmb{1}\big)-\pmb{1}}  + \frac{1}{2}\sin 2\theta \hbar\omega \pmb{1} \,.
\end{equation}
Now we evaluate $\langle E(\theta) \rangle$ in both high and low temperature limit. In high temperature limit $k_B T >> \hbar\omega$, we have 
\begin{equation}
\langle E(\theta)\rangle \approx \frac{\hbar\omega \sin 2\theta \pmb{1}}{\pmb{1} + \sin 2\theta n \beta \hbar{\omega}\pmb{1}-\pmb{1}}  + \frac{1}{2}\sin 2\theta \hbar\omega \pmb{1} = \frac{1}{\beta}= k_B T = \text{constant}\,,
\end{equation}
which is independent of the phase $\theta$.

In low temperature limit $\hbar{\omega} >> k_B T$ , or $T\rightarrow 0$, $\exp\big( \sin 2\theta  n \beta\hbar\omega \pmb{1}\big) \rightarrow \infty$, hence
\begin{equation}
\langle E(\theta)\rangle = \frac{1}{2}\sin 2\theta \hbar\omega \pmb{1}\,.
\end{equation} 
This is just the same as the average ground state energy in \ref{eq:groundstateenergy} for $n=0$ case. That means at $T = 0K$, all states are at their ground state.

\section{Conclusion}
In this paper, we demonstrate how the fundamental concept of duality can be represented by rigorous mathematical and quantum physics formalism. The symmetry of duality and multi-duality are represented by the groups $\mathbb{Z}_2$ and $\mathbb{Z}_2 \otimes \cdots \otimes \mathbb{Z}_2 $. The 4-duality group
$\mathbb{Z}_2 \otimes \mathbb{Z}_2$ plays an important role in various dual structures, and can be represented by
a 4-tableau. The concepts of dual pairs and dual invariant numbers are also introduced. We have developed quantum field theory with duality and multi-duality symmetry and demonstrated how duality is embedded in the interaction terms. We have discovered that dual invariance implies topological property, for which the Chern-Simons action and the chiral anomaly action is a topological invariant. This profound result establishes a relation between dual invariance and topological invariance. Finally and most importantly, we exhibit how the oscillation of two dual states would produce a duality wave, and upon field quantization this produces a quasi-particle quanta called the dualiton. This is a new exotic particle produced by duality. A physical practice would be local continuous spin flip of an electron that produces a wave signal, and such wave is a particle quantum mechanically. This new particle is expected to have various contributions in particle physics and condensed matter physics.

\subsection*{Conflict of interest}
There is no conflict of interest for this paper.


\begin{thebibliography}{}

\bibitem{DiracAntiMatter}
Dirac, P. A. M.: A Theory of Electrons and Protons. Proceedings of the Royal Society A. 126 (801): 360–365. (1930)

\bibitem{Dirac}
Dirac, P. A. M.: The Quantum Theory of the Electron. Proceedings of the Royal Society A: Mathematical, Physical and Engineering Sciences. 117 (778): 610–624. (1928)

\bibitem{Dirac2}
Dirac, P. A. M.: A Theory of Electrons and Protons. Proceedings of the Royal Society A: Mathematical, Physical and Engineering Sciences. 126 (801): 360–365. (1930)

\bibitem{MAMasymmetry}
A. D. Sakharov.: "Violation of CP invariance, C asymmetry, and baryon asymmetry of the universe". Journal of Experimental and Theoretical Physics Letters. 5: 24–27. (1967)

\bibitem{parityviolation}
Lee, T. D.; Yang, C. N. (1956). "Question of Parity Conservation in Weak Interactions". Physical Review. 104 (1): 254–258.

\bibitem{Wu}
Wu, C. S.; Ambler, E.; Hayward, R. W.; Hoppes, D. D.; Hudson, R. P. "Experimental Test of Parity Conservation in Beta Decay". Physical Review. 105 (4): 1413–1415. (1957)

\bibitem{Higgs1}
P.~W. Higgs.: Broken symmetries, massless particles and gauge fields,
\textit{Phys. Lett.} \textbf{12} (1964) 132.

\bibitem{Higgs2}
P.~W. Higgs.: Broken Symmetries and the Masses of Gauge Bosons.
\textit{Phys. Rev. Lett.} \textbf{13} (1964) 508.

\bibitem{Englert}
F. Englert, R. Brout.: Broken Symmetry and the Mass of Gauge Vector Mesons.
\textit{Phys. Rev. Lett.} \textbf{13}:\,321--323. Aug. 1964.

\bibitem{Weinberg}
S. Weinberg.: A Model of Leptons. \textit{Phys. Rev. Lett.} \textbf{19} (1967) 1264.

\bibitem{Salam}
A.  Salam.:  Weak  and  electromagnetic  interactions, in:  N.  Svartholm  (Ed.),  Elementary  Particle Physics: Relativistic Groups and Analyticity. \textit{Proceedings of the Eighth Nobel Symposium, Almquvist and Wiskell}, 1968, p. 367.

\bibitem{Kibble}
T.W.B. Kibble, Symmetry breaking in non-Abelian gauge theories. \textit{Phys. Rev}. 155 (1967) 1554.

\bibitem{string1}
K. Becker, M. Becker and J. H. Schwarz.: String theory and M-theory-A modern introduction. \textit{Cambridge University Press} (2007)

\bibitem{string2}
J. Polchinski.: String Theory-Introduction to the bosonic string (volume 1).  \textit{Cambridge University Press} (2007)


\bibitem{Feyn1}
R. P. Feynman. Space-Time Approach to Non-Relativistic Quantum Mechanics. \textit{Reviews of Modern Physics.} \textbf{20} (2). 1948.

\bibitem{Feyn2}
R. P. Feynman and A. R. Hibbs. Quantum mechanics and Path Integrals. New York: McGraw-Hill. 1965.


\bibitem{Peskin}
P. Schroeder and D. V. Schroeder, An Introduction to Quantum Field Theory. \textit{ABP}, 1995.


\bibitem{Rydle}
L. H. Ryder.: Quantum Field Theory. \textit{Cambridge University Press}. (1985)

\bibitem{Shannon1}
E. C. Shannon.: A Mathematical Theory of Communication.  \textit{Bell System Technical Journal}. \textbf{27} (3): 379–423. Jul 1948.

\bibitem{Shannon2}
E. C. Shannon.: A Mathematical Theory of Communication.  \textit{Bell System Technical Journal}. \textbf{27} (4):  623–656. Oct 1948.



\end{thebibliography}
\end{document}